\documentclass{aa}  

\usepackage{graphicx}
\usepackage{txfonts}
\begin{document}

   \title{PENELLOPE IV. A comparison between optical forbidden lines and $\rm H_2$ UV lines in the Orion OB1b and $\sigma$-Ori associations\thanks{Based on data obtained within ESO programme 106.20Z8}}
   \subtitle{}
\titlerunning{PENELLOPE IV}
   
   \author{M. Gangi                 \inst{1,2}
          \and B. Nisini            \inst{2}
          \and C. F. Manara         \inst{3}     
          \and K. France            \inst{4}
          \and S. Antoniucci        \inst{2}
          \and K. Biazzo            \inst{2}
          \and T. Giannini          \inst{2}
          \and G. J. Herczeg        \inst{5,6}
          \and J. M. Alcal\'a       \inst{7}          
          \and A. Frasca            \inst{8}   
          \and K. Maucó             \inst{3}
          \and J. Campbell-White    \inst{3}
          \and M. Siwak             \inst{9,10}
          \and L. Venuti            \inst{11}
          \and P. C. Schneider      \inst{12}
          \and \'A. K\'osp\'al      \inst{9,10, 13, 14}
          \and A. Caratti o Garatti \inst{7}
          \and E. Fiorellino        \inst{7}
          \and E. Rigliaco          \inst{15}
          \and R. K. Yadav          \inst{16}
}

   \institute{ ASI, Italian Space Agency, Via del Politecnico snc, 00133 Rome, Italy \\
              \email{manuele.gangi@asi.it}
        \and  INAF - Osservatorio Astronomico di Roma, Via Frascati 33, I-00078 Monte Porzio Catone, Italy
        \and  European Southern Observatory, Karl-Schwarzschild-Strasse 2, 85748 Garching bei M\"unchen, Germany 
        \and  Laboratory for Atmospheric and Space Physics, University of Colorado Boulder, Boulder, CO 80303, USA
        \and  Kavli Institute for Astronomy and Astrophysics, Peking University, Yiheyuan 5, Haidian Qu, 100871 Beijing, People’s Republic of China 
        \and  Department of Astronomy, Peking University, Yiheyuan 5, Haidian Qu, 100871 Beijing, People’s Republic of China 
        \and  INAF - Osservatorio Astronomico di Capodimonte - Salita Moiariello 16, 80131 Napoli, Italy 
        \and  INAF - Osservatorio Astrofisico di Catania - Via S. Sofia 78, 95123 Catania, Italy 
        \and Konkoly Observatory, Research Centre for Astronomy and Earth Sciences, Eötvös Loránd Research Network (ELKH), KonkolyThege Miklós út 15-17, 1121 Budapest, Hungary
        \and CSFK, MTA Centre of Excellence, Konkoly-Thege Miklós út 15-17, 1121 Budapest, Hungary 
        \and SETI Institute, 339 Bernardo Ave, Suite 200, Mountain View, CA 94043, USA
        \and Hamburg Observatory, Gojenbergsweg 11, 21029 Hamburg, Germany
        \and Max Planck Institute for Astronomy, K\"onigstuhl 17, 69117 Heidelberg, Germany
        \and ELTE E\"otv\"os Lor\'and University, Institute of Physics, P\'azm\'any P\'eter s\'et\'any 1/A, 1117 Budapest, Hungary
        \and INAF - Osservatorio Astronomico di Padova, Vicolo dell’osservatorio 5, 35122 Padova, Italy
        \and National Astronomical Research Institute of Thailand (NARIT), Sirindhorn AstroPark, 260 Moo 4, T. Donkaew, A. Maerim, Chiangmai 50180, Thailand}
   \date{Received ; Accepted }

 
  \abstract
   {Observing the spatial distribution and excitation processes of atomic and molecular gas in the inner regions (< 20 au) of young (< 10 Myr) protoplanetary disks helps us to understand the conditions for the formation and evolution of planetary systems.}
   {In the framework of the PENELLOPE and ULLYSES projects, we aim to characterize the atomic and molecular component of protoplanetary disks in a sample of 11 Classical T Tauri Stars (CTTs) of the Orion OB1 and $\sigma$-Orionis associations.}
   {We analyzed the flux-calibrated optical-forbidden lines and the fluorescent ultraviolet $\rm H_2$ progressions using spectra acquired with ESPRESSO at VLT, UVES at VLT and HST-COS. Line morphologies were characterized through Gaussian decomposition. We then focused on the properties of the narrow low-velocity (FWHM < 40 $km$ $s^{-1}$ and |$v_p$| < 30 $km$ $s^{-1}$) component (NLVC) of the [\ion{O}{i}] 630 nm line, compared with the properties of the UV-$\rm H_2$ lines.}
   {We found that the [\ion{O}{i}]630 NLVC and the UV-$\rm H_2$ lines are strongly correlated in terms of peak velocities, full width at half maximum, and luminosity. Assuming that the line width is dominated by Keplerian broadening, the [\ion{O}{i}]630 NLVC originates from a disk region between 0.5 and 3.5 au, while that of UV-$\rm H_2$ in a region from 0.05 and 1 au. The luminosities of the [\ion{O}{i}]630 NLVC and UV-$\rm H_2$ correlate with the accretion luminosity with a similar slope, as well as with the luminosity of the \ion{C}{iv}154.8, 155 nm doublet. We discuss such correlations in the framework of the currently suggested excitation processes for the [\ion{O}{i}]630 NLVC.}
   {Our results can be interpreted in a scenario in which the [\ion{O}{i}]630 NLVC and UV-$\rm H_2$ have a common disk origin with a partially overlapped radial extension. We also suggest that the excitation of the [\ion{O}{i}] NLVC is mainly induced by stellar FUV continuum photons more than being of thermal origin. This study demonstrates the potential of contemporaneous wide-band high-resolution spectroscopy in linking different tracers of protoplanetary disks.}


\maketitle
%

\begin{table*}
\center
\caption{\label{tab:target_list} List of sources with their stellar and accretion properties, and disk inclination when available.}
\begin{tabular}{lc|cccccc|c}
\hline
\hline
Name    & $\rm Dist.^{(a)}$ 			& $ \rm SpT^{(b)}$    & $\rm L_{\star}^{(b)}$ 				&	$\rm M_{\star}^{(b)}$ 					& $\rm Av^{(b)}$ & $\rm log\ L_{acc}^{(b)}$				& $\rm log\ \dot{M}_{acc}^{(b)}$ 	& $\rm  i_{disk}$ \\
        & [pc] 					&  		 & [$\rm L_{\sun}$] &	[$\rm M_{\sun}$]	&    &	[$\rm L_{\odot}$]		& $\rm [M_{\odot}$ $\rm yr^{-1}]$ & [deg]      \\
\hline
\noalign{\smallskip}
CVSO58	& $349.0 \pm 2.8$ 		& K7   & 0.32 & 0.81 & 0.8 & -1.12 & -8.37 &  ...$\rm ^{(c)}$ \\
\noalign{\smallskip}
CVSO90	& $338.7^{+3.8}_{-3.7}$ 	& M0.5 & 0.13 & 0.62 & 0.1 & -1.34 & -8.61 &  ...$\rm ^{(c)}$ \\
\noalign{\smallskip}
CVSO104	& $360.7^{+3.9}_{-3.8}$	& M2   & 0.37 & 0.37 & 0.2 & -1.73 & -8.49 &  37$\rm ^{(e)}$\\
\noalign{\smallskip}
CVSO107	& $330.4 \pm 2.5$			& M0.5 & 0.32 & 0.53 & 0.3 & -1.30 & -7.30 &  22.3 $\pm$ 3.9$\rm ^{(d)}$ \\
\noalign{\smallskip}
CVSO109	& 400					& M0.5 & 0.92 & 0.46 & 0.1 & -0.77 & -7.49 &  9.2 $\pm$ 2.7$\rm ^{(d)}$  \\ 
\noalign{\smallskip}
CVSO146	& $332.0 \pm 1.7$			& K6   & 0.80 & 0.86 & 0.6 & -1.46 & -8.57 &  25.8 $\pm$ 4.8$\rm ^{(d)}$ \\ 
\noalign{\smallskip}
CVSO165	& 400					& K6   & 0.98 & 0.84 & 0.2 & -2.05 & -9.10 &  45.85 $\pm$ 4.07$\rm ^{(e)}$ \\
\noalign{\smallskip}
CVSO176	& $302.4^{+2.9}_{-2.8}$	& M3.5 & 0.34 & 0.25 & 1.0 & -1.27 & -7.84 &  52.4 $\pm$ 7.8$\rm ^{(d)}$ \\
\noalign{\smallskip}
\hline
\hline
\noalign{\smallskip}
SO518	& $392.3^{+3.9}_{-3.8}$	& K7 & 0.24 & 0.81 & 1.0 & -1.22 & -8.53 &  78$\rm ^{(f)}$ \\
\noalign{\smallskip}
SO583	& 385					& K5 & 3.61 & 1.09 & 0.4 & -0.30 & -7.21 &  ... \\
\noalign{\smallskip}
SO1153	& $390.3^{+4.1}_{-4.0}$	& K7 & 0.17 & 0.76 & 0.1 & -0.88 & -8.24 &  29.49 \\
\noalign{\smallskip}
\hline
\end{tabular}
\begin{quotation}                  
\textbf{Notes.} 

$\rm ^{(a)}$ Computed from Gaia EDR3 parallaxes with reliable astronometric solutions \citep{Gaia2021}. For those without reliable astronometric solutions (i.e. CVSO109, CVSO165 and SO583), the mean distance of the association with an uncertainty of 10\% is assumed. $\rm ^{(b)}$ Computed in \citet{Manara2021}. $\rm ^{(c)}$ Inclination assumed to be $\rm 60^\circ$. $\rm ^{(d)}$ Computed in \citet{Pittman2022} from $TESS$ stellar rotation periods and $\rm v \sin{i}$ from \citet{Manara2021} and \citet{Kounkel2019}. $\rm ^{(e)}$ Binary system, evidences of circumbinary disk \citep{Frasca2021}. Inclination computed as the average of the inclinations of the components. $\rm ^{(f)}$ Computed in Maucò et al. (in prep.).

\end{quotation}
\end{table*}

\section{Introduction}
Protoplanetary disks of young stellar objects (YSOs) are the birth site of planets. They undergo important processes leading to mass accretion onto the star, ejection of outflows, and photo-evaporated disk winds \citep{Hartmann2016, Ercolano2017, Pascucci2022}. These processes combine to the disk dissipation on timescales of a few million years \citep[see review by][]{Manara2022}.
In this framework, the composition and spatial distribution of atomic and molecular gas in the inner regions (i.e. few au) of protoplanetary disks is one of the key ingredients for the formation and evolution of planetary systems. 

The atomic protoplanetary disk components can be spectroscopically traced by atomic or weakly ionized forbidden lines in the optical and infrared (IR) spectral range. These lines usually present a composite profile, with high-velocity components (HVC, |$\rm v_p$| > 30 $\rm km$ $\rm s^{-1}$), attributed to extended collimated jets, and low-velocity components (LVC, |$\rm v_p$| < 30 $\rm km$ $\rm s^{-1}$), associated with compact (0.5-10 au) protoplanetary disk winds \citep[e.g.,][]{Hartigan1995}. When observed at medium- or high-spectral resolution, the LVC often shows a contribution from a broad component (BLVC, FWHM > 40 $\rm km$ $\rm s^{-1}$) and a narrow component \citep[NLVC, FWHM < 40 $\rm km$ $\rm s^{-1}$, e.g.,][]{Rigliaco2013, Simon2016, Banzatti2019, Giannini2019, Gangi2020}. On the other hand, the molecular content in the inner disk region (< 10 au) can be characterized through specific emission and absorption bands like those of the $\rm H_2$, CO, $\rm H_2$O, and OH, observed from the UV to the IR \citep[e.g.,][]{Najita2007, France2012, Banzatti2022}.

One of the open questions concern the excitation mechanism of the oxygen LVC. Thermal excitation of the atomic oxygen (i.e. collisional excitation with hydrogen atoms or electrons) was suggested as the major mechanism to the line emissions \citep[e.g.,][]{Hartigan1995}. Indeed, the [\ion{O}{i}]557\footnote{Hereafter, where not otherwise specified, all the lines are identified with the wavelength expressed in nm}/[\ion{O}{i}]630 and [\ion{S}{ii}]406/[\ion{O}{i}]630 line ratios are usually interpreted in terms of diagnostic models involving collisionally-excited processes to set constraints on the gas temperature and density \citep[e.g.,][]{Natta2014, Giannini2015, Fang2018, Giannini2019}. This scenario is classically supported by the fact that the kinematic components of the [\ion{S}{ii}]406 line are found to be similar to those of the [\ion{O}{i}]630 and the [\ion{S}{ii}]406 lines being collisionally excited \citep[e.g.][]{Fang2018}. However, the large spread observed in the [\ion{O}{i}]557/[\ion{O}{i}]630 and [\ion{S}{ii}]406/[\ion{O}{i}]630 line ratios also suggest non-thermal contributions \citep[e.g.][Nisini et al., in prep.]{Rigliaco2013}.

A non-thermal contribution to the excitation of \ion{O}{} may arise from the action of FUV photons (1400 - 1700 \AA) through two distinct processes: (i) photo-dissociation of the OH molecule \citep{Gorti2011,Rigliaco2013} and (ii) FUV pumping \citep{Nemer2020}. OH molecules can be dissociated into the $^1D$ and $^1S$ fine structure levels of the ground state of \ion{O}{} by FUV photons reaching the disk surface. They eventually decay to the fundamental level along the [\ion{O}{i}]630 and [\ion{O}{i}]557 transitions. As pointed out by \citet{Rigliaco2013}, this photo-dissociated layer may have both a bound component in Keplerian rotation and an unbound component at larger scales ($\geq$ 10 au), making it kinematically indistinguishable from a photoevaporative wind. FUV pumping consists of excitation to higher \ion{O}{} levels by absorption of FUV photons with a subsequent cascade of radiative or collisional de-excitations towards the upper levels of the [\ion{O}{i}] lines. \citet{Nemer2020} found that this process can appreciably contribute to the excitation of the [\ion{O}{i}]557 and [\ion{O}{i}]630 lines both in X-ray driven photoevaporative wind models \citep{Owen2010} and in magnetothermal models \citep{Wang2019}, dominating the lines emission up to a fraction of $\sim$ 90\%. Along with stellar UV photons, an additional source of excitation may arise in the so-called external photoevaporative winds \citep[e.g.,][ and references therein]{Winter2022}, where the action of UV-dominated irradiation from close members of the star forming region, like the OB stars, can contribute to the dissociation of OH. This would lead to an important increase of the [\ion{O}{i}]630 line luminosity, as recently discussed in \citet{Ballabio2023}.

From an observational point of view, we can investigate the role of the aforementioned processes by linking together the properties of the [\ion{O}{i}] LVC and those of the dipole-allowed electronic transitions of the $\rm H_2$ molecules in the UV (hereafter UV-$\rm H_2$). The UV-$\rm H_2$ are in fact photo-excited by Ly$\alpha$ photons \citep[e.g.,][]{France2012} and therefore they can be used as an indirect probe of the role that a FUV-continuum field may have in exciting the \ion{O}{} species. This multi-wavelength approach requires medium or high resolution spectroscopic investigation of large samples of YSOs with a wide and simultaneous spectral coverage, from the UV to the NIR. The simultaneity between different spectral bands is essential to avoid biases induced by variability. 

In this framework, the Hubble Space Telescope (HST) Director's Discretionary Time ULLYSES program \citep[][]{Roman-Duval2020} is devoted to acquire UV spectra for about 70 low-mass ($\sim$0.1-2 $M_{\odot}$) YSOs with ages from 1 to 10 Myrs. This project is flanked by the public ESO VLT large programme PENELLOPE \citep[][]{Manara2021}, to obtain contemporaneous high-resolution optical and NIR spectra. With contemporaneous observations that minimize the spectral changes caused by variability, these complementary surveys offer a once-in-a-lifetime opportunity to make significant advances in the study of the physics of low-mass YSOs.

In this work we add new observational constraints on the \ion{O}{} excitation mechanism by comparing the properties of optical forbidden lines with those of the UV-$\rm H_2$ in a sample of 11 CTTs observed in the framework of the ULLYSES and PENELLOPE collaborations. This is the first study on the PENELLOPE series devoted to the analysis of forbidden lines, and it is based on the sample of the Orion SFR presented in \citet{Manara2021}. The forbidden line properties of the other SFRs observed within the ULLYSES and PENELLOPE collaborations will be analysed in forthcoming works. The paper is organized as follows. In Sect. 2 we present the sample and the data, while the analysis of the optical and UV spectra is reported in Sect. 3. Results are shown in Sect. 4 and the correlations between the properties of the different lines are reported in Sect. 5. Discussion and conclusion are finally presented in Sect. 6 and 7, respectively.

\section{Targets and data}
Our sample consists of 8 CTTs of the Orion OB1 and 3 CTTs of the $\rm \sigma$-Orionis associations. Spectra from the FUV to the NIR, were acquired in the framework of the PENELLOPE Large Program and the ULLYSES public survey. The list of the sources and their basic stellar and accretion parameters, characterized in \citet{Manara2021}\footnote{Accretion parameters were determined by \citet{Manara2021} through multi-components accretion flows \citep{Manara2013} and they are adopted here for homogeneity with previous works. However, other determination is also provided by \citet{Pittman2022}.}, are reported in Table \ref{tab:target_list}. Masses are in the range between 0.25 and 1.09 $\rm M_{\odot}$, spectral types between M0.5 and K7 and luminosities between 0.13 and 3.61 $\rm L_{\odot}$.

In this work, for each target we use three high-resolution ESPRESSO \citep[R=140000, $\rm \lambda\ 380-788\ nm$, ][]{Pepe2021} or UVES \citep[R=70000, $\rm \lambda\ \sim 330-450, 480-680\ nm$,][]{Dekker2000} spectra, acquired with a one day cadence, and a contemporaneous flux-calibrated medium-resolution X-Shoother \citep[R=5400-18400, $\rm \lambda\ \sim 300-2500\ nm,$ ][]{Vernet2011} spectrum. We also include medium-resolution (R $\rm \sim 15000$) HST-COS spectra covering the FUV region ($\rm 136\ nm < \lambda < 177\ nm$).

Details on the observational strategy, standard data reduction and analysis are reported in \citet{Manara2021} and \citet{Espaillat2022}. In the following we describe the method used in this work.

\begin{table}
\caption{\label{tab:line_atomic_list} Relevant parameters of the observed atomic lines}
\begin{tabular}{lccccc}
\hline
\hline
\noalign{\smallskip}
\multicolumn{6}{c}{ATOMIC SPECIES (OPTICAL)} 	\\
\hline
ID			   &	$\rm \lambda$	& Upper			& Lower					&	$\rm E_u$	&	$\rm E_l$ 	\\
			&	[nm]			& level			& level					&	[eV]		&	[eV]	 	\\
\hline
$[\ion{O}{I}]$	&	630.0304		& $\rm {}^1D_2$		& $\rm {}^3P_2$		&	1.97		&	0.00		\\
$[\ion{O}{I}]$	&	557.7339		& $\rm {}^1S_0$ 	& $\rm {}^1D_2$ 	&	4.19		&	1.97		\\
$[\ion{S}{II}]$	&	673.0816		& $\rm {}^2D_{3/2}$	& $\rm {}^4S_{3/2}$	&	1.84		&	0.00		\\
$[\ion{S}{II}]$	&	406.8600		& $\rm {}^2P_{3/2}$	& $\rm {}^4S_{3/2}$	&	3.05		&	0.00		\\
$[\ion{N}{II}]$	&	658.3450		& $\rm {}^1D_2$		& $\rm {}^3P_2$		&	1.90		&	0.02		\\
\hline
\hline
\end{tabular}
\end{table} 
 
\begin{table}
\caption{\label{tab:line_mol_list} Relevant parameters of the observed molecular lines}
\begin{tabular}{lccc}
\hline
\hline
\noalign{\smallskip}
\multicolumn{4}{c}{$\rm H_2$ MOLECULAR SPECIES (UV)} 	\\
\hline
Line ID		& $\rm \lambda$	    & $\rm B_{mn}^{(a)}$ & $\rm [v', J']^{(b)}$	\\
		& [nm]			    &                    & 					     \\
\hline
(1-6) R(3)	& 143.101			& 0.058              & [1,4]				 \\
(1-6) P(5)	& 144.612			& 0.083              & [1,4]				 \\
(1-7) R(3)	& 148.957			& 0.094              & [1,4]				 \\
(1-7) P(5)	& 150.476			& 0.115              & [1,4]				 \\
\noalign{\smallskip}
(1-6) P(8)	& 146.708			& 0.080              & [1,7]				 \\
(1-7) R(6)	& 150.045			& 0.101              & [1,7]				 \\
(1-7) P(8)	& 152.465			& 0.111              & [1,7]				 \\
(1-8) R(6)	& 155.687			& 0.074              & [1,7]			     \\
\hline
\hline
\end{tabular}
\begin{quotation}
(a) Branching ratio, defined as the ratio of the line transition probability to the total transition probability out of state $\rm [v', J']$ (see \citet{France2012} for details).

(b) The quantum numbers v' and J' denote the vibrational and rotational quantum numbers in the excited ($\rm B^1\Sigma_u^+$) $\rm H_2$ electronic state.
\end{quotation}
\end{table}

\section{Data analysis}\label{sec:data_analysis}
\subsection{Optical spectra}
We focus on the five brightest optical forbidden transitions, i.e. the [\ion{O}{i}] lines at 630 and 557 nm, the [\ion{S}{ii}] at 673 and 406 nm and the [\ion{N}{ii}] at 658 nm. Their relevant atomic parameters retrieved from the \textsc{nist}\footnote{https://www.nist.gov/pml/atomic-spectra-database} database are summarized in Table \ref{tab:line_atomic_list}.

\begin{figure*}
\center
\includegraphics[trim=0 150 0 100,width=0.7\columnwidth, angle=0]{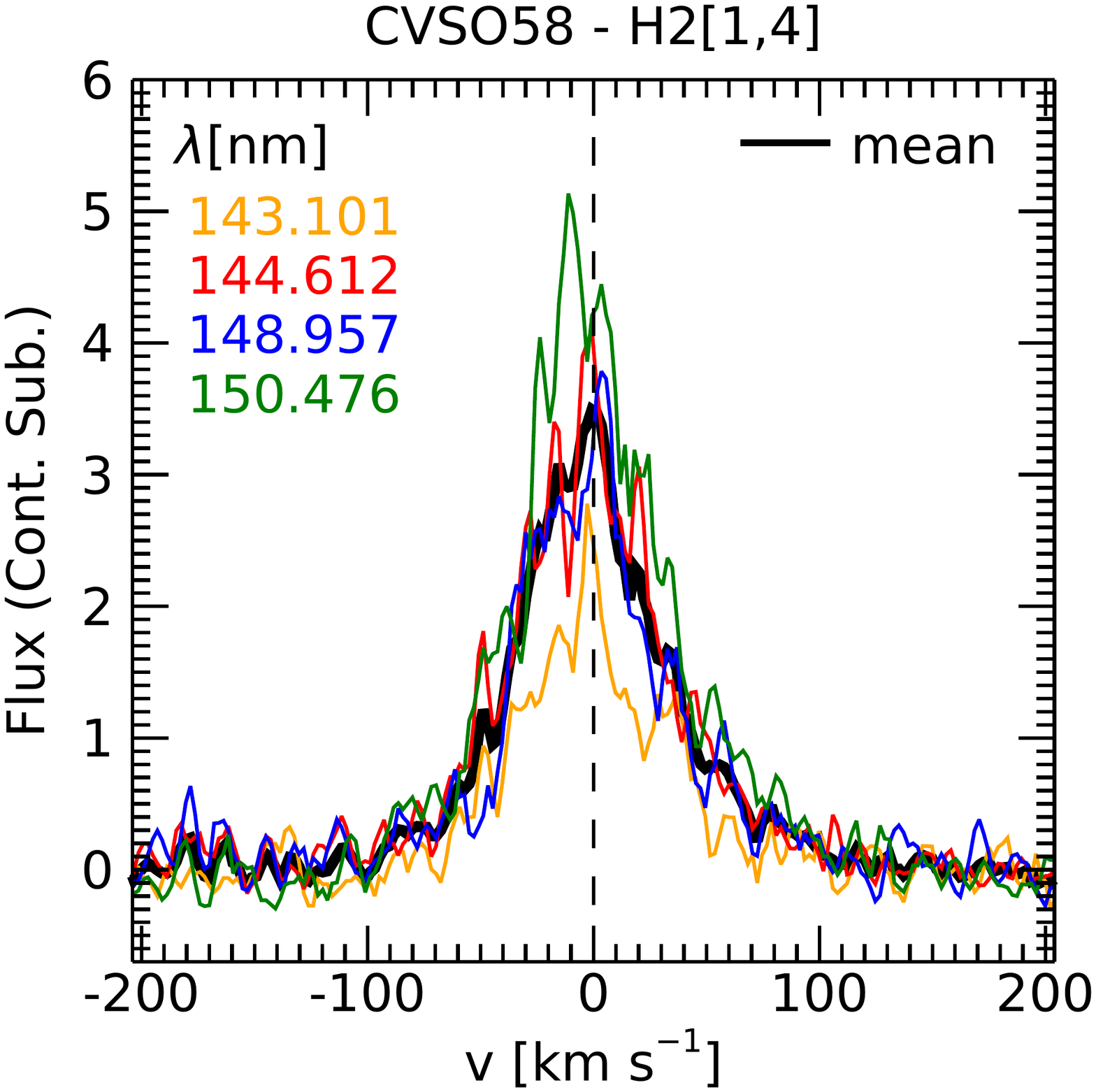}
\includegraphics[trim=0 150 0 100,width=0.7\columnwidth, angle=0]{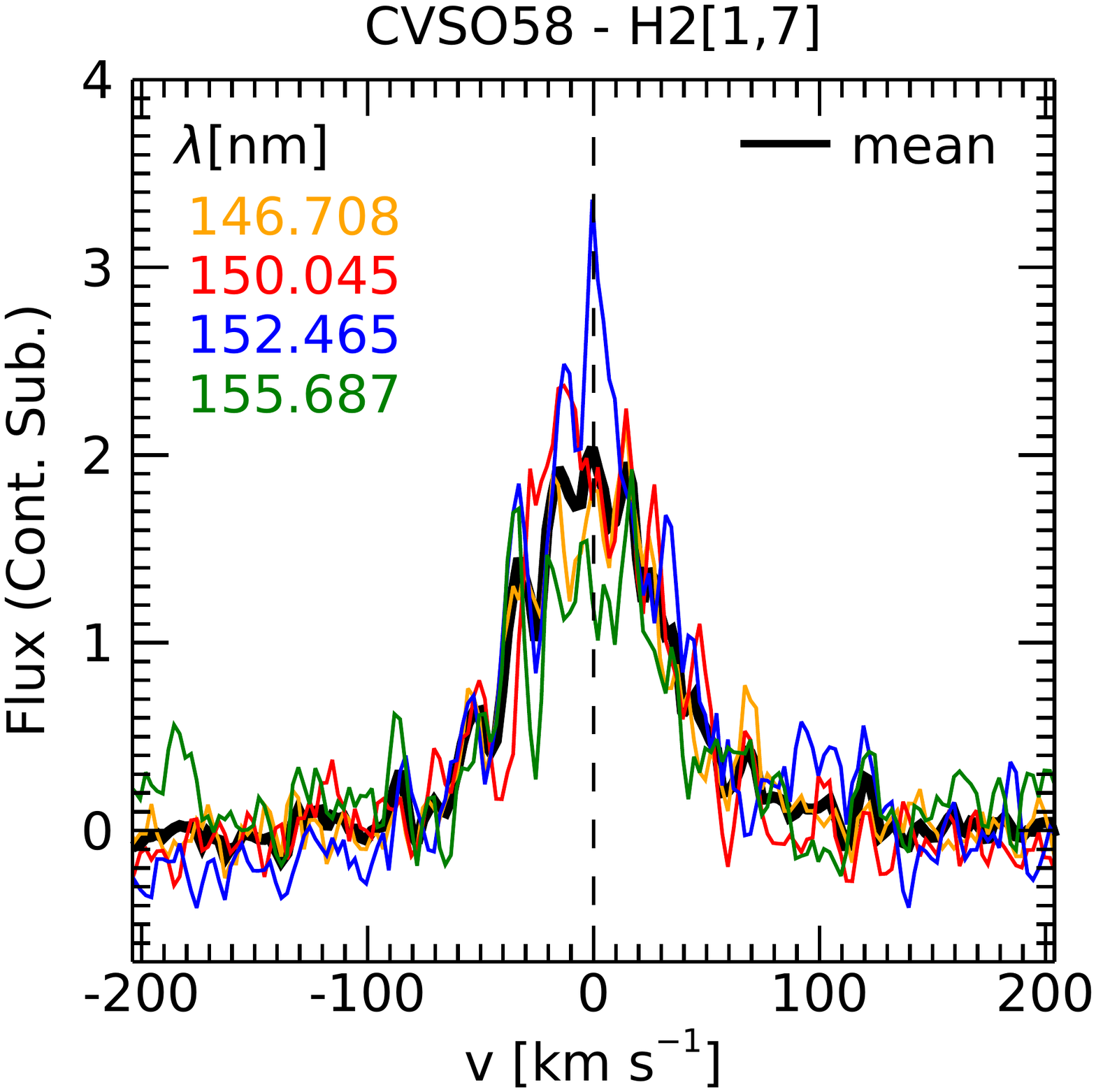}
\begin{center}\caption{\label{fig:H2_combine} Example of $\rm H_2$ [1,4] (left) and $\rm H_2$ [1,7] (right) progression lines profile. The averaged profile is indicated in black. Flux units are $\rm 10^{-15} ergs^{-1} cm^{-2} \AA^{-1}$.}\end{center}
\end{figure*}

For each epoch, we retrieved the five UVES/ESPRESSO spectral segments ($\sim 100$ \AA\ larger) containing the emission profiles. The photospheric contribution was subtracted by the PENELLOPE collaboration, as described in \citet{Manara2021}. We first corrected spectral segments for the radial velocity computed by \citet{Manara2021} by means of cross-correlation of appropriate template and target spectrum. The typical estimated wavelength accuracy of our calibration is about 0.5 $\rm km$ $\rm s^{-1}$. Each spectral profile was then normalized to the local continuum and flux calibrated on the basis on the local continuum extracted from the contemporaneous medium resolution X-Shooter spectra. After checking that the multi-epochs profiles show no appreciable line variability, we obtained the final profiles as the median of the available multi-epochs spectra. This allowed us to increase the signal-to-noise ratio ($SNR$) to a sufficient level for the morphological analysis in the majority of cases.

Gaussian decomposition was performed employing an \textsc{idl} procedure to fit multi-component optical/infrared high-resolution profiles \citep{Gangi2020, Gangi2021}. In short, this procedure is based on $\rm \chi^2$ minimization and provides for each component the width, peak velocity, and peak intensity values. The total number of components was determined following the criteria adopted in \citet{Banzatti2019}, i.e. as the minimum number of Gaussians that yields a $\rm \chi^2$ stable at 20\% of its minimum value. To estimate errors on fit parameters, we simulated for each profile $10^4$ data-sets with random gaussian distribution having as central value the observed spectral points and standard deviation the local $SNR$. Each simulated profile was then decomposed and the errors were determined as the sigma of the fit parameter distributions. To limit the high level of degeneracy involved in this kind of analysis, for each star we first applied the procedure to the highest $SNR$ line profile, namely the [\ion{O}{i}]630. We then used the obtained kinematic solutions as initial parameters for the decomposition of the other lines, since it is known that individual kinematic components share similar profiles among the different forbidden lines \citep[e.g.,][]{Fang2018}. Despite this, however, the high noise present in these profiles made it impossible to distinguish between broad and narrow low-velocity components for most cases. A direct comparison with the respective low-velocity [\ion{O}{i}]630 components must be then taken with caution.

The FWHMs were deconvolved by the instrumental width, $\rm \sigma_{instr}$ assuming a Gaussian profile with $\rm \sigma_{instr} = 0.09\ \AA$ for UVES and $\rm \sigma_{instr} = 0.04\ \AA$ for ESPRESSO. Finally, for each component we have derived the line flux and luminosity, corrected for extinction assuming the Av values computed by \citet{Manara2021}, reported here in Table \ref{tab:target_list} for completeness, and the reddening law by \citet{Cardelli1989} with $R_v =$ 3.1. In cases where a line was not detected, we estimated a 3-$\rm \sigma$ upper limit as $\rm 3 \times$ \textsc{rms} $\times \Delta \lambda$, with \textsc{rms} the local flux-noise and $\rm \Delta \lambda$ the expected line width. The latter was estimated from the other detected lines and the typical values range from 1 to 4 \AA.

\subsection{UV spectra}\label{subsec:UV_spectra}
We restrict our analysis to the brightest 4 $\rm H_2$ emission lines in both of the progressions [$v'$,$J'$] = [1,7] and [1,4], for a total of 8 lines detected in all of the sources (Table \ref{tab:line_mol_list}). We first correct spectral profiles for the radial velocity adopting the values reported by \citet{Manara2021}. Peak velocities are found to be roughly consistent with stellar velocities, with differences well below of the wavelength solution accuracy of COS ($\rm \sim 15$ $\rm  km$ $\rm  s^{-1}$). For each progression we averaged the four spectral profiles obtaining single lines to which we refer as $\rm H_2$ [1,4] and $\rm H_2$ [1,7] (Fig. \ref{fig:H2_combine}).

We perform a Gaussian fit of such profiles taking into account the line broadening introduced by the COS line-spread-function (LSF). This latter depends on both the Telescope position and the wavelength range\footnote{this is due to the polishing errors on the HST primary and secondary mirrors: https://www.stsci.edu/hst/instrumentation/cos/performance/spectral-resolution} and it is approximately a Lorentzian profile. In addition to the instrumental broadening, the broad LSF wings can substantially alter the line profile and mimic a broad low-velocity component. For this reason it is therefore particularly important to correct for this contribution. For this purpose, we followed the approach of \citet{France2012}, in which a Gaussian component with infinite resolution is convolved with the appropriate LSF. We choose the COS LSF G160M/1611, corresponding to the Telescope LifeTime position 4, and average the LSFs at 1467, 1500, 1524, 1556 \AA\ and 1431, 1446, 1489, 1504 \AA\ for the $\rm H_2$[1,7] and $\rm H_2$[1,4] line profiles, respectively. For each averaged profile we then derive the peak intensity, FWHM and peak velocity. 

Finally, from a given progression we compute the total line flux as

\begin{eqnarray}
\rm F_m = \frac{1}{N} \sum \frac{F_{mn}}{B_{mn}},
\end{eqnarray}

where $F_{mn}$ is the integrated flux of the specific line from the rovibrational state $m$ to the electronic state $n$, $B_{mn}$ is the corresponding branching ratio and $N$ is the number of emission lines of the progression \citep{France2012}. Individual line fluxes ($F_{mn}$) were corrected for the extinction using the Av reported in \citet{Manara2021} and assuming the extinction law of \citet{Whittet2004} towards HD29647. This latter was shown to be better suited for use in the NUV than that of \citet{Cardelli1989} in the case of Orion OB1 \citep{Pittman2022}. The total line fluxes were then converted into luminosity by adopting the distances reported in Table \ref{tab:target_list}.

\begin{figure*}
\includegraphics[trim=20 140 0 140,width=.68\columnwidth, angle=0]{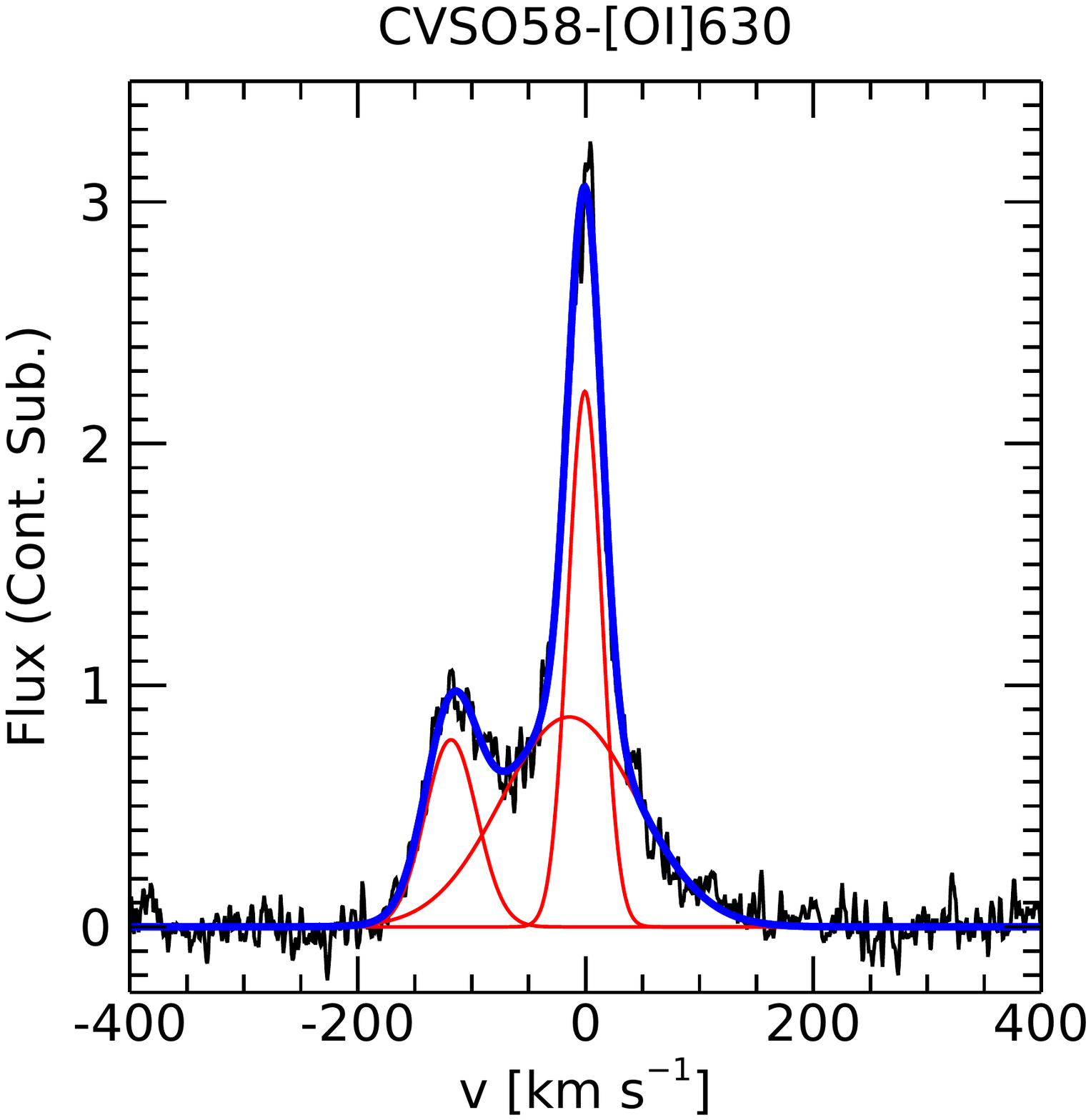}
\includegraphics[trim=20 140 0 140,width=.68\columnwidth, angle=0]{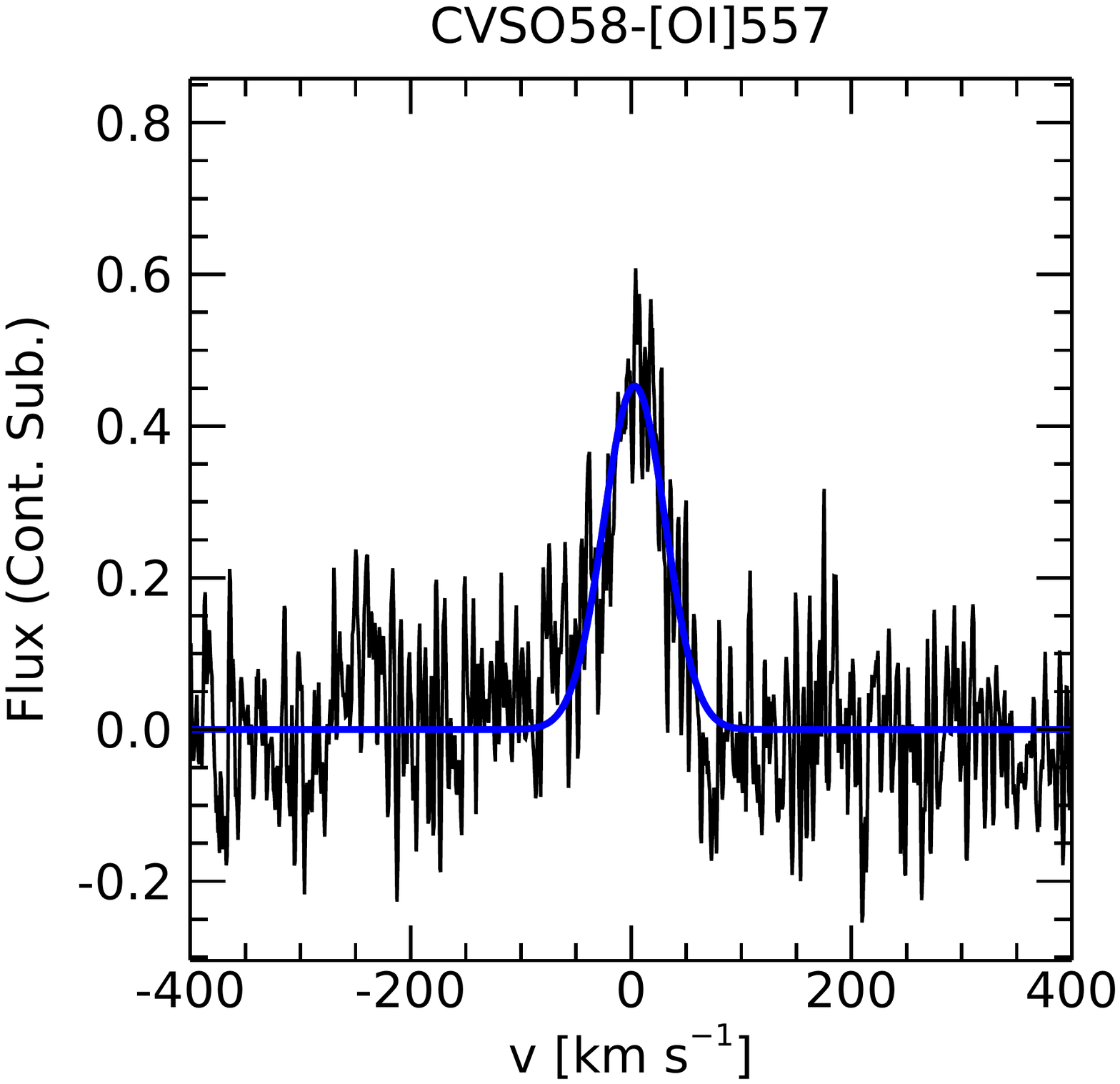}
\includegraphics[trim=20 140 0 140,width=.68\columnwidth, angle=0]{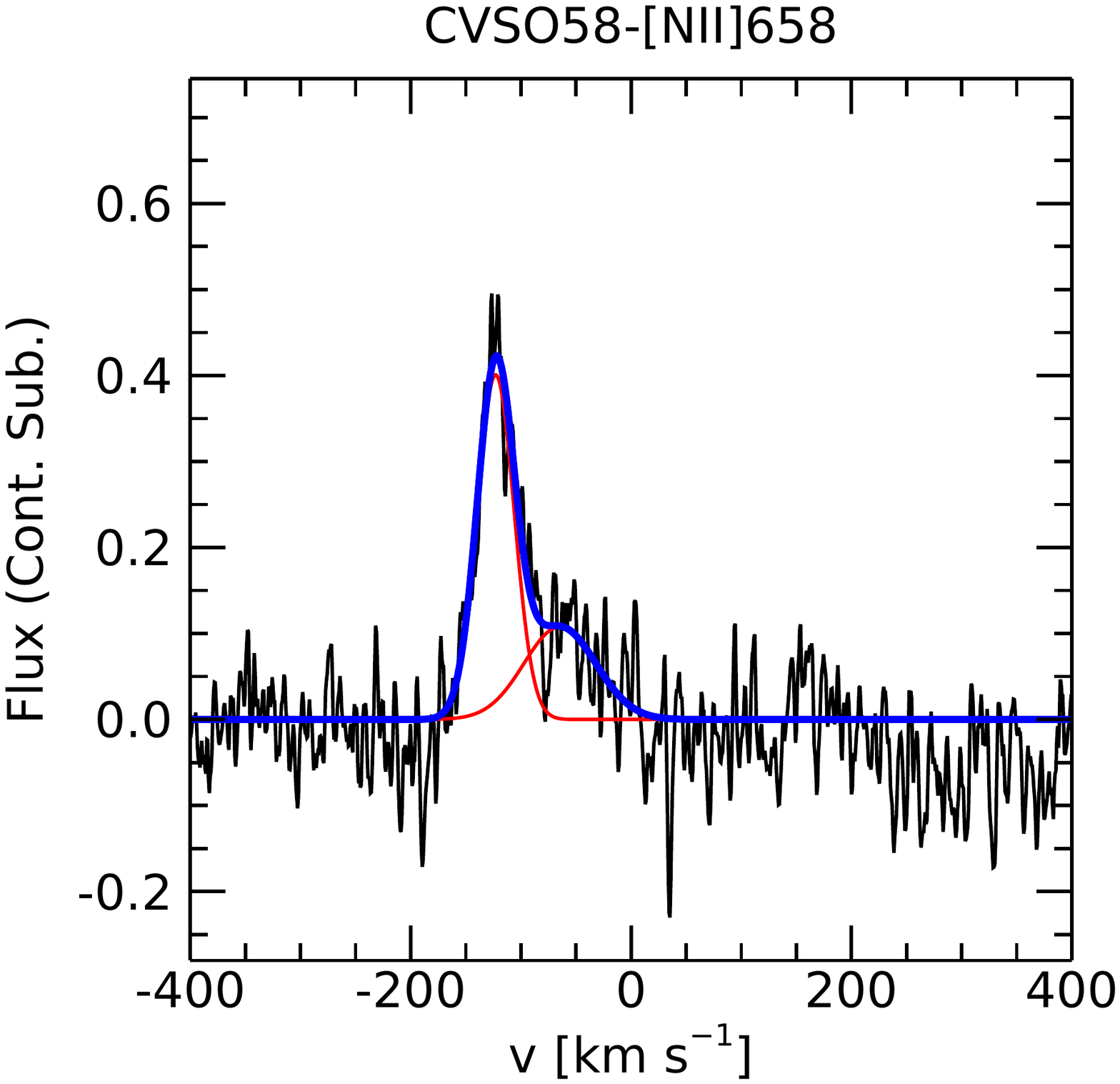}
\includegraphics[trim=20 180 0 140,width=.68\columnwidth, angle=0]{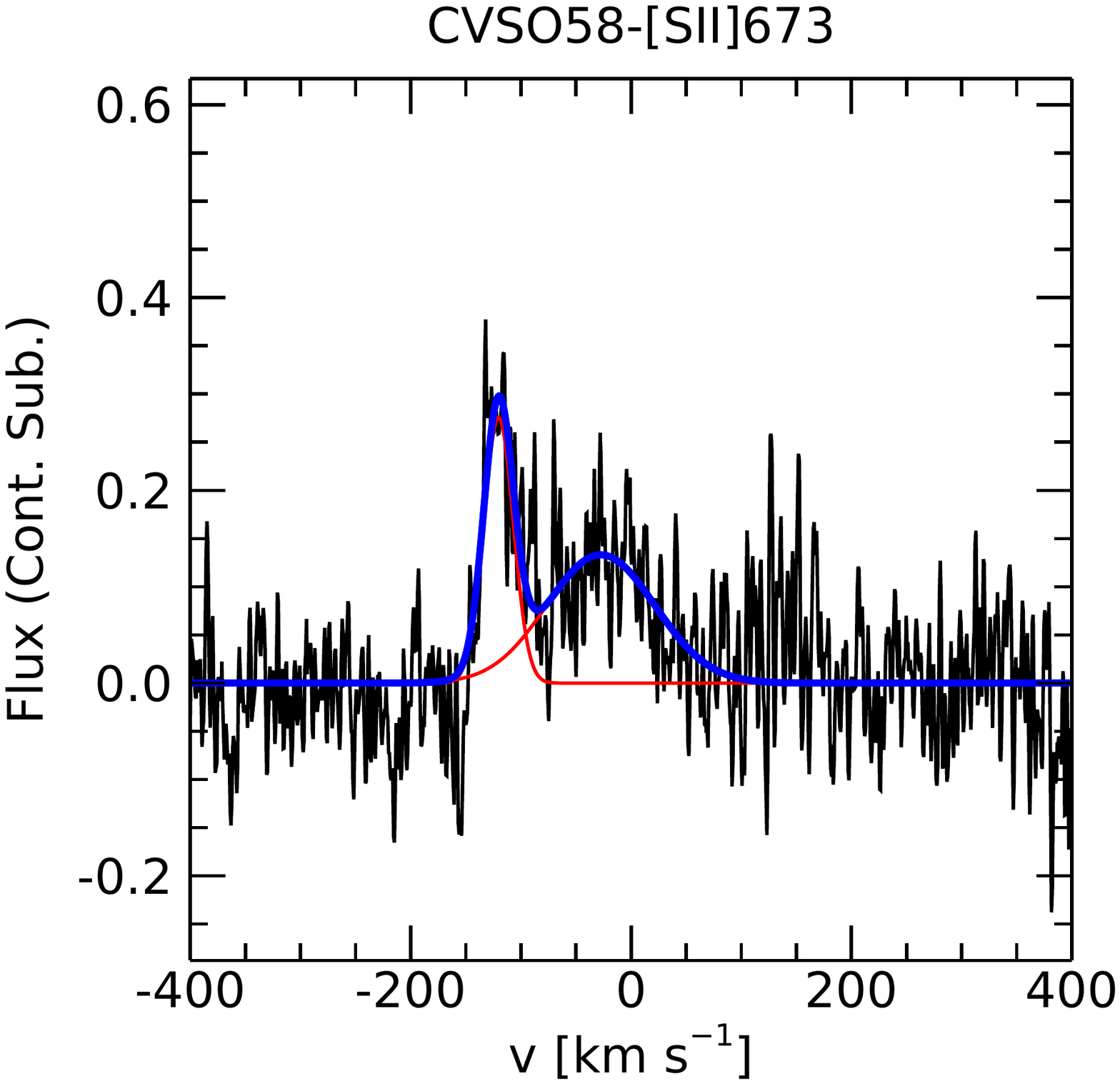}
\includegraphics[trim=20 180 0 140,width=.68\columnwidth, angle=0]{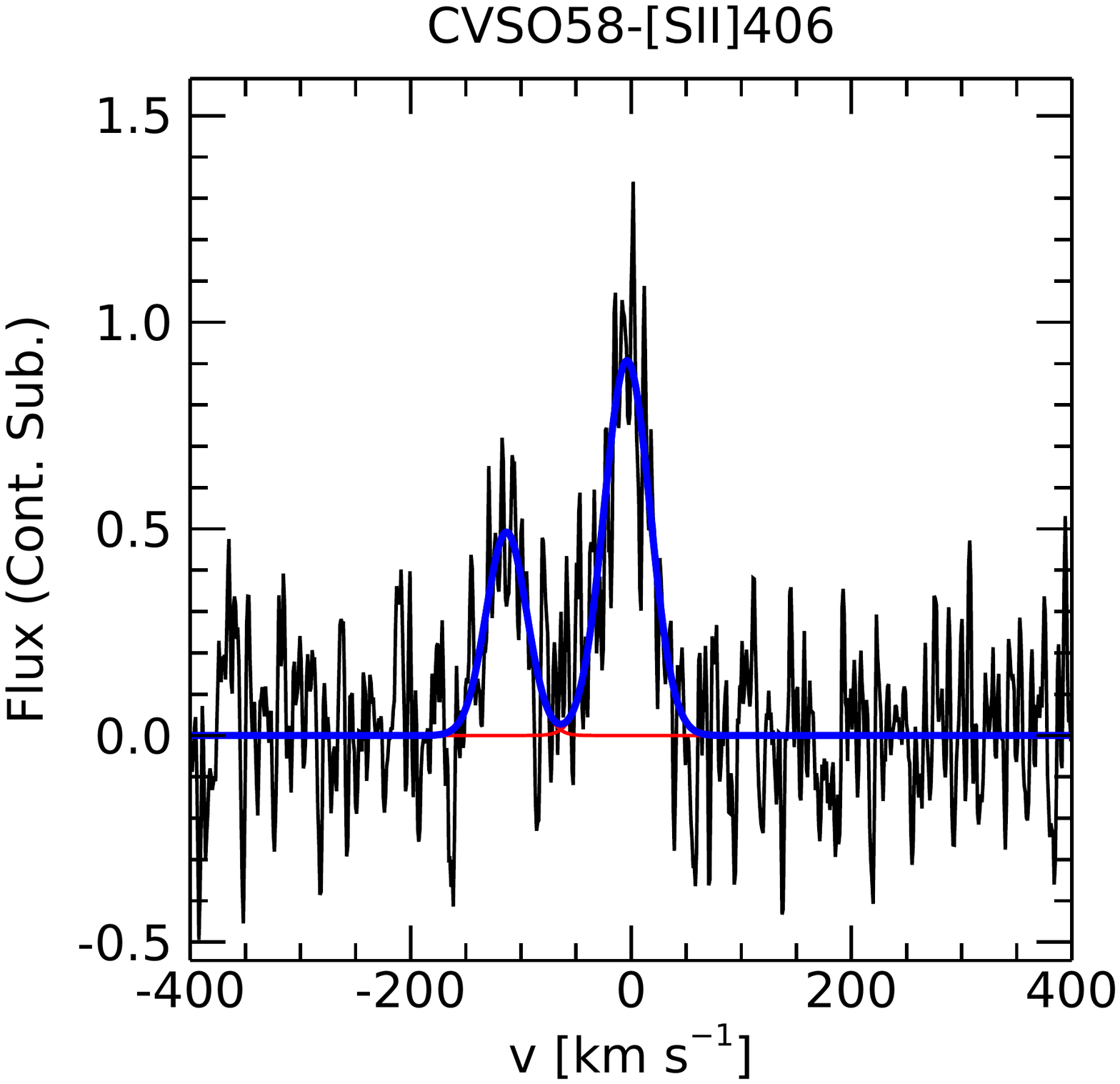}
\begin{center}\caption{\label{fig:example_opt_lines} Example of Gaussian decomposition. The continuum-subtracted forbidden line profiles are shown with black lines. In blue we plot the fit to the profile while individual components are shown in red lines. Flux units are $\rm 10^{-15}$ $\rm ergs^{-1}$ $\rm cm^{-2}$ $\rm \AA^{-1}$. For each panel, target name and line diagnostics are indicated. The complete sample is reported in Fig. \ref{fig:complete_sample}}\end{center}
\end{figure*}

\begin{figure*}
\begin{center}
\includegraphics[trim=20 40 0 0,width=.68\columnwidth, angle=0]{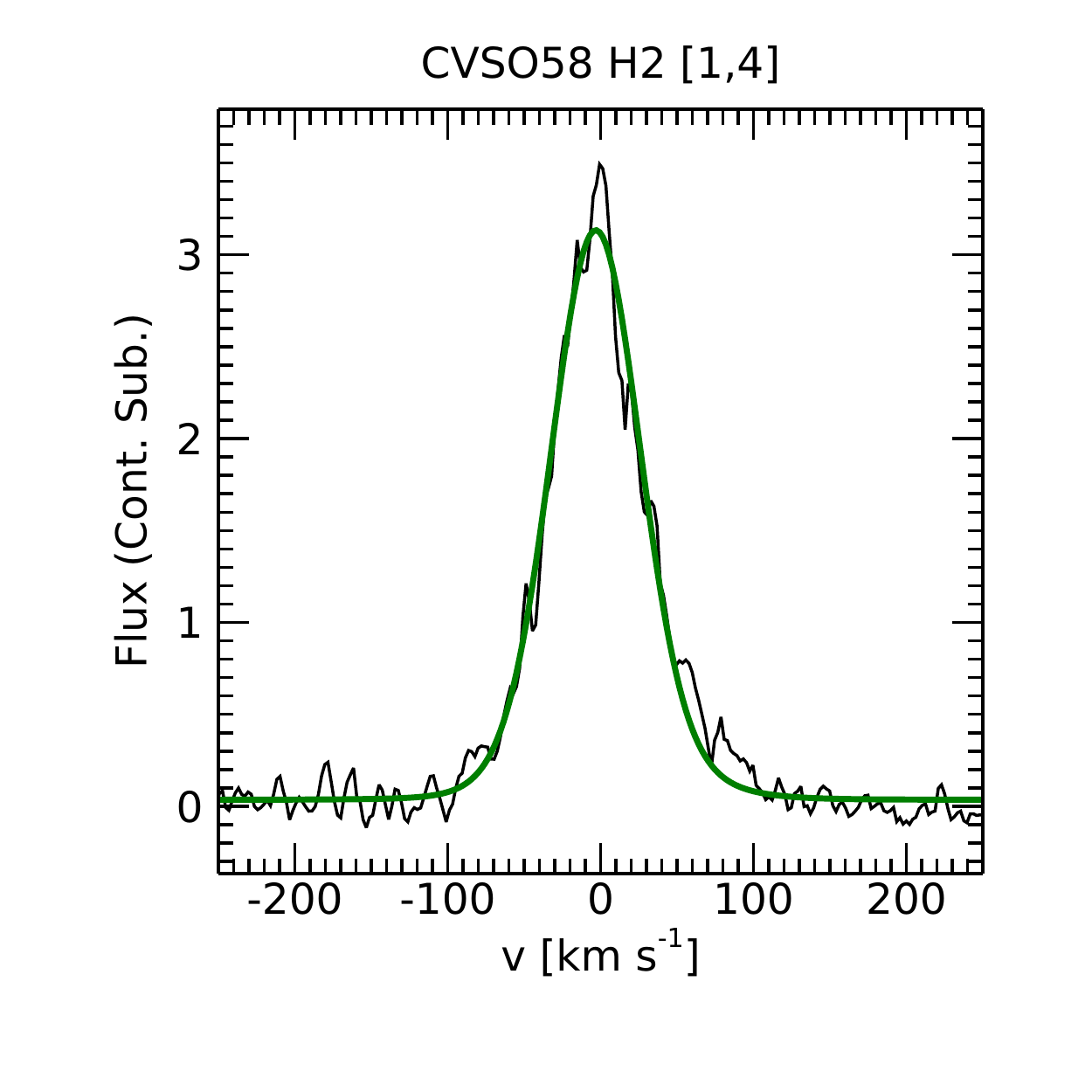}
\includegraphics[trim=20 40 0 0,width=.68\columnwidth, angle=0]{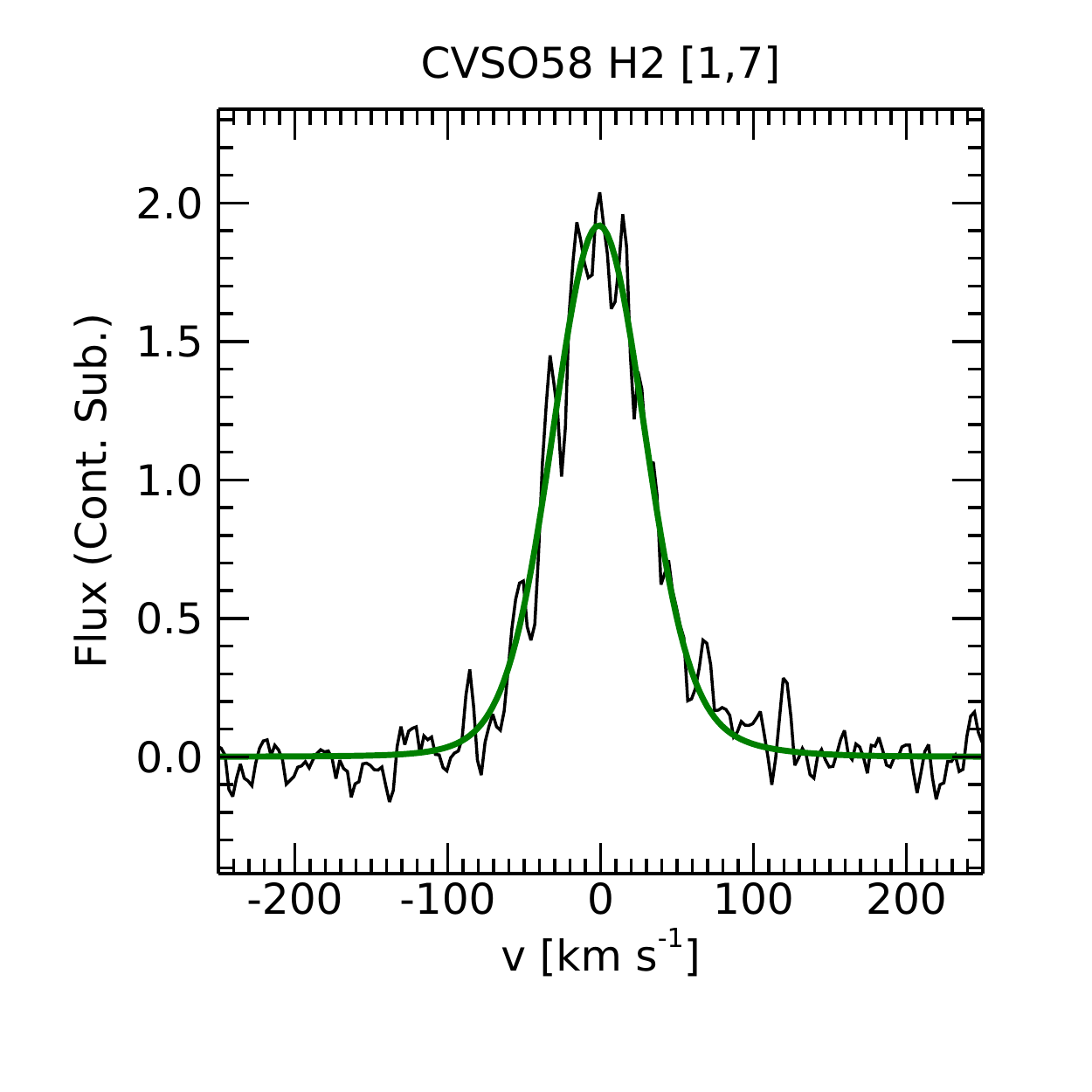}
\begin{center}\caption{\label{fig:example_H2_lines} Example of Gaussian fitting of continuum-subtracted $\rm H_2$ [1,4] and [1,7] averaged line profiles. Flux units are $\rm 10^{-15} ergs^{-1} cm^{-2} \AA^{-1}$. For each panel, target name and line profile are indicated. The complete sample is reported in Fig. 
\ref{fig:complete_sample_H2_14} and \ref{fig:complete_sample_H2_17}}\end{center}
\end{center}
\end{figure*}

\begin{figure}
\includegraphics[trim=10 130 50 50,width=0.97\columnwidth, angle=0]{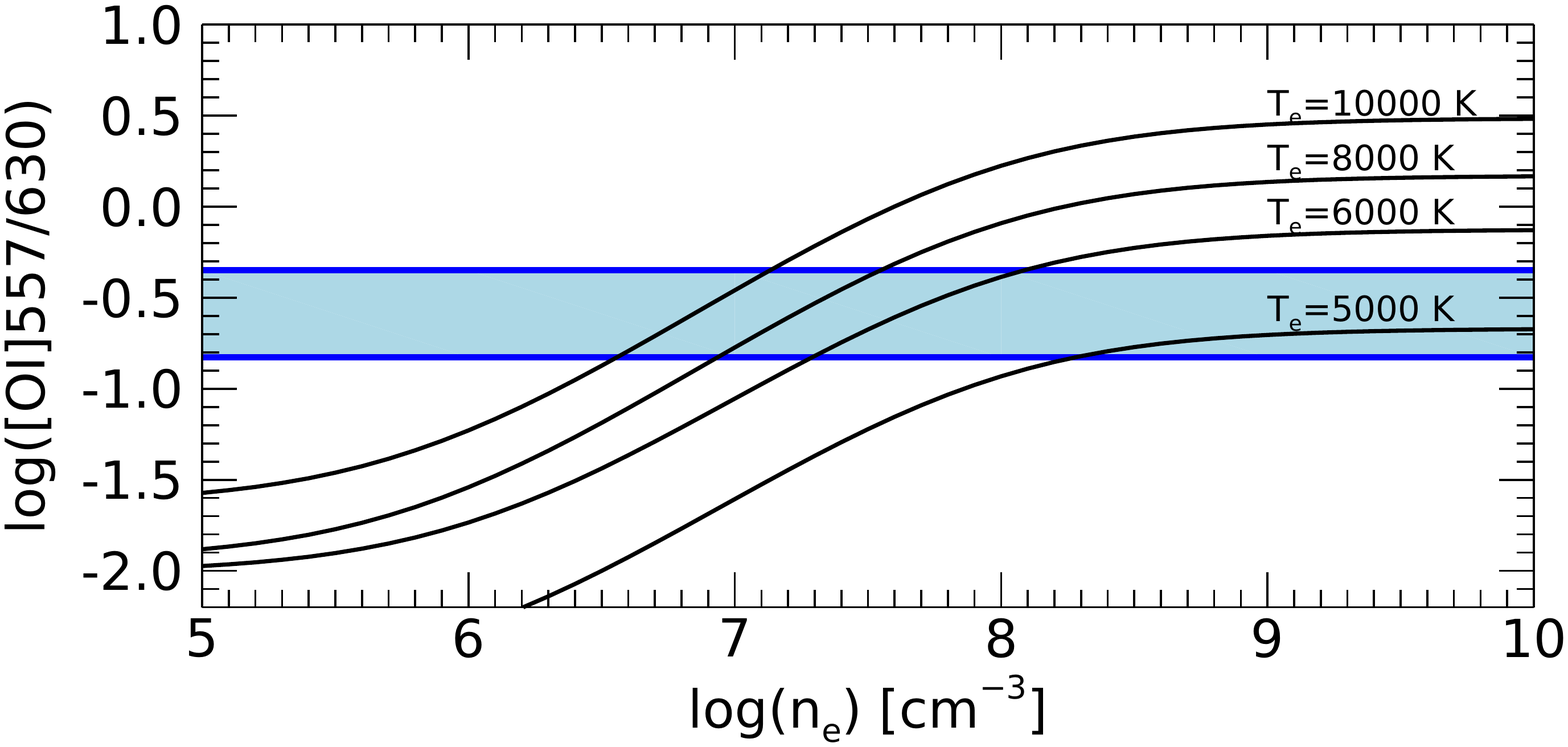}
\includegraphics[trim=10 130 50 50,width=0.97\columnwidth, angle=0]{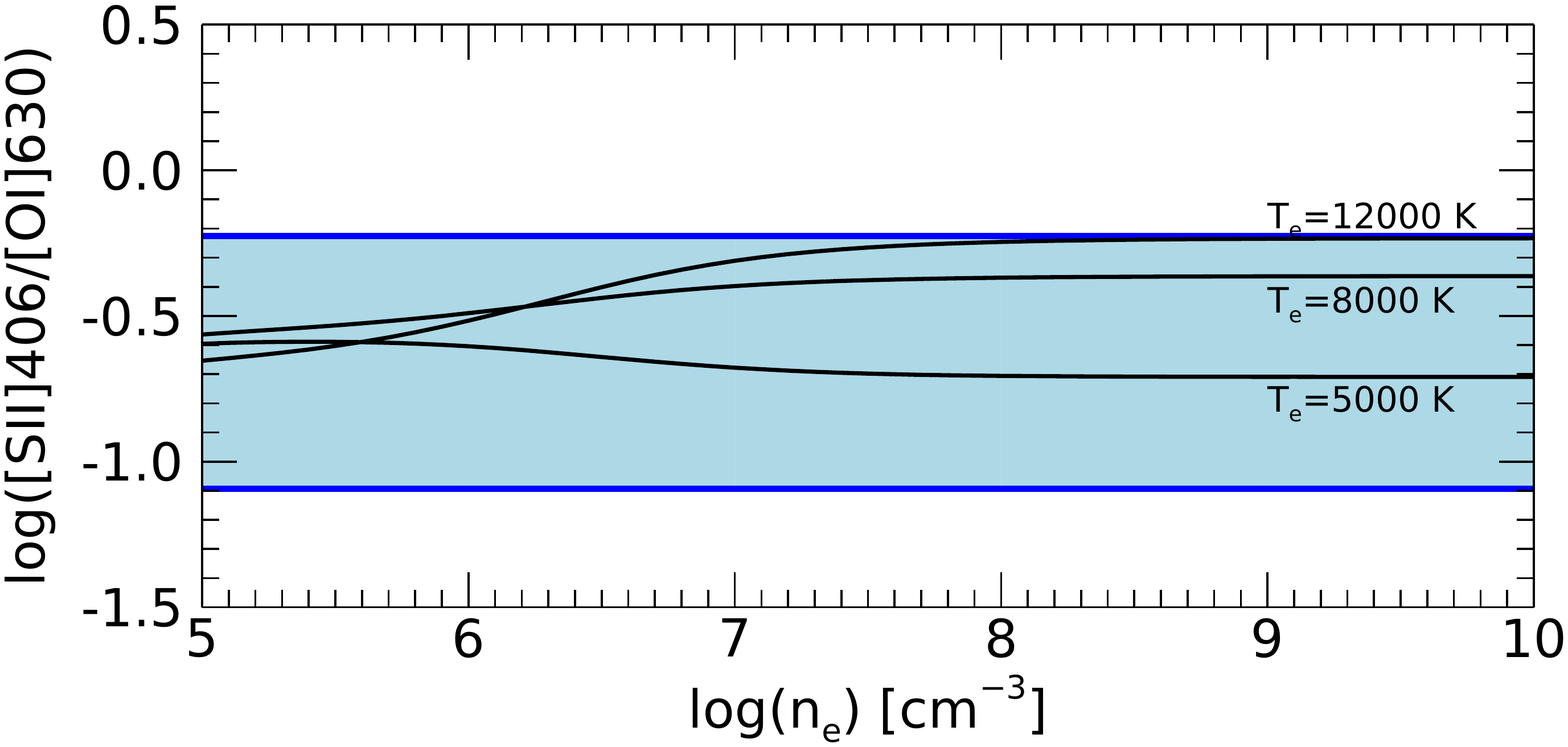}
\begin{center}\caption{\label{fig:diagnostic_diagrams} Diagnostic diagrams of line luminosity ratios due to thermal excitation as a function of density and temperature as computed by \citet{Giannini2019}. Solar abundance is assumed for all involved species. Blue regions indicate the range of values measured in this work for the LVCs.}
\end{center}
\end{figure}

\begin{figure*}
\begin{center}
\includegraphics[trim=120 130 80 100,width=1.5\columnwidth, angle=0]{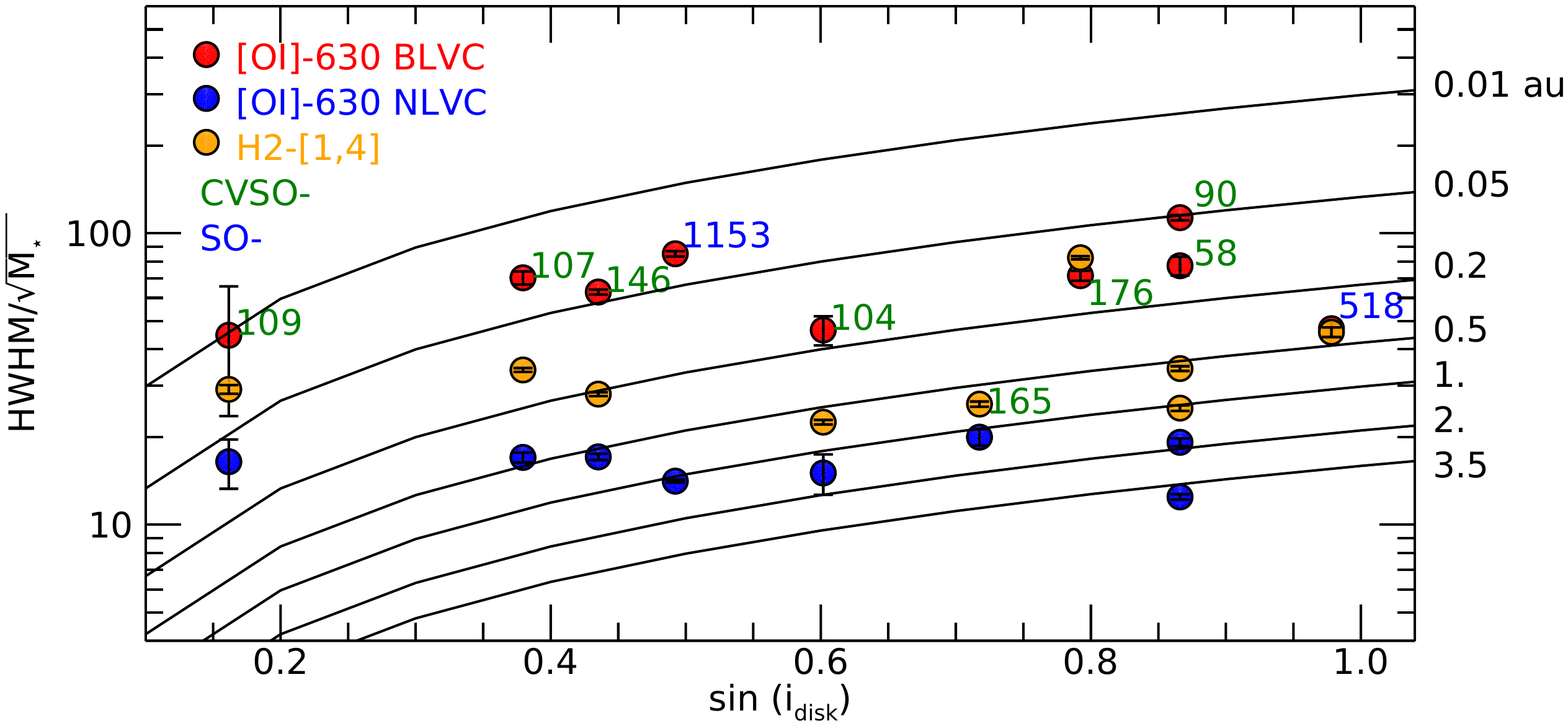}
\caption{\label{fig:emitting_sizes} Half-width at half maximum divided by the square root of the stellar mass for the $\rm H_2$ (orange), [\ion{O}{i}] NLVC (blue) and BLVC (red) as a function of the sine of the disk inclination. Keplerian models for gas emitted from disk radii of 0.01, 0.05, 0.2, 0.5, 1, 2, and 3.5 au are shown as solid black lines.} \end{center}
\end{figure*}

\begin{figure*}
\includegraphics[trim=20 190 0 70,width=.68\columnwidth, angle=0]{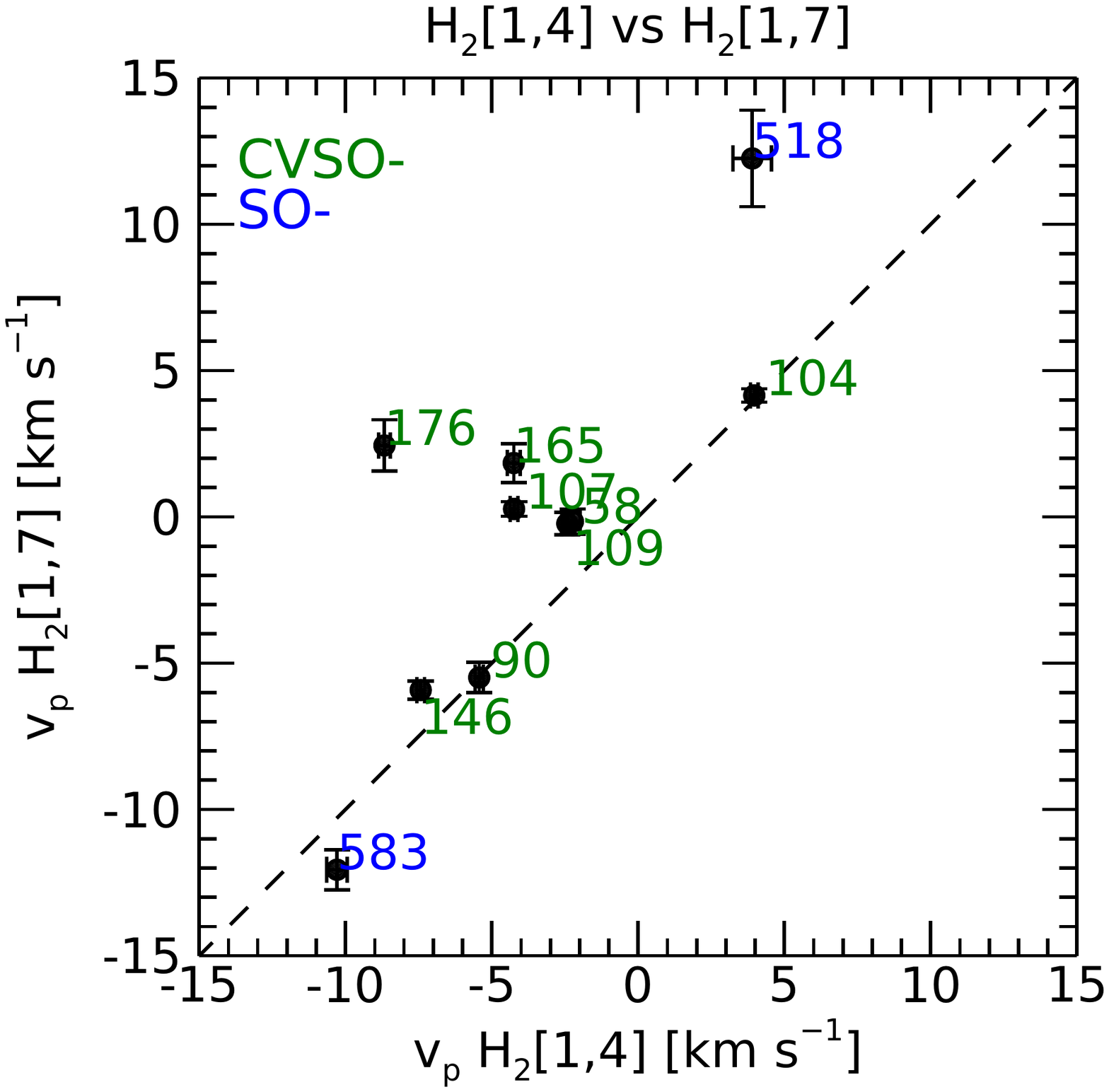}
\includegraphics[trim=20 190 0 70,width=.68\columnwidth, angle=0]{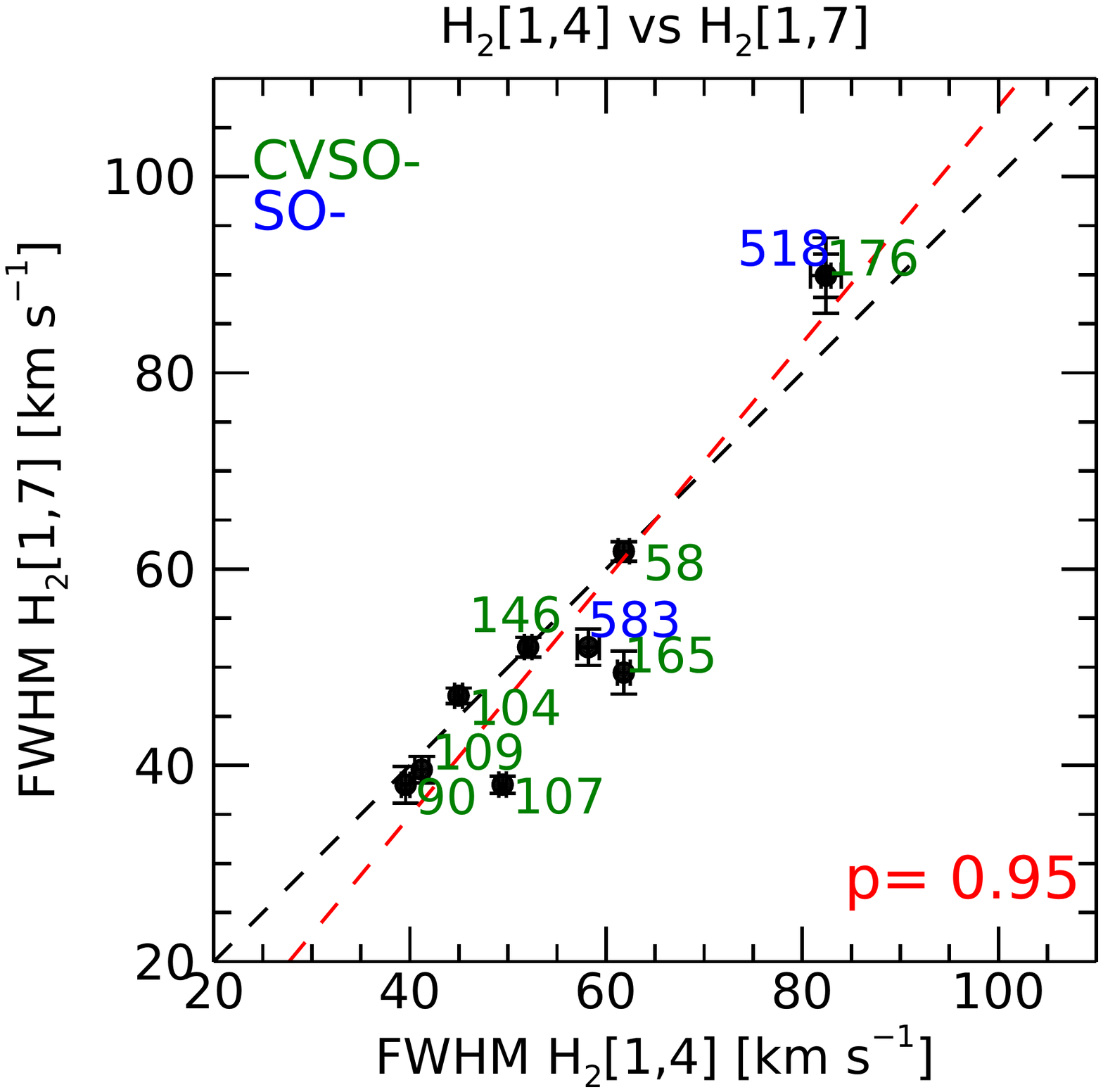}
\includegraphics[trim=20 190 0 70,width=.68\columnwidth, angle=0]{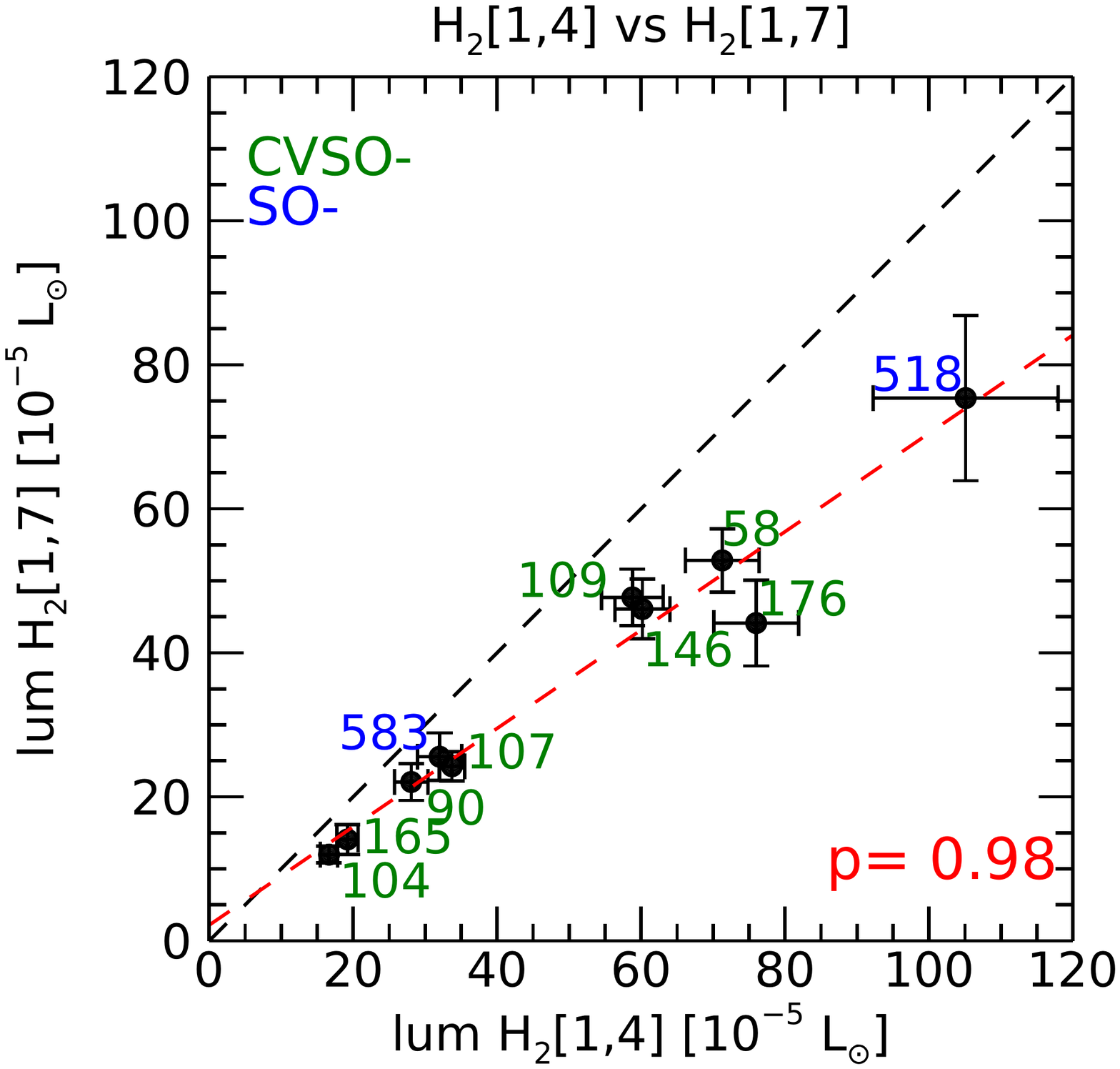}
\includegraphics[trim=20 190 0 70,width=.68\columnwidth, angle=0]{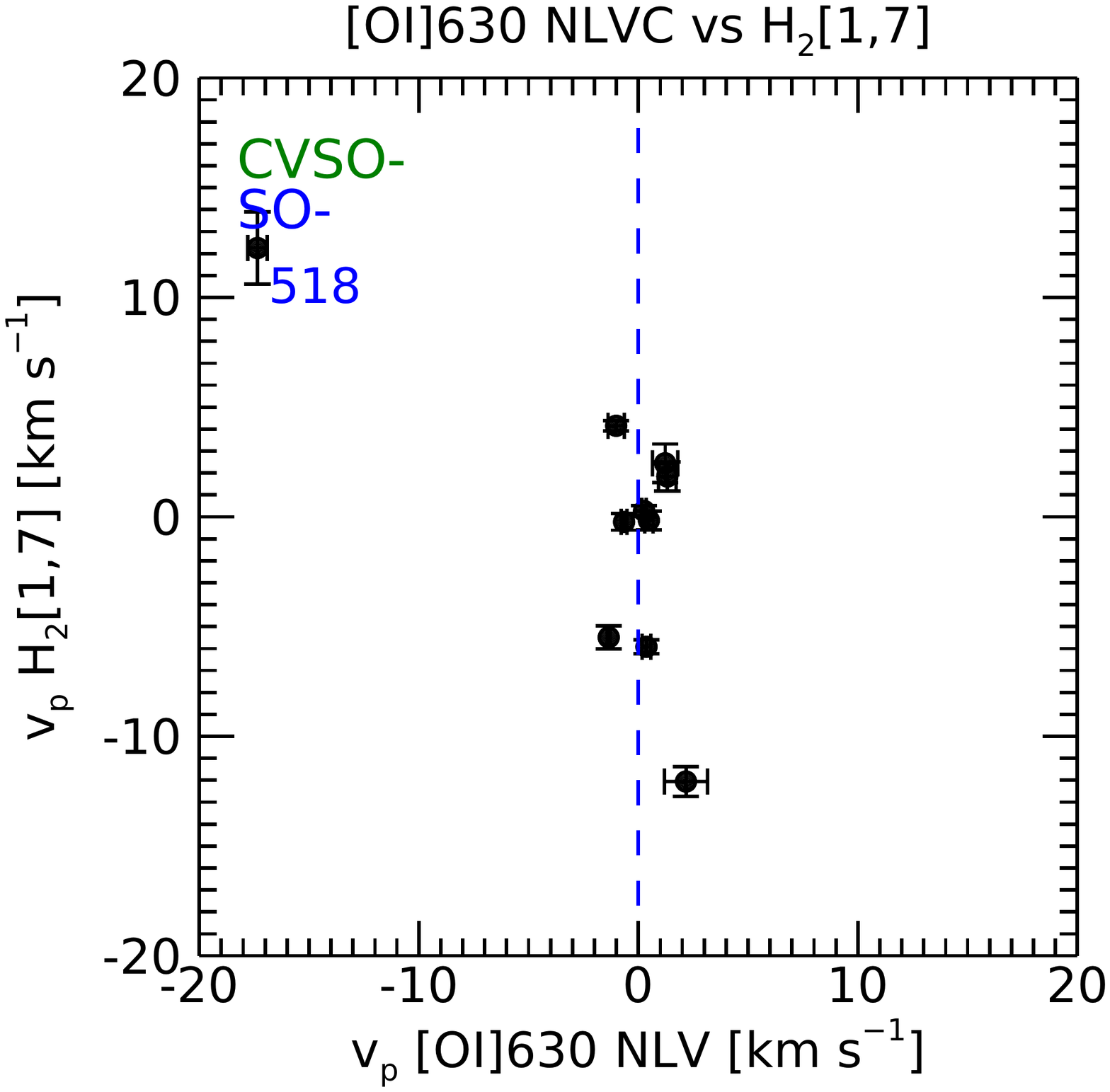}
\includegraphics[trim=20 190 0 70,width=.68\columnwidth, angle=0]{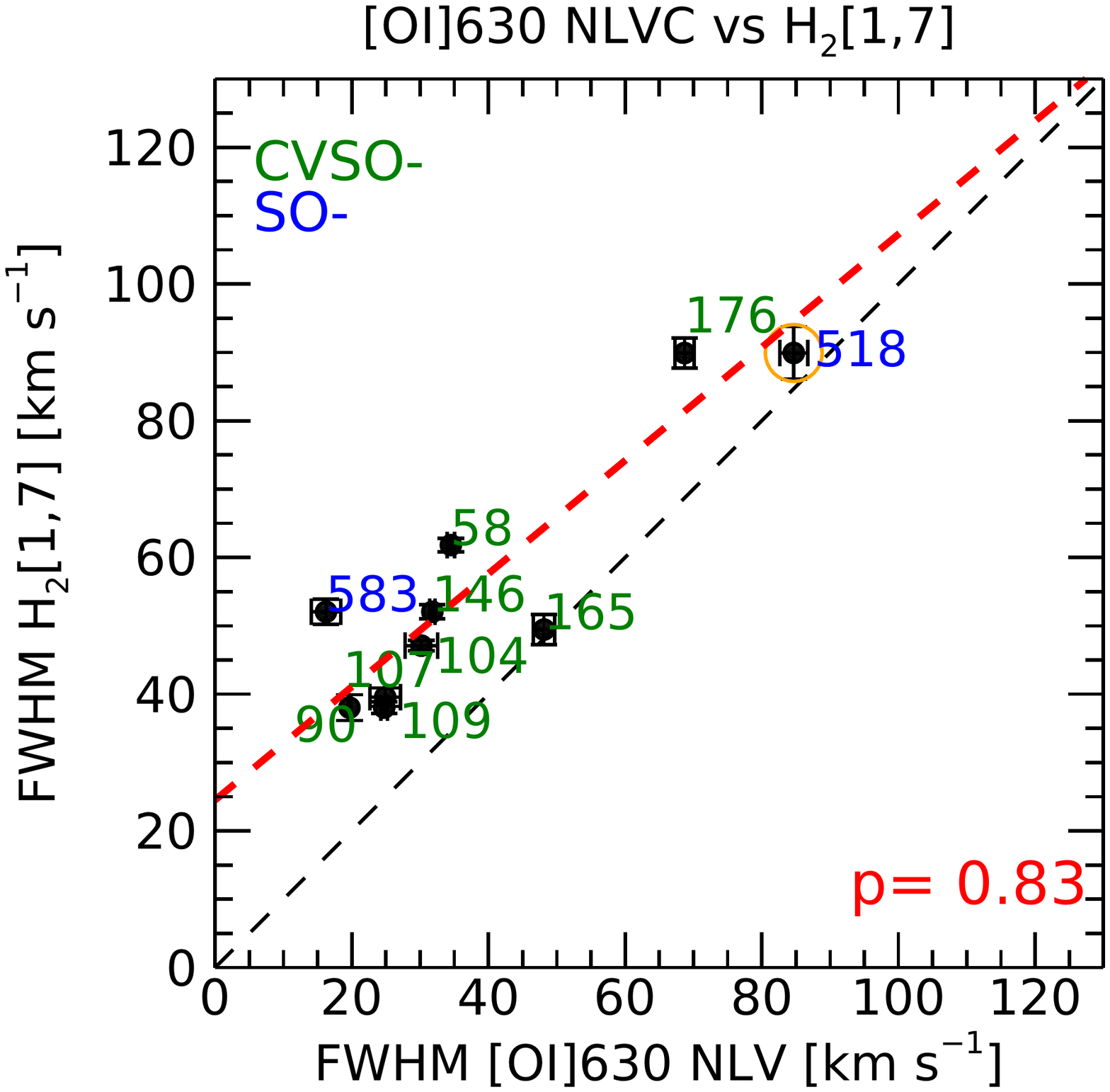}
\includegraphics[trim=20 190 0 70,width=.68\columnwidth, angle=0]{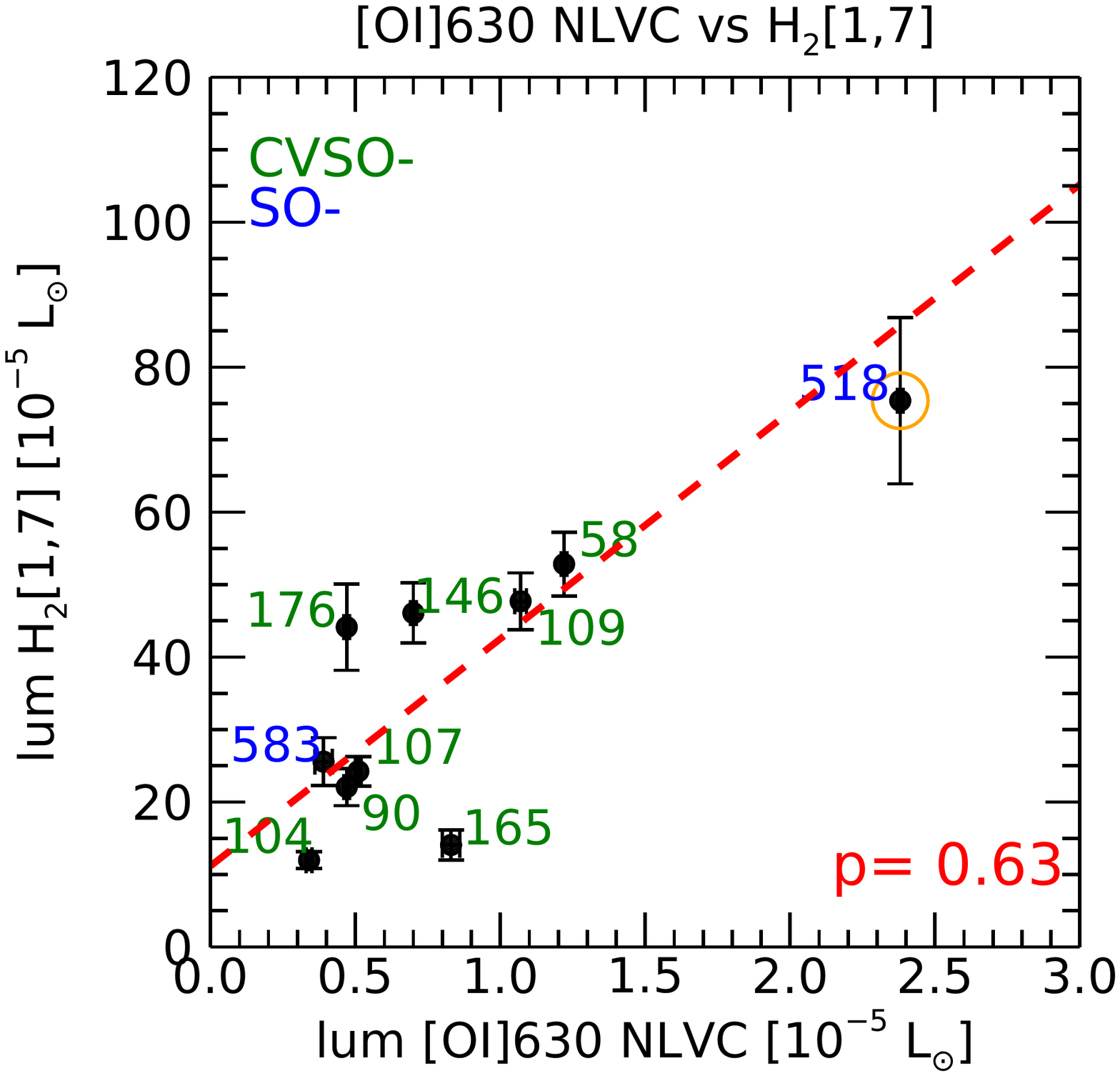}
\includegraphics[trim=20 170 0 70,width=.68\columnwidth, angle=0]{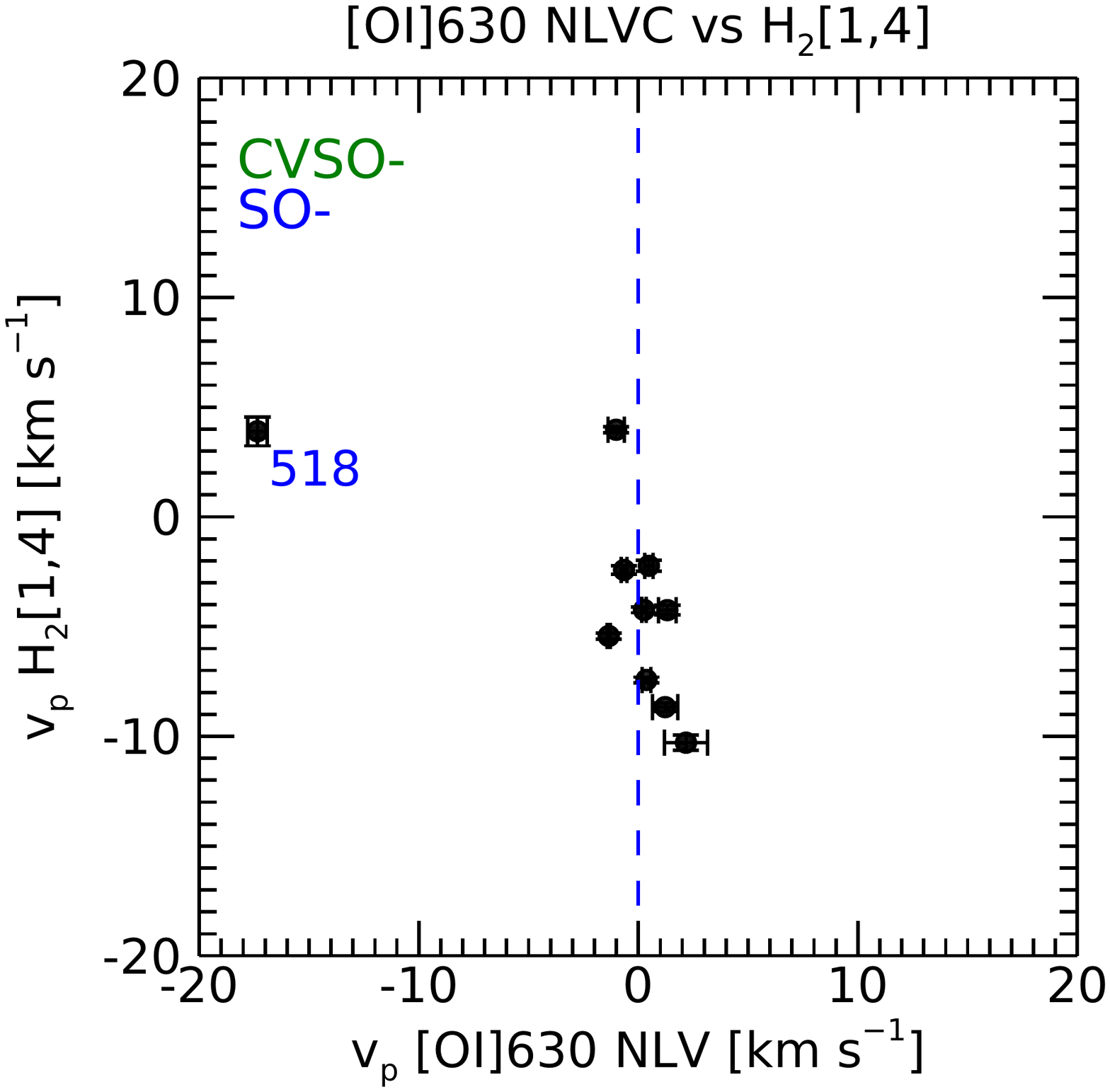}
\includegraphics[trim=20 170 0 70,width=.68\columnwidth, angle=0]{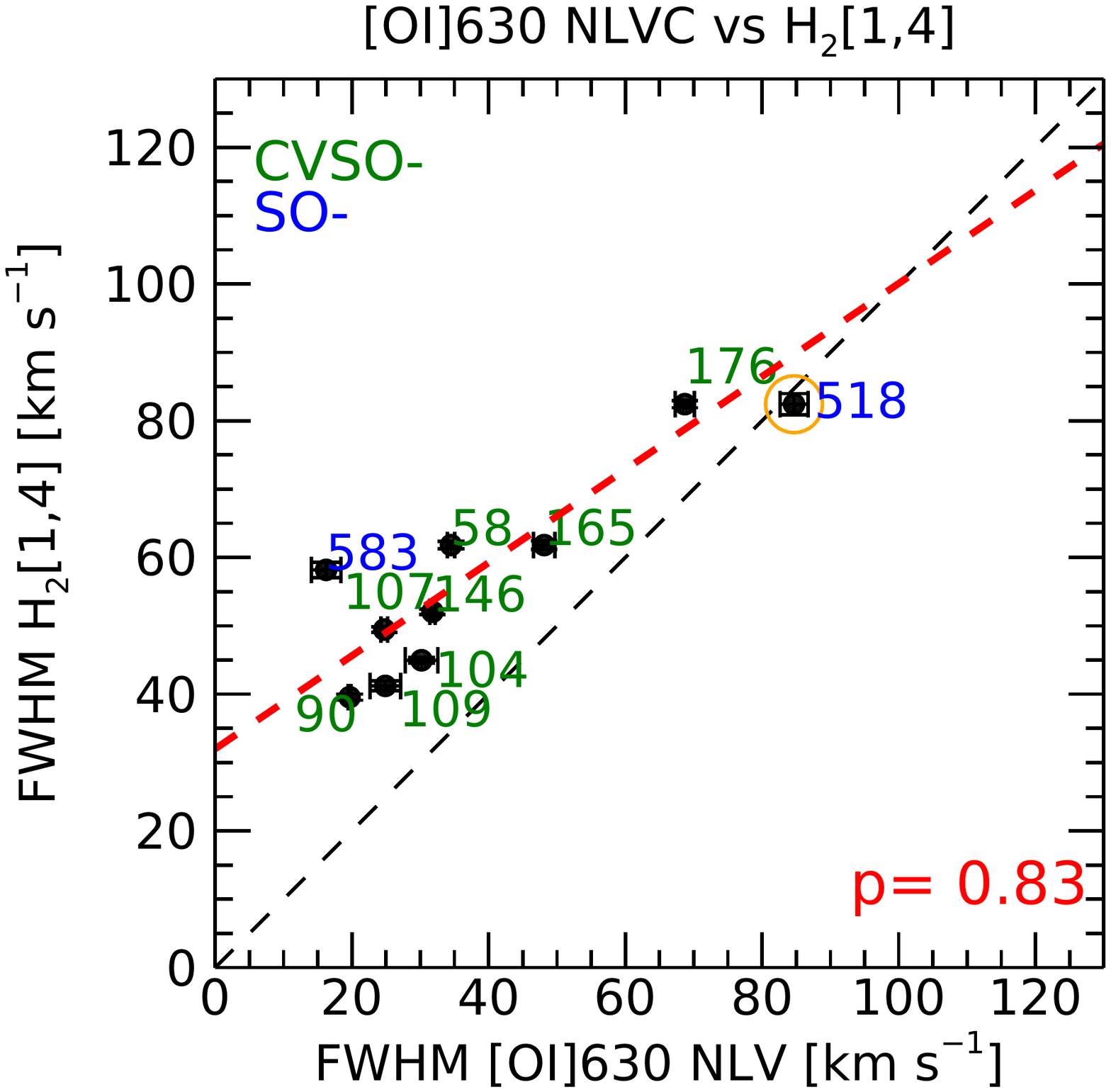}
\includegraphics[trim=20 170 0 70,width=.68\columnwidth, angle=0]{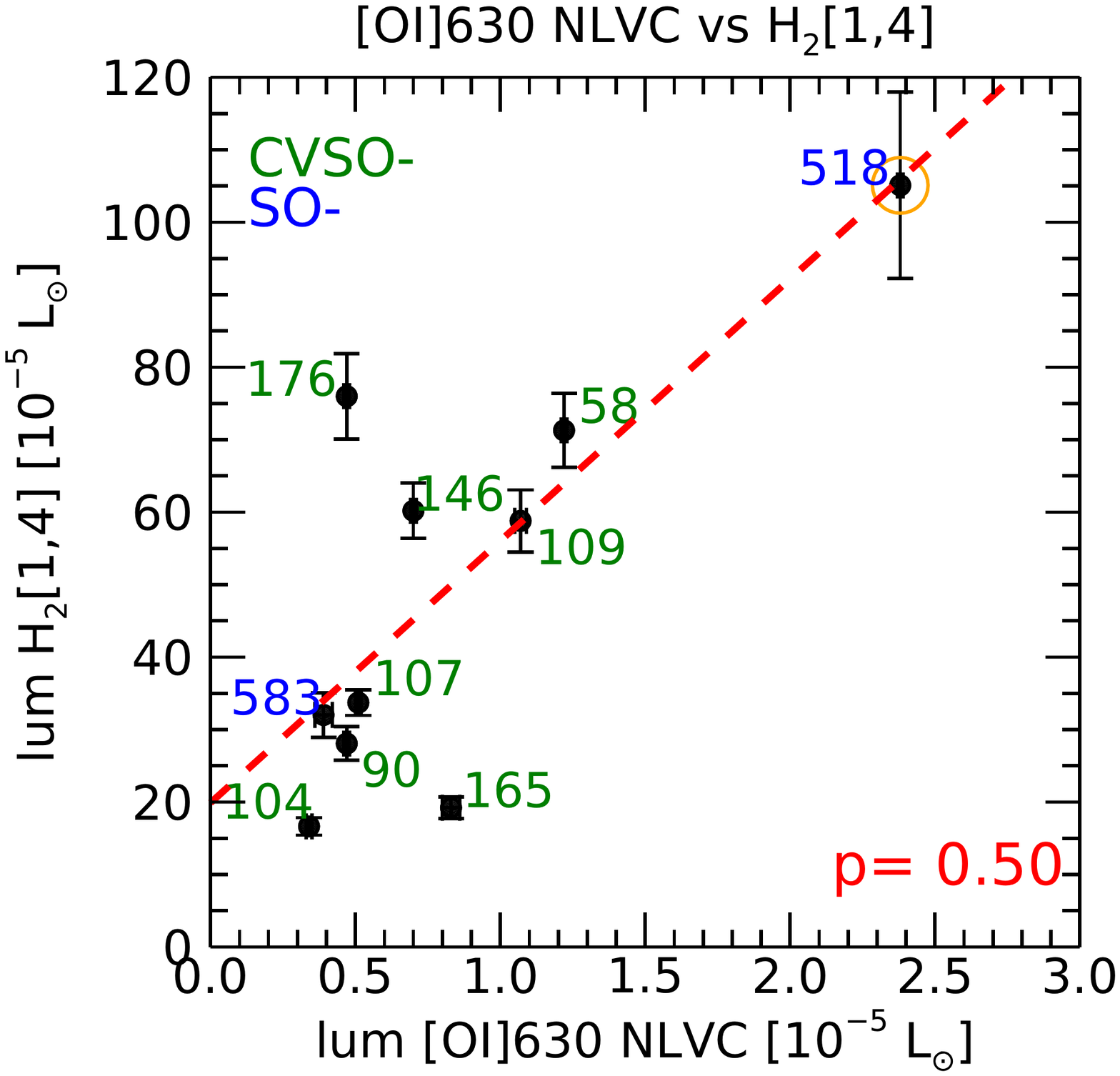}
\begin{center}\caption{\label{fig:correlations} Correlations between kinematic and luminosity properties of the [\ion{O}{i}]630 NLVC and UV $\rm H_2$ lines. From top to bottom: $\rm H_2$ [1,7] versus $\rm H_2$ [1,4],  $\rm H_2$ [1,7] versus [\ion{O}{i}] and $\rm H_2$ [1,4] versus [\ion{O}{i}]. The dashed black lines represent the one-to-one correlations, while linear fits are shown as dashed red lines. The SO518 source (orange circled point) was excluded from the fit (see Sect. \ref{sec:kin_corr} for details). The Pearson coefficients are also reported. The $\rm v_p$ errors on $\rm H_2$-UV lines reported in these plots are those resulting from the Gaussian decomposition procedure (Sect \ref{sec:data_analysis}). The estimated $\rm \sim 15$ $\rm km$ $\rm s^{-1}$ uncertainty due to the wavelength calibration of COS should be added in quadrature.} \end{center}
\end{figure*}

\section{Results}

Fig. \ref{fig:example_opt_lines} and \ref{fig:example_H2_lines} show examples of the observed atomic and $\rm H_2$ lines and their Gaussian analysis for the CVSO58 source, while the complete sample is reported in Fig. \ref{fig:complete_sample}, \ref{fig:complete_sample_H2_14} and \ref{fig:complete_sample_H2_17}. Fitted parameters (i.e. peak velocity, FWHM, flux and luminosity) are reported in Tables \ref{tab:GaussDecompRes} and \ref{tab:GaussDecompRes_H2_14}. 

\begin{table*}
\center
\caption{\label{tab:det_tatistics} Detection of kinematical components drawn from Gaussian decomposition.}
\begin{tabular}{llllll}
\hline
\hline
Name       & [\ion{O}{i}]630      & [\ion{O}{i}]557  & [\ion{N}{ii}]658 & [\ion{S}{ii}]673 & [\ion{S}{ii}]406 \\
\hline
\noalign{\smallskip}
CVSO58	&  NL, BL, Hb              & BL               &  Hb           & BL, Hb      & NL, Hb      \\    
\noalign{\smallskip}
CVSO90	&  NL, BL, Hr, Hb         & NL, BL        &  Hb, Hr     & Hb, Hr      & BL, Hb, Hr \\ 
\noalign{\smallskip}
CVSO104	&  NL, BL                     & BL               & -                & -               & NL             \\ 
\noalign{\smallskip}
CVSO107	&  NL, BL, Hb              & NL, BL         & -               & BL, Hb     & NL, BL, Hb \\ 
\noalign{\smallskip}
CVSO109	&  NL, BL                     & NL               & -               & -               & NL              \\  
\noalign{\smallskip}
CVSO146	&  NL, BL                     & BL               & -               & Hb            & NL, BL        \\ 
\noalign{\smallskip}
CVSO165	&  BL, Hb, Hr               & BL               & -               & Hb, Hr      & -                 \\ 
\noalign{\smallskip}
CVSO176	& BL, Hb                      & BL               & -               & -               & BL              \\ 
\noalign{\smallskip}
\hline
\hline
\noalign{\smallskip}
SO518	& BL, Hb, Hr                & BL               & Hb            &  Hb           & BL, Hb       \\ 
\noalign{\smallskip}
SO583	&  NL, BL                     & BL               & -               &  BL           & -                 \\ 
\noalign{\smallskip}
SO1153	&  NL, Hb                     & NL, Hb        & Hb           & Hb            & Hb              \\ 
\noalign{\smallskip}
\hline
\end{tabular}
\begin{quotation}                  
\textbf{Notes.}
\textbf{NL}: NLVC (|$\rm v_p$| < 30 $\rm km$ $\rm s^{-1}$, FWHM < 40 $\rm km$ $\rm s^{-1}$), \textbf{BL}: BLVC (|$\rm v_p$| < 30 $\rm km$ $\rm s^{-1}$, FWHM > 40 $\rm km$ $\rm s^{-1}$), \textbf{Hr}: redshifted HVC ($\rm v_p$ > 30 $\rm km$ $\rm s^{-1}$), \textbf{Hb}: blueshifted HVC  ($\rm v_p$ < -30 $\rm km$ $\rm s^{-1}$).
\end{quotation}
\end{table*}  

\subsection{Detection statistics}
The UV-$\rm H_2$ lines were detected in all the sources; their profiles can be always reproduced with a single Gaussian component, with the current spectral resolution and $SNR$. In contrast, the optical forbidden lines present composite profiles, where components at different velocities and FWHMs can be identified, in line with previous similar studies. A detailed view of the different detected components is reported in Table \ref{tab:det_tatistics}. In short, the two [\ion{O}{i}] lines were detected in all the sources, with the [\ion{O}{i}]630 the most structured one. They always show LVCs peaked at zero velocities, except for 4 sources showing a slight blue-shift values compatible with slow disk winds.

The [\ion{S}{ii}]637 and [\ion{S}{ii}]406 lines were detected in 8 and 9 sources out of 11, respectively. Compared to the [\ion{O}{i}] lines, the [\ion{S}{ii}] HVCs were detected more often, a behavior already observed in high excitation or ionization lines \citep[e.g.,][]{Natta2014}. Finally, the [\ion{N}{ii}]658 line was detected in 4 sources and only in the HVCs, as expected given the high ionization needed to have an appreciable \ion{N}{ii} abundance (e.g., Nisini et al. in prep).

\subsection{Line ratios}\label{sec:line_lum}
We investigate here the line luminosity ratios under the assumption that the emission of the atomic components have a thermal origin. Fig. \ref{fig:diagnostic_diagrams} shows the diagnostic diagrams of line luminosity ratios based on the excitation model of \citet{Giannini2015,Giannini2019}. The latter assumes an NLTE approximation for the line emission, with population of levels determined by assuming equilibrium between collisional excitation and de-excitation with electrons, and radiative decay. Different curves correspond to different temperatures, as labelled, while the range of ratios computed from the extinction-corrected LVCs luminosities are depicted in blue boxes.

From the top panel of Fig. \ref{fig:diagnostic_diagrams}, we find that the [\ion{O}{i}]557/[\ion{O}{i}]630 ratios are consistent with density $\rm n_e \gtrsim 10^{6.5}$ $\rm cm^{-3}$ and temperature $\rm 5000 \leq T \leq 10 000$ K, in agreement with previous similar studies \citep[e.g.,][]{Natta2014, Fang2018, Giannini2019}. The [\ion{S}{ii}]406/[\ion{O}{i}]630 ratio, is expected to have a very little dependence with the density and temperature, since both the [\ion{S}{ii}]406 and [\ion{O}{i}]630 lines have a similar critical density (bottom panel of Fig. \ref{fig:diagnostic_diagrams}). However, the observed [\ion{S}{ii}]406/[\ion{O}{i}]630 ratios are consistent with the range of densities and temperatures indicated by the [\ion{O}{i}]557/[\ion{O}{i}]630 ratios, with the exception of 3 sources (i.e. CVSO90, CVSO104 and SO1153), whose [\ion{S}{ii}]406/[\ion{O}{i}]630 ratios lies at around 0.1. This value would be compatible with a temperature of about 3000 K that is not enough to sufficiently populate the \ion{S}{ii} atomic sub-levels to give the observed line fluxes \citep[e.g.,][]{Giannini2019}.

\subsection{Emitting sizes}\label{sec:em_sizes}
Average emitting sizes can be deduced from the line width under the assumption that this latter is dominated by the bulk motion of Keplerian rotation. In such case the average emitting size can be expressed as
\begin{eqnarray}
\rm R_K = \Biggl{(}\frac{\rm \sin (i_{disk})}{\rm HWHM}\Biggr{)}^2 \times \rm G \times M_{\star},
\end{eqnarray}

where $\rm i_{disk}$ is the disk inclination angle and $\rm HWHM$ is the line half-width at half maximum.

The assumption of a purely Keplerian broadening has been extensively applied for both the NLVC and BLVC of the [\ion{O}{i}]630 line \citep[][]{Simon2016, Fang2018, McGinnis2018, Banzatti2019, Gangi2020}, and it is usually assumed to be valid if the component originates in slow disk winds, where the wind poloidal velocity is small and the broadening is still dominated by the Keplerian motion. However, as discussed in \citet{Weber2020}, vertical velocity gradients might contribute to the line broadening, particularly at low inclinations (i.e. $\rm \lesssim$ $20^{\circ}$). 

Regarding the UV-$\rm H_2$ lines, their widths can be reasonably considered to be dominated by Keplerian broadening, since significant thermal broadening would require temperatures higher than the dissociation temperature of $\rm H_2$ \citep{Lepp1983} so the broadening induced by the turbulence of the disk can be negligible \citep{France2012}.

In Fig. \ref{fig:emitting_sizes} we report the HWHM divided by the square root of the stellar mass as a function of $\rm \sin(i_{disk})$ for the [\ion{O}{i}]630 NLVC and BLVC and for the $\rm H_2$ [1,4] line. From the over-plotted Keplerian models of constant radius we can infer that the [\ion{O}{i}]630 NLVC is emitted by regions with size between $\sim$0.5 and $\sim$3.5 au while the [\ion{O}{i}]630 BLVC traces an innermost region between $\sim$0.01 and $\sim$0.2 au. On the contrary, the UV-$\rm H_2$ emission appear to originate from an intermediate region, partially superimposed on the previous two, namely between $\sim$0.05 and $\sim$1 au.

\section{Correlations}

\subsection{[\ion{O}{I}]630 versus UV-$\rm H_2$ line properties}\label{sec:kin_corr}
Fig. \ref{fig:correlations} shows the correlations between $\rm v_p$, FWHM and line luminosity of the [\ion{O}{I}]630 NLVC, $\rm H_2$ [1,4] and [1,7] lines.

The $\rm H_2$ [1,4] and [1,7] lines are well correlated in terms of both kinematics and luminosity. In particular, in six sources the $\rm v_p$ follows a one-to-one correlation and in the other four the $\rm H_2$ [1,4] appears to be blueshifted but always within the velocity accuracy of COS ($\rm \sim$ 15 $\rm km$ $\rm s^{-1}$). The FWHMs show a one-to-one relation and the line luminosities correlate with a slope lower than one, i.e. lines from the [1,4] progression are brighter than the corresponding lines of the [1,7] progression.

Regarding the [\ion{O}{I}]630 NLVC, we find a similar behaviour in the correlations with the $\rm H_2$ [1,7] and [1,4]. In particular, with the exception of SO518, the [\ion{O}{I}] peak velocities are always consistent with zero. The SO518 source has a high disk inclination angle ($\rm i_{disk} = 78^{\circ}$, Table \ref{tab:target_list}), thus the LVC properties may be heavily contaminated by the HVC. For this reason, we have excluded this source in all of the  correlations concerning the atomic forbidden lines. A high spread around the $\rm v_p$ of both the $\rm H_2$ lines is present. Although this spread shows a predominance of blue-shifted velocity, it is comparable to the velocity accuracy of COS. We have also checked that the larger blue-shifted peaks (up to $\rm -10$ $\rm km$ $\rm s^{-1}$) are not associated with sources having a broad or a HVC in the [\ion{O}{I}]630 emission, which could have contaminated the $\rm H_2$ line profile. The FWHMs are highly correlated, and for most cases, the FHWM of the [\ion{O}{I}]630 is smaller than that of the $\rm H_2$ species. Finally, the line luminosities show a steep relation, with the luminosity of the $\rm H_2$ emission a factor 30 to 40 higher than that of the [\ion{O}{I}]630. 

\begin{figure}
\includegraphics[trim=100 50 100 35,width=1.\columnwidth, angle=0]{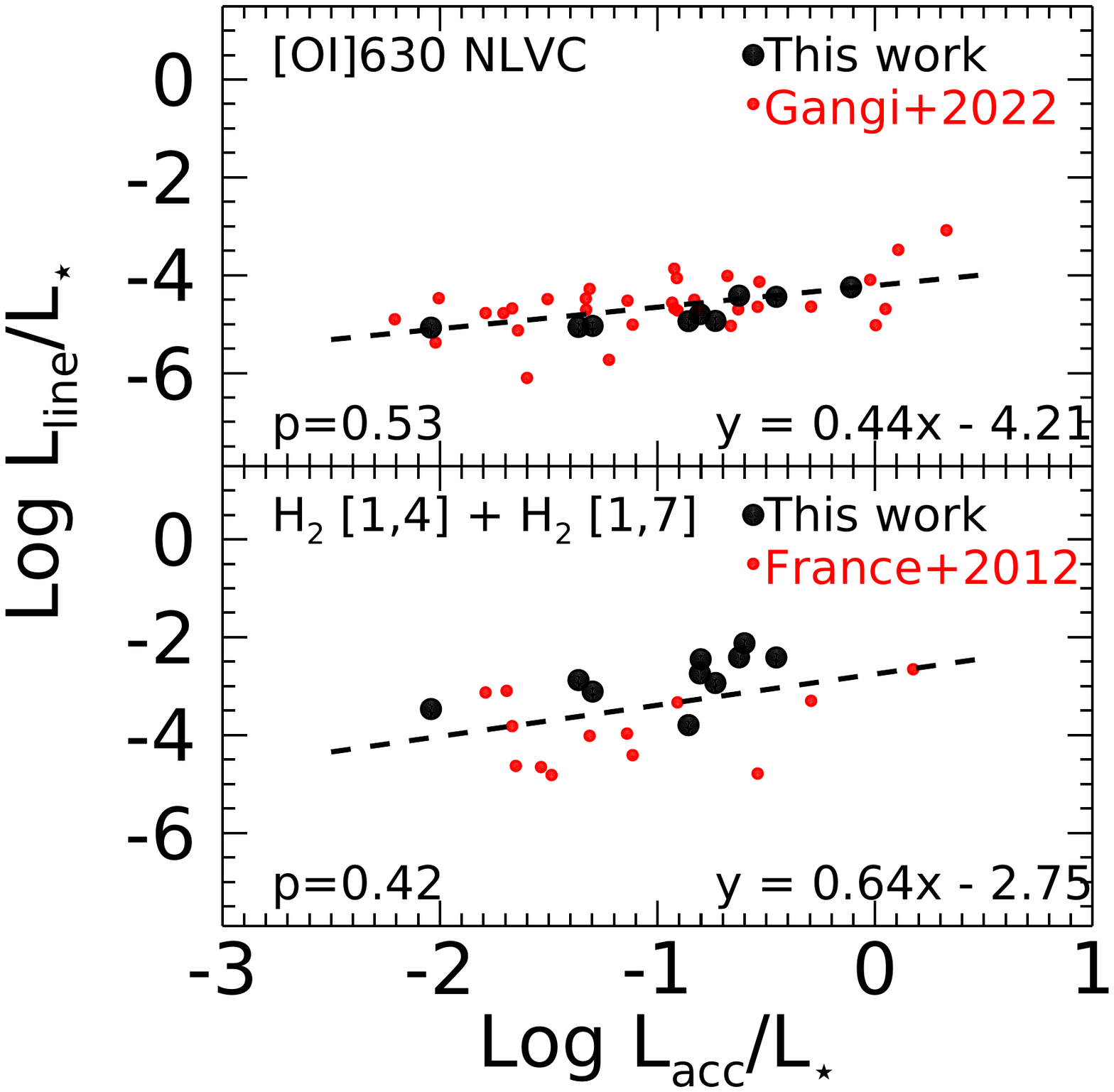}
\begin{center}\caption{\label{fig:lline_vs_lacc} Line luminosity as a function of accretion luminosity. Both quantities are normalized to the stellar luminosity. From top to bottom: [\ion{O}{I}]630 NLVC and $\rm H_2$ [1,4] and [1,7] sum. Red dots indicate values found in \citet{Gangi2022} and \citet{France2012} for a sample of CTTs in the Taurus-Auriga association. Linear fits to the all data points are marked by dashed lines. Analytical solution and Pearson coefficients are also labelled.}\end{center}
\end{figure}

\subsection{[\ion{O}{I}]630 and UV-$\rm H_2$ line luminosity versus accretion luminosity}\label{sec:lum_vs_acc}
We can obtain insights on how the physical origin of the different line components is connected to the accretion mechanisms by looking at the correlation between $\rm L_{line}$ and $\rm L_{acc}$. Indeed, the [\ion{O}{i}]630 LVC and the HVC correlate with $\rm L_{acc}$ \citep[e.g., ][]{Rigliaco2013, Natta2014, Simon2016, Nisini2018, Gangi2022}. Moreover, \citet{Rigliaco2013} and \citet{Gangi2022} found out a slightly different slope in the $\rm L_{line} - L_{acc}$ relation of these two components, which might suggest that they originate from distinct mechanisms, but both are related to accretion processes.

Fig. \ref{fig:lline_vs_lacc} (top panel) shows the $\rm L_{line} - L_{acc}$ correlation for the [\ion{O}{i}]630 NLVC, compared with the distribution found in \citet{Gangi2022} for a sample of CTTs of the Taurus-Auriga association. Both the $\rm L_{line}$ and the $\rm L_{acc}$ are normalized to the stellar luminosity to exclude the correlation between $\rm L_{line} - L_{\star}$ and $\rm L_{acc} - L_{\star}$ \citep{Mendigutia2015}. We find that the NLVC luminosity correlates with $\rm L_{acc}$, in good agreement with \citet{Gangi2022}. A best linear regression fit in log scale yields:
\begin{eqnarray}
\rm log{\frac{L_{\ion{[O}{i]},NLVC}}{L_{\star}}} = 0.44 (\pm 0.12)\ log{ \frac{L_{acc}}{L_{\star}}} -4.21 (\pm 0.12).
\end{eqnarray}

In the bottom panel of Fig. \ref{fig:lline_vs_lacc} we report the $\rm L_{line} - L_{acc}$ distribution of the UV-$\rm H_2$ lines. To increase the statistics we also included $\rm L_{line}$ values from \citet{France2012} and the corresponding $\rm L_{acc}$ from \citet{Gangi2022}. We find a good correlation between the two quantities. More interestingly, the slope of the distribution is compatible, within errors, with that of [\ion{O}{i}]630 NLVC. A best linear regression fit of the total sample yields:

\begin{eqnarray}
\rm log{\frac{L_{H_2}}{L_{\star}}} = 0.64 (\pm 0.30)\ log{ \frac{L_{acc}}{L_{\star}}} -2.75 (\pm 0.36).
\end{eqnarray}

\subsection{[\ion{O}{I}]630 versus [\ion{S}{II}]406 line properties}\label{sec:OI_vs_SII}
\begin{figure*}
\begin{center}
\includegraphics[trim=20 170 0 160,width=0.85\columnwidth, angle=0]{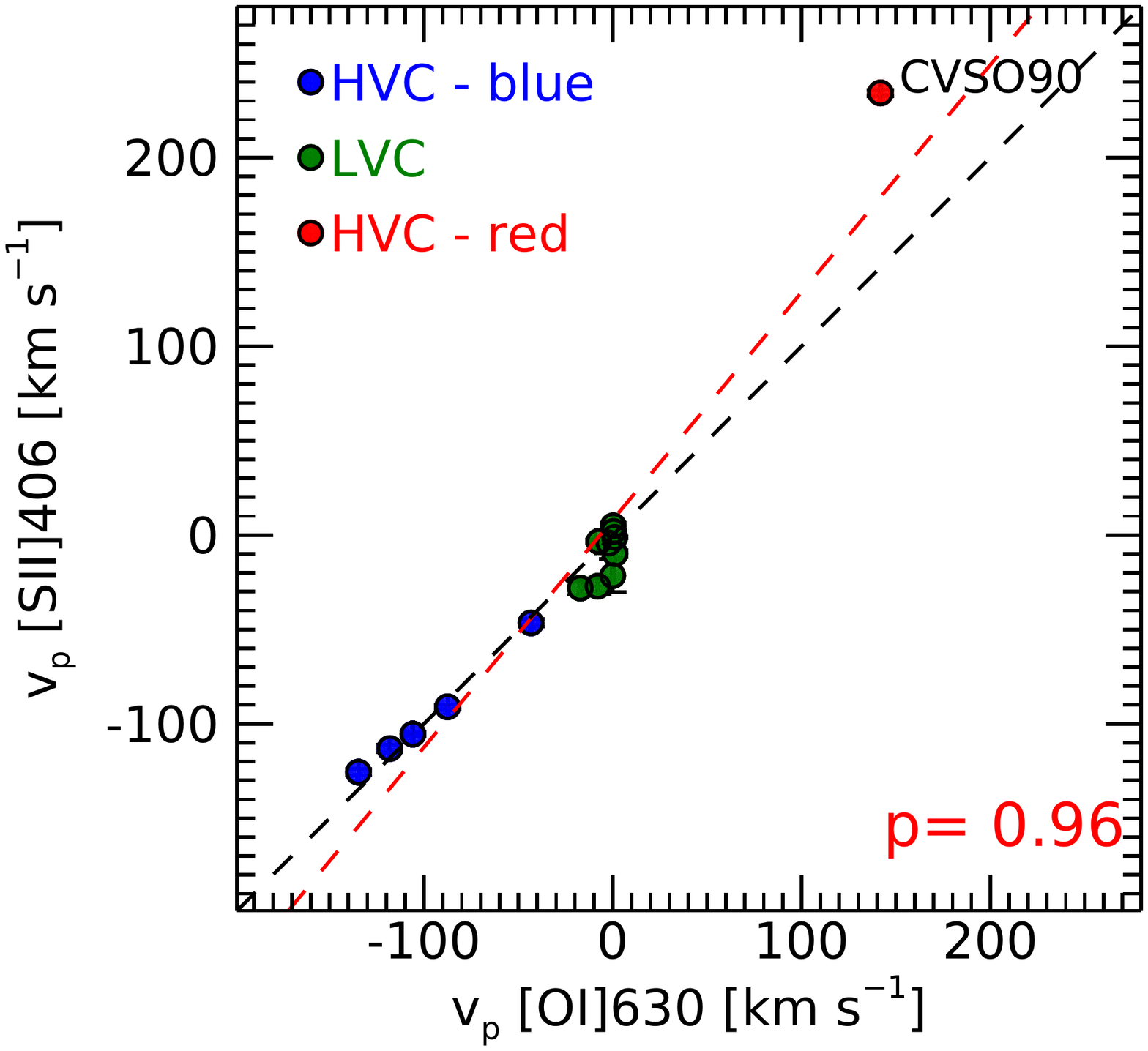}
\includegraphics[trim=20 170 0 160,width=0.85\columnwidth, angle=0]{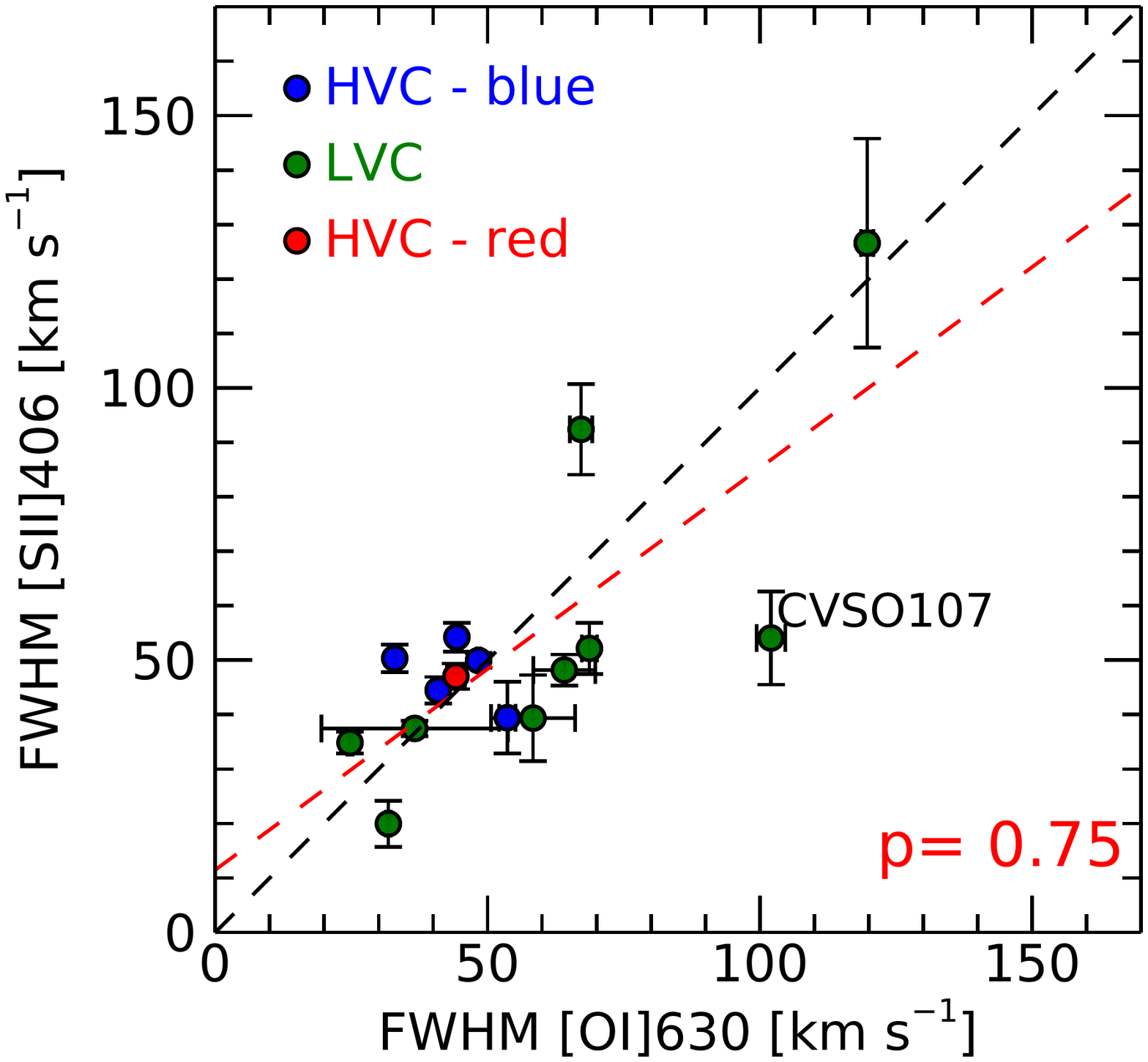}
\caption{\label{fig:compare_components} Peak velocities (left) and FWHMs (right) of individual [\ion{S}{ii}]406 and [\ion{O}{i}]630 components. Different colors refer to different velocity components, as labelled in the panels.}
\end{center}
\end{figure*}

Fig. \ref{fig:compare_components} shows the correlations between $\rm v_p$ and FWHMs of the [\ion{O}{i}]630 and [\ion{S}{ii}]406 obtained from the Gaussian decomposition of the profiles as explained in Sect. \ref{sec:data_analysis}. When it was possible to distinguish NLVC and BLVC in the [\ion{O}{i}]630 line but not in the [\ion{S}{ii}]406 line, we averaged the $\rm v_p$ and FWHMs of the NLVC and BLVC, using the peak intensities as weights. Overall, we find a good correlation between the two quantities but also note that few sources (i.e. CVSO58, CVSO90, CVSO104 and CVSO107), whose [\ion{O}{i}]630 LVC is centered at zero velocity, show slight blue-shifts in their [\ion{S}{ii}]406 LVCs (see Table \ref{tab:GaussDecompRes}). This might indicate different emitting regions for the two species.

\subsection{[\ion{O}{I}]630, [\ion{S}{II}]406 and UV-$\rm H_2$ versus UV emission}\label{sec:flux_vs_UV}
\begin{figure}
\begin{center} 
\includegraphics[trim=30 160 50 160,width=.7\columnwidth, angle=0]{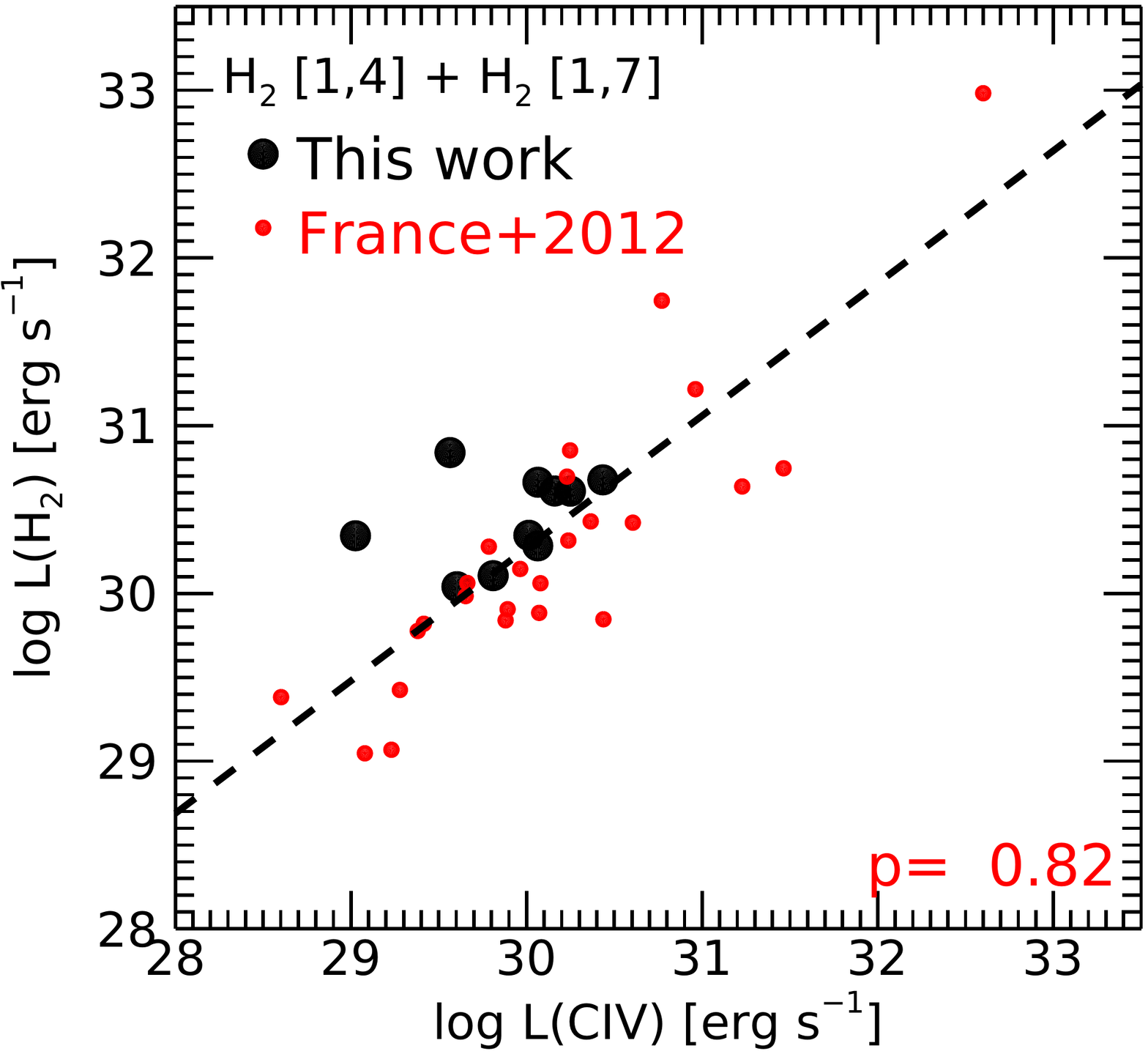}
\caption{\label{fig:CIV_H2_corr} Correlations between extinction-corrected line luminosities of $\rm H_2$ [1,4], [1,7] and \ion{C}{iv} at 155 nm (black points) compared with that of a CTTs sample reported in \citet{France2012} (red points). Linear fit to the total distribution is shown as dashed black line. The Pearson coefficient is also reported.}\end{center}
\end{figure}

\begin{figure*}
\begin{center} 
\includegraphics[trim=20 160 0 160,width=0.8\columnwidth, angle=0]{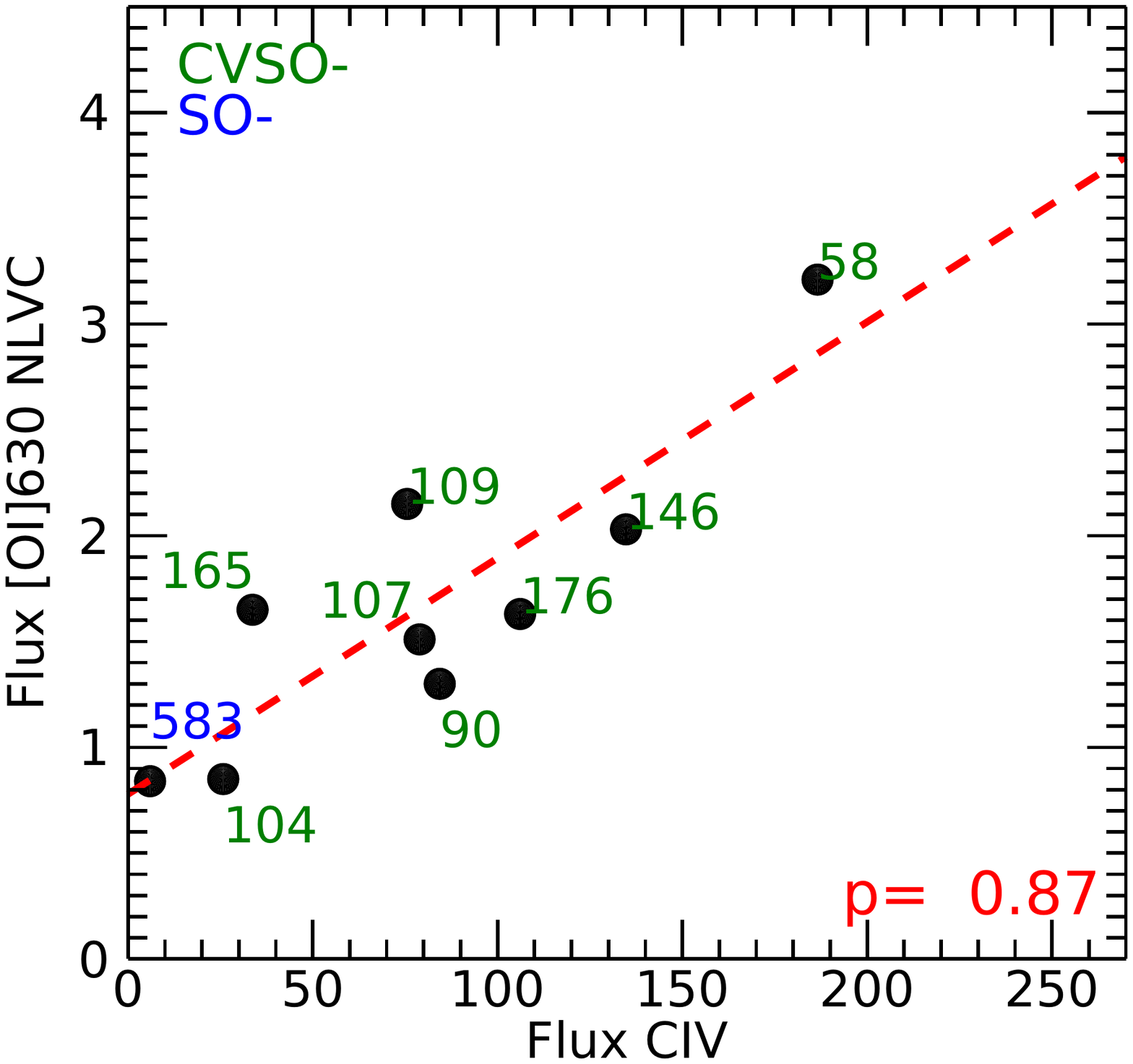}
\includegraphics[trim=20 160 0 160,width=0.8\columnwidth, angle=0]{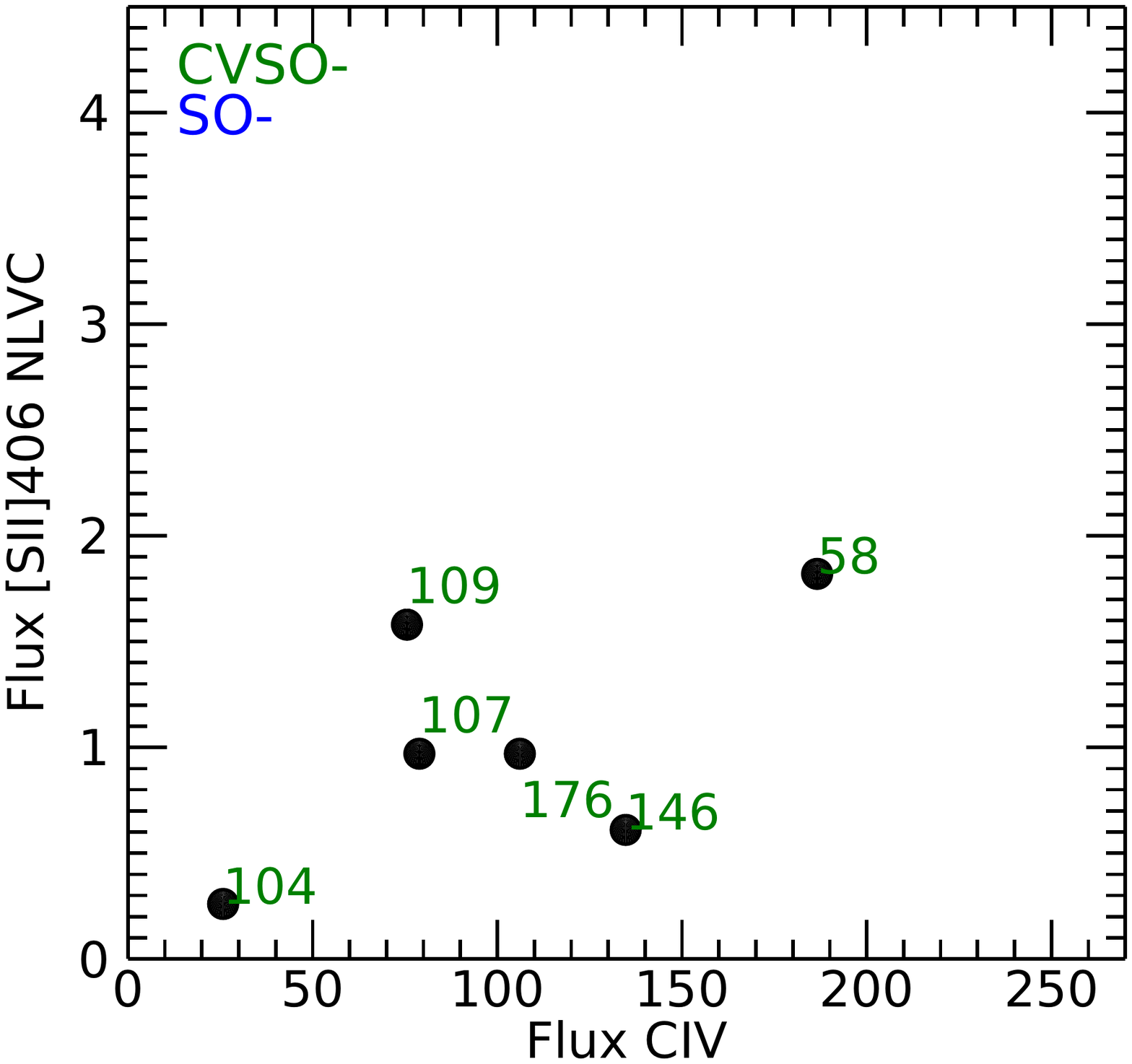}
\caption{\label{fig:CIV_correlation} Correlations between extinction-corrected line fluxes of [\ion{O}{i}]630 NLVC, [\ion{S}{ii}]406 NLVC and \ion{C}{iv} at 155 nm. Flux units are $\rm 10^{-15} ergs^{-1} cm^{-2}$. The SO518 source was excluded (see Sect. \ref{sec:kin_corr} for details).} Linear fit is shown as dashed red line.\end{center}
\end{figure*}

We look here at a possible direct dependency of the line fluxes from the UV photons by means of the \ion{C}{IV}-154.8, 155 nm doublet, which has been found to correlate with both the FUV continuum and the UV-$\rm H_2$ fluxes \citep[][]{France2012, France2014}. Fig. \ref{fig:CIV_H2_corr} correlates the luminosities of the UV-$\rm H_2$ with that of the \ion{C}{IV} doublet. To increase the statistics, we also included values from the sample of CTTs reported in \citet{France2012}. As expected, the UV-$\rm H_2$ luminosities are well correlated with those of the \ion{C}{IV} and follow the same trend as shown in \citet{France2012}. In Fig. \ref{fig:CIV_correlation}, we report the correlation between the line fluxes of the [\ion{O}{i}]630 and [\ion{S}{ii}]406 with those of the \ion{C}{IV} doublet. We find a good correlation for the [\ion{O}{i}]630 line, while no evidence of a trend is present for the [\ion{S}{ii}]406 line.

\section{Discussion}

\subsection{On the origin of the [\ion{O}{i}]630 NLVC}
As described in the introduction, the excitation mechanisms of the atomic oxygen component is still controversial. The similarities in the kinematical properties of [\ion{O}{i}] and [\ion{S}{ii}], together with the consistency of the line ratios with simple thermal models, have supported a thermal origin for the oxygen component \citep[see][]{Natta2014, Fang2018}, in contrast to suggestions of an origin from OH dissociation due to FUV photons \citep[][]{Rigliaco2013, Gorti2011} or from FUV pumping \citep[][]{Nemer2020}.

In Sect. \ref{sec:OI_vs_SII} we have found a kinematic correlation between the different components of [\ion{O}{i}]630 and [\ion{S}{ii}]406. However, for a few sources the [\ion{S}{ii}] LVC peak velocity is more blueshifted than the [\ion{O}{i}] LVC, suggesting a different emitting regions for the two species. This evidence, together with the correlations between the [\ion{O}{i}]630 NLVC and UV-$\rm H_2$ line transition properties (Sect. \ref{sec:kin_corr}), can significantly alter the above scenario.

In particular, the positive correlation between the FWHMs of the [\ion{O}{i}]630 NLVC and UV-$\rm H_2$ components presented in Sect. \ref{sec:kin_corr} may indicate that the two species are spatially connected. Under the assumption that the line broadening is dominated by Keplerian rotation, we have estimated in Sect. \ref{sec:em_sizes} average emitting sizes between $\sim0.5$ and $\sim3.5$ au for the [\ion{O}{i}]630 NLVC and between $\sim0.05$ and $\sim1$ au for the UV-$\rm H_2$, a result in good agreement with previous studies of other CTTs samples \citep[e.g.,][]{France2012, Fang2018, McGinnis2018, Gangi2020} and also consistent with the [\ion{O}{i}] spatial size of TW Hya as recently measured by \citet{Fang2023}. In Sect. \ref{sec:kin_corr}, we also point out that the peak velocities of the [\ion{O}{i}]630 NLVC are always centered at zero velocity. The same occurs for the UV-$\rm H_2$ lines, within the $\sim15$ $\rm km$ $\rm s^{-1}$ wavelength resolution accuracy of COS \citep{France2012}. However, we stress that the decomposition of the [\ion{O}{i}]630 LVC into a broad and narrow component, in contrast with a single Gaussian for the UV-$\rm H_2$ profile, may reflect the different spectral resolution and $SNR$ in the optical and UV spectral regions. For example, previous studies at higher $SNR$ have found that multi-components or double-picked morphology for the UV-$\rm H_2$ profiles can be reproduced with simple disk emission models \citep[e.g.,][]{Hoadley2015,Schneider2015}. With this caveat in mind, these two pieces of evidence suggest that the atomic and molecular components originate from regions that might overlap and that both species mostly have a bound disk origin. 

In Sect. \ref{sec:kin_corr}, we have also shown that the line luminosity of the [\ion{O}{i}]630 NLVC and UV-$\rm H_2$ components are correlated, with the latter a factor 30-40 higher than that of the [\ion{O}{i}]. Line luminosities are also correlated with the accretion luminosities (Sect. \ref{sec:lum_vs_acc}). The $\rm L_{line}-L_{acc}$ dependence found in this study  for the [\ion{O}{i}]630 NLVC line is consistent with that measured in other samples of YSOs \citep[][]{Rigliaco2013, Natta2014, Simon2016, Nisini2018, Gangi2022}. More interestingly, we have found that the $\rm L_{line}-L_{acc}$ relation also persists for the UV-$\rm H_2$ species and, within the limit of the low statistics of this study, it shows the same dependence as that of [\ion{O}{i}]630 NLVC. The UV-$\rm H_2$ $\rm L_{line}-L_{acc}$ correlation can be naturally explained by the fact that the $\rm H_2$ species are photo-excited by Ly$\alpha$ photons which, in turn, originate from the accretion processes \citep[e.g.,][]{Hoadley2015, Arula2023}. These results suggest a non-negligible role of the FUV photons in exciting the [\ion{O}{i}]630 NLVC for the sample analysed in this work. 

 This latter conclusion is further investigated by looking at a possible link between line fluxes and FUV continuum. In Sect. \ref{sec:flux_vs_UV} we found that the [\ion{O}{i}]630 NLVC fluxes are well correlated with those of the \ion{C}{IV}-154.8, 155 nm doublet, a well known tracer of UV continuum, while the [\ion{S}{ii}]406 NLVC does not show any particular trend. This different behaviour supports the scenario in which the [\ion{O}{i}]630 and [\ion{S}{ii}]406 NLVCs could have different dominant excitation mechanisms.

In conclusion, the above evidence point towards a framework where the [\ion{O}{i}]630 NLVC and UV-$\rm H_2$ have a common disk origin with a partially overlapping region. An important contribution for the excitation of the [\ion{O}{i}]630 NLVC component might be compatible with a non-thermal process due to the action of FUV photons. 

\subsection{Ly$\rm \alpha$ and FUV-continuum penetration depths}

\begin{figure}
\begin{center}
\includegraphics[trim=0 0 0 0,width=1.\columnwidth, angle=0]{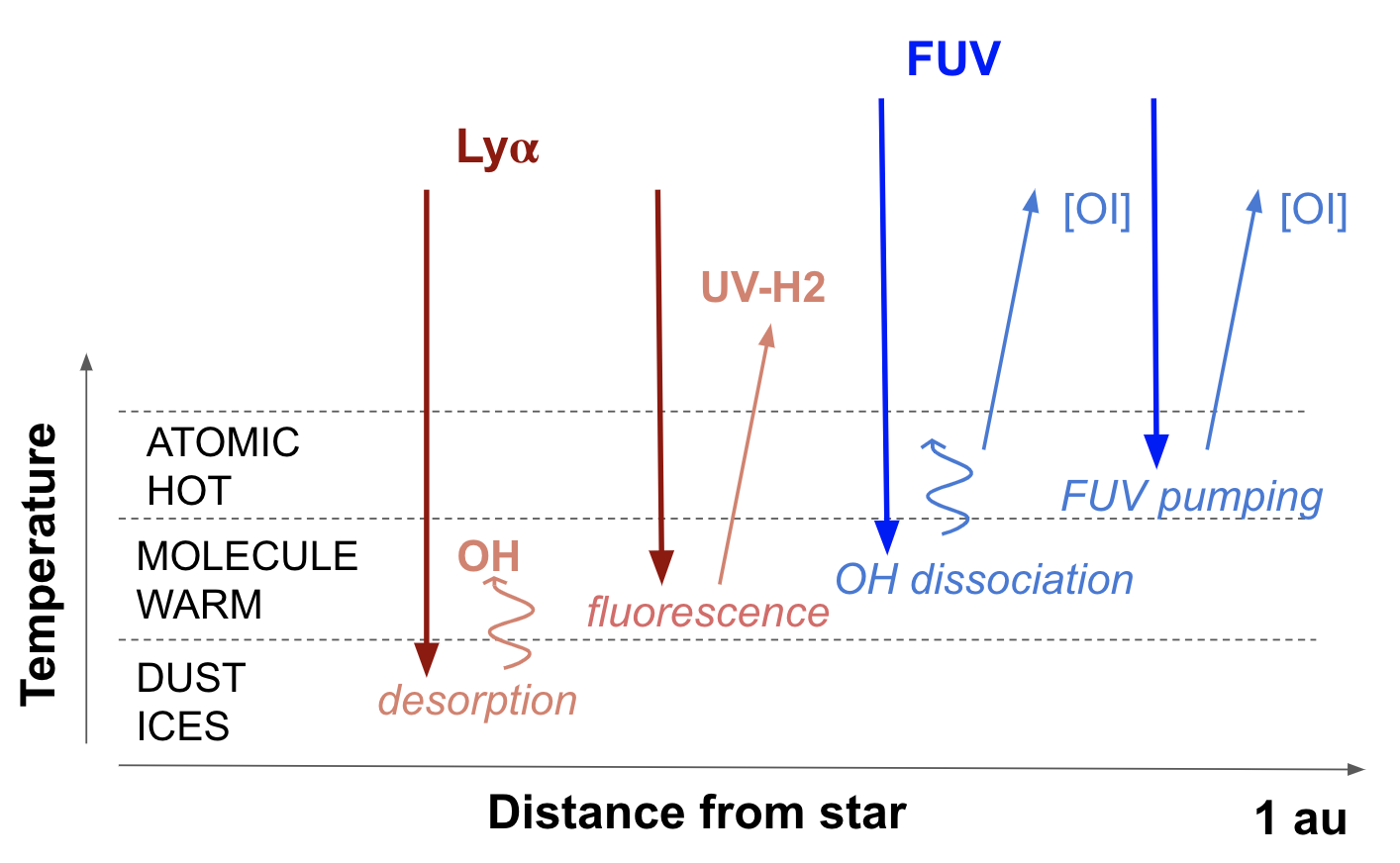}
\end{center}
\begin{center}\caption{\label{fig:schema_test} Schematic view of the disk vertical stratification and penetration depths of Ly$\rm \alpha$ and FUV-continuum photons.}
\end{center}
\end{figure}

We discuss here whether the above conclusion might be compatible with the capabilities of the exciting radiation to impact on different disk layers. 
Fig. \ref{fig:schema_test} shows a schematic view of the disk vertical stratification and penetration depths of Ly$\rm \alpha$ and FUV-continuum. The dust tends to settle in the inner layer, where the temperature is low enough to freeze out molecules on the surface of dust grains. In contrast, in the upper layer the temperature is sufficiently warm to prevent freezing and molecular species can form, while close to the disk surface only the atomic species can survive. In this framework, the FUV continuum dominates in the atomic layer while the Ly$\rm \alpha$ dominates in the molecular surface \citep{Bethell2011}. Such a behaviour is particularly true in the inner disk regions ($\sim$1 au), where the bulk of UV-$\rm H_2$ and [\ion{O}{i}]630 NLVC emission arise. Models have shown that, together with the fluorescence of $\rm H_2$, the Ly$\rm \alpha$ radiation is also responsible for the photo-desorption of different molecular species, such as $\rm H_2O$ and $\rm OH$, trapped in the ice layer of dusts \citep{Fogel2011}. The molecular layer can be then enriched with OH species, which in turn might be efficiently photo-dissociated by the action of FUV photons reaching the warm disk layer. In addition, the penetration efficiency of UV photons has been found to increase as the dust settling increases \citep{Dullemond2004}, further facilitating the OH photodissociation.

Therefore, the correlation between the UV-$\rm H_2$ and [\ion{O}{i}]630 NLVC we found and the scenario in which the UV field could give a contribution to the oxygen atomic excitation appears to be consistent with the picture depicted here.
However, we cannot discriminate between the OH photo-dissociation mechanism and the FUV pumping as major UV contributor to the \ion{O}{} excitation, neither evaluate their relative contribution. Thermo-chemical models that simultaneously take into account the molecular and atomic species are necessary for that purpose. Similarly, and due to the low statistics of our data, we cannot point out any trend associated with a contribution from external UV fields. In particular, we note that the $\rm \sigma-$Orionis sources of our sample might have an important external contribution due to their proximity to OB stars. Complete samples in different star forming regions are necessary to shed light on this latter point.

\subsection{Comparison with near-IR $H_2$ emission}
In a previous work \citep{Gangi2020}, we have looked at the link between the [\ion{O}{i}]630 NLVC and the ro-vibrational hydrogen 1-2 S(1) transition at 2.12 $\mu$m (hereafter NIR-$\rm H_2$) with a homogeneous and simultaneous dataset of 36 CTTs of the Taurus-Auriga star forming region, in the framework of the GHOsT project \citep{Alcala2021}. We found: (i) a strong kinematical link between the two species, both in terms of peak velocity and FWHM and (ii) a weak correlation between the luminosities, with the NIR-$\rm H_2$ showing on average lower luminosity than that of the [\ion{O}{i}]630. The kinematic link was interpreted in the framework where the neutral atomic and near-infrared molecular components are part of the same disk wind, with the latter tracing a more external region with sizes between 2 and 20 au.

The weak correlation between the line luminosities was later examined by \citet{Rab2022}, who interpreted the trend in terms of photo-evaporative disk-wind coupled with thermo-chemical models. These latter included the UV pumping and the OH photo-dissociation for the calculation of the NLTE level populations of the \ion{O}{} species and the pumping induced by dust grains for the NIR-$\rm H_2$. Their models are able to populate reasonably well the distribution of the [\ion{O}{i}]630 luminosity, confirming the non-negligible role of UV photons. On the other hand, the NIR-$\rm H_2$ line luminosities are found to be under predicted up to an order of magnitude. This might suggest that the dissociation mechanisms for this component could also depend on a detailed inclusion of the dynamical processes linked to the disk structure and wind radial extension.

Finally, in a subsequent work \citep{Gangi2022}, we found no significant correlation between the luminosity of the NIR-$\rm H_2$ line and $\rm L_{acc}$. Again, this was explained in the framework where the NIR-$\rm H_2$ dissociation is strongly affected by several processes not necessarily connected with accretion.

\begin{figure*}
\begin{center}
\includegraphics[trim=0 60 0 0,width=1.7\columnwidth, angle=0]{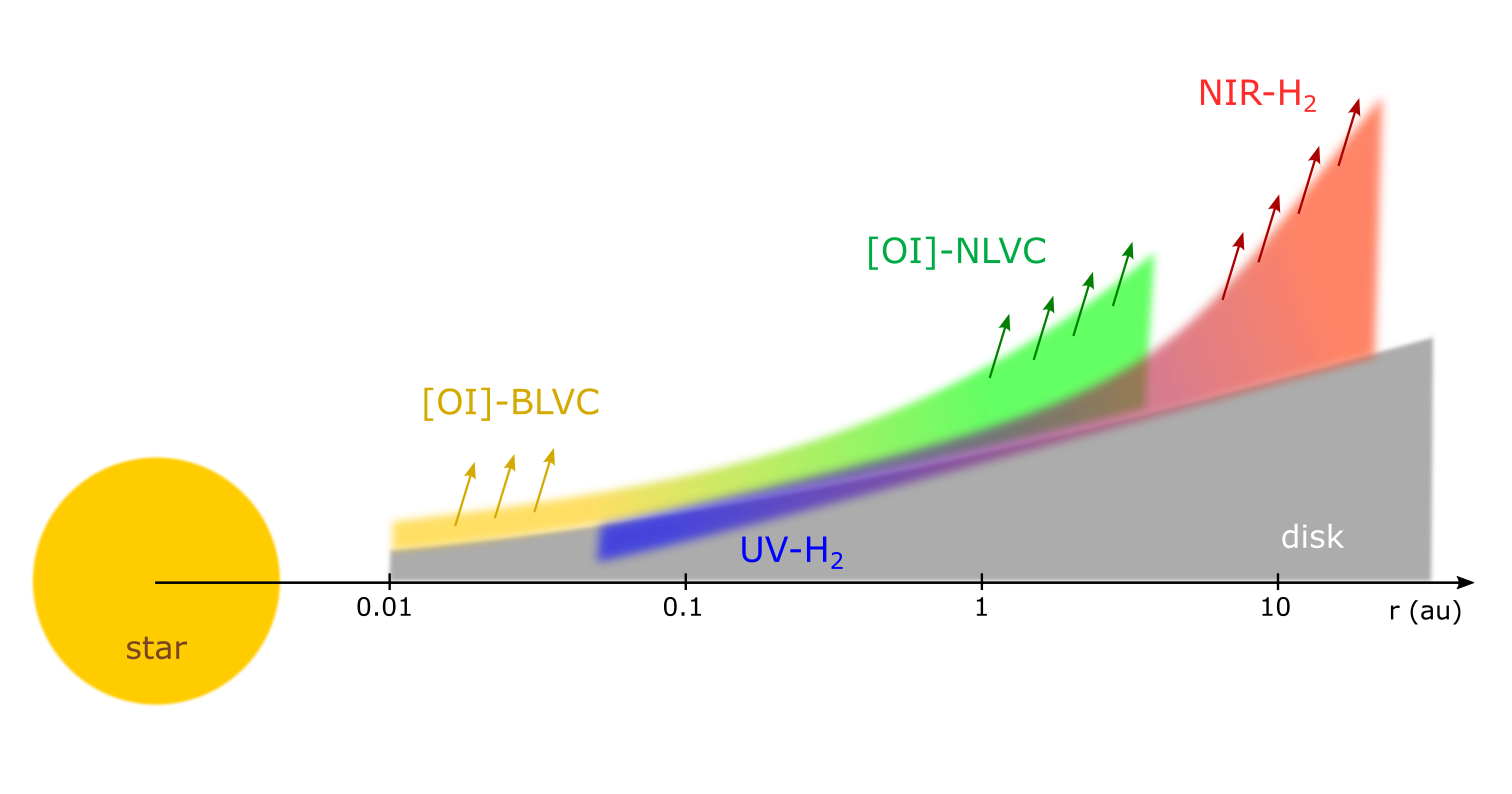}
\end{center}
\begin{center}\caption{\label{fig:final_sketch} Sketch (not in scale) showing a possible distribution of the atomic [\ion{O}{i}] and molecular $\rm H_2$ components in protoplanetary disk. Horizontal axis is in log scale.}
\end{center}
\end{figure*}

\subsection{A schematic view of the \ion{O}{i} and $\rm H_2$ protoplanetary disk components}
If the scenario discussed in the previous subsection can be extended to the sample analyzed in this work, we can trace a global picture of the atomic [\ion{O}{i}]630 NLVC, molecular UV-$\rm H_2$ and NIR-$\rm H_2$ emissions as schematized in Fig. \ref{fig:final_sketch}. 

First, we found evidence from line widths that all the three species are spatially connected, with the UV-$\rm H_2$ and the NIR-$\rm H_2$ regions partially overlapping to that of the [OI]630 LVC. The UV-$\rm H_2$ component traces the inner ($\lesssim$ 1 au) molecular protoplanetary disk, where stellar FUV photons are able to reach the warm disk and photo-excite the species. The NIR-$\rm H_2$ is the more extended component, with sizes up to $\sim$20 au, but it is not present in the inner regions ($\lesssim$ 1-2 au). 

On the other hand, the atomic \ion{O}{} component is emitted from a more internal region, which reaches up fractions of a few au. The differences we found in the FWHMs between the BLVC and NLVC are consistent with the classical interpretation of the BLVC as tracer of the innermost atomic protoplanetary disk region, in contrast to the corresponding NLVC \citep[e.g.,][]{Ercolano2017}. In this framework, we stress that the spatial overlapping with the UV-$\rm H_2$ region supports the scenario in which the FUV photons provide a substantial contribution to the \ion{O}{} line excitation.

Finally, we conclude that, although the blue-shifted [\ion{O}{i}]630 LVC and NIR-$\rm H_2$ lines show that they can be both associated with slow winds, a contribution to the line emission from gas bound in the disk cannot be ruled out. For this purpose we highlight the importance of an accurate wavelength calibration and a precise correction for the stellar radial velocity \citep[e.g.,][]{Campbell2023}.

\section{Summary and conclusion}
In the framework of the PENELLOPE and ULLYSES projects we have presented a study of the atomic and molecular protoplanetary disk components in a sample of 11 CTTs of the Orion OB1 and $\rm \sigma$-Orionis associations. We analyzed contemporaneous high-resolution optical and ultraviolet ESPRESSO at VLT, UVES at VLT and HST-COS spectra, focusing on the five brightest optical forbidden lines and on the fluorescent ultraviolet $\rm H_2$ [1,4], [1,7] progressions. We applied a Gaussian decomposition of the line profiles to separate different kinematic components. The optical forbidden lines were deconvolved into components at different velocities, in line with previous high-resolution studies of CTTs. On the contrary, the $\rm H_2$ [1,4], [1,7] line progressions, detected in all sources, were fit as a single Gaussian. We then focused on the comparison between the [\ion{O}{i}]630 narrow-low-velocity component (NLVC, |$\rm v_p$| < 30 $\rm km$ $\rm s^{-1}$, FWHM < 40 $\rm km$ $\rm s^{-1}$) and the $\rm H_2$ line progressions, with the aim of investigating the \ion{O}{} excitation mechanisms. The main results of our study are summarized below.

We found a strong kinematic link between the [\ion{O}{i}]630 NLVC and the UV-$\rm H_2$. In particular, the FWHMs of the two components are tightly correlated while the peak velocities are consistent with zero velocity.

Assuming that the line width is dominated by Keplerian broadening, we measured the average radius of the disk region where the emission originates. We found that the [\ion{O}{i}]630 NLVC originates from radii in between 0.5 au and 3.5 au and UV-$\rm H_2$ from 0.05 au and 1 au.

We found a strong correlation between the line luminosities ($\rm L_{line}$) of the [\ion{O}{i}]630 NLVC and UV-$\rm H_2$, as well as between $\rm L_{line}$ and the accretion luminosities ($\rm L_{acc}$). In particular, the $\rm L_{line}$-$\rm L_{acc}$ relations for the two species have a similar slope of the linear fit.

The UV-$\rm H_2$ $\rm L_{line}$ correlates with the luminosity of \ion{C}{iv}-154.8, 155 doublet, in agreement with the results of \citet{France2012}. In addition, we also found that such correlation is also valid for the [\ion{O}{i}]630 NLVC, while no correlation is found for the [\ion{S}{ii}]406 LVC.

We interpreted these results in terms of a common disk origin for the [\ion{O}{i}]630 NLVC and UV-$\rm H_2$ species, which partially overlap in space. We suggest a possible dominant role of the FUV photons in exciting the \ion{O}{i} species, at variance with a thermal origin. Finally, we proposed a global picture of the distributions of atomic and molecular species in protoplanetary disks including the near-infrared $\rm H_2$-2.12 $\mu$m emission properties from previous similar studies.

This work highlights the potential of contemporaneous wide-band high-resolution spectroscopy to provide insights on the complex physical processes in the environments around YSOs. At the same time we stress that a future observational strategy probing statistically significant samples in different star forming regions, coupled with advanced thermo-chemical models, is essential to confirm the inferred gas distribution and the interaction between the exciting radiation and the inner disk structure. This can be achieved thanks to the effort of a large community like that involved in the PENELLOPE at VLT and ULLYSES at HST projects. 

\begin{acknowledgements}
This work has been supported by the projects PRIN-INAF 2019 "Spectroscopically Tracing the Disk Dispersal Evolution (STRADE)", PRIN-INAF 2019 "Planetary systems at young ages (PLATEA)" and by the Large Grant INAF 2022 YODA (YSOs Outflows, Disks and Accretion: towards a global framework for the evolution of planet forming systems). This work also benefited from discussions with the ODYSSEUS team (HST AR-16129), \url{https://sites.bu.edu/odysseus/}. Funded by the European Union under the European Union’s Horizon Europe Research \& Innovation Programme 101039452 (WANDA) and 716155 (SACCRED). Views and opinions expressed are however those of the author(s) only and do not necessarily reflect those of the European Union or the European Research Council. Neither the European Union nor the granting authority can be held responsible for them.      
\end{acknowledgements}

\clearpage

\begin{appendix}\section{Additional tables and images}
\begin{figure*}[!h]
\includegraphics[trim=20 210 0 70,width=.6\columnwidth, angle=0]{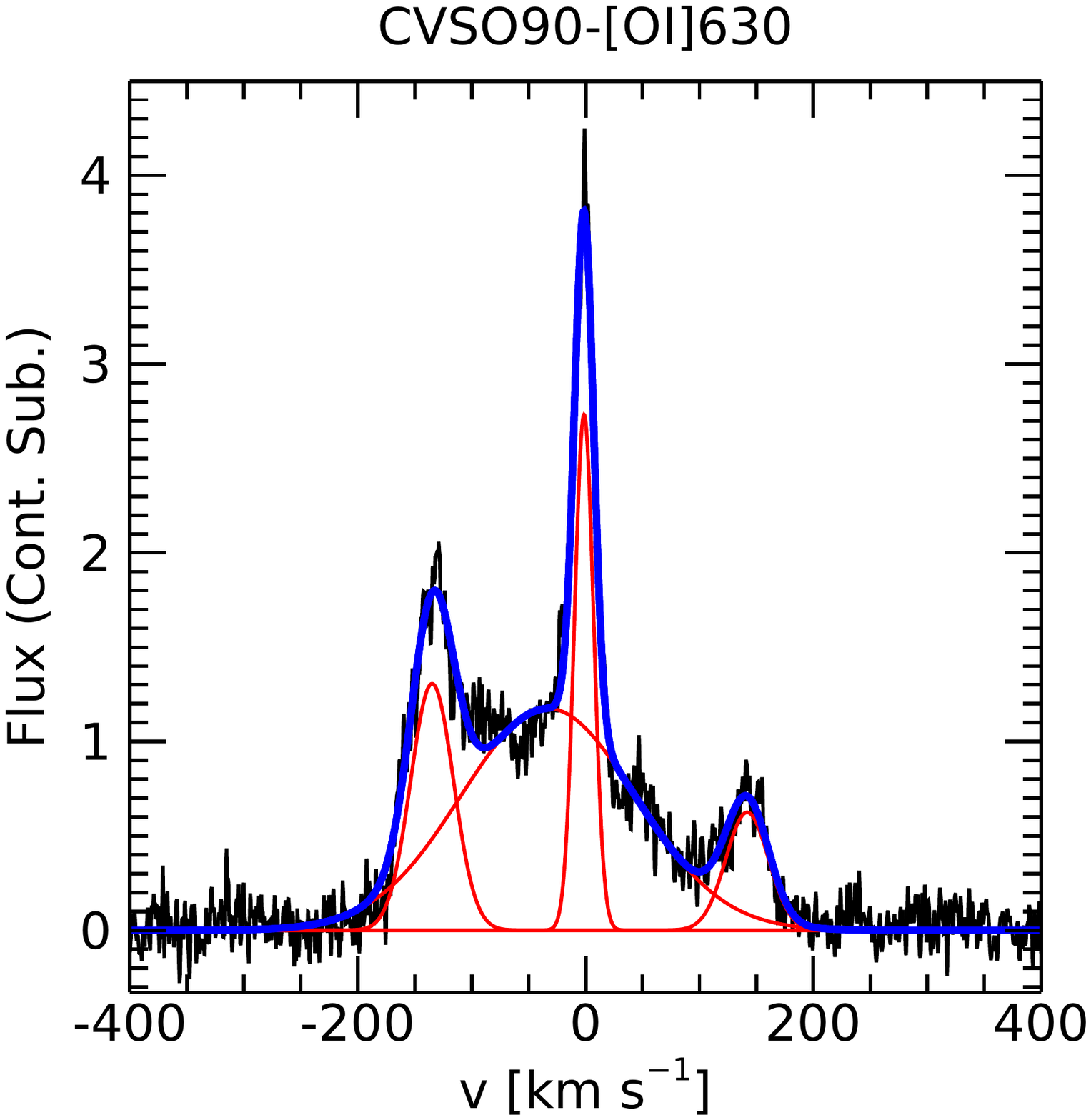}
\includegraphics[trim=20 210 0 70,width=.6\columnwidth, angle=0]{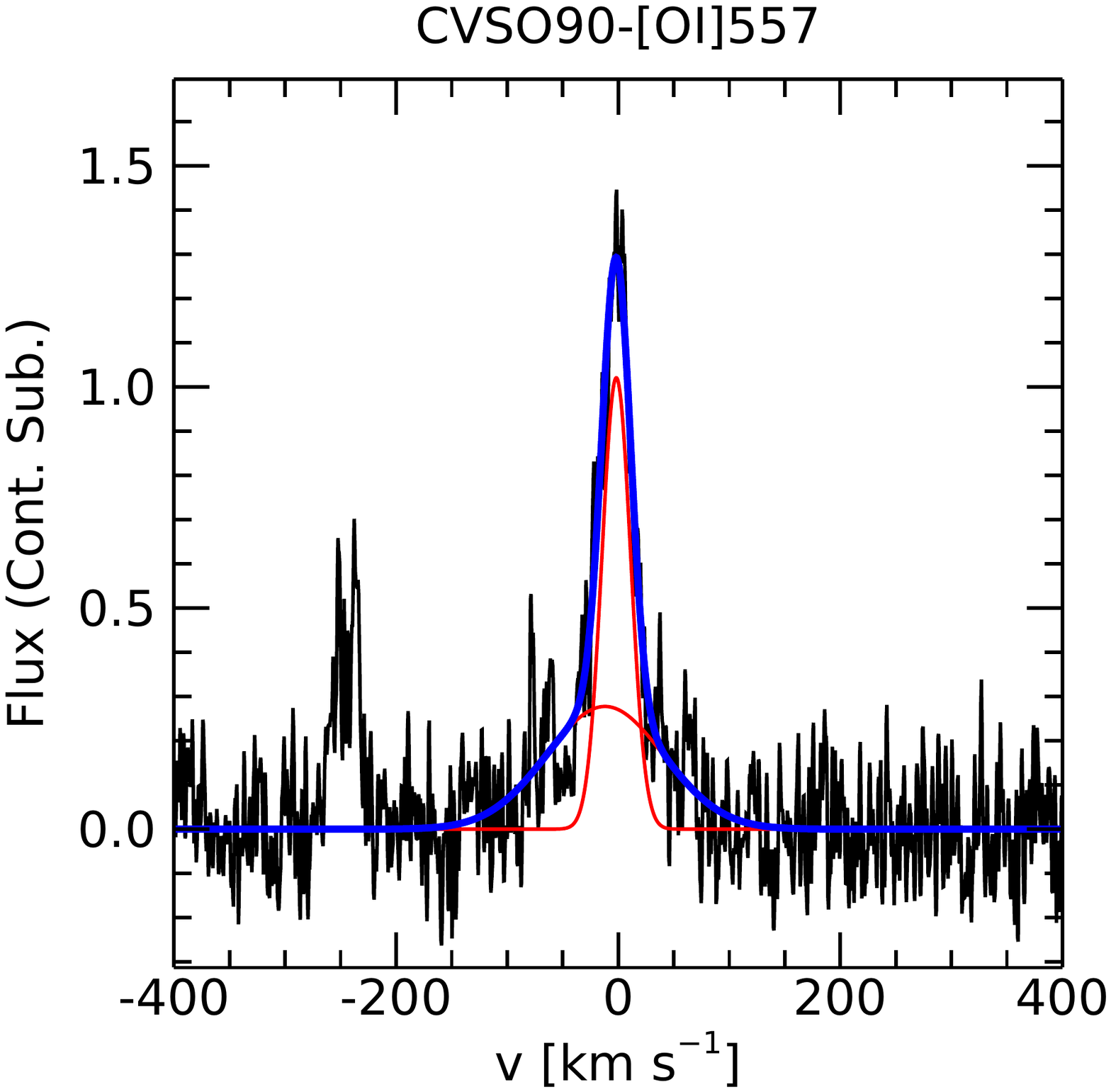}
\includegraphics[trim=20 210 0 70,width=.6\columnwidth, angle=0]{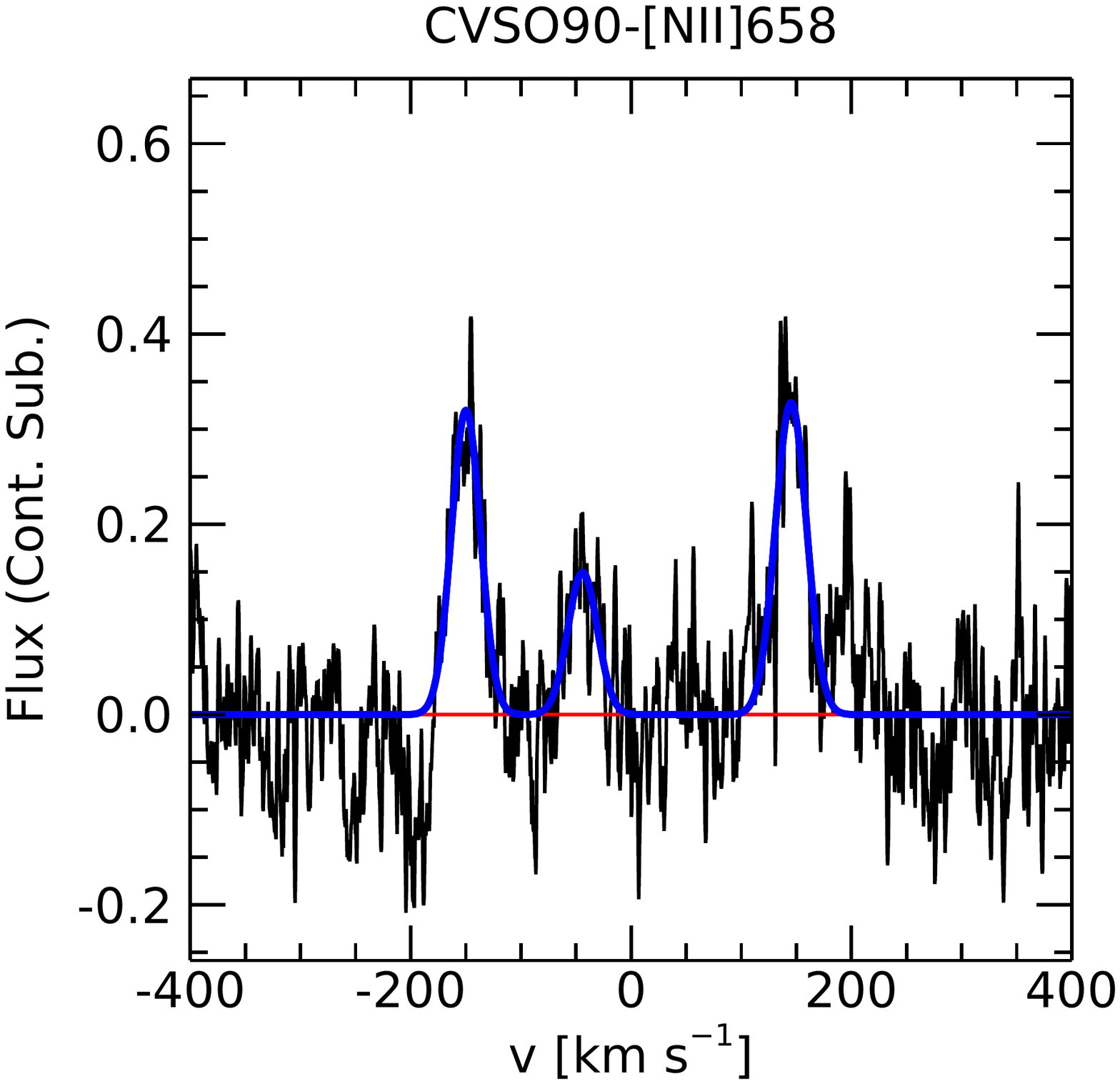}
\includegraphics[trim=20 210 0 70,width=.6\columnwidth, angle=0]{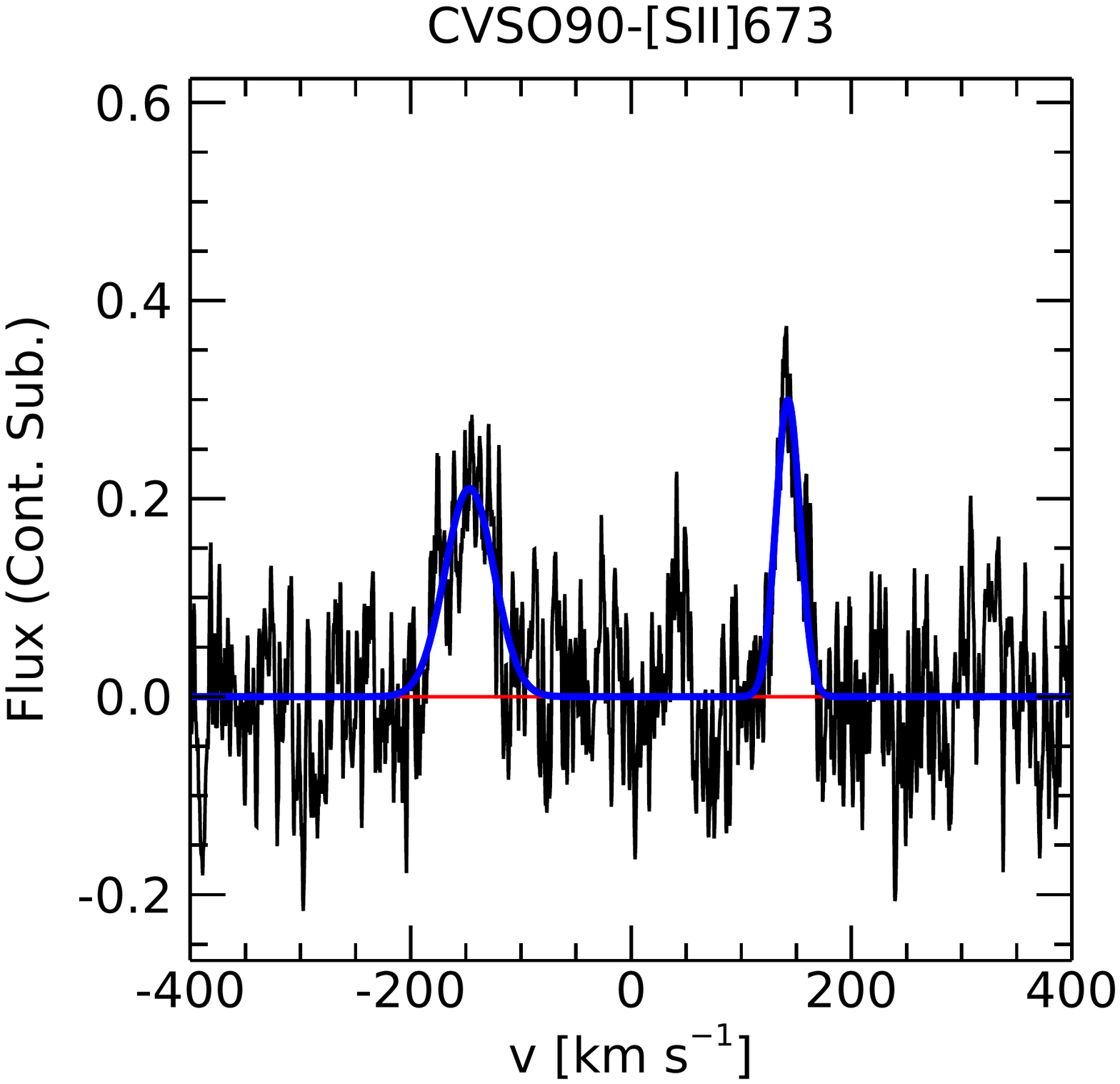}
\includegraphics[trim=20 210 0 70,width=.6\columnwidth, angle=0]{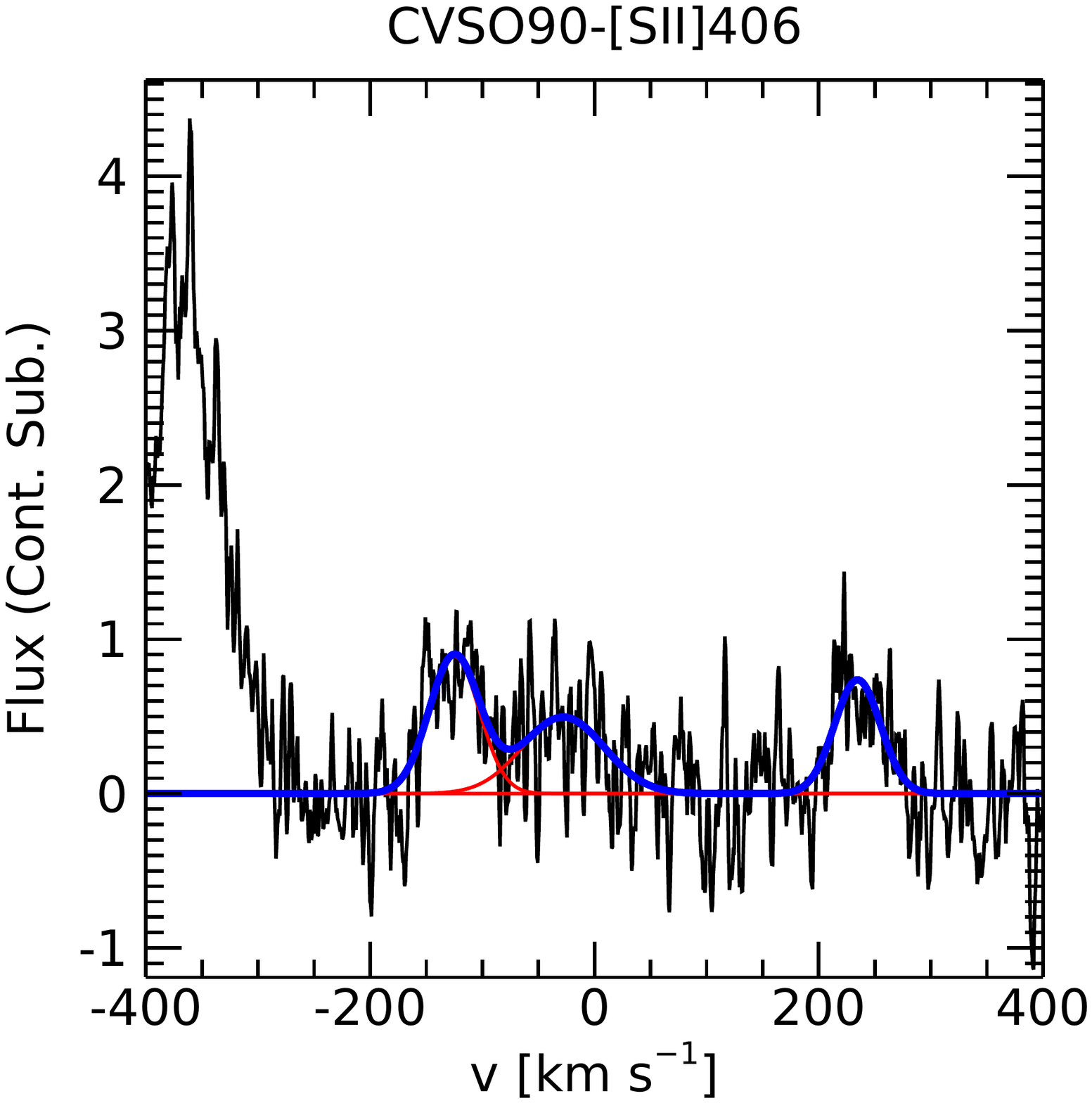}
\end{figure*}

\begin{figure*}[!h]
\includegraphics[trim=20 210 0 70,width=.6\columnwidth, angle=0]{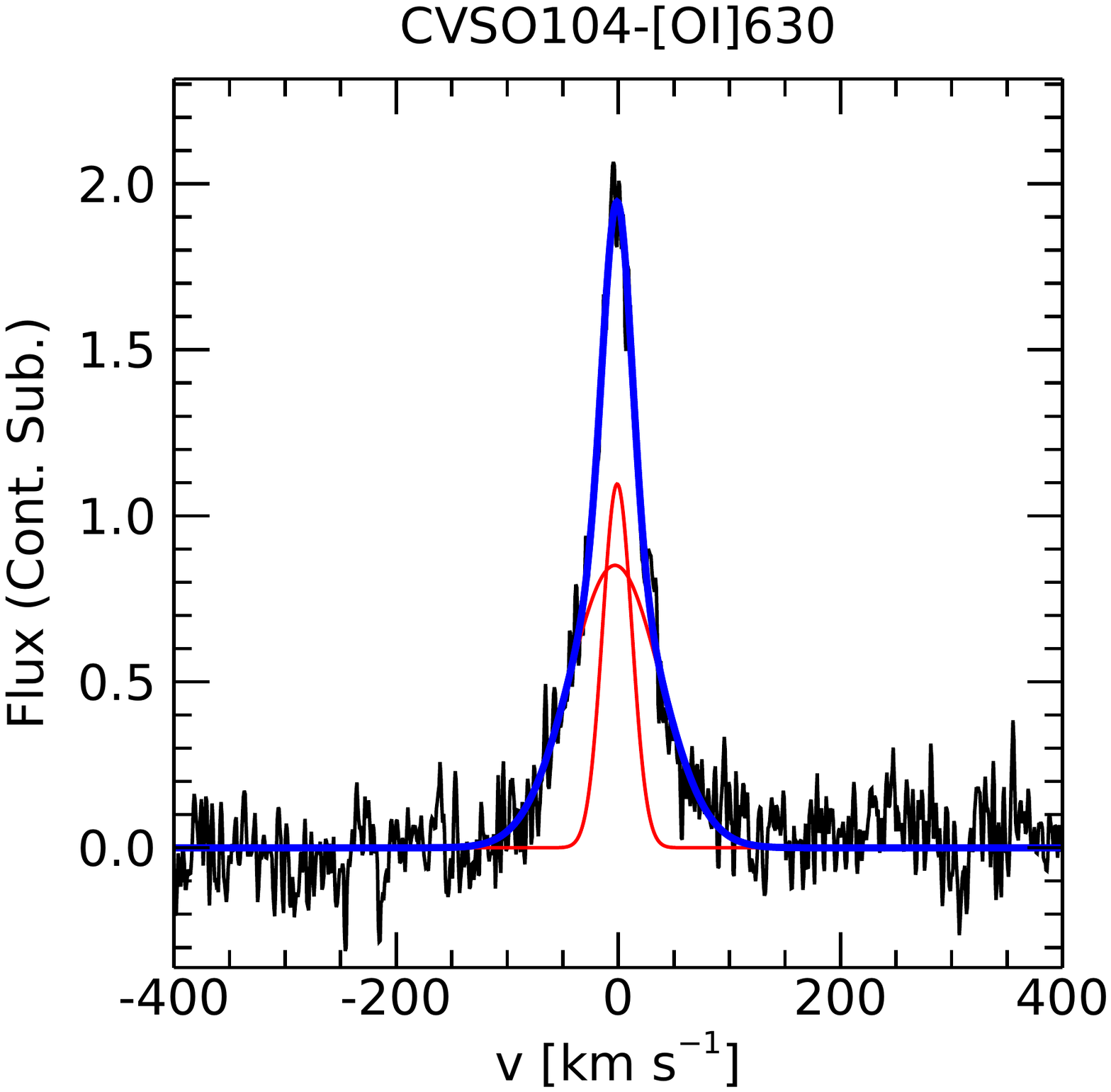}
\includegraphics[trim=20 210 0 70,width=.6\columnwidth, angle=0]{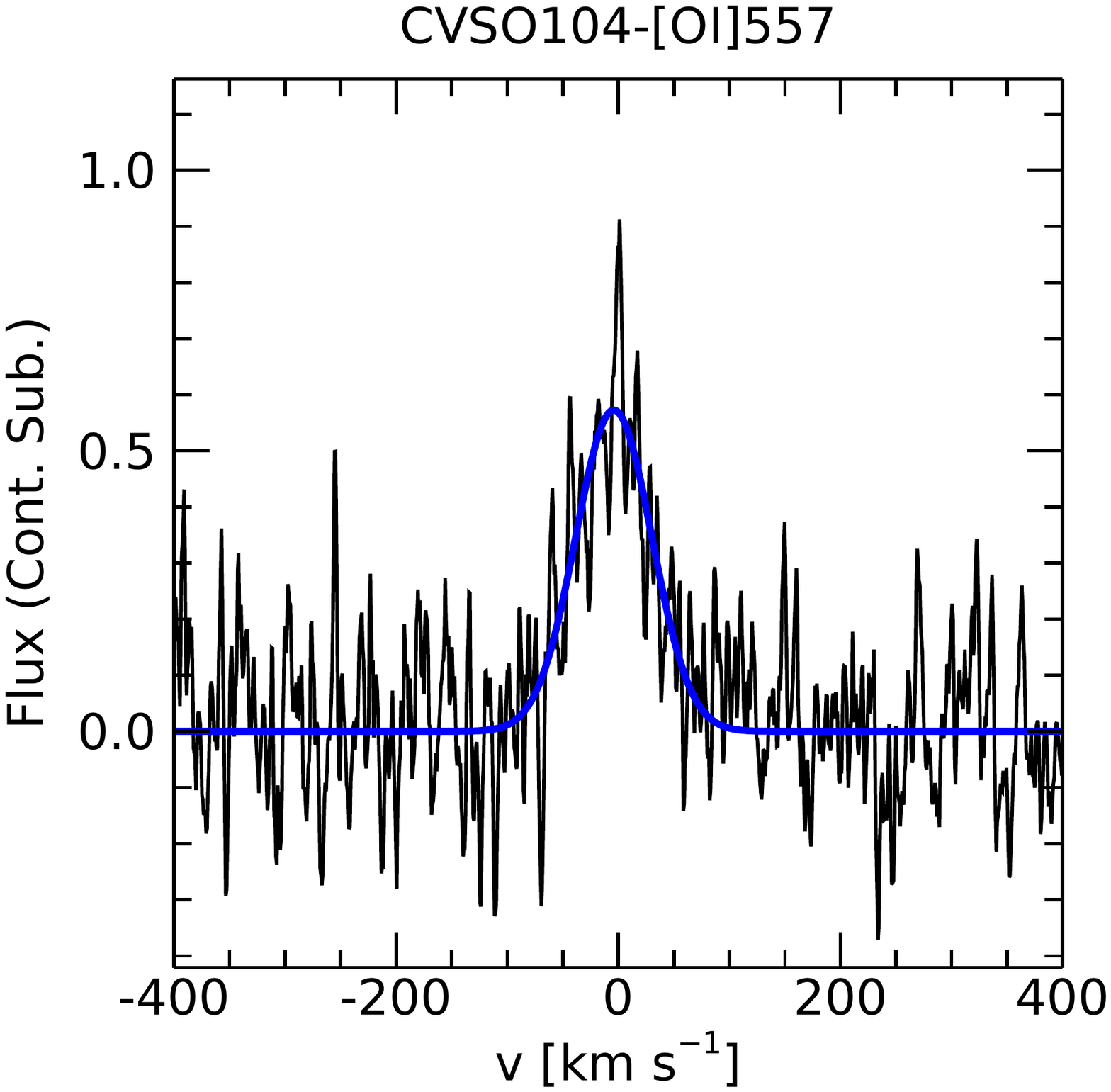}
\includegraphics[trim=20 210 0 70,width=.6\columnwidth, angle=0]{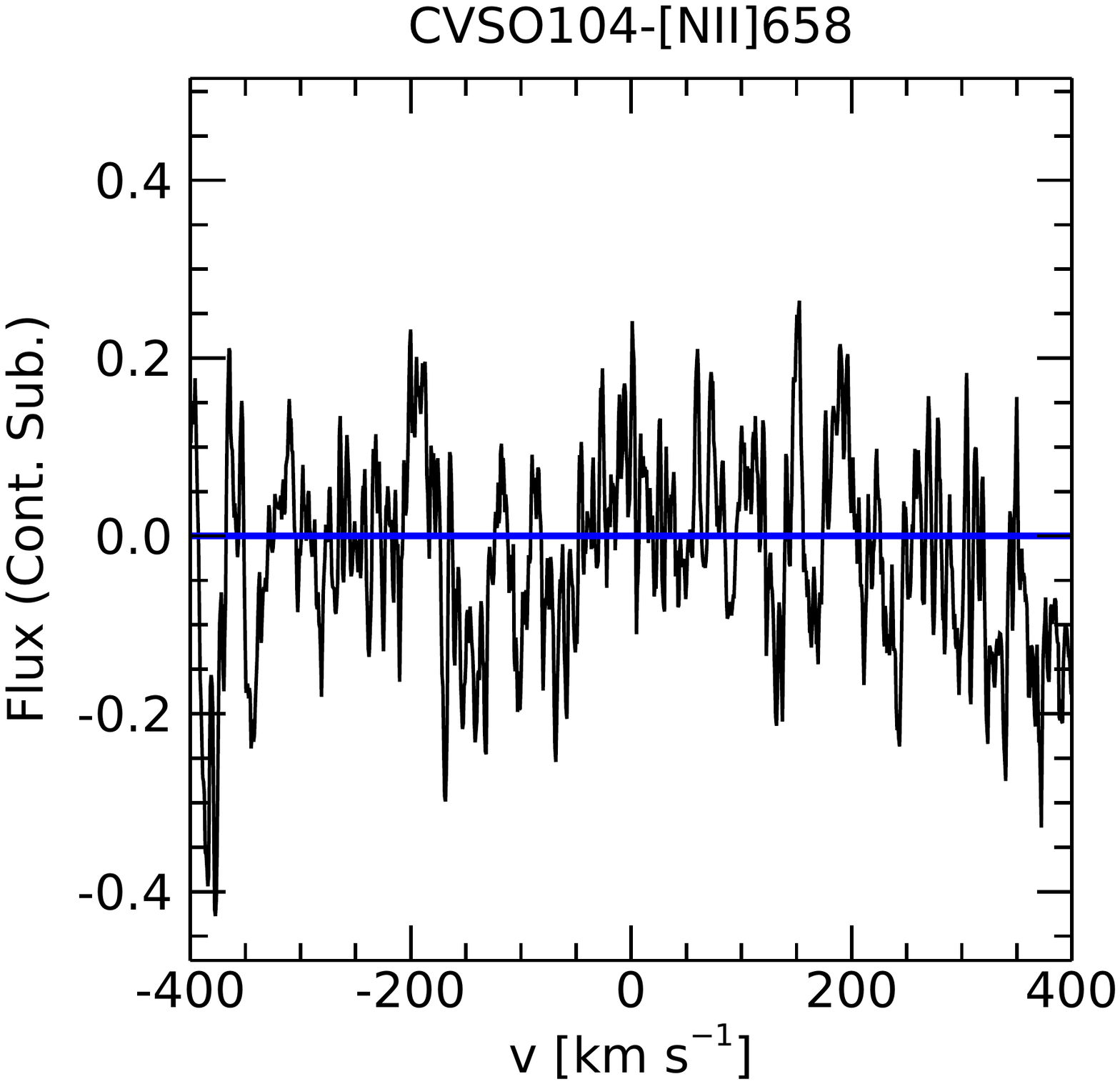}
\includegraphics[trim=20 210 0 70,width=.6\columnwidth, angle=0]{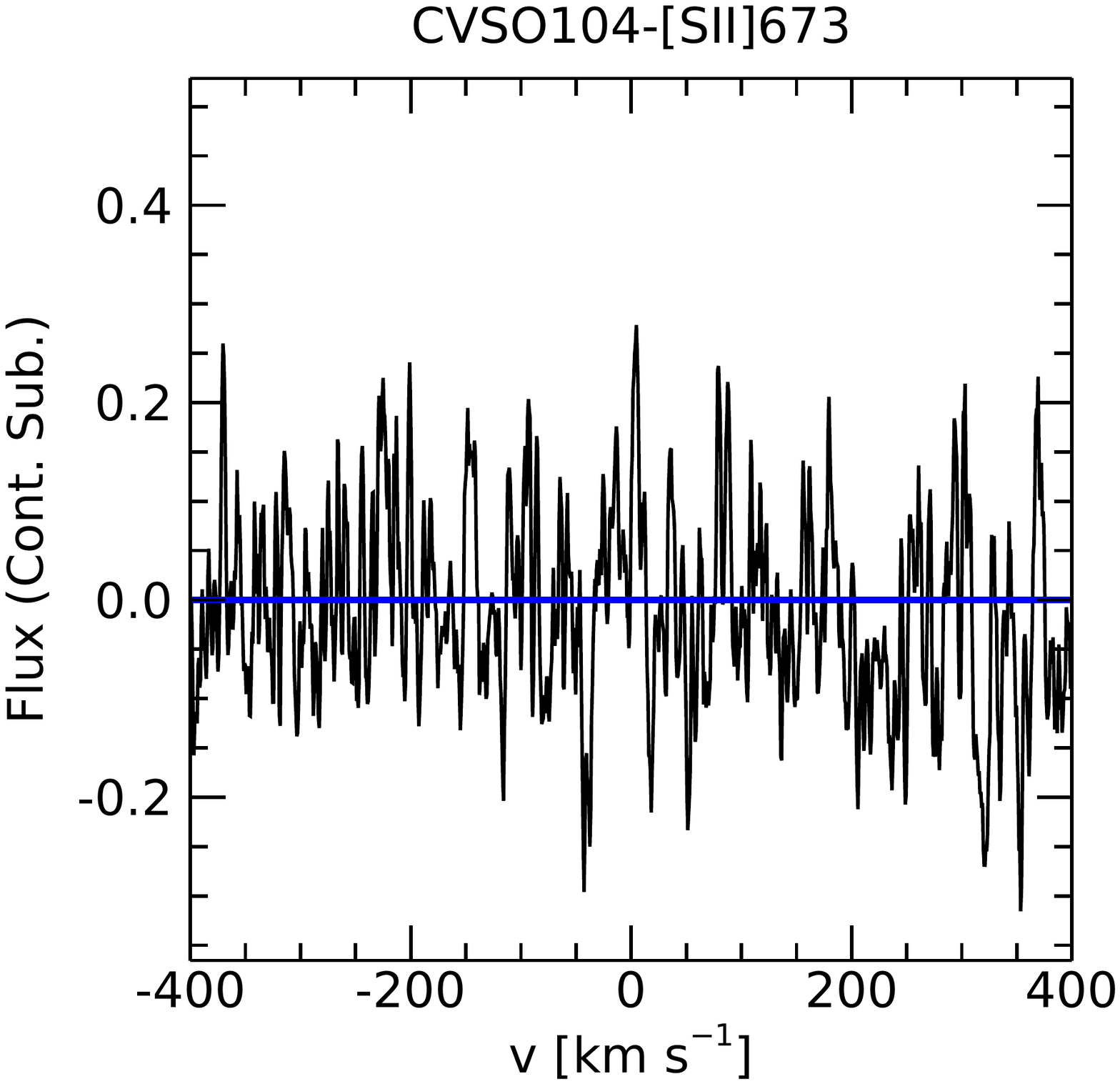}
\includegraphics[trim=20 210 0 70,width=.6\columnwidth, angle=0]{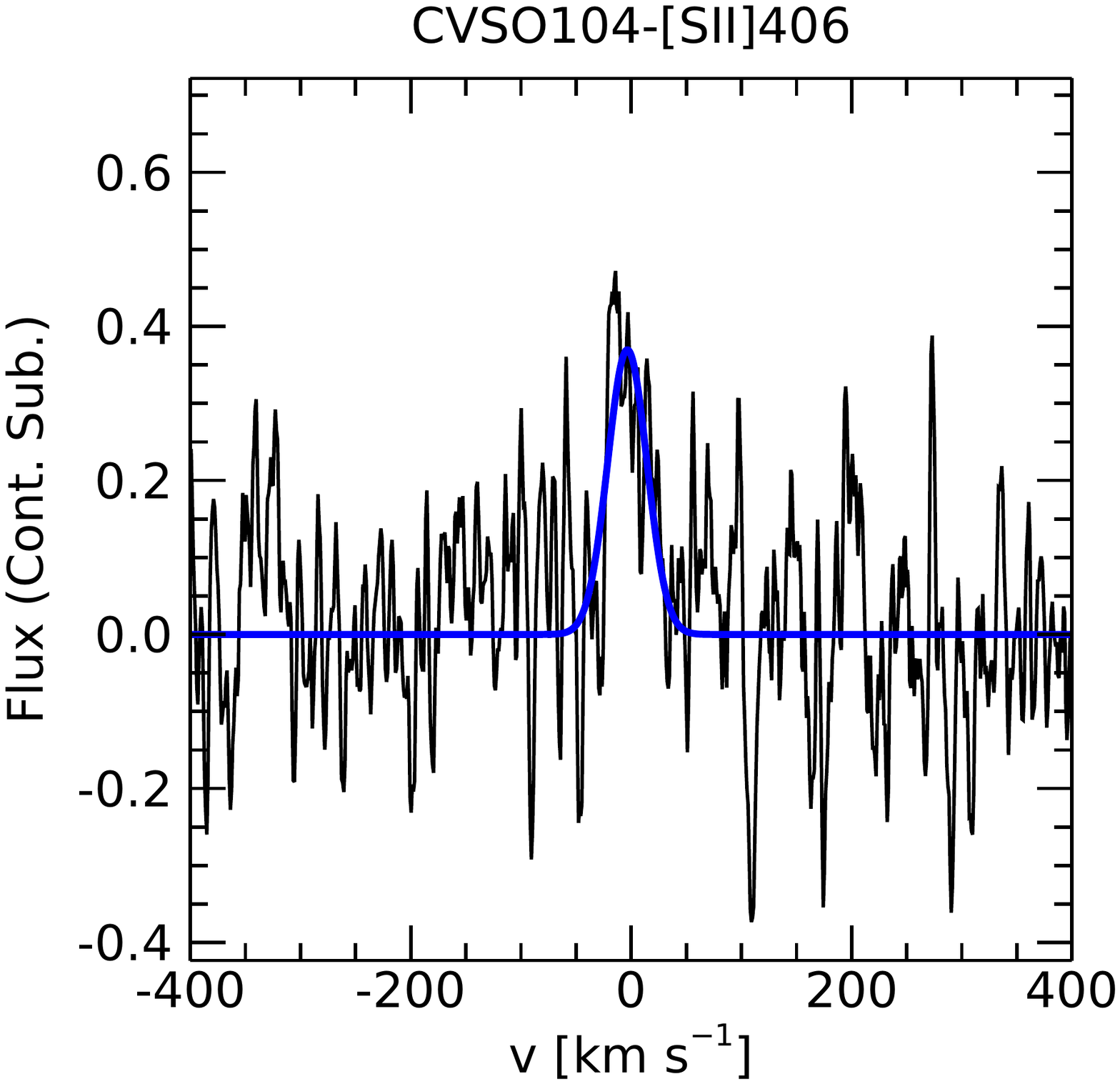}
\begin{center}\caption{\label{fig:complete_sample}Continuum-subtracted optical forbidden line profiles (black). In blue we plot the fit to the profile, obtained by adding single or multiple Gaussians (red lines). Flux units are $\rm 10^{-15} ergs^{-1} cm^{-2} \AA^{-1}$. For each panel we indicate the target name and the line diagnostics.}\end{center}
\end{figure*}

\begin{figure*}[!h]
\includegraphics[trim=20 210 0 70,width=.63\columnwidth, angle=0]{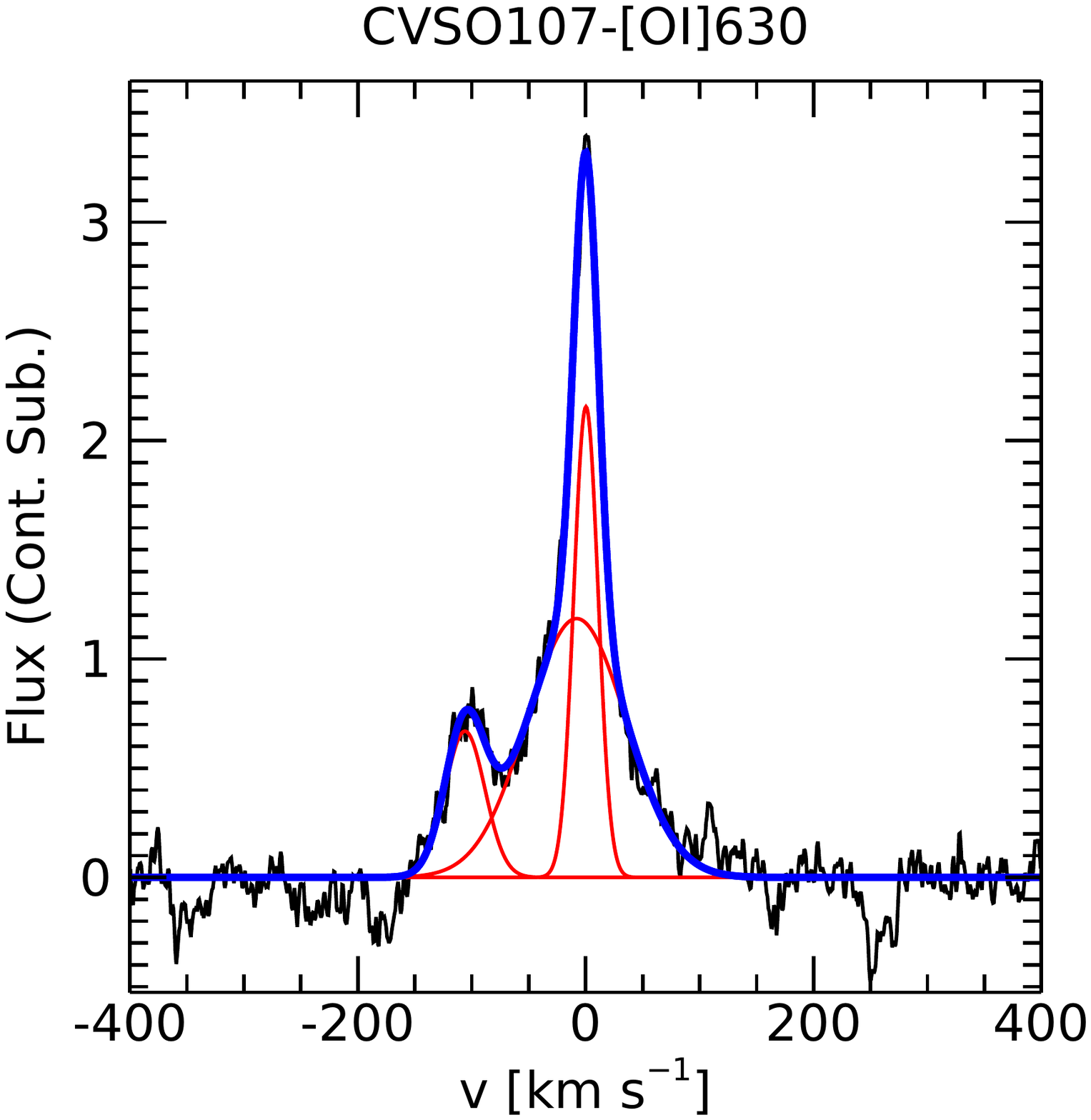}
\includegraphics[trim=20 210 0 70,width=.63\columnwidth, angle=0]{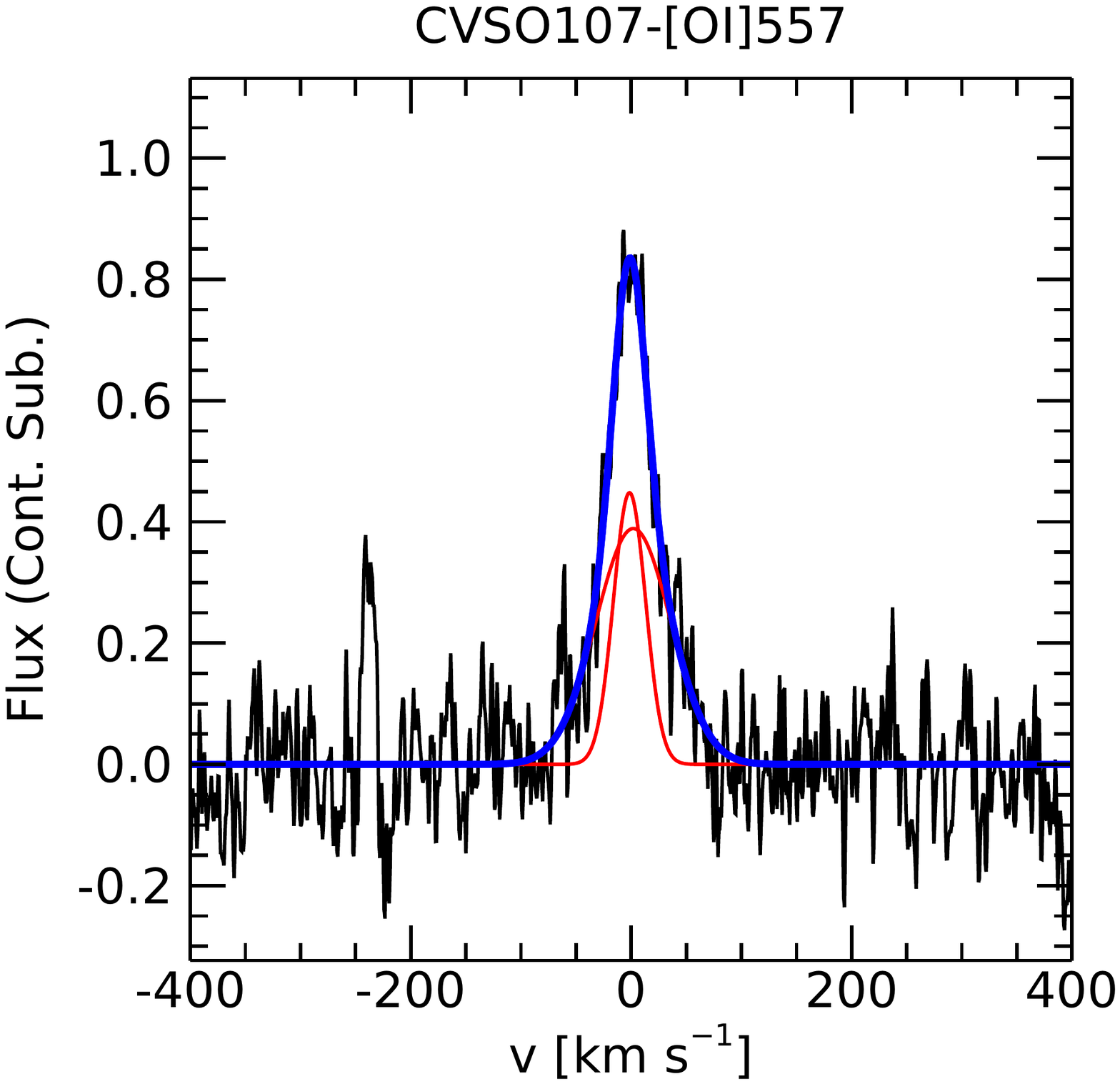}
\includegraphics[trim=20 210 0 70,width=.63\columnwidth, angle=0]{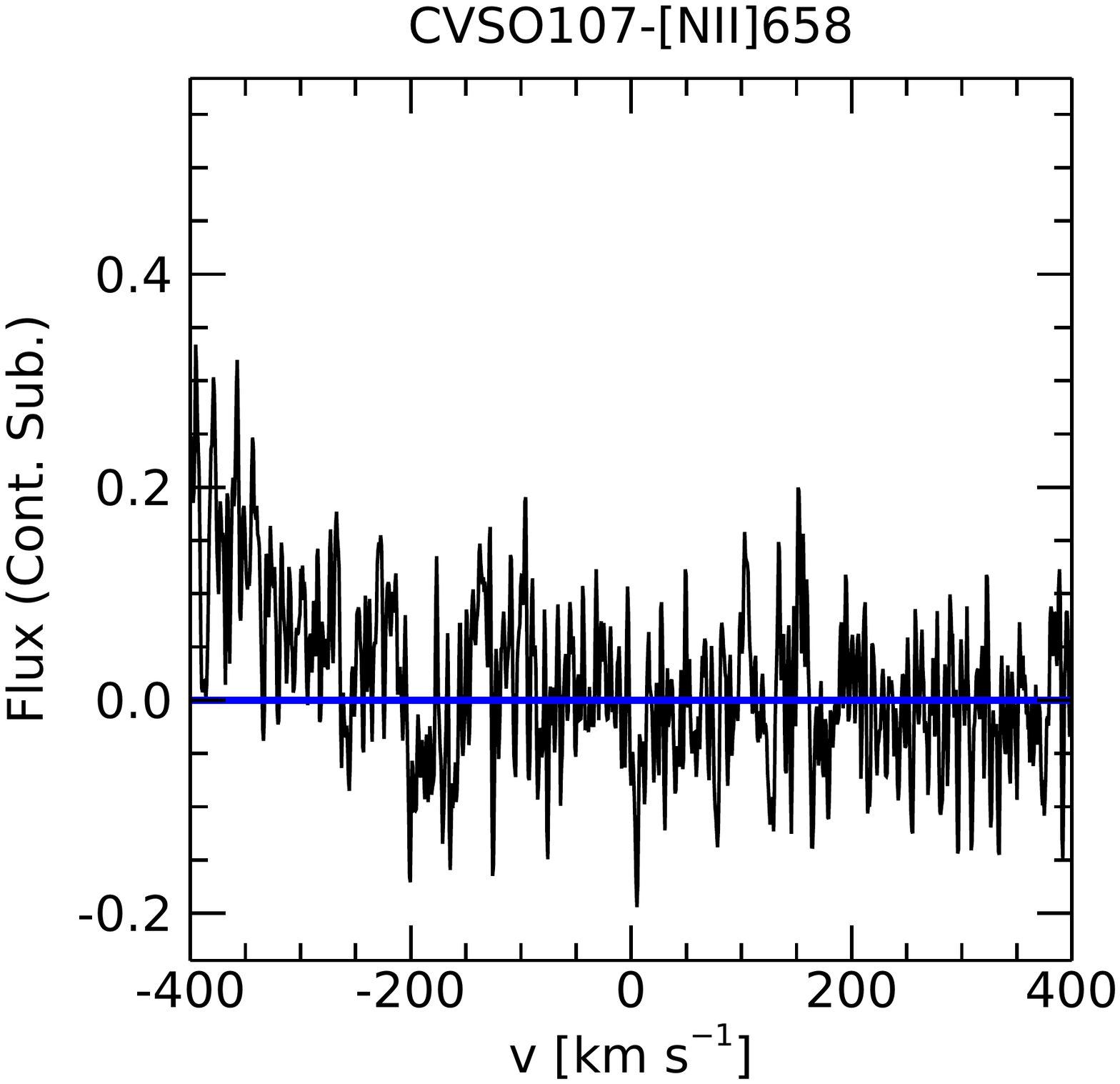}
\includegraphics[trim=20 210 0 70,width=.63\columnwidth, angle=0]{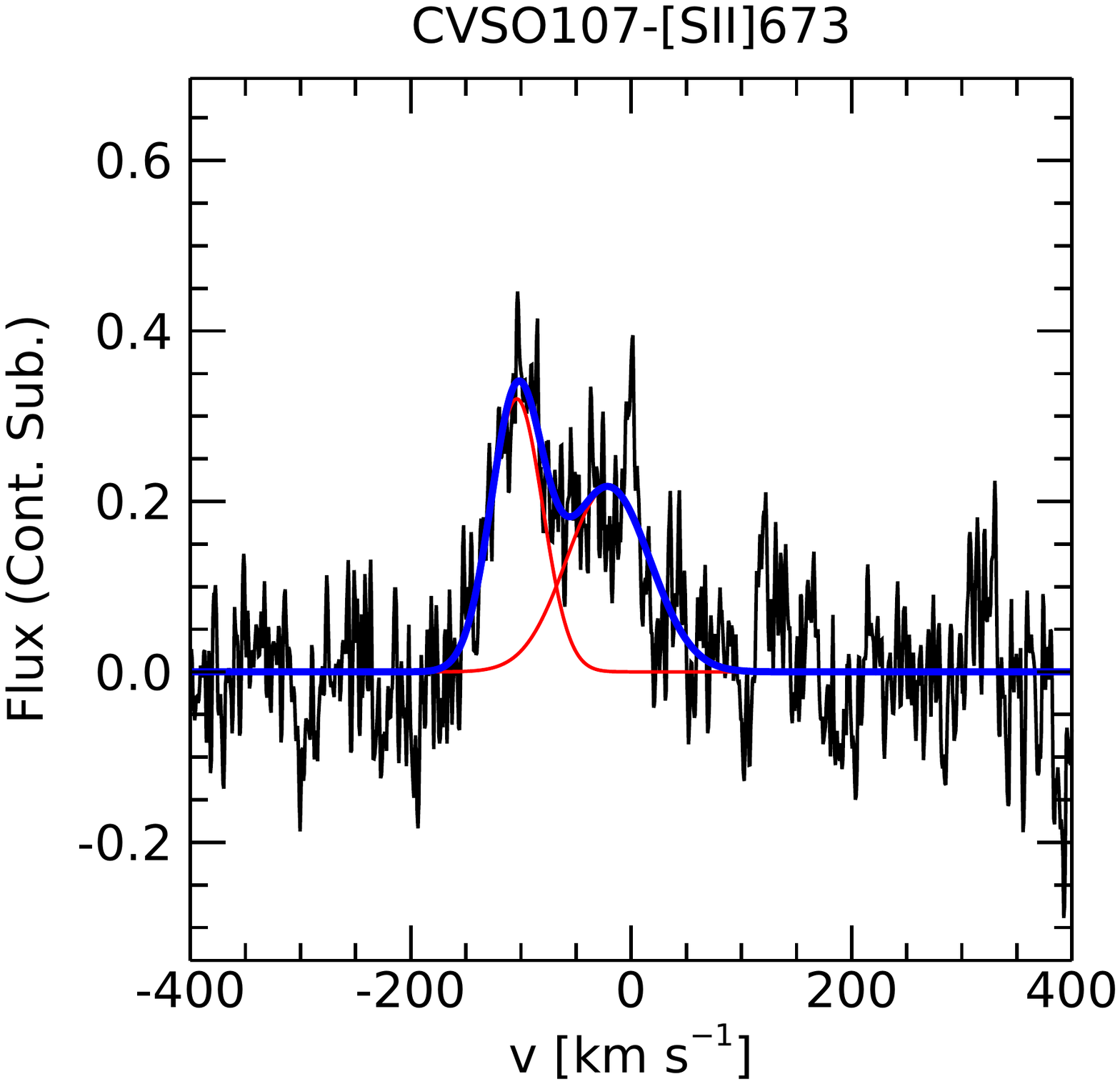}
\includegraphics[trim=20 210 0 70,width=.63\columnwidth, angle=0]{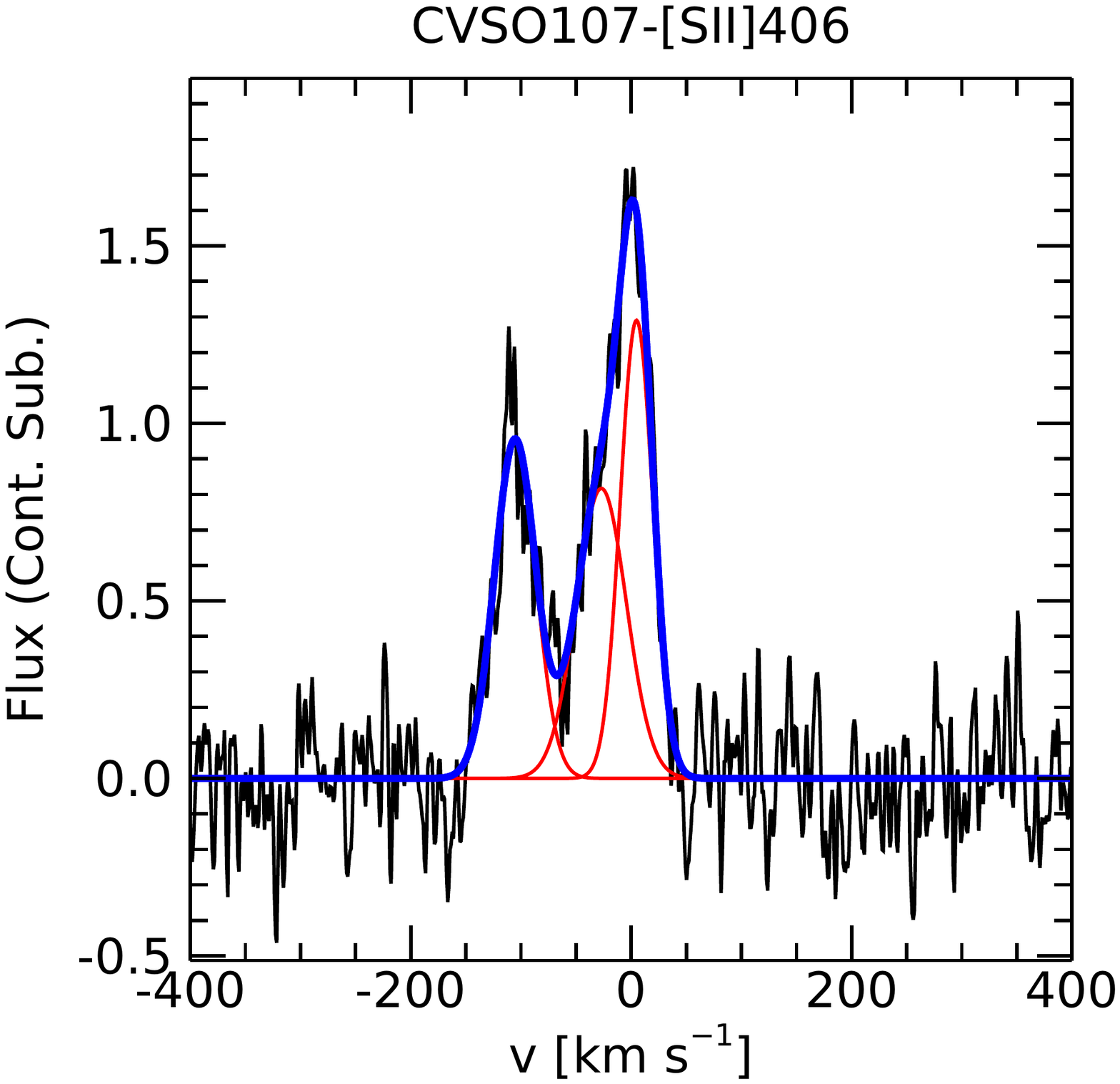}
\end{figure*}

\begin{figure*}[!h]
\includegraphics[trim=20 210 0 70,width=.63\columnwidth, angle=0]{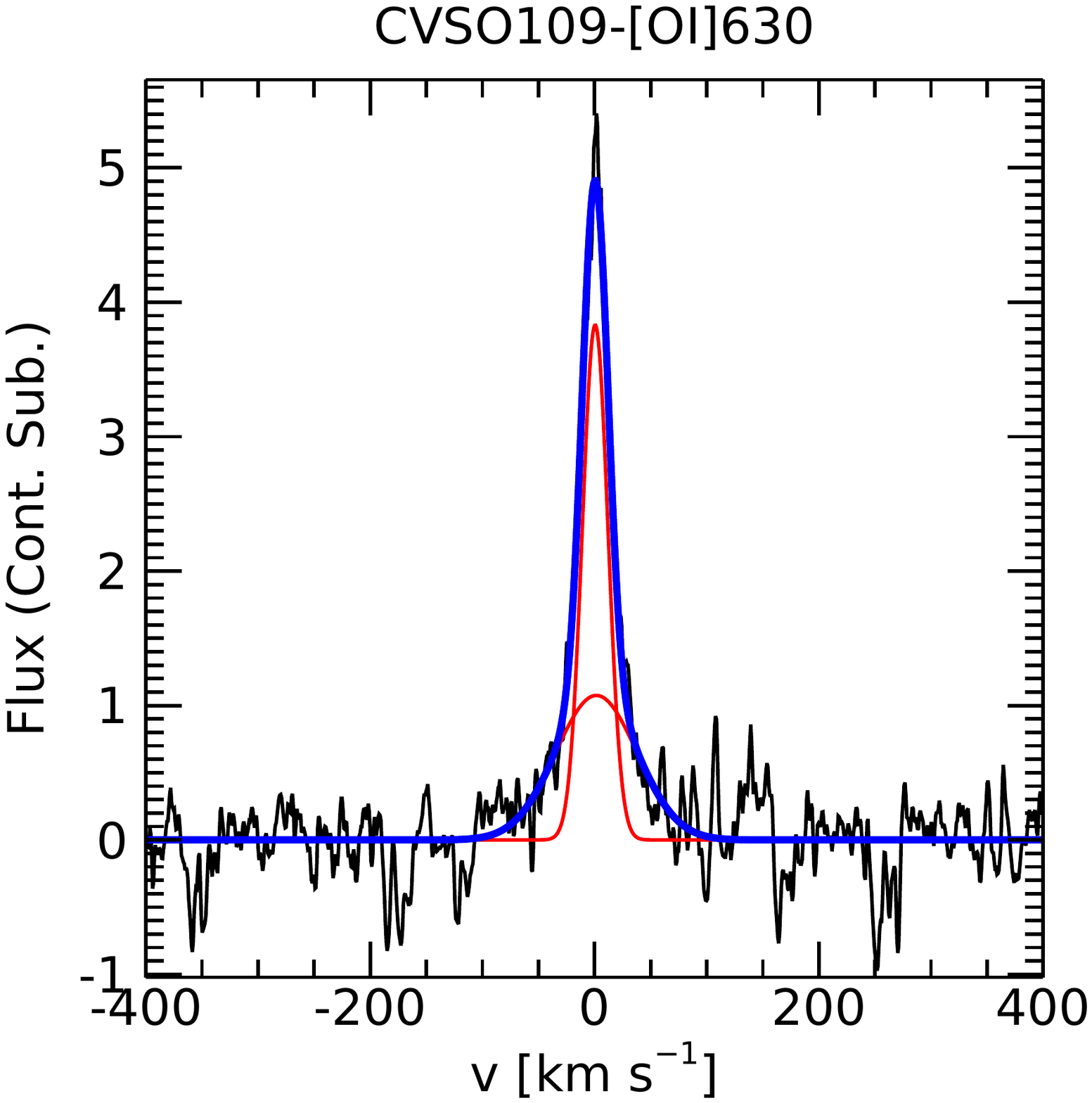}
\includegraphics[trim=20 210 0 70,width=.63\columnwidth, angle=0]{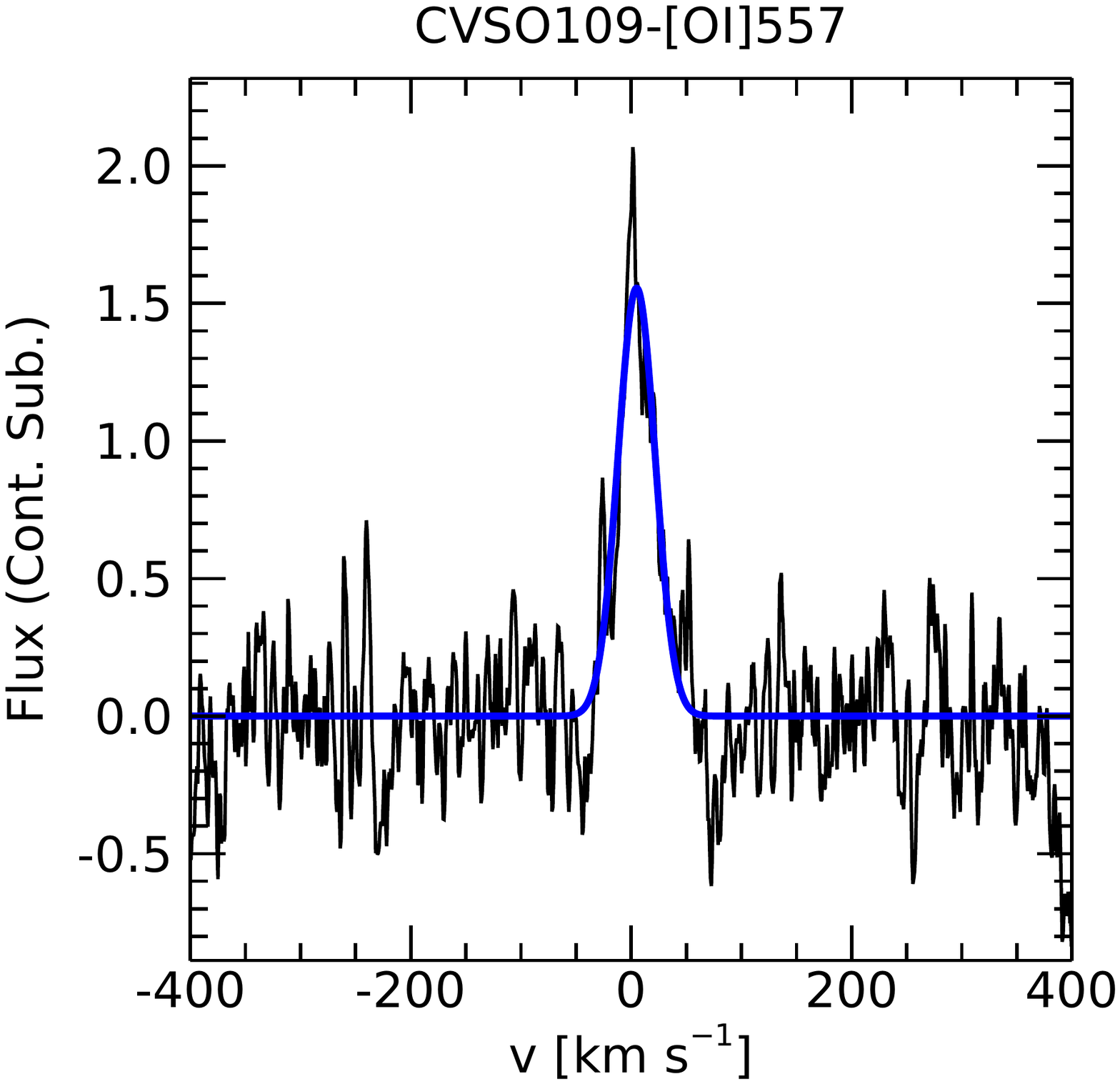}
\includegraphics[trim=20 210 0 70,width=.63\columnwidth, angle=0]{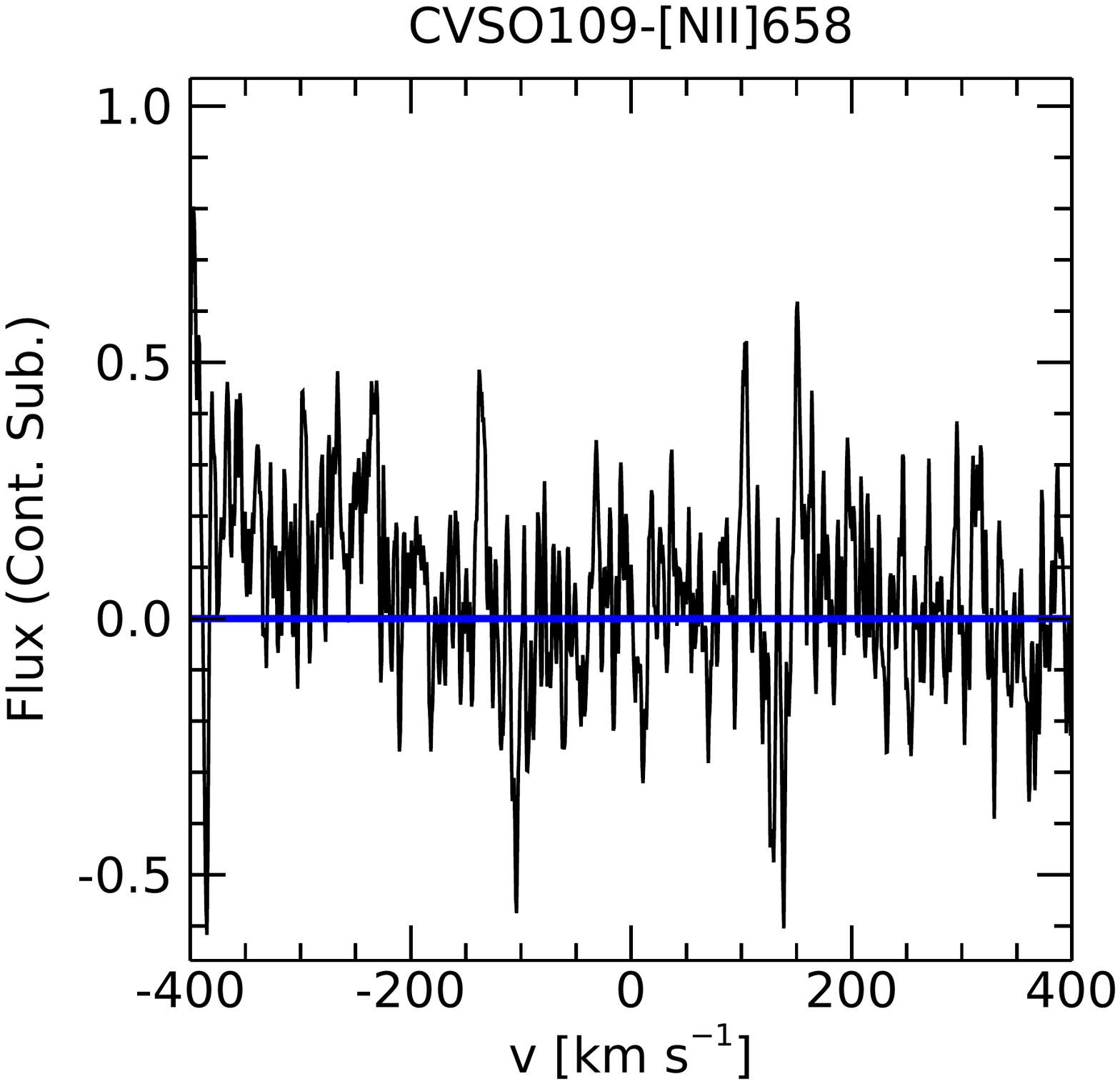}
\includegraphics[trim=20 210 0 70,width=.63\columnwidth, angle=0]{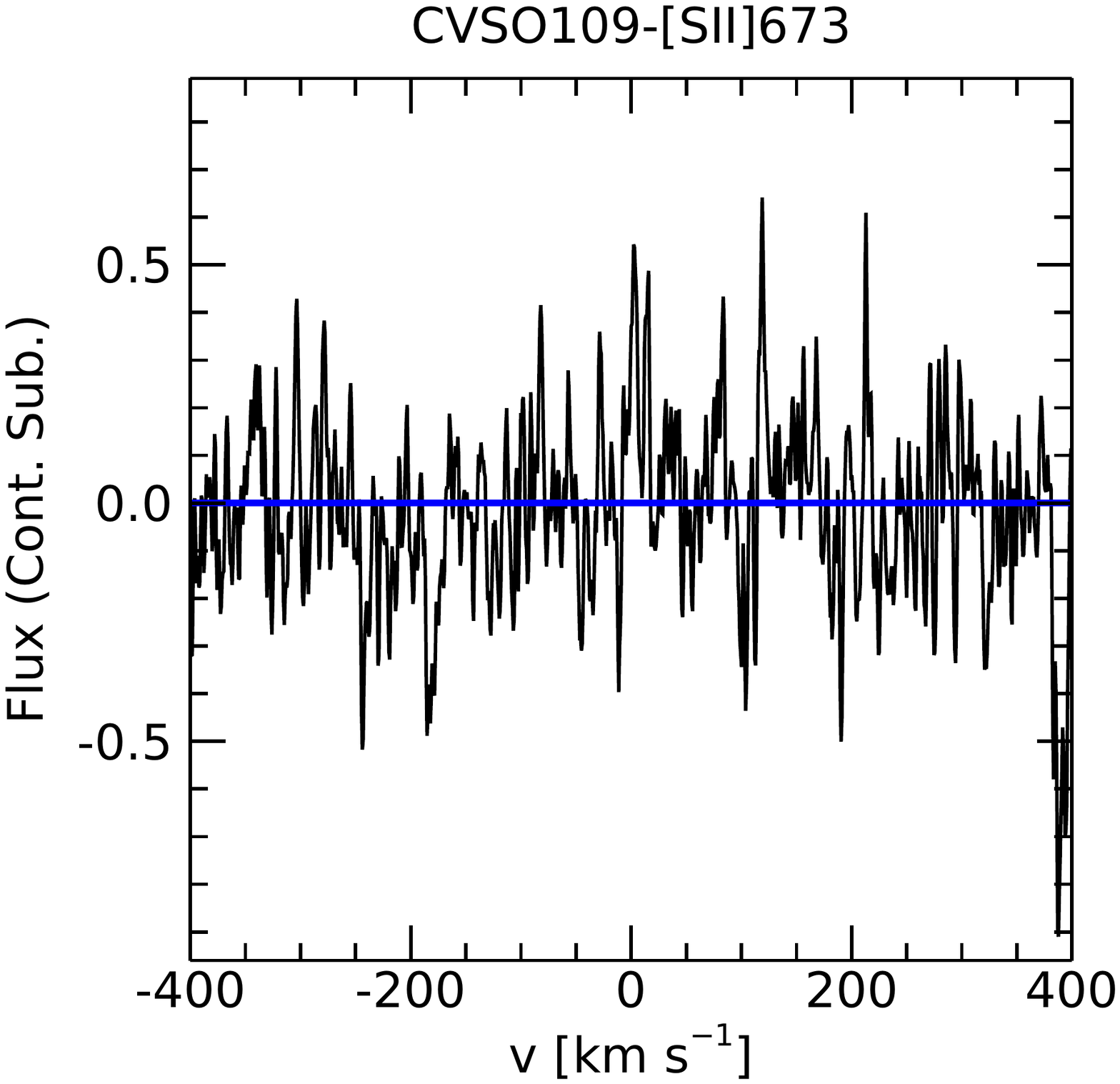}
\includegraphics[trim=20 210 0 70,width=.63\columnwidth, angle=0]{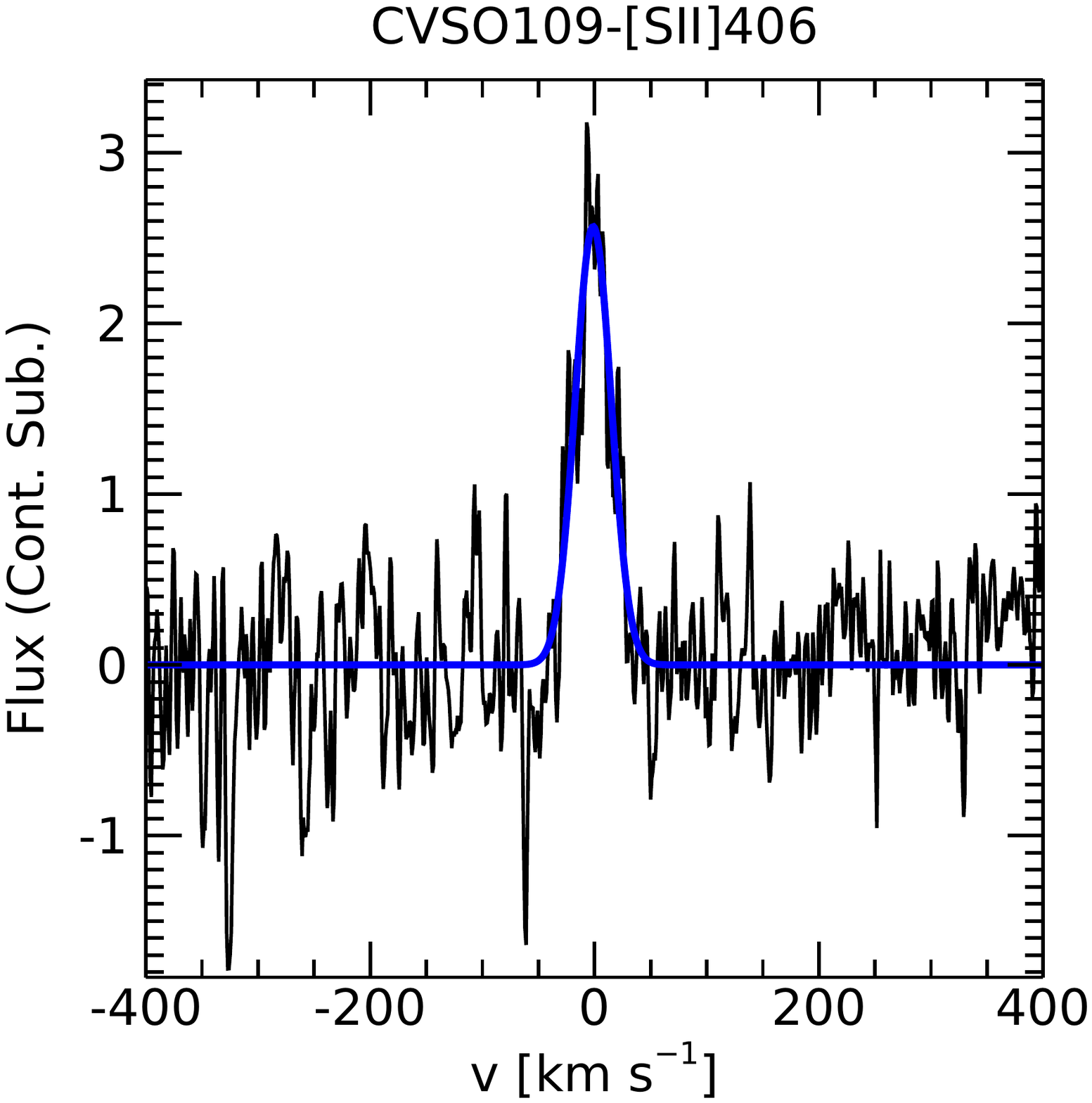}
\begin{center} \textbf{Fig. A.1.} continued.\end{center}
\end{figure*}

\begin{figure*}[!h]
\includegraphics[trim=20 210 0 70,width=.63\columnwidth, angle=0]{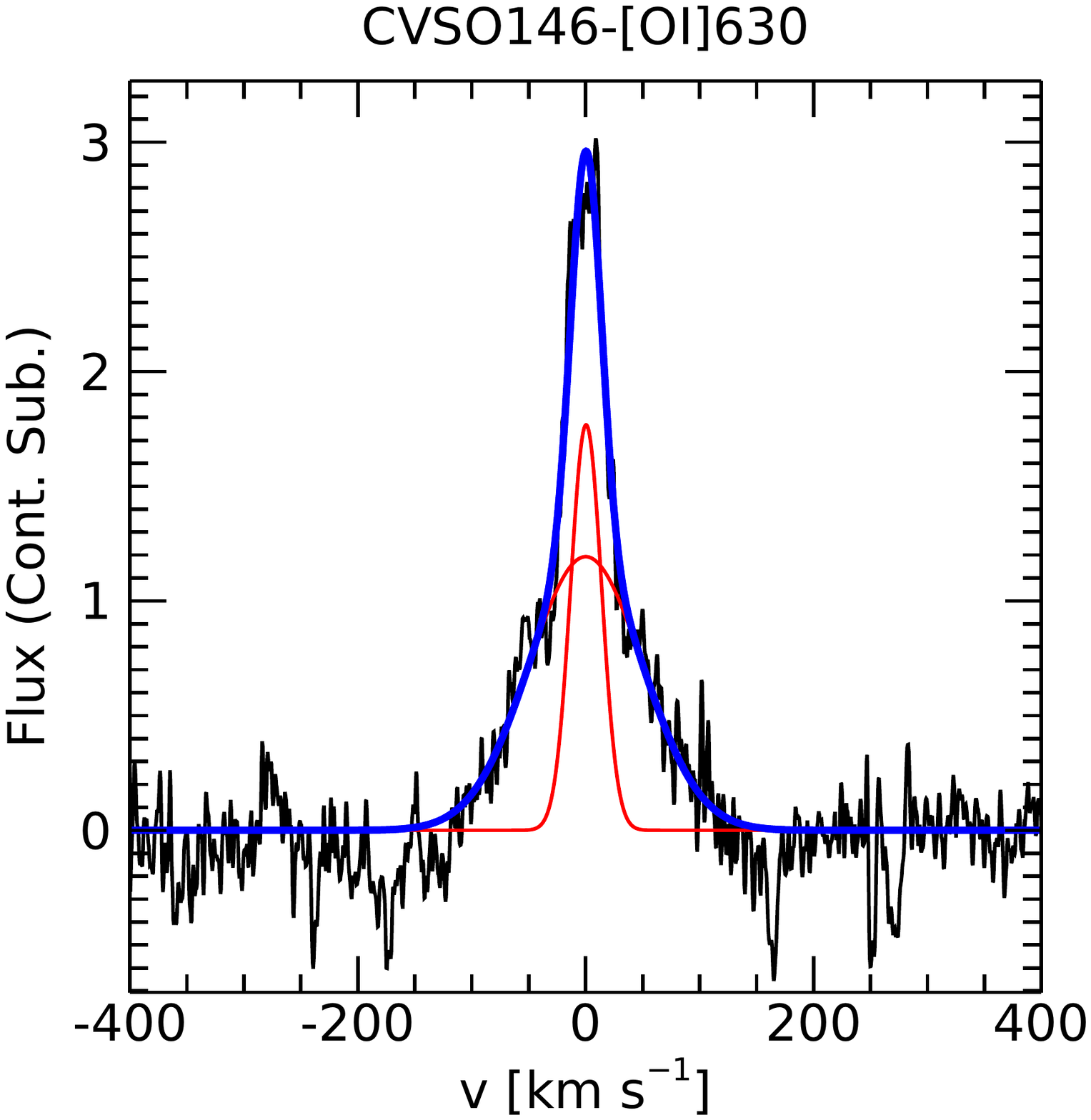}
\includegraphics[trim=20 210 0 70,width=.63\columnwidth, angle=0]{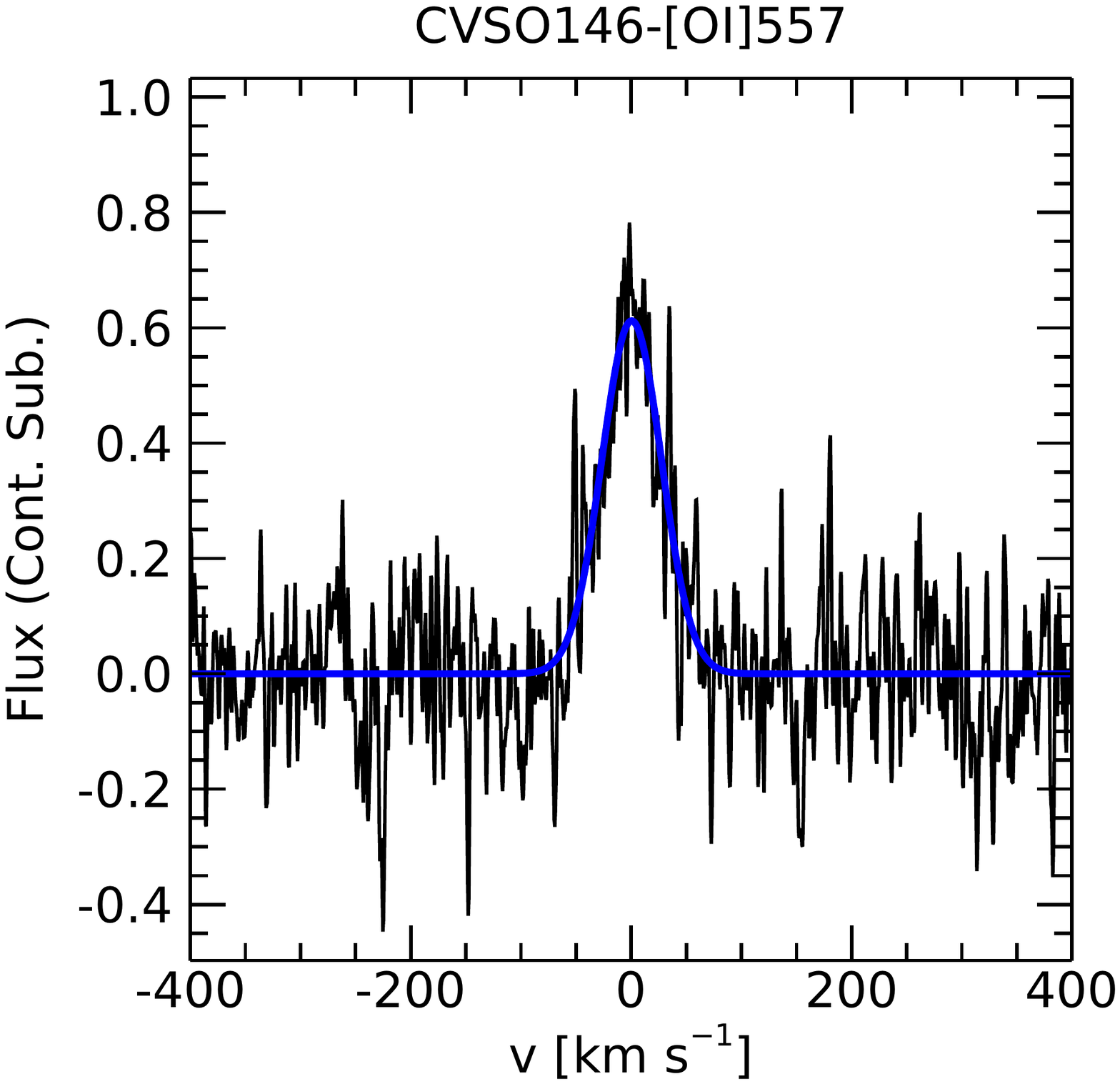}
\includegraphics[trim=20 210 0 70,width=.63\columnwidth, angle=0]{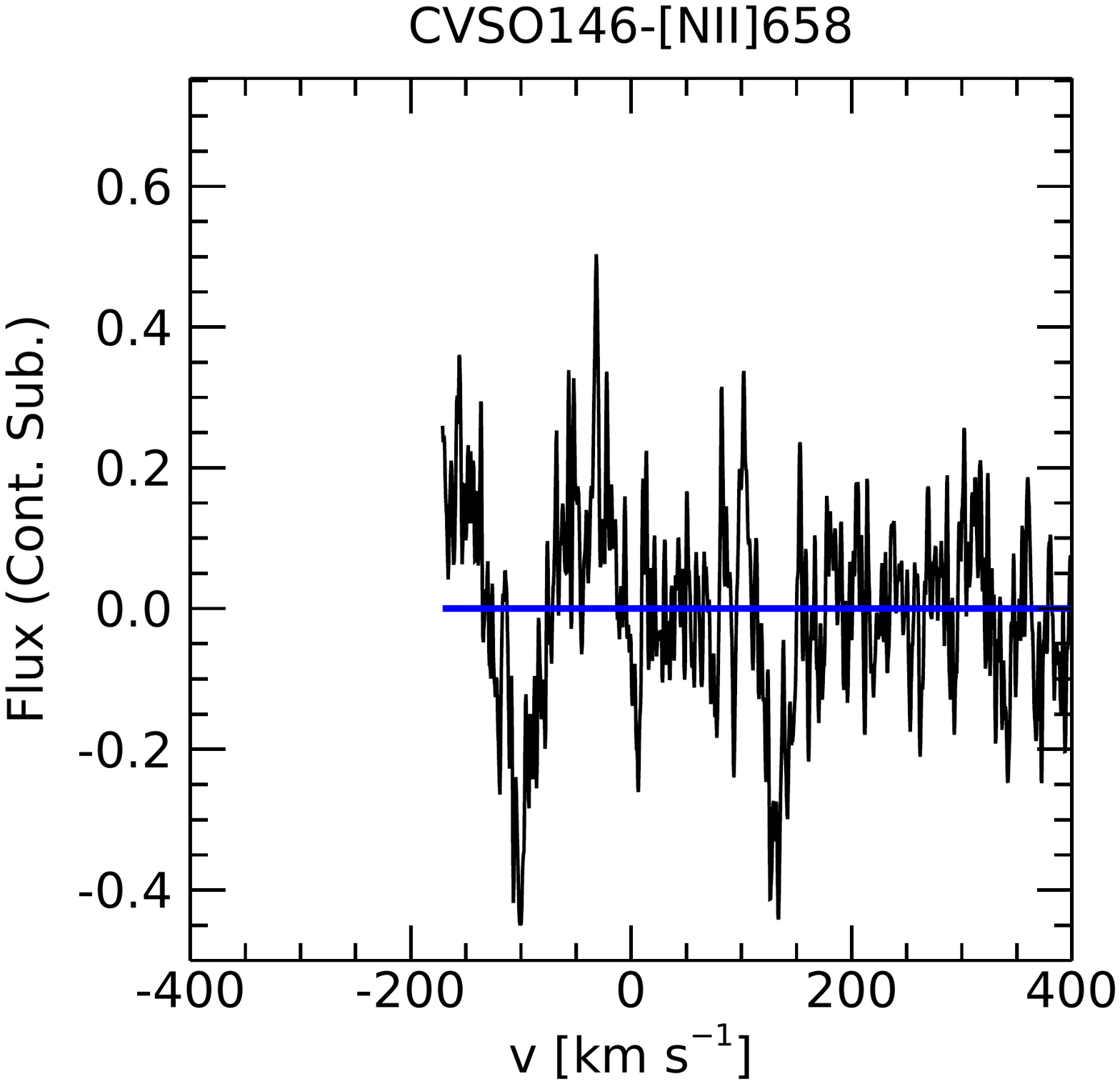}
\includegraphics[trim=20 210 0 70,width=.63\columnwidth, angle=0]{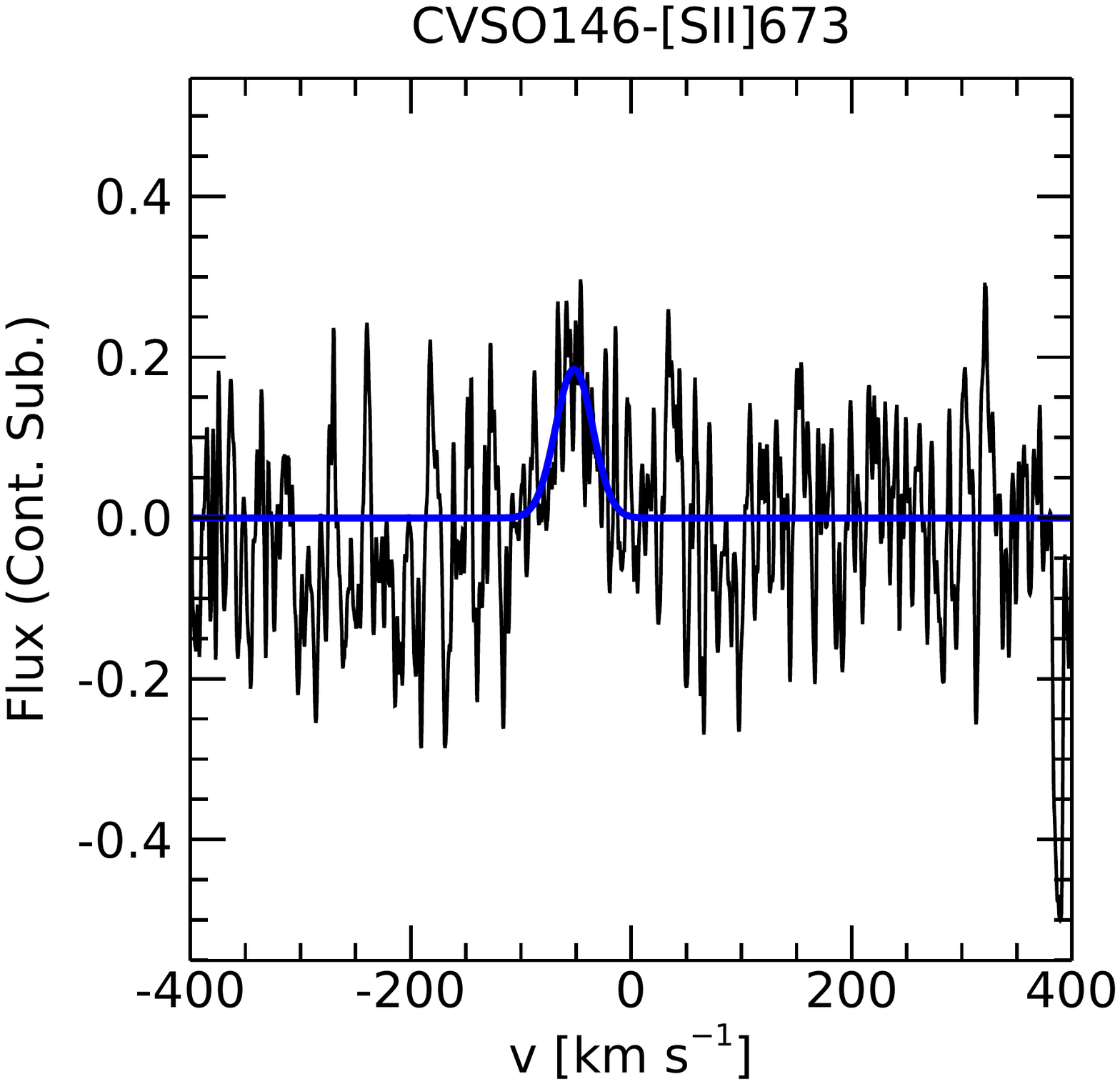}
\includegraphics[trim=20 210 0 70,width=.63\columnwidth, angle=0]{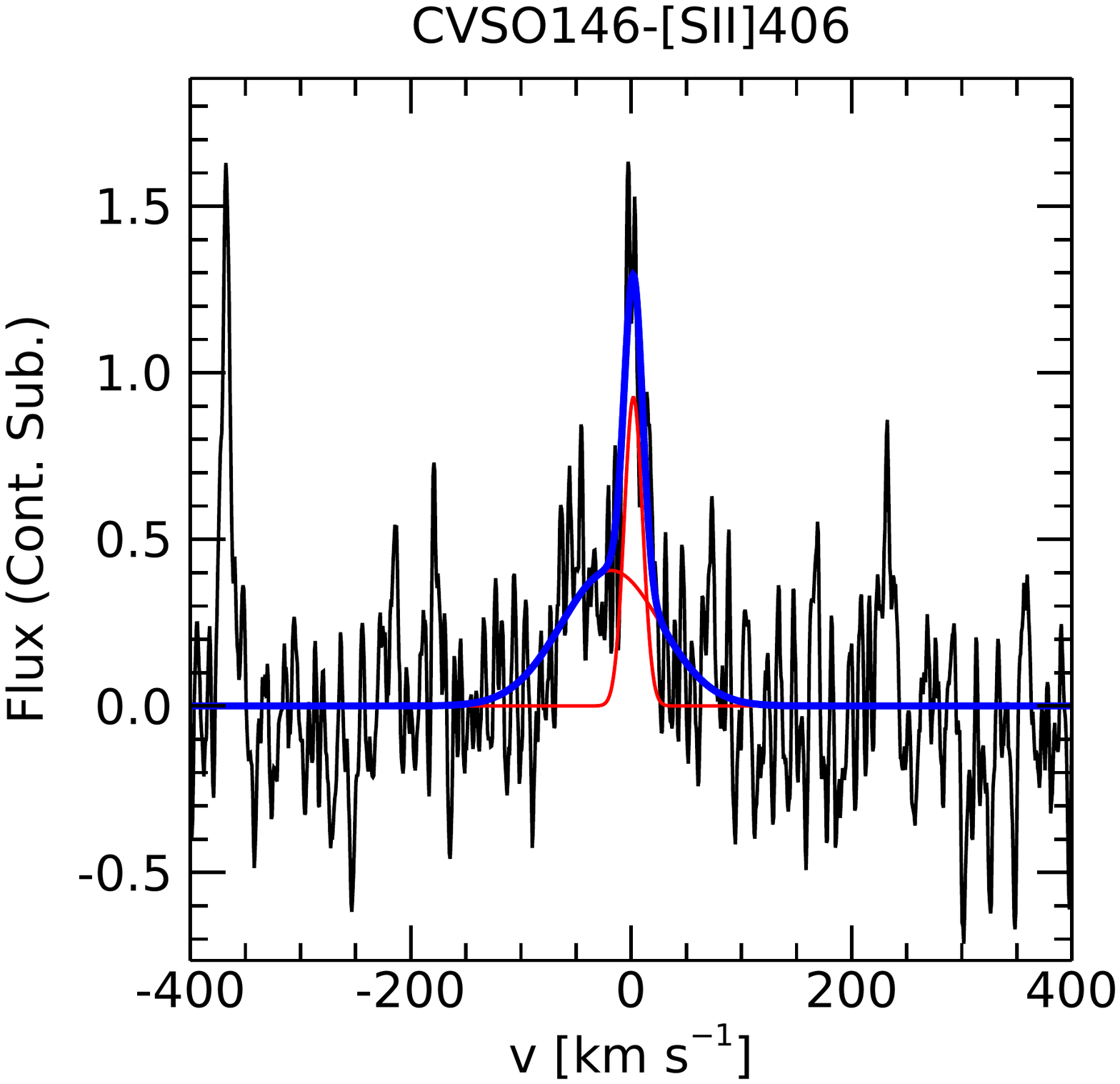}
\end{figure*}

\begin{figure*}[!h]
\includegraphics[trim=20 210 0 70,width=.63\columnwidth, angle=0]{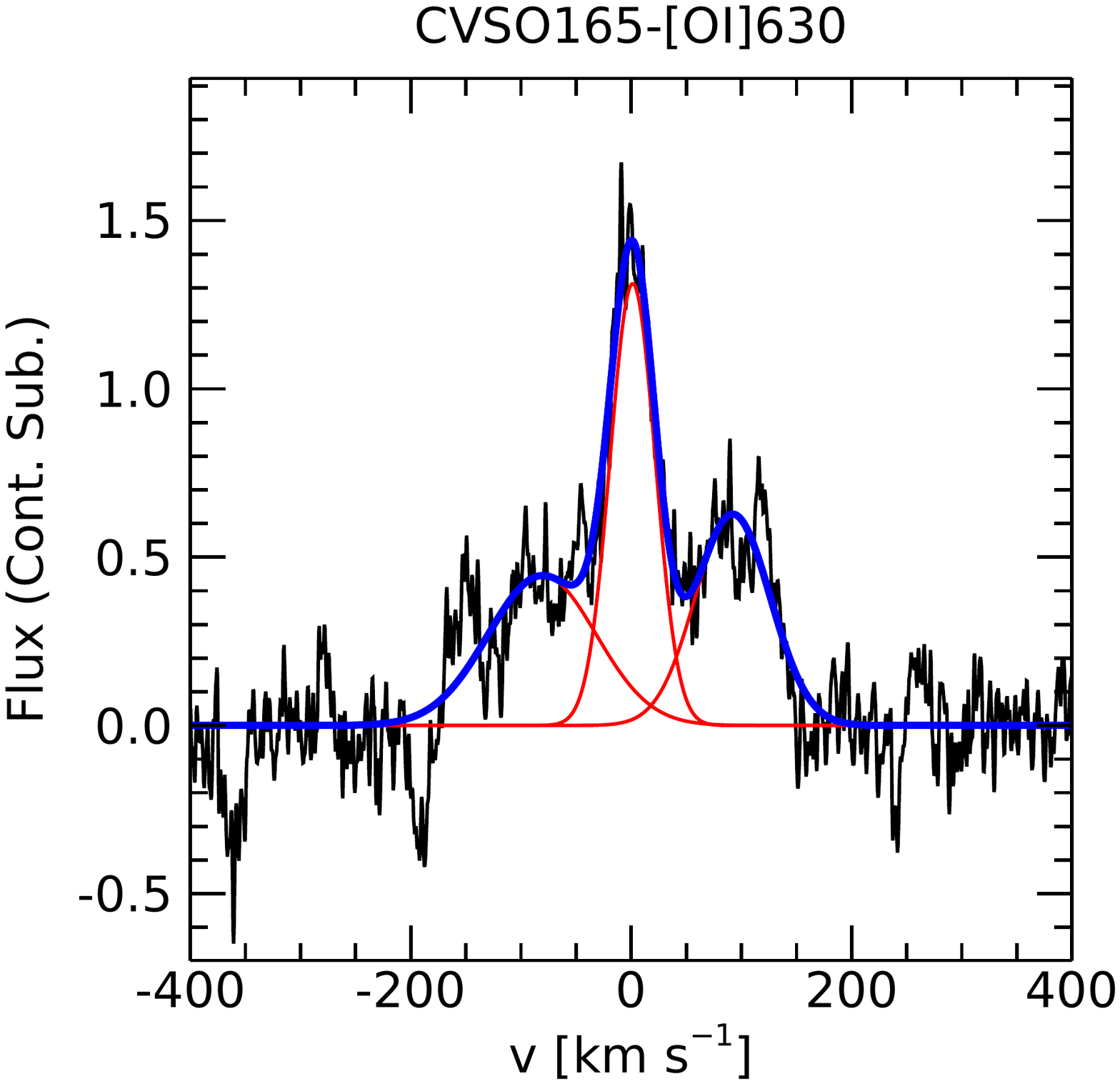}
\includegraphics[trim=20 210 0 70,width=.63\columnwidth, angle=0]{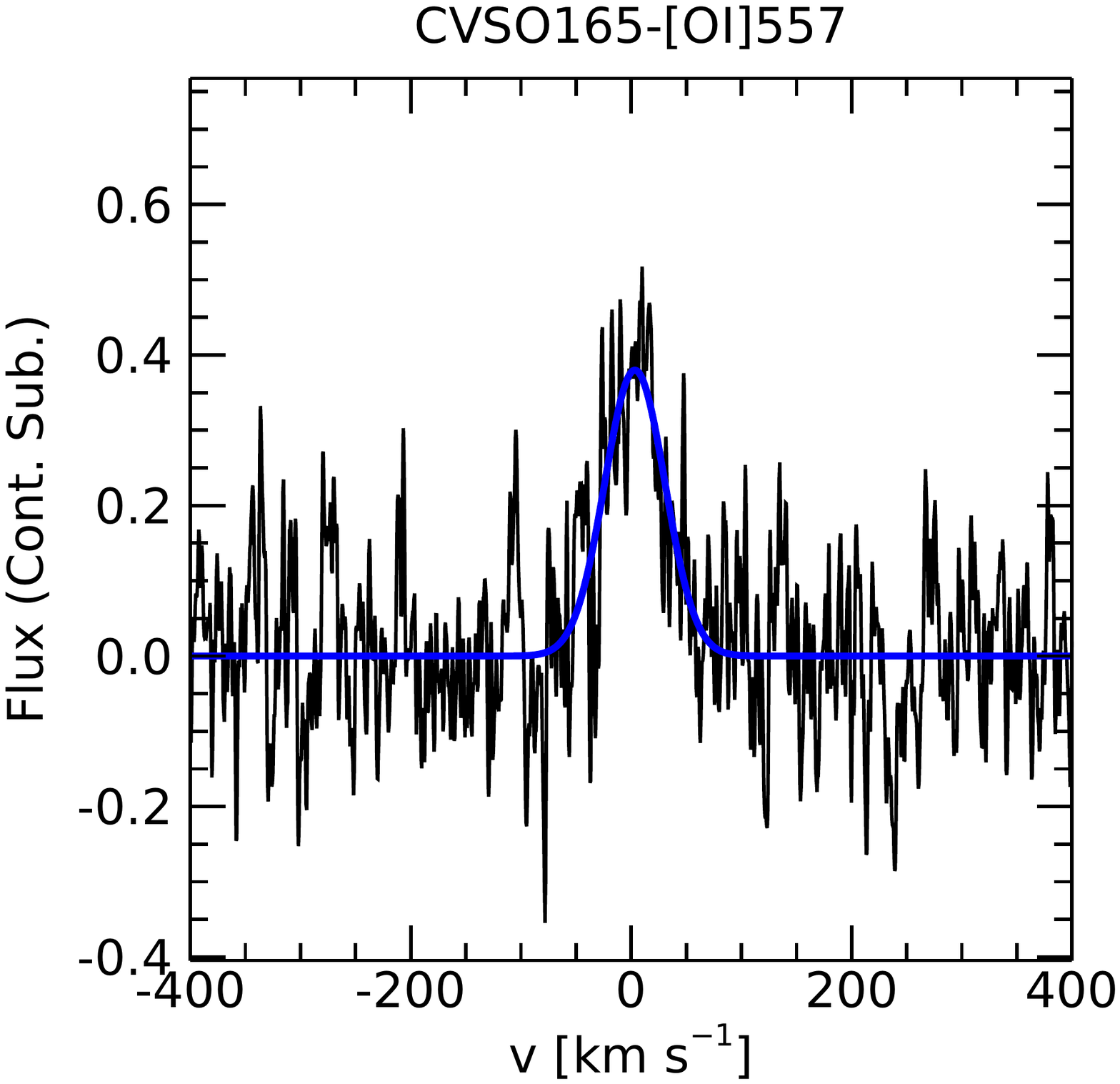}
\includegraphics[trim=20 210 0 70,width=.63\columnwidth, angle=0]{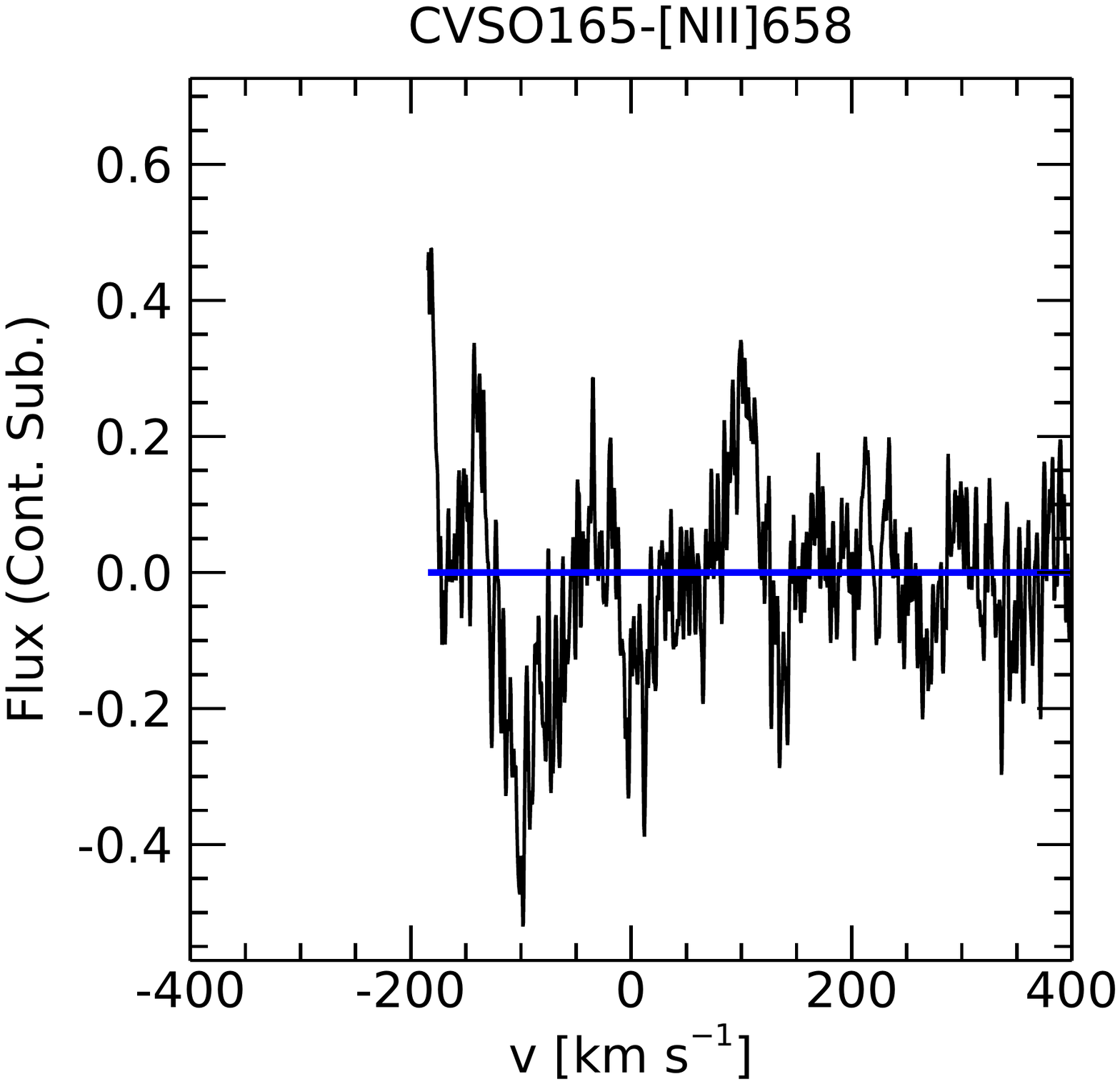}
\includegraphics[trim=20 210 0 70,width=.63\columnwidth, angle=0]{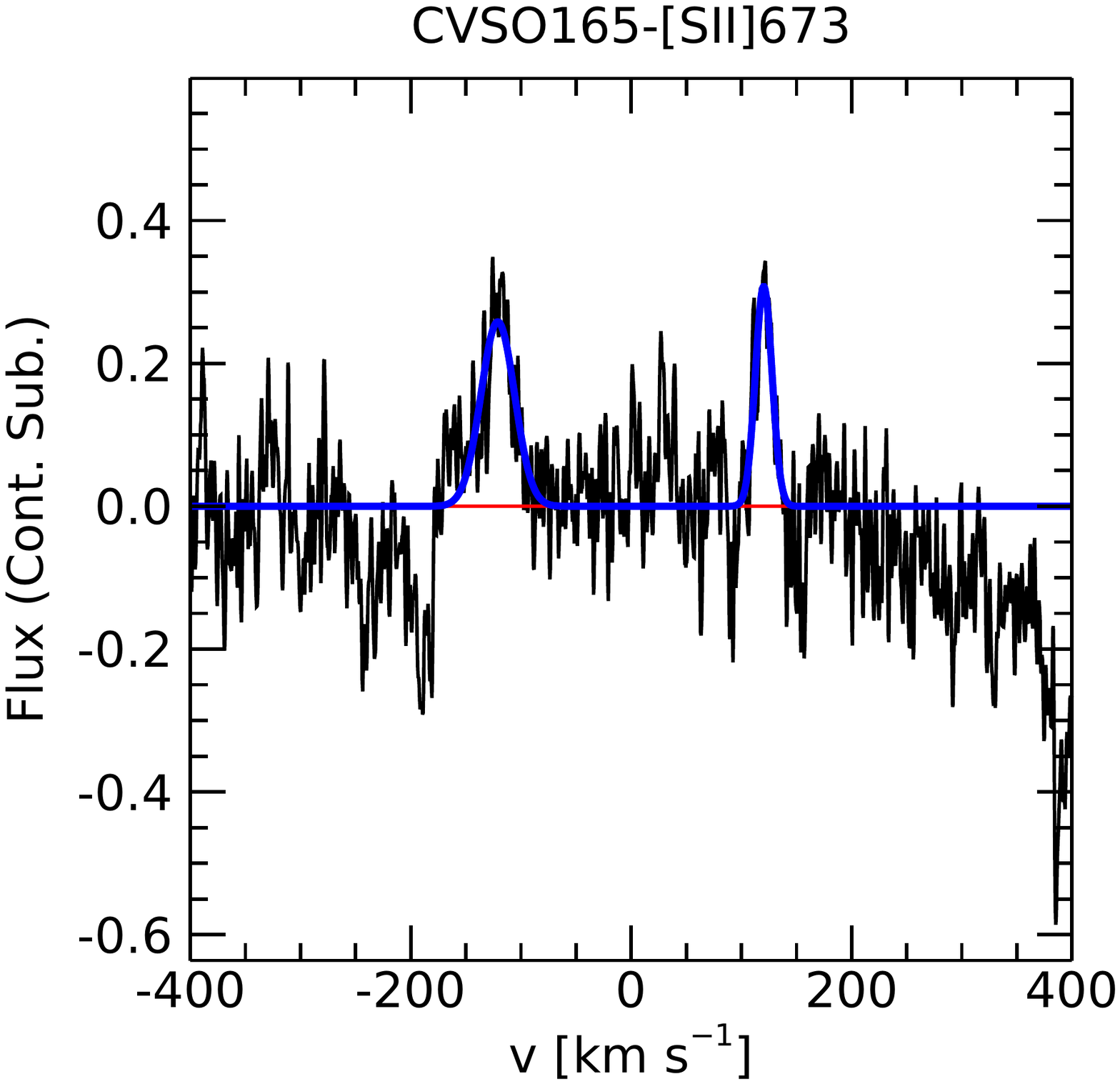}
\includegraphics[trim=20 210 0 70,width=.63\columnwidth, angle=0]{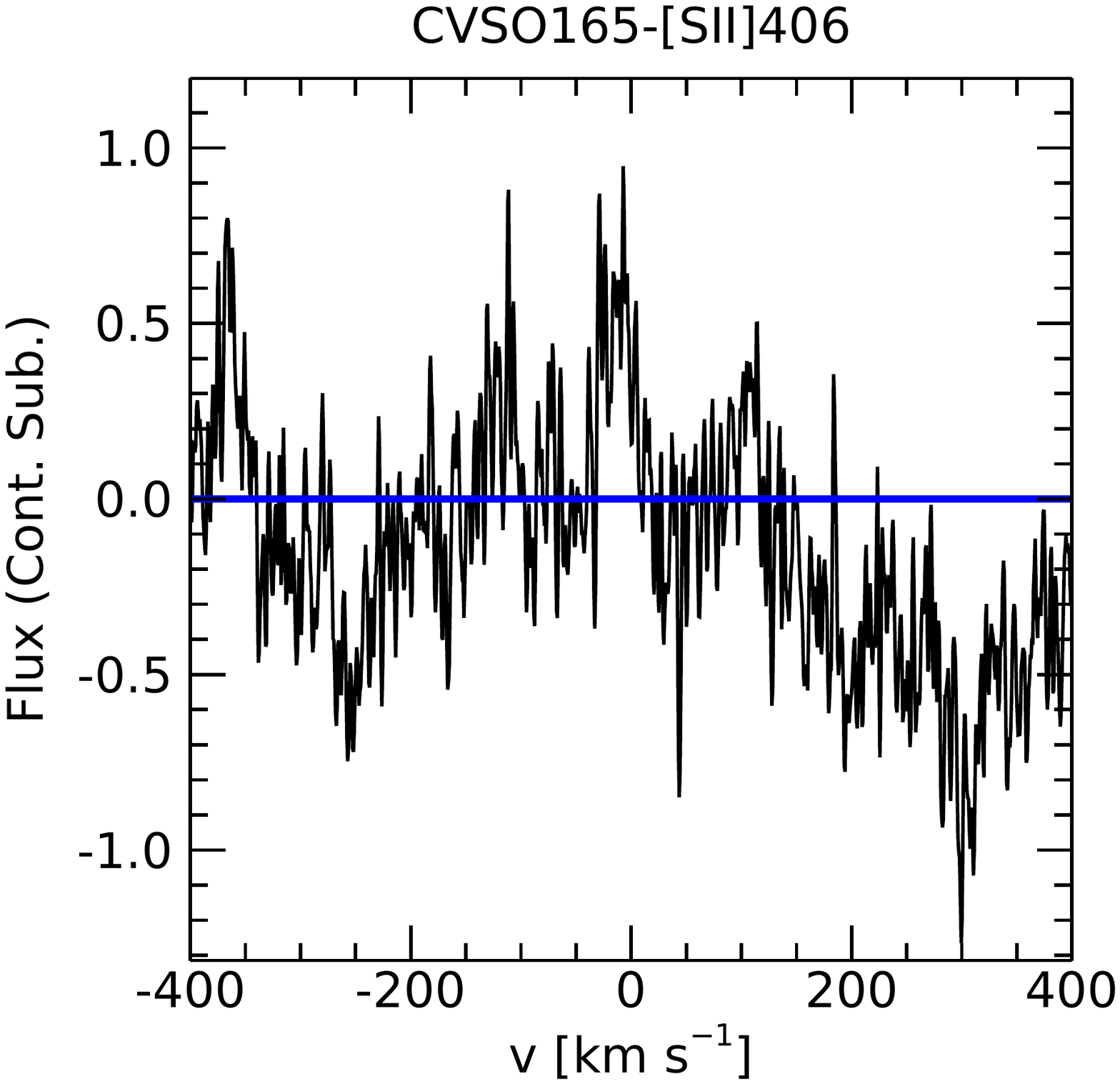}
\begin{center} \textbf{Fig. A.1.} continued.\end{center}
\end{figure*}

\begin{figure*}[!h]
\includegraphics[trim=20 210 0 70,width=.63\columnwidth, angle=0]{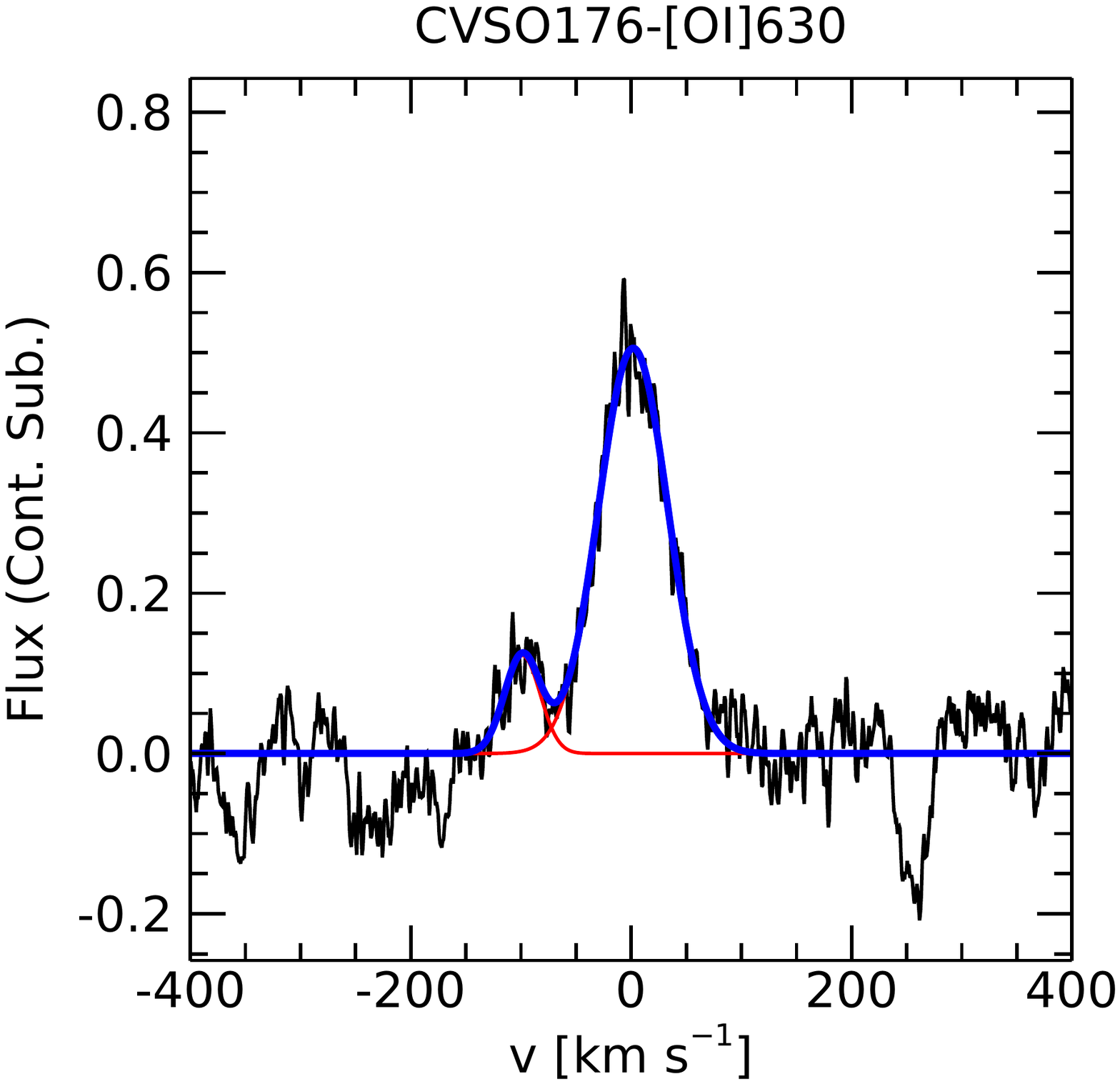}
\includegraphics[trim=20 210 0 70,width=.63\columnwidth, angle=0]{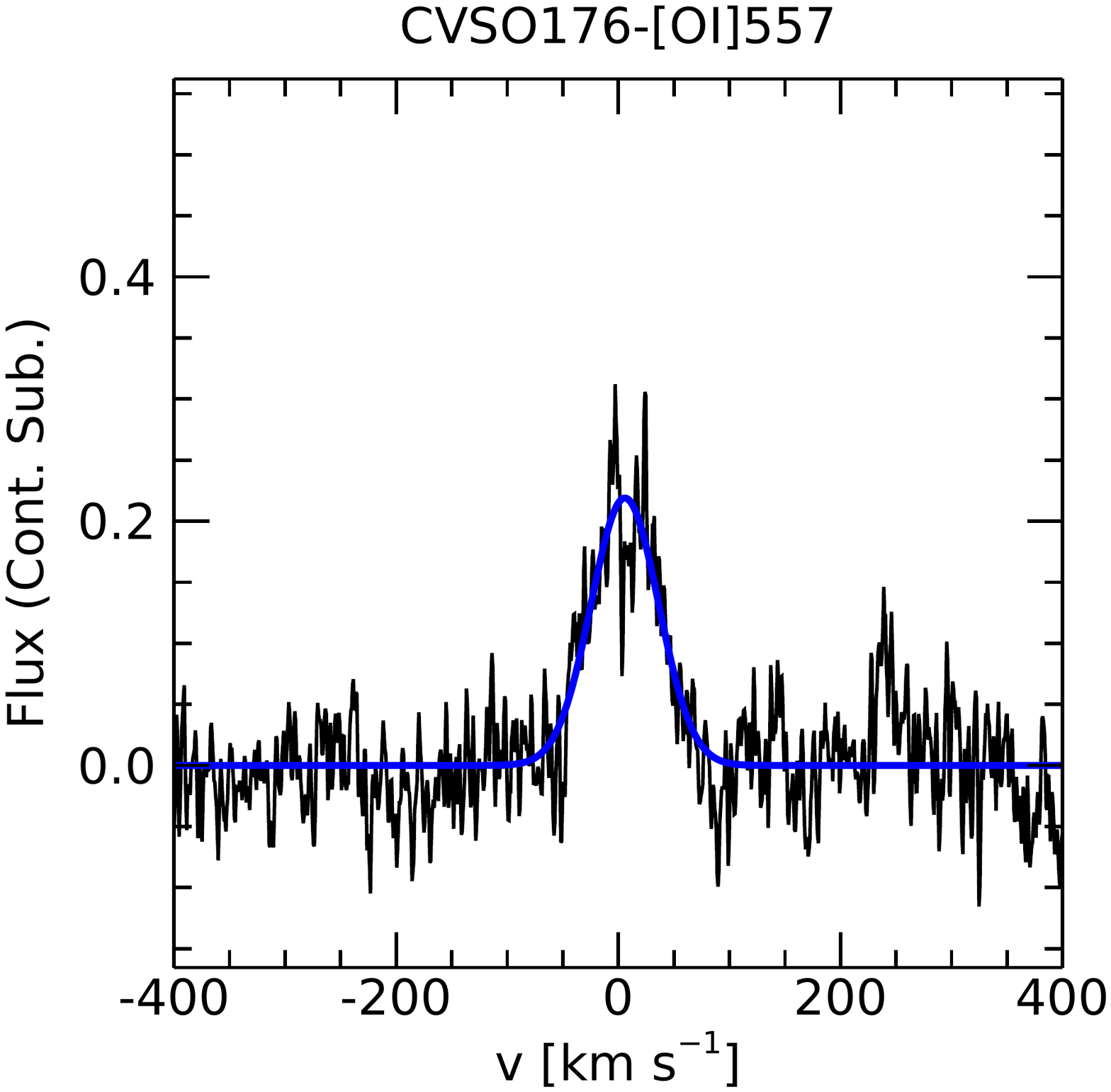}
\includegraphics[trim=20 210 0 70,width=.63\columnwidth, angle=0]{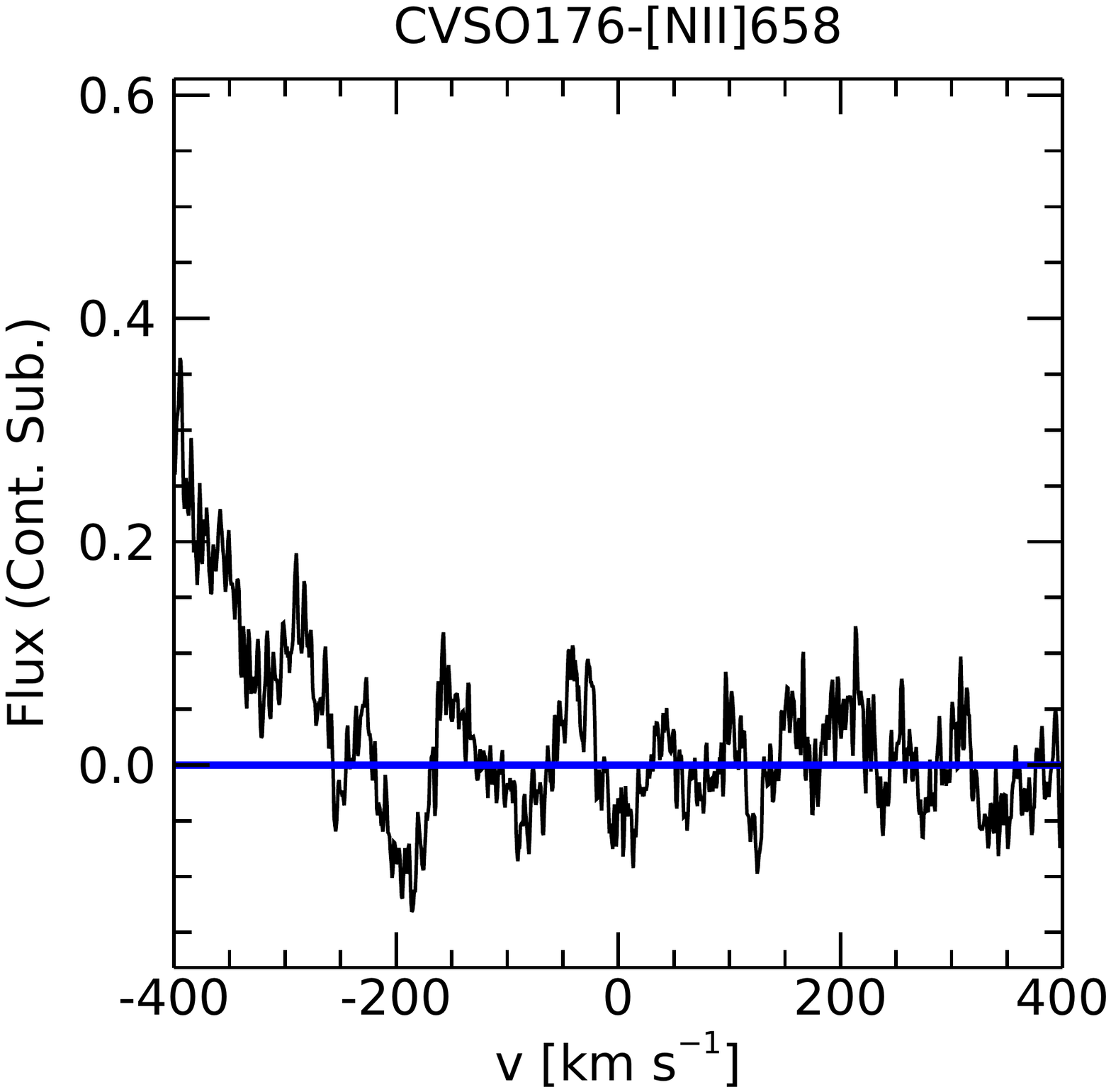}
\includegraphics[trim=20 210 0 70,width=.63\columnwidth, angle=0]{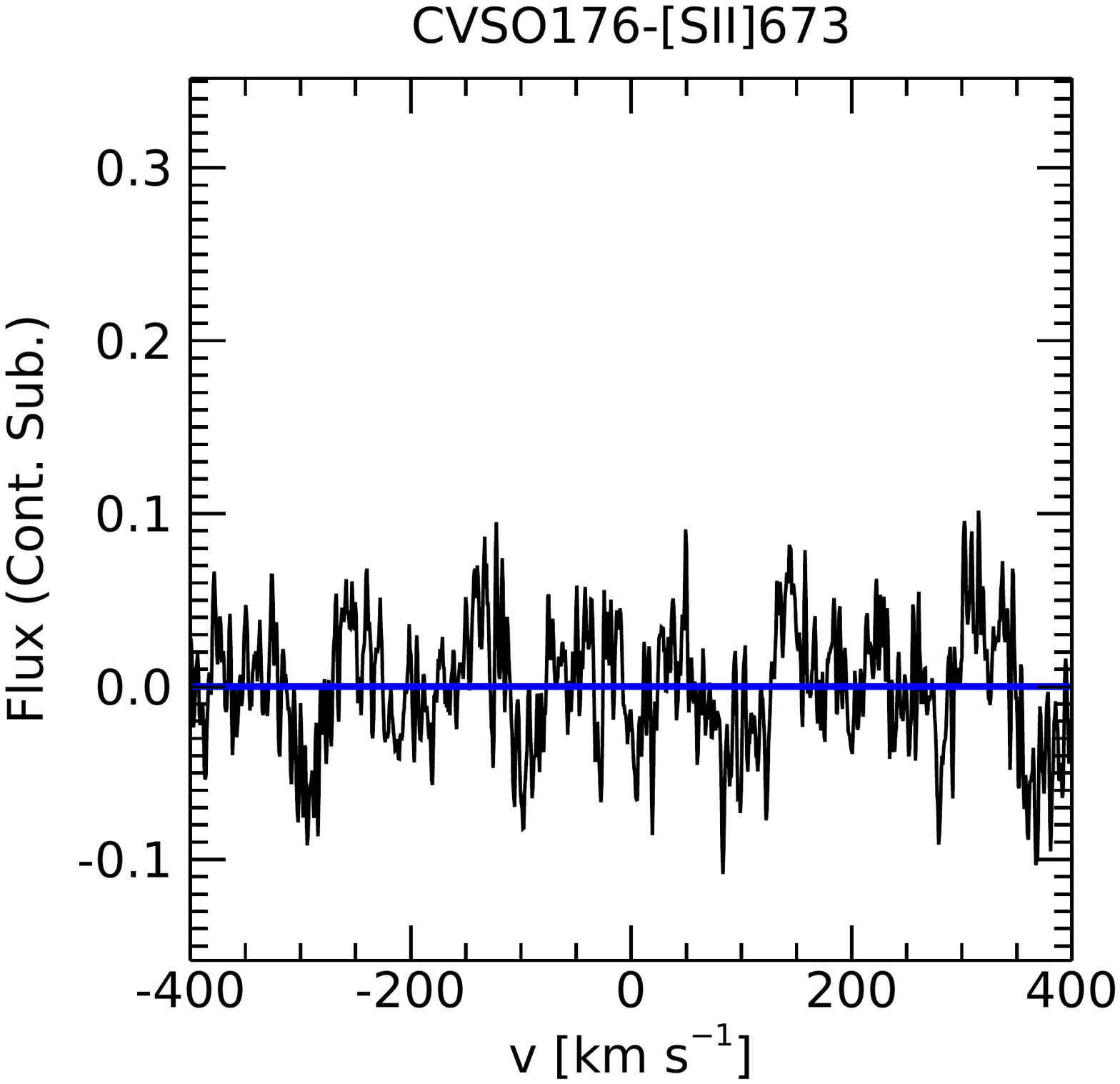}
\includegraphics[trim=20 210 0 70,width=.63\columnwidth, angle=0]{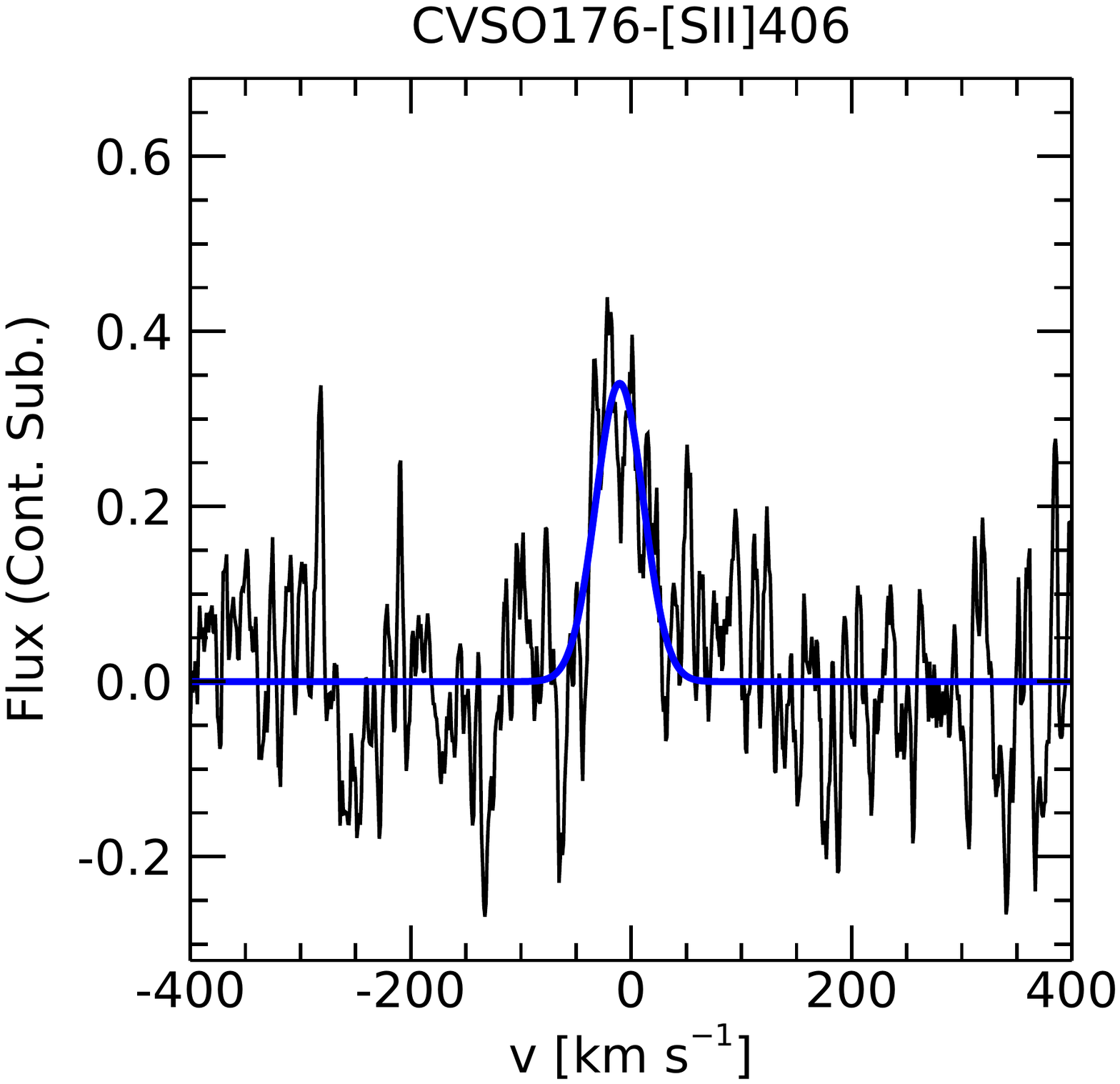}
\end{figure*}

\begin{figure*}[!h]
\includegraphics[trim=20 210 0 70,width=.63\columnwidth, angle=0]{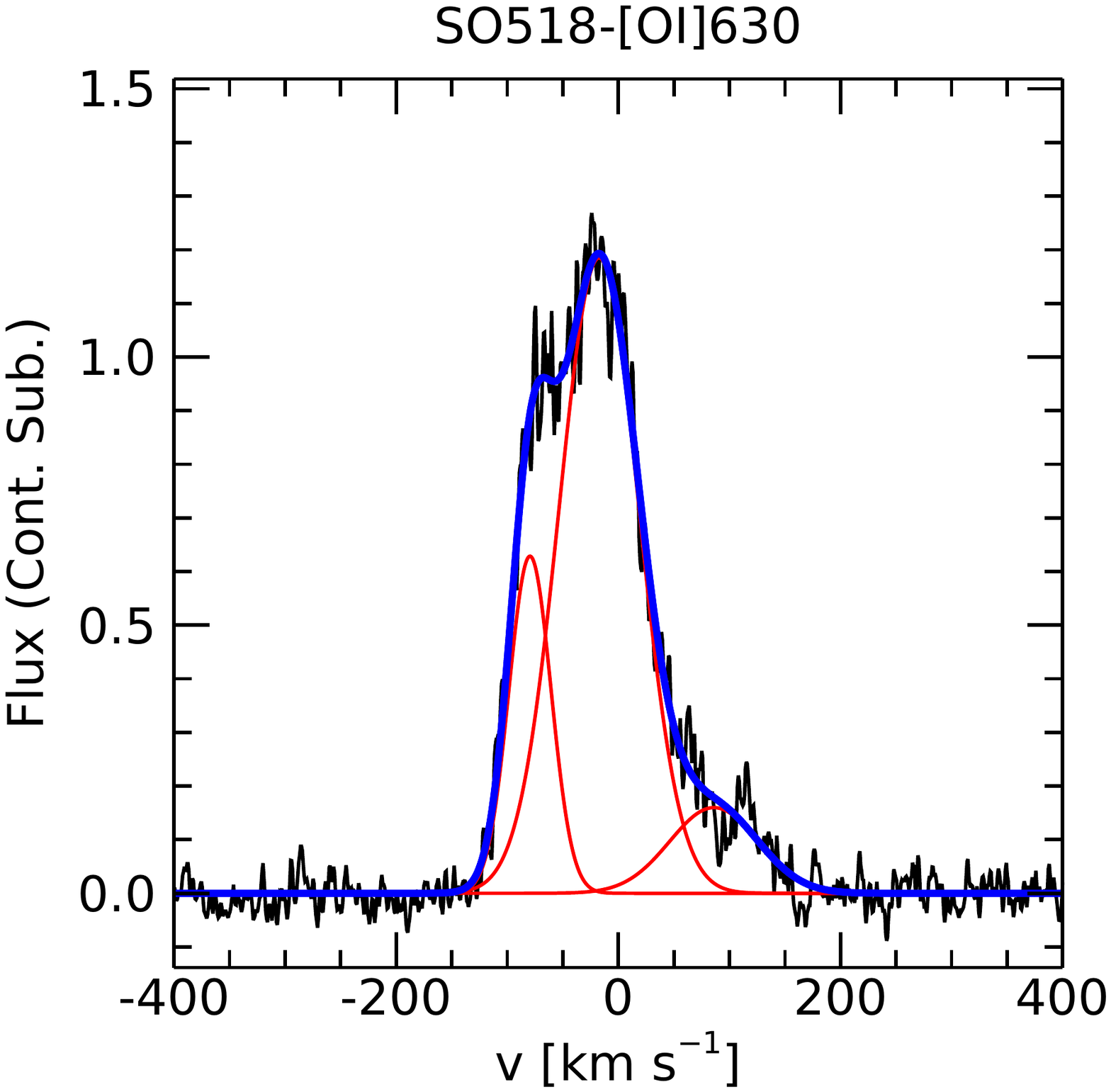}
\includegraphics[trim=20 210 0 70,width=.63\columnwidth, angle=0]{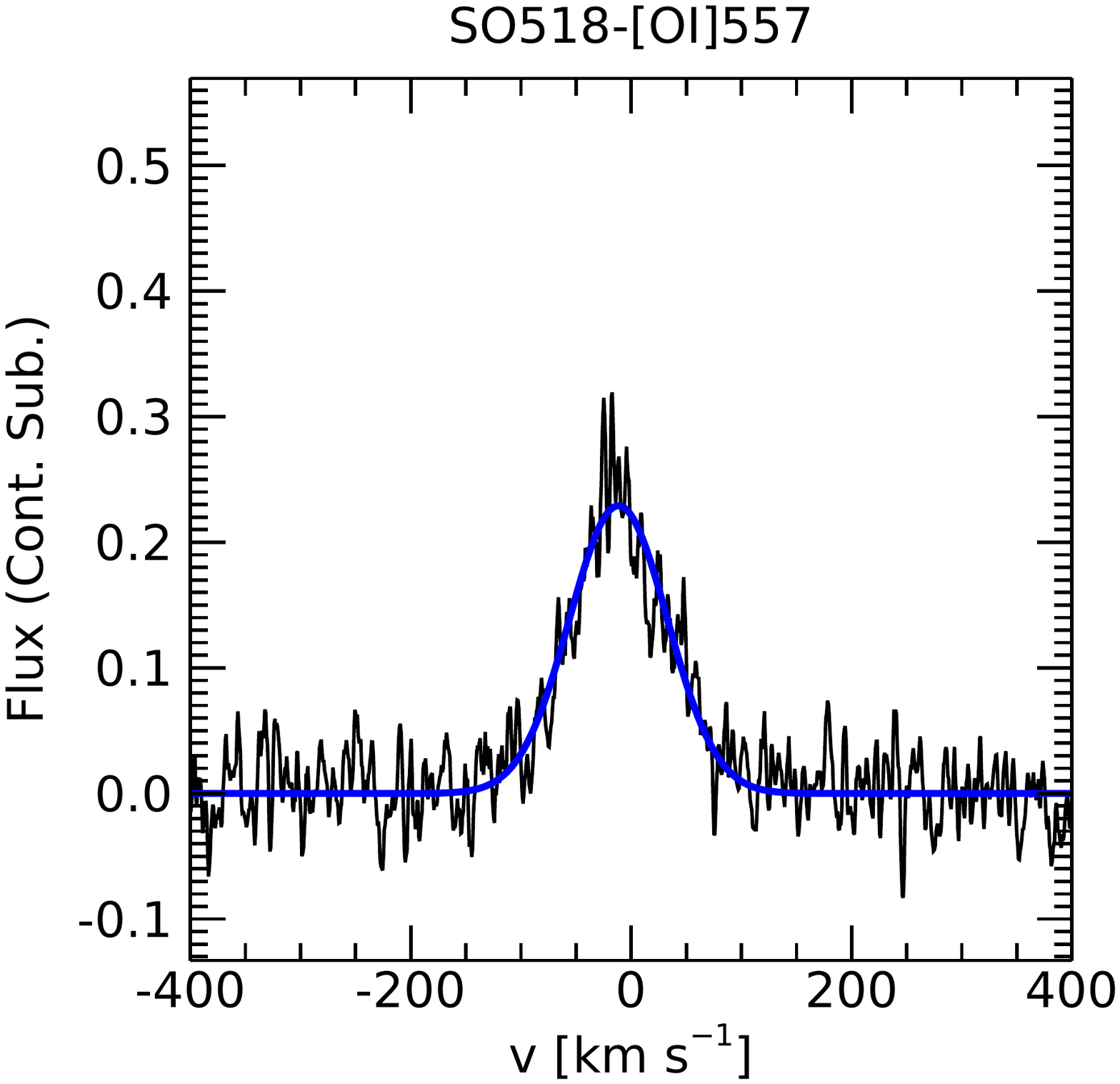}
\includegraphics[trim=20 210 0 70,width=.63\columnwidth, angle=0]{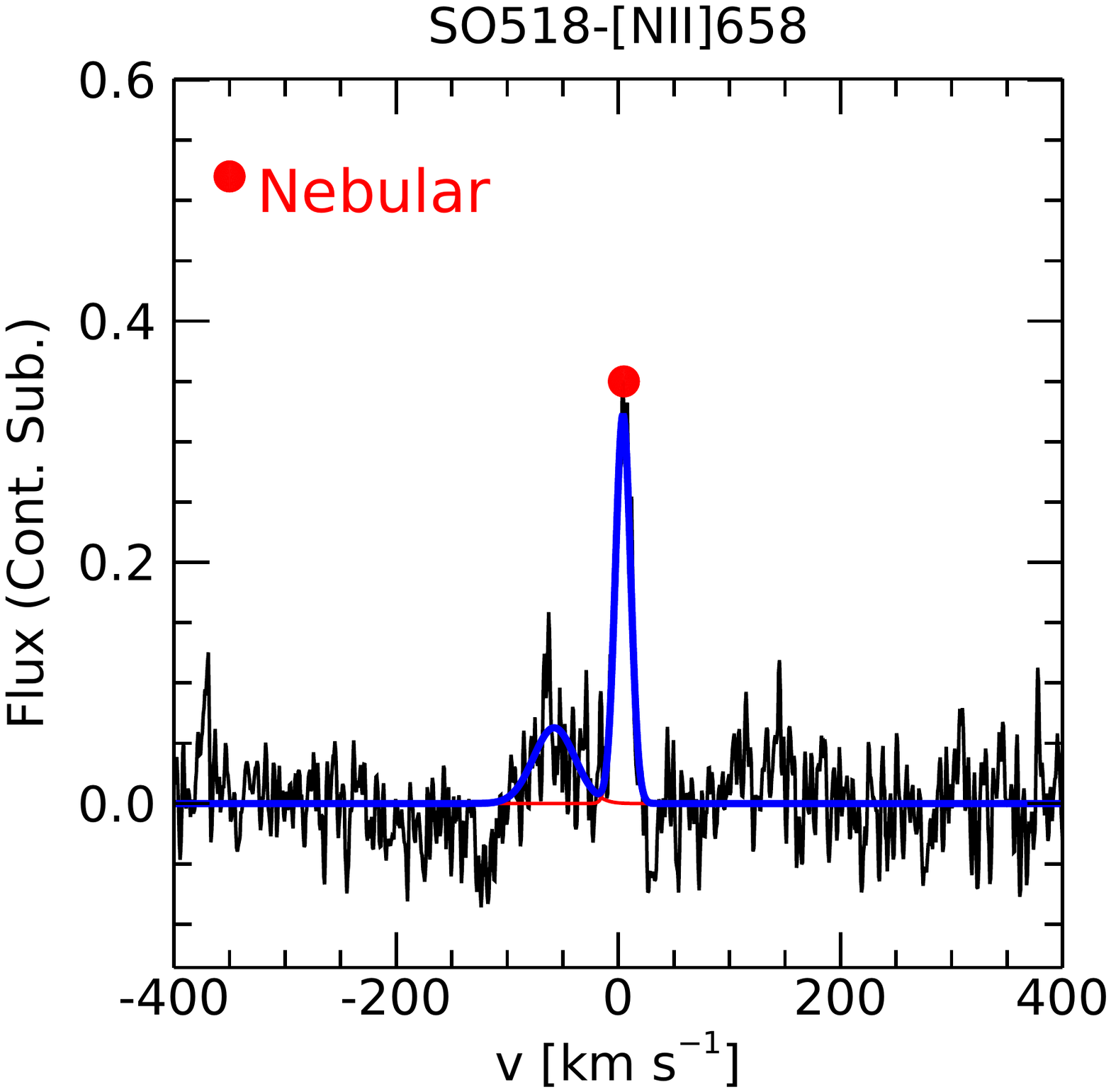}
\includegraphics[trim=20 210 0 70,width=.63\columnwidth, angle=0]{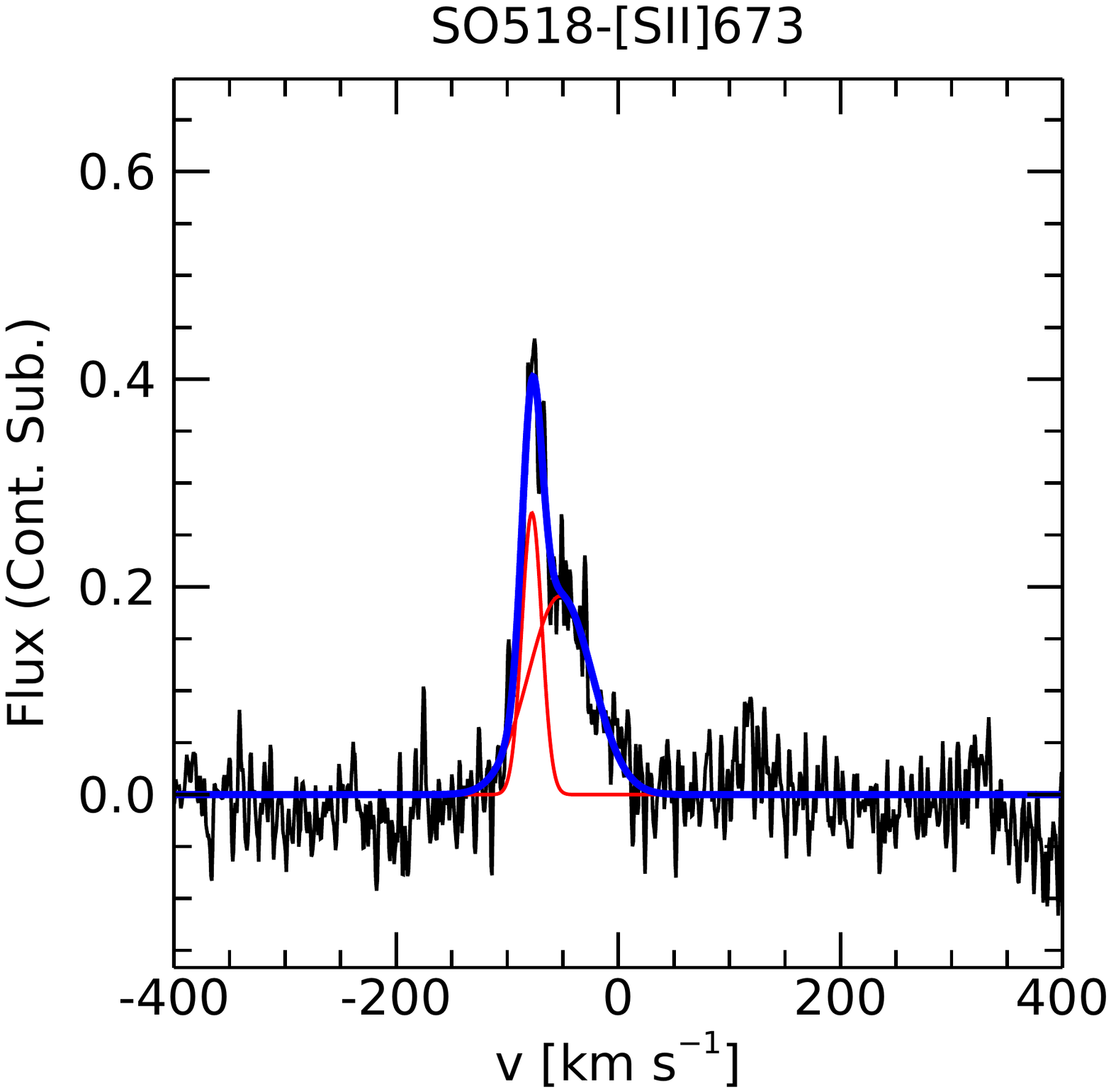}
\includegraphics[trim=20 210 0 70,width=.63\columnwidth, angle=0]{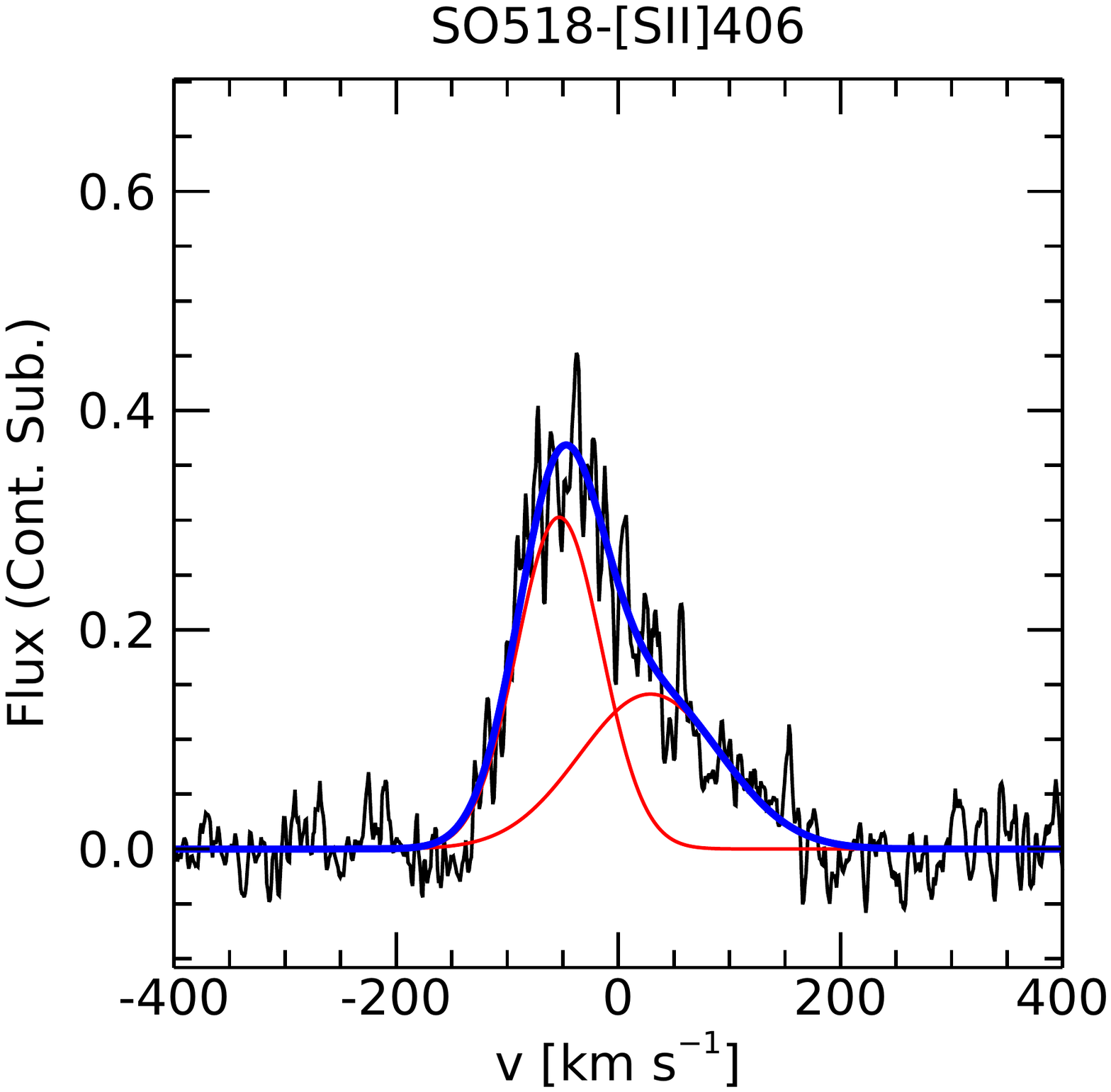}
\begin{center} \textbf{Fig. A.1.} continued.\end{center}
\end{figure*}

\begin{figure*}[!h]
\includegraphics[trim=20 210 0 70,width=.63\columnwidth, angle=0]{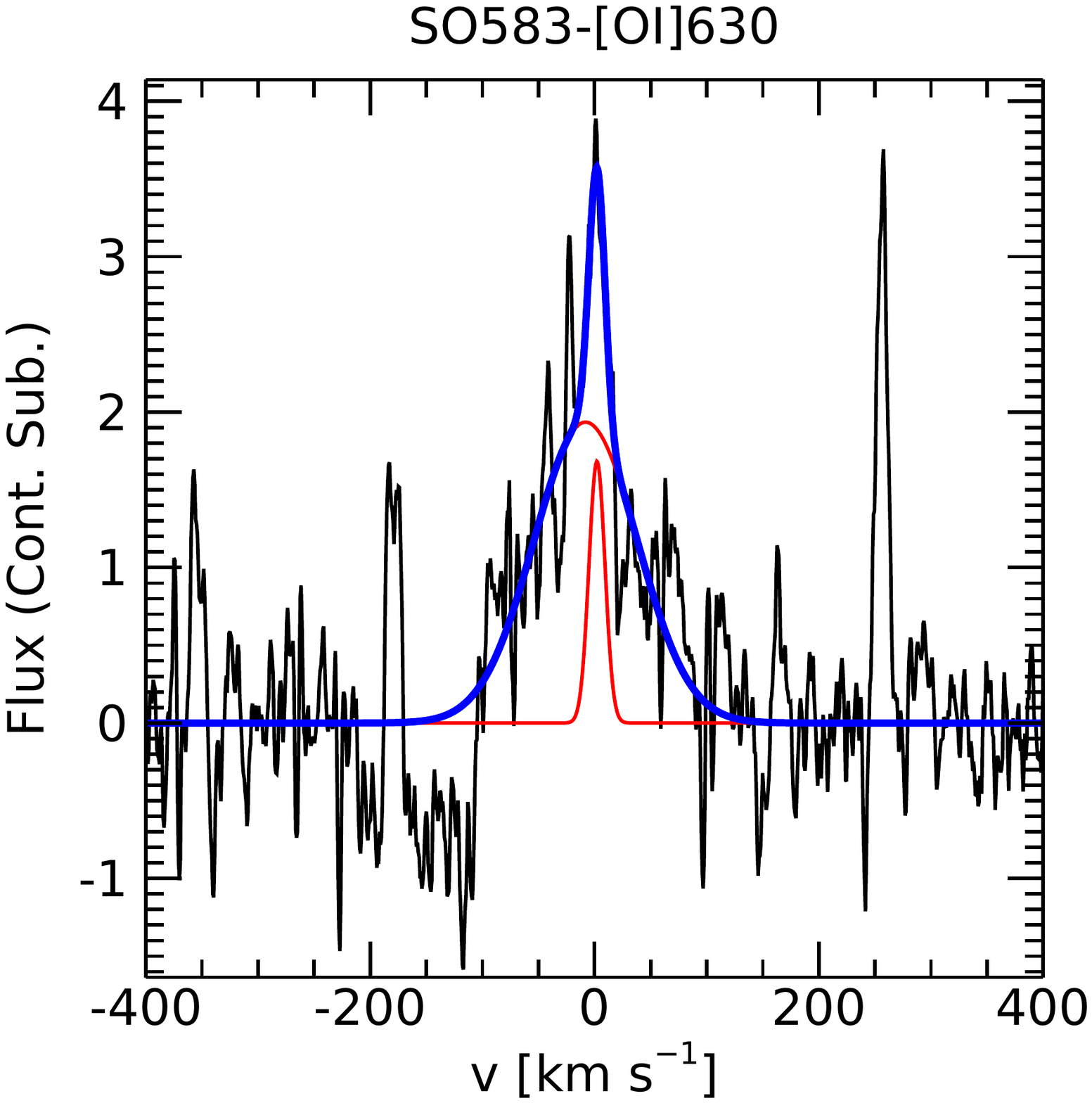}
\includegraphics[trim=20 210 0 70,width=.63\columnwidth, angle=0]{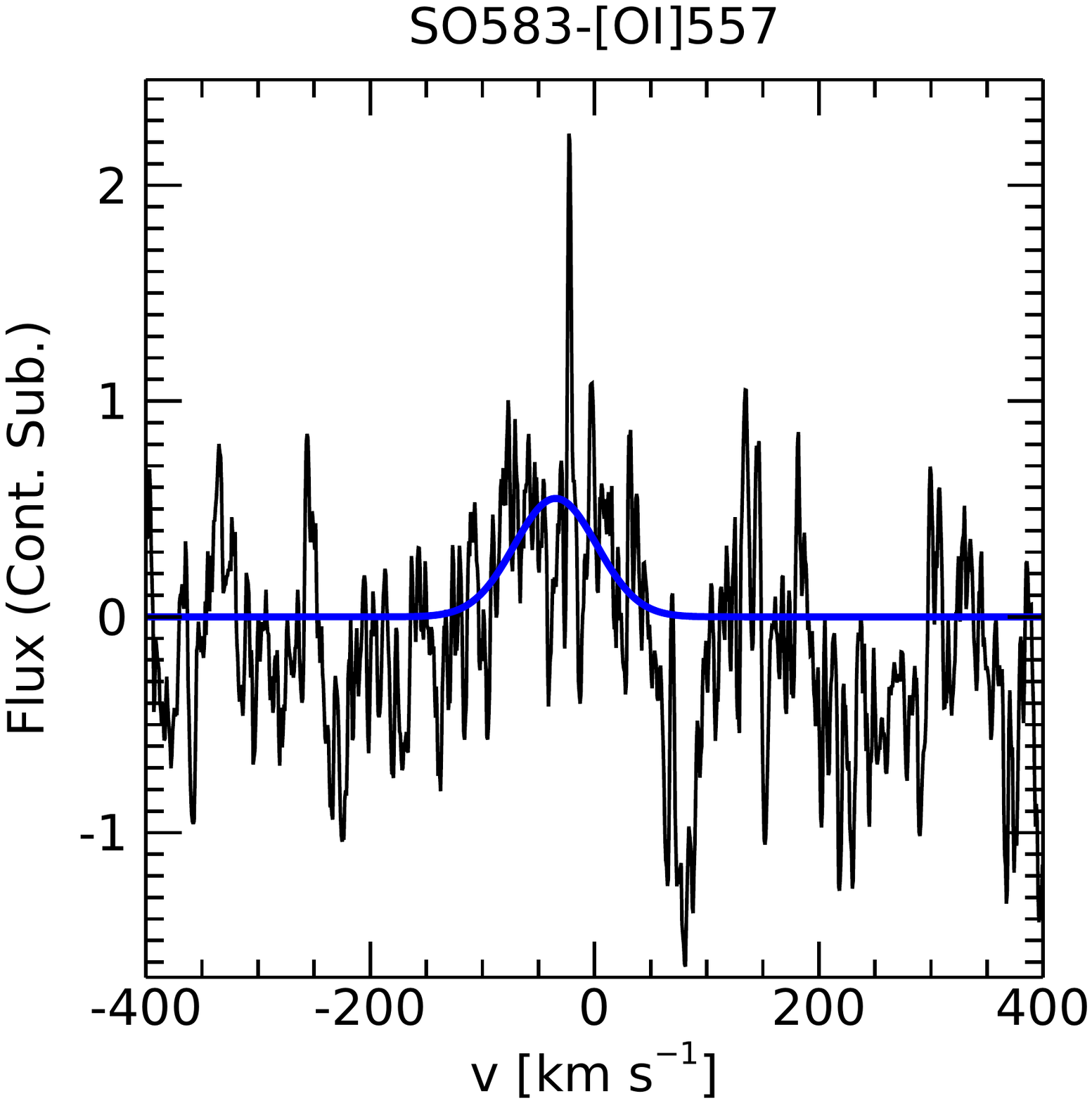}
\includegraphics[trim=20 210 0 70,width=.63\columnwidth, angle=0]{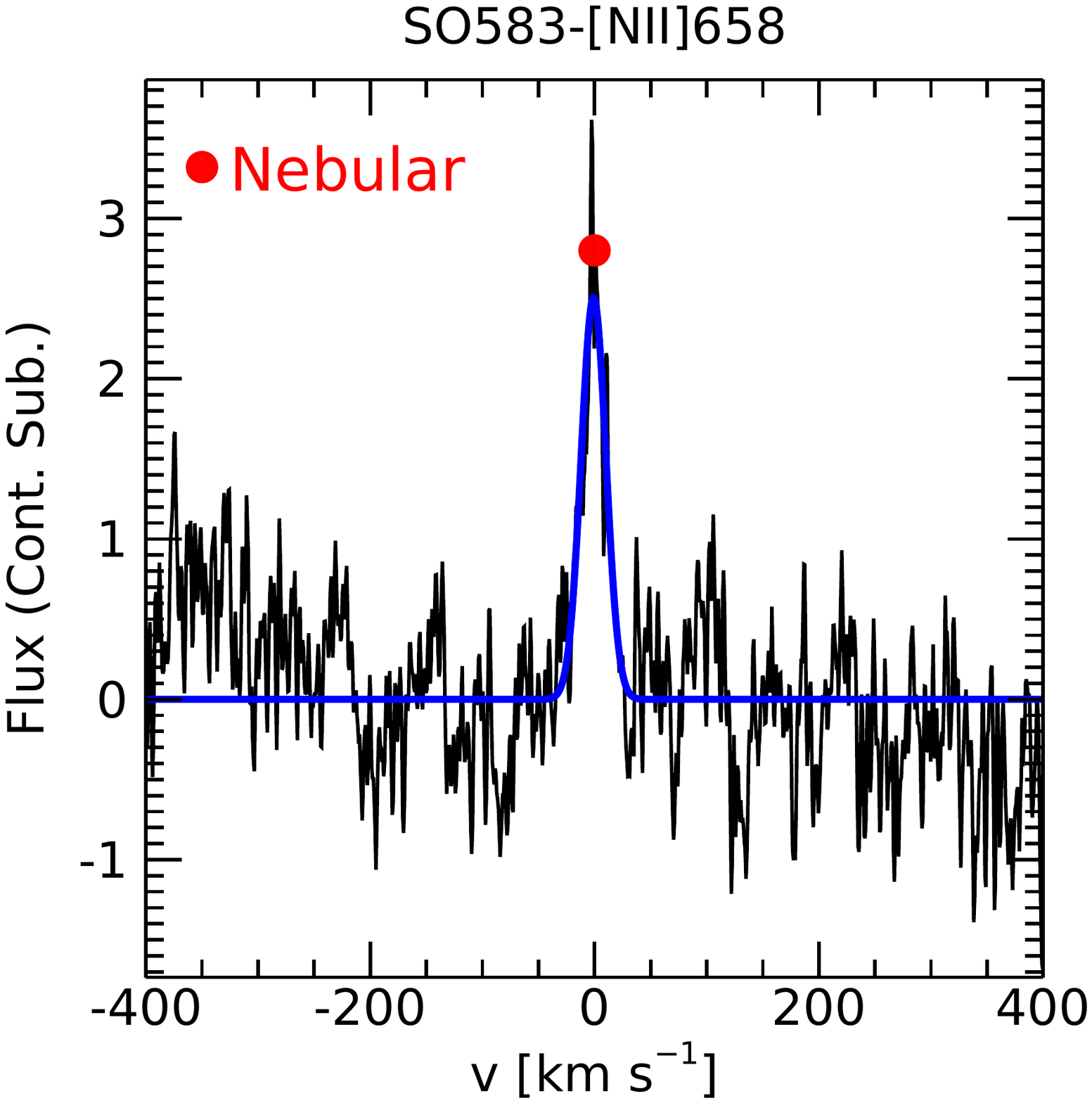}
\includegraphics[trim=20 210 0 70,width=.63\columnwidth, angle=0]{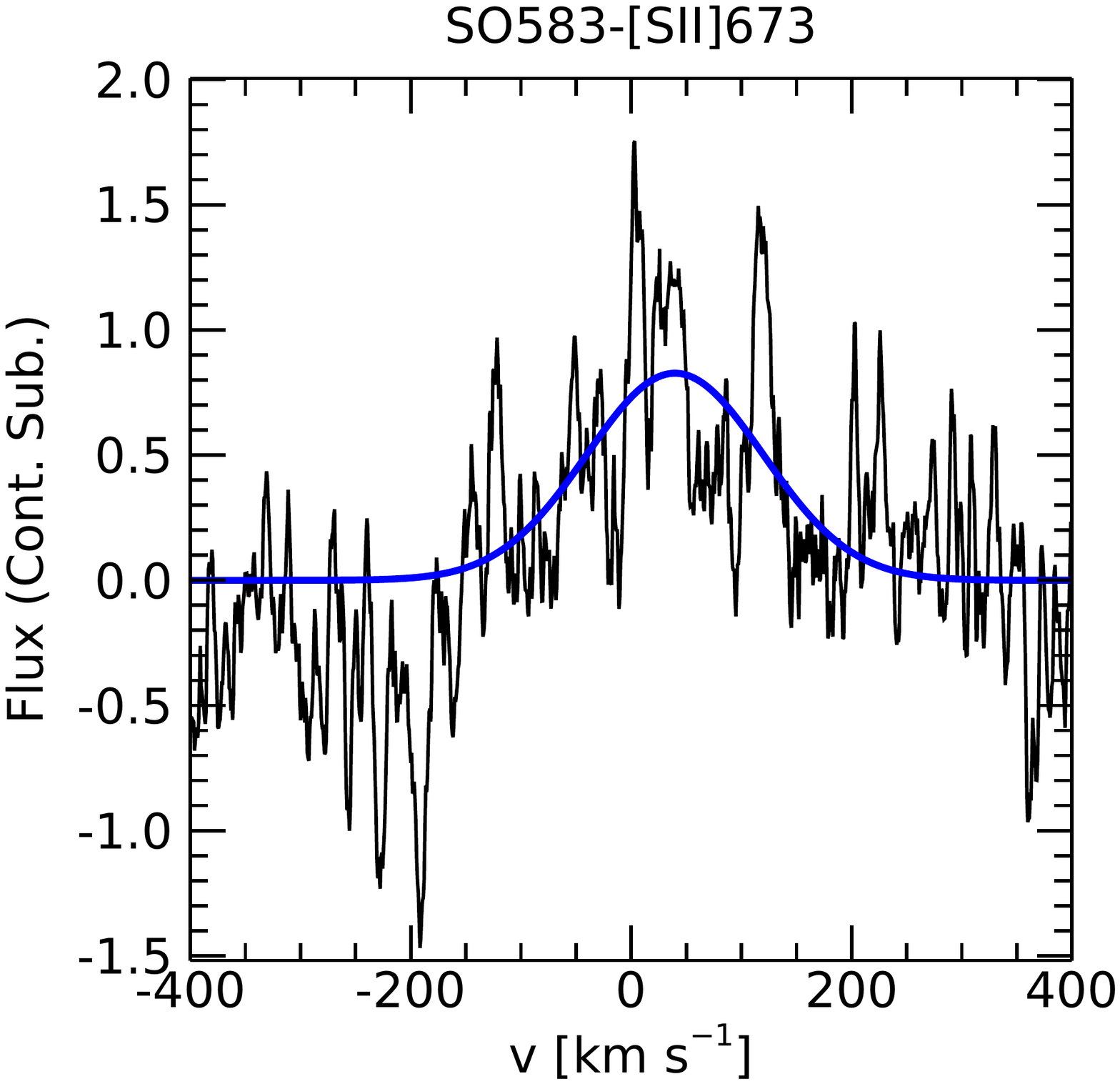}
\includegraphics[trim=20 210 0 70,width=.63\columnwidth, angle=0]{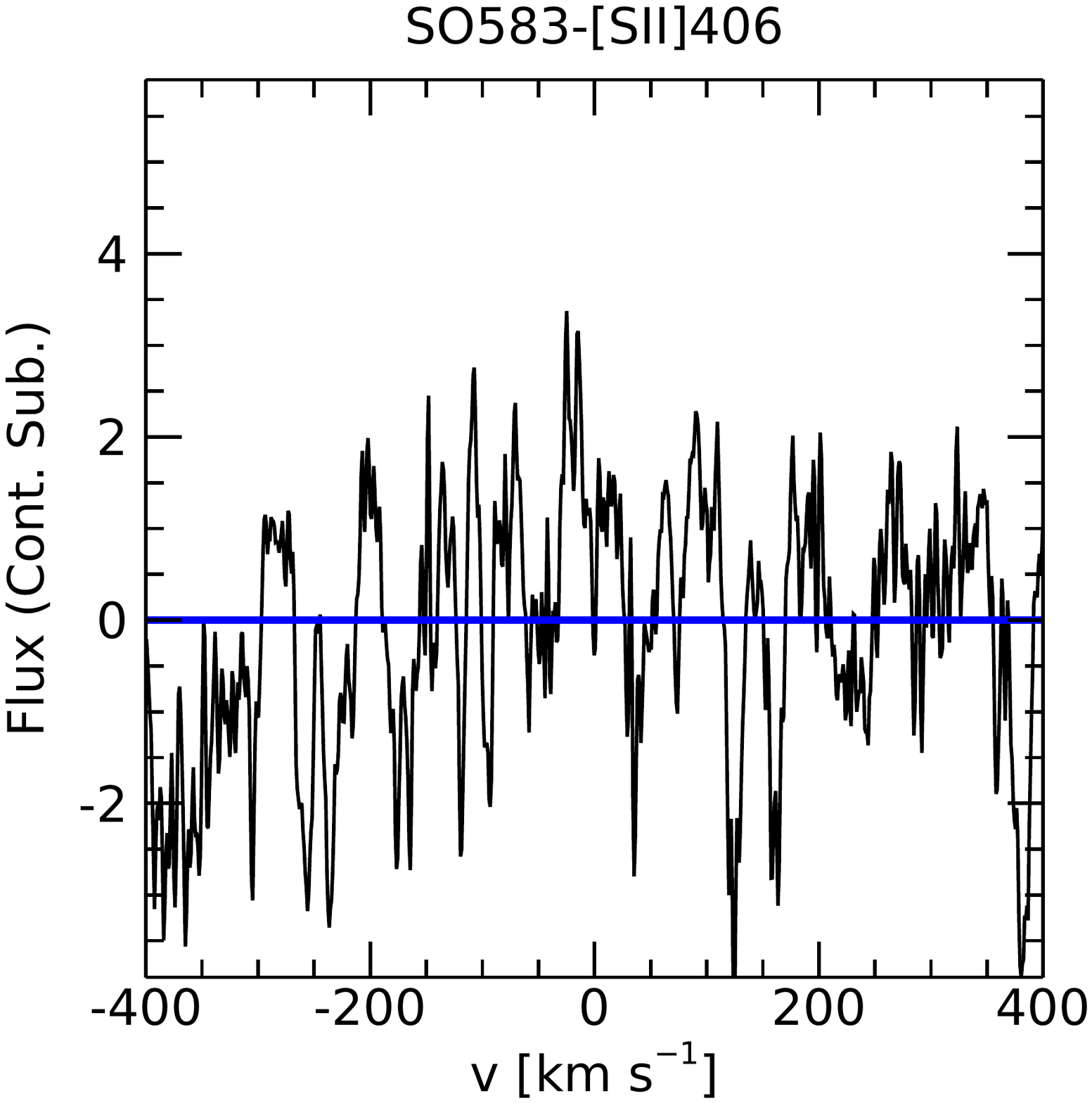}
\end{figure*}

\begin{figure*}[!h]
\includegraphics[trim=20 210 0 70,width=.63\columnwidth, angle=0]{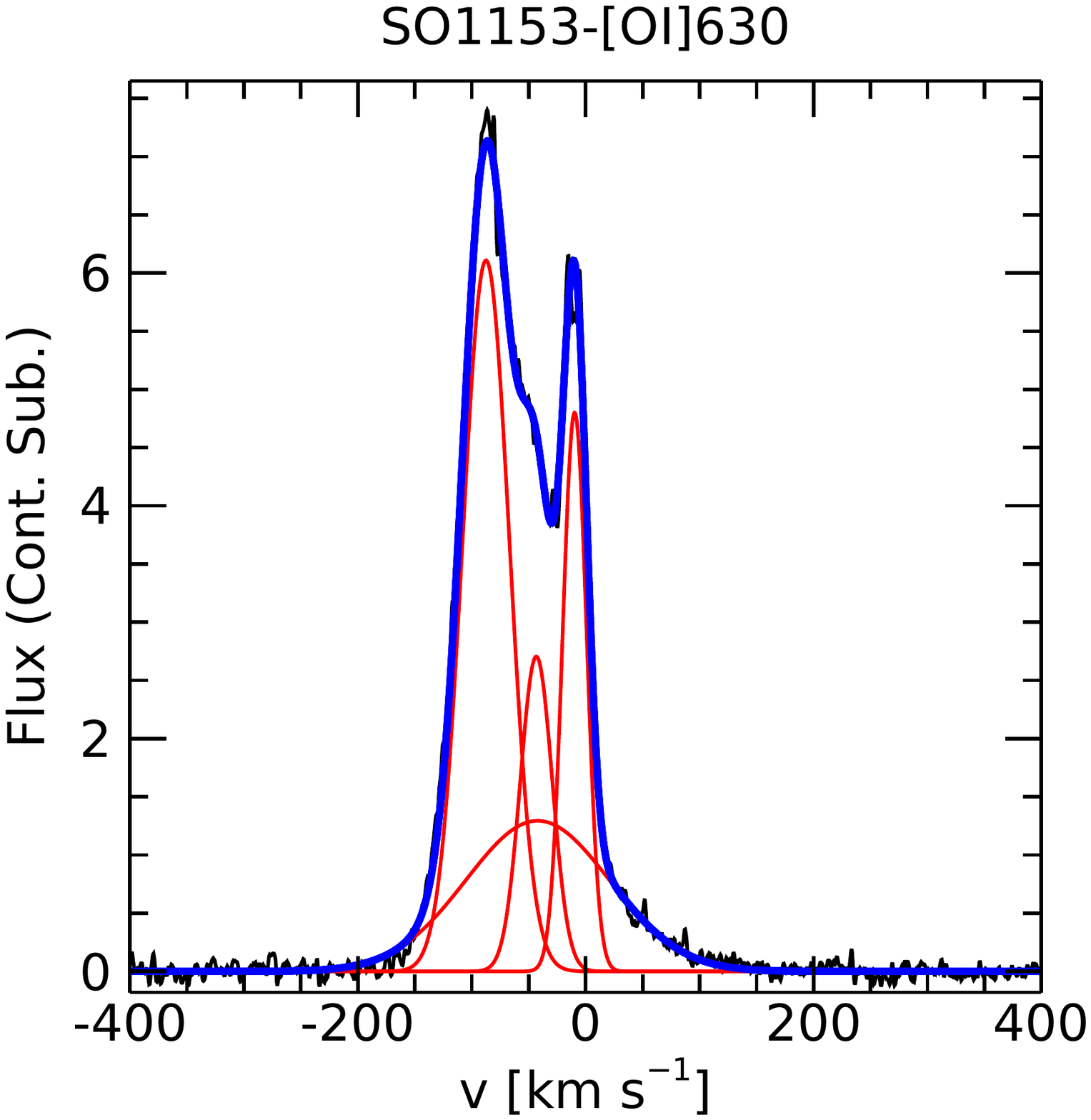}
\includegraphics[trim=20 210 0 70,width=.63\columnwidth, angle=0]{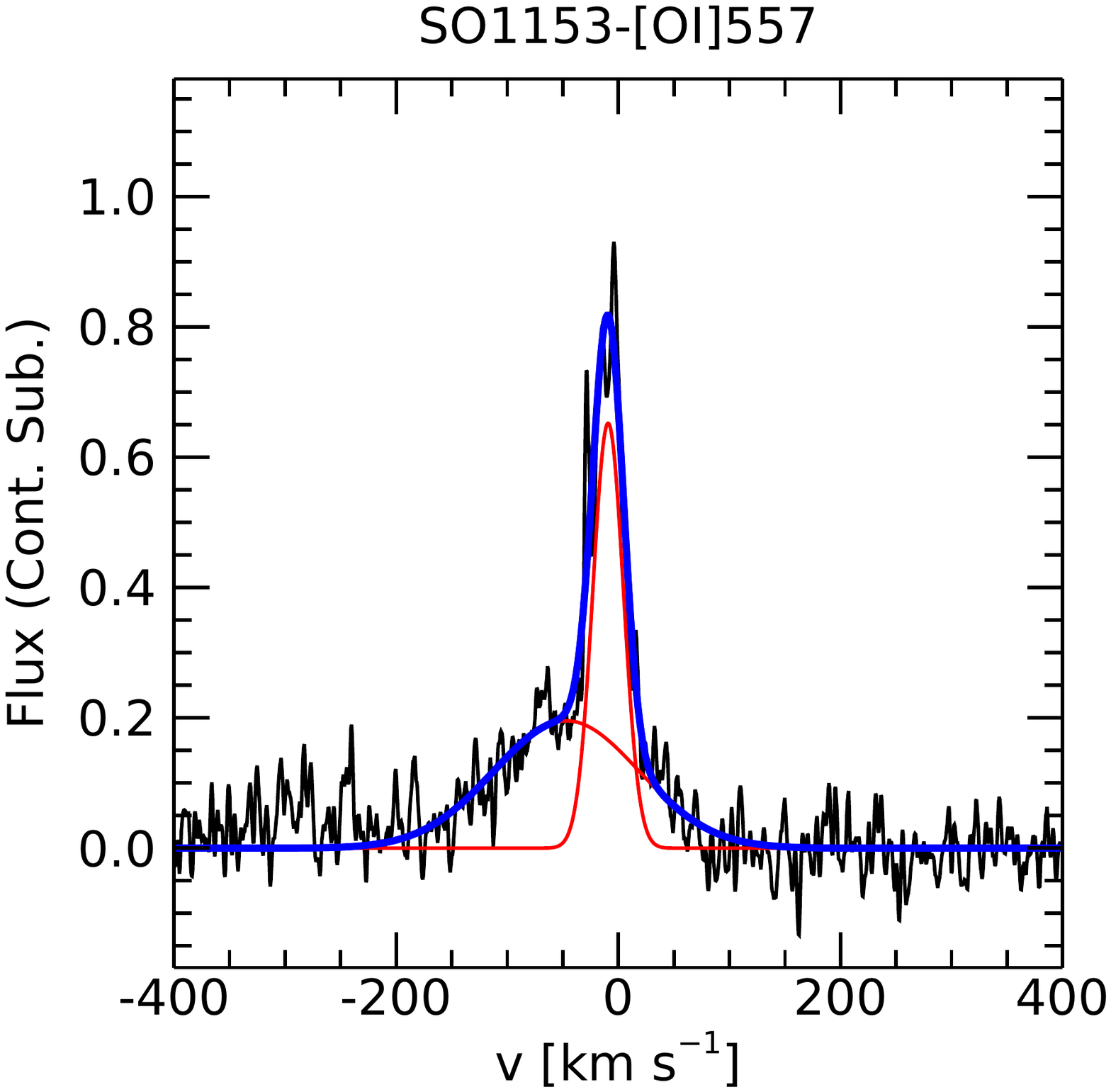}
\includegraphics[trim=20 210 0 70,width=.63\columnwidth, angle=0]{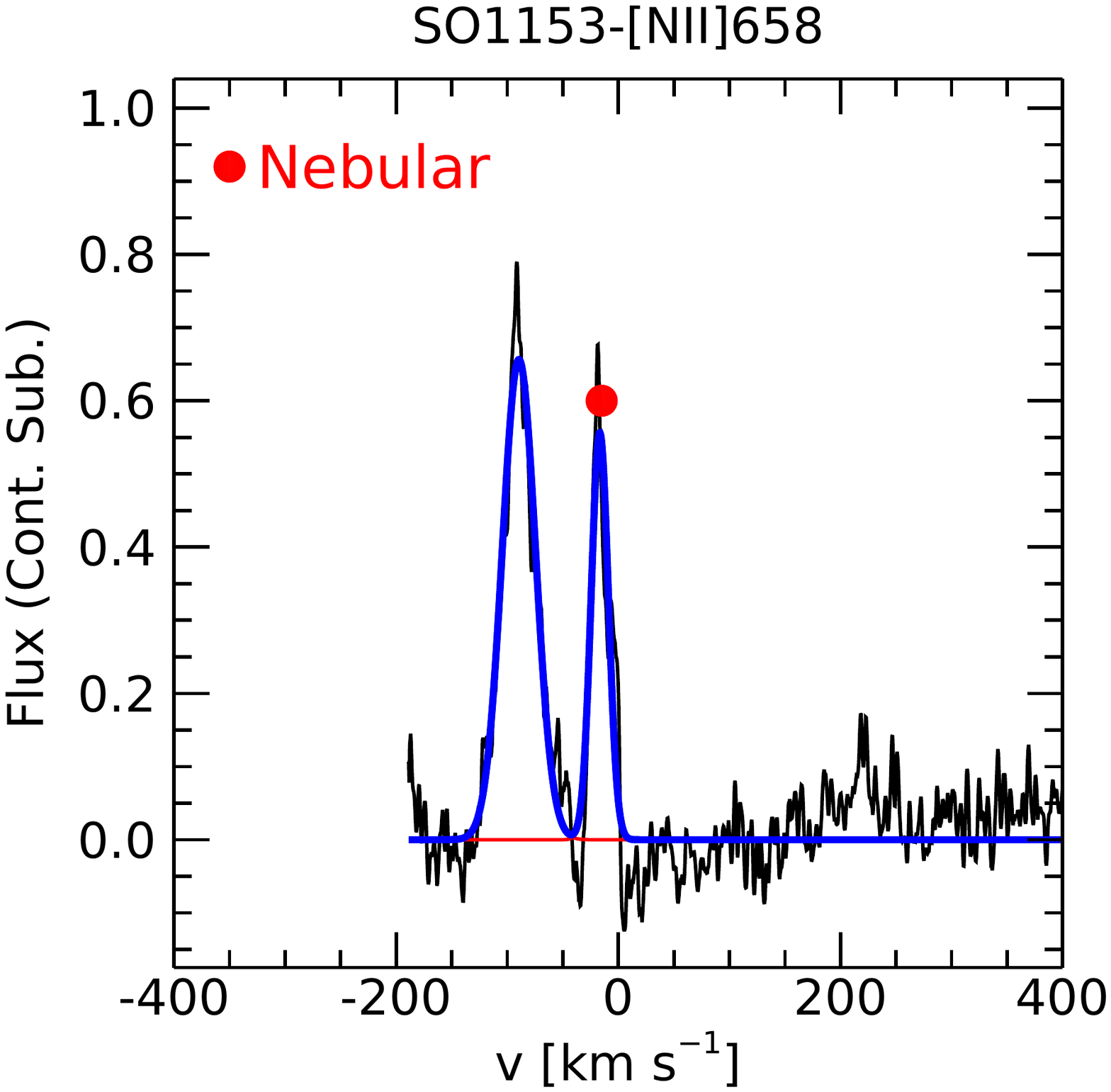}
\includegraphics[trim=20 210 0 70,width=.63\columnwidth, angle=0]{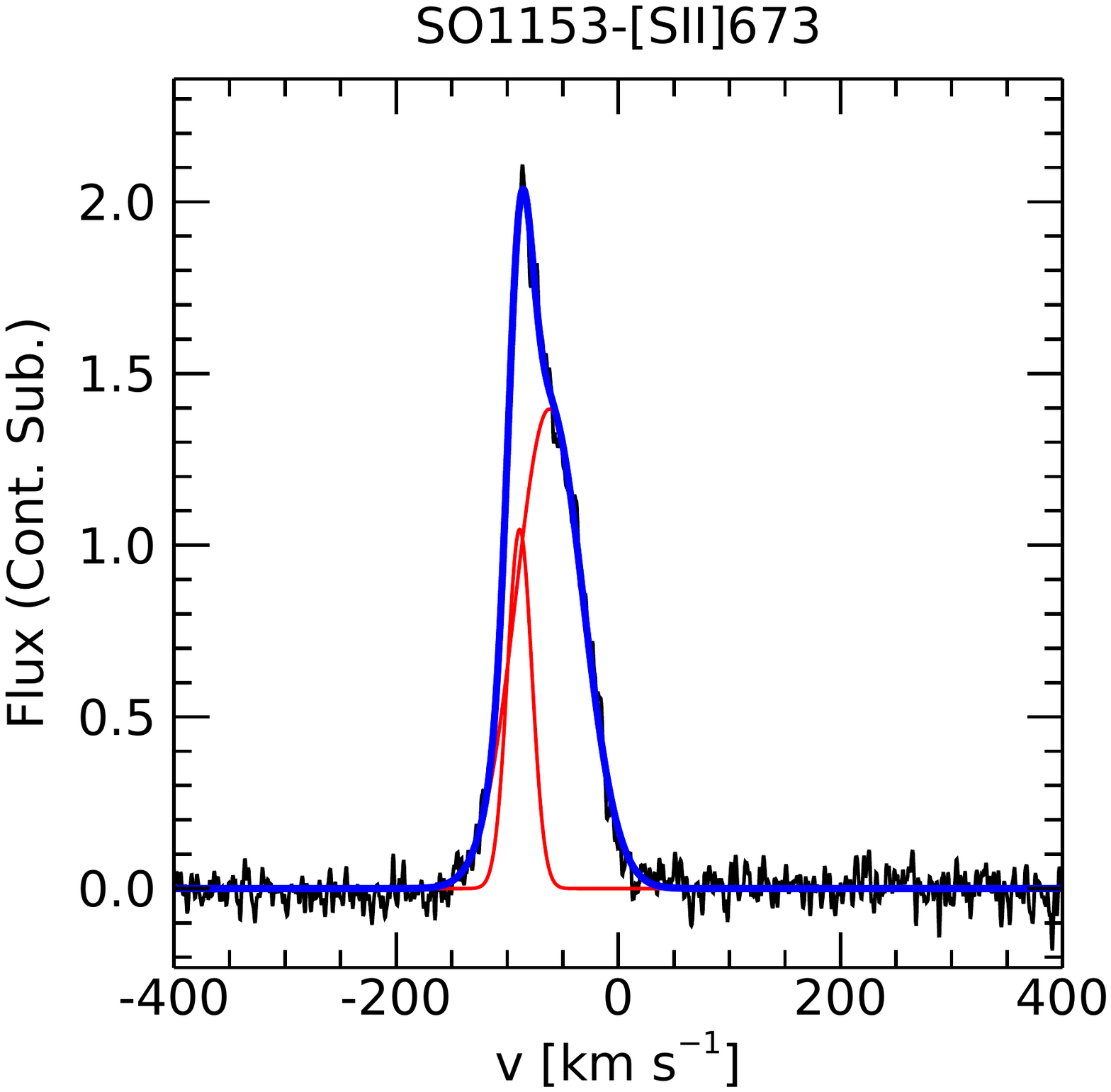}
\includegraphics[trim=20 210 0 70,width=.63\columnwidth, angle=0]{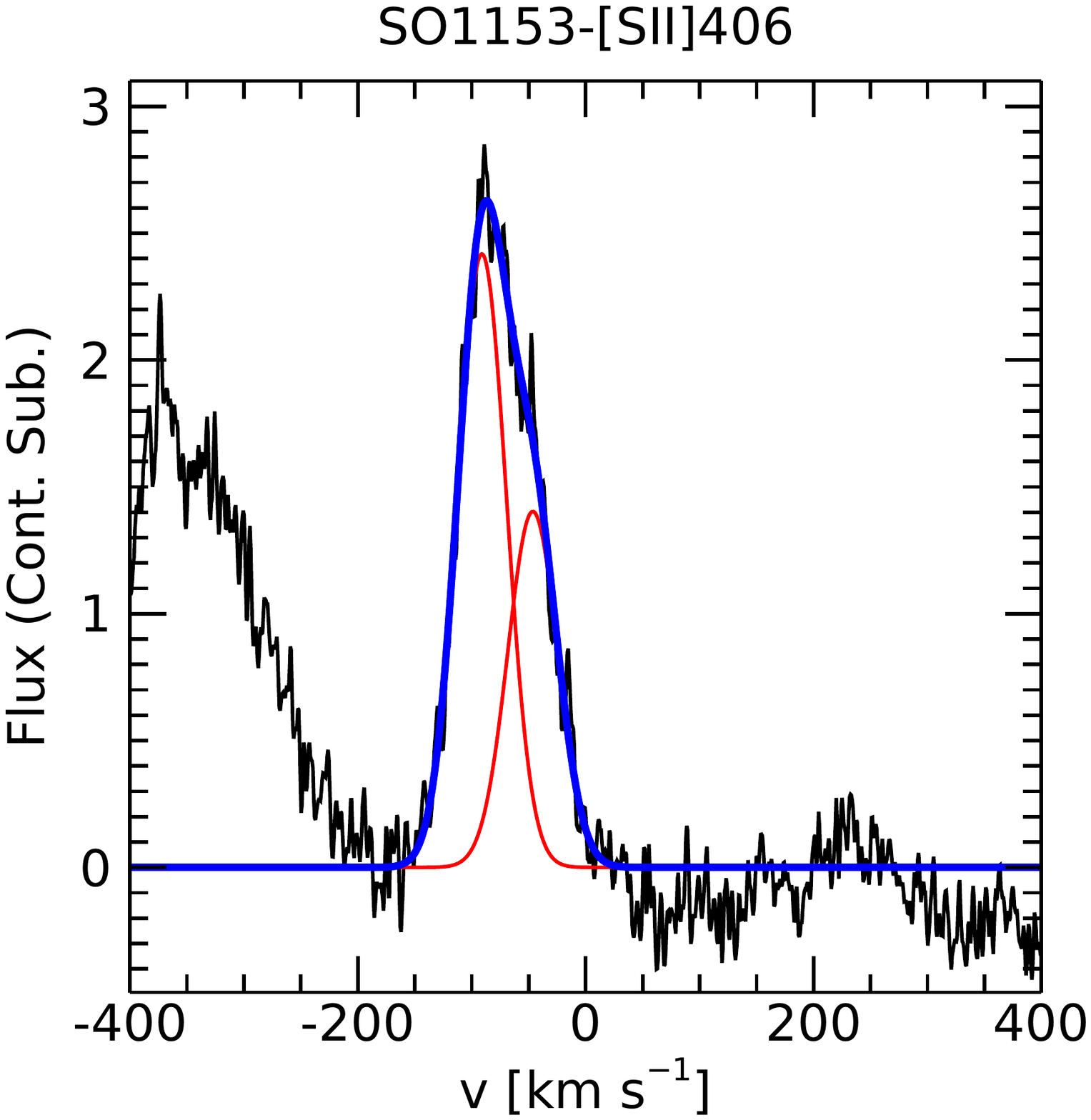}
\begin{center} \textbf{Fig. A.1.} continued.\end{center}
\end{figure*}

\begin{figure*}[!h]
\includegraphics[trim=20 40 0 0,width=.65\columnwidth, angle=0]{Figures/Gauss_decomp_H2/CVSO58_H2_14_final_fit.pdf}
\includegraphics[trim=20 40 0 0,width=.65\columnwidth, angle=0]{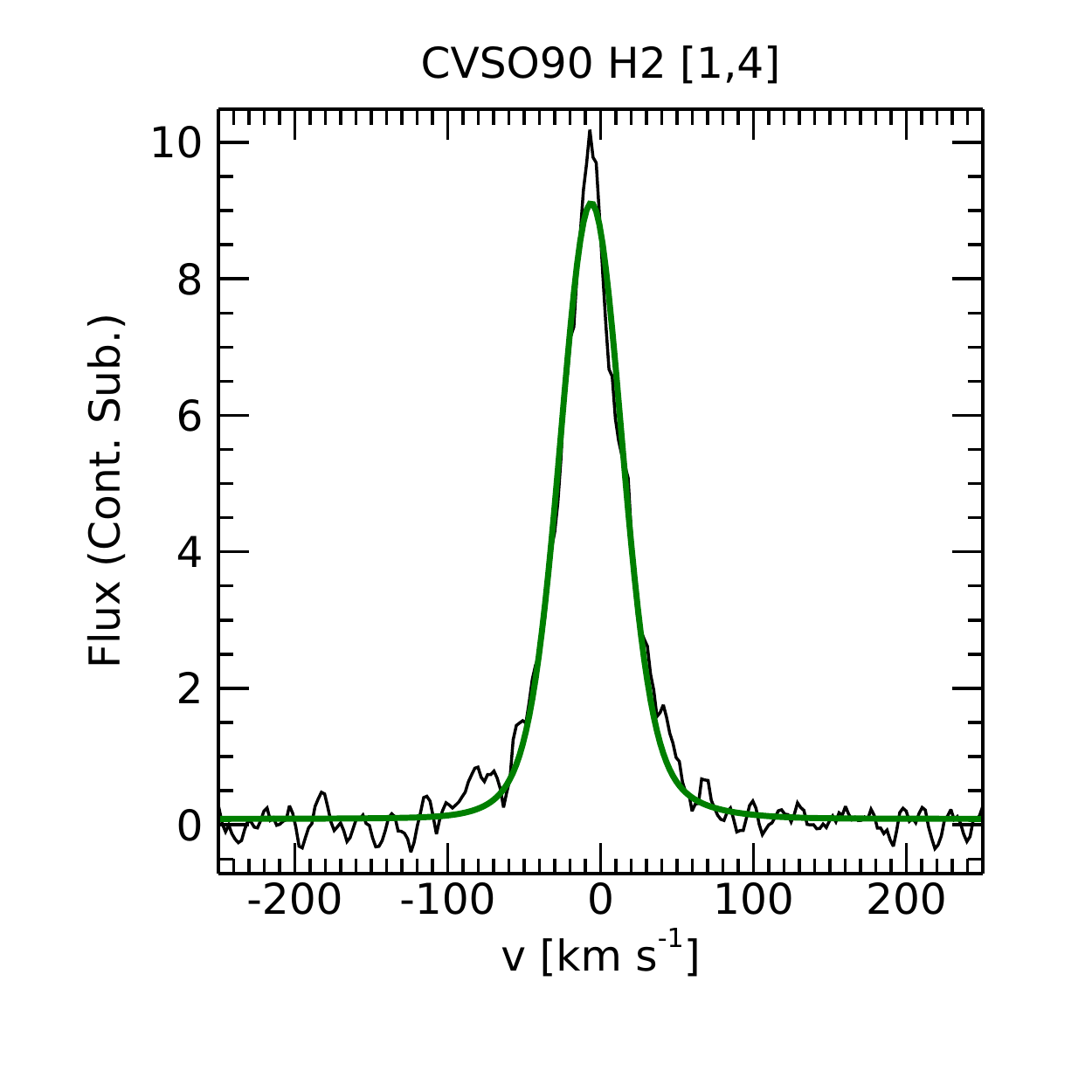}
\includegraphics[trim=20 40 0 0,width=.65\columnwidth, angle=0]{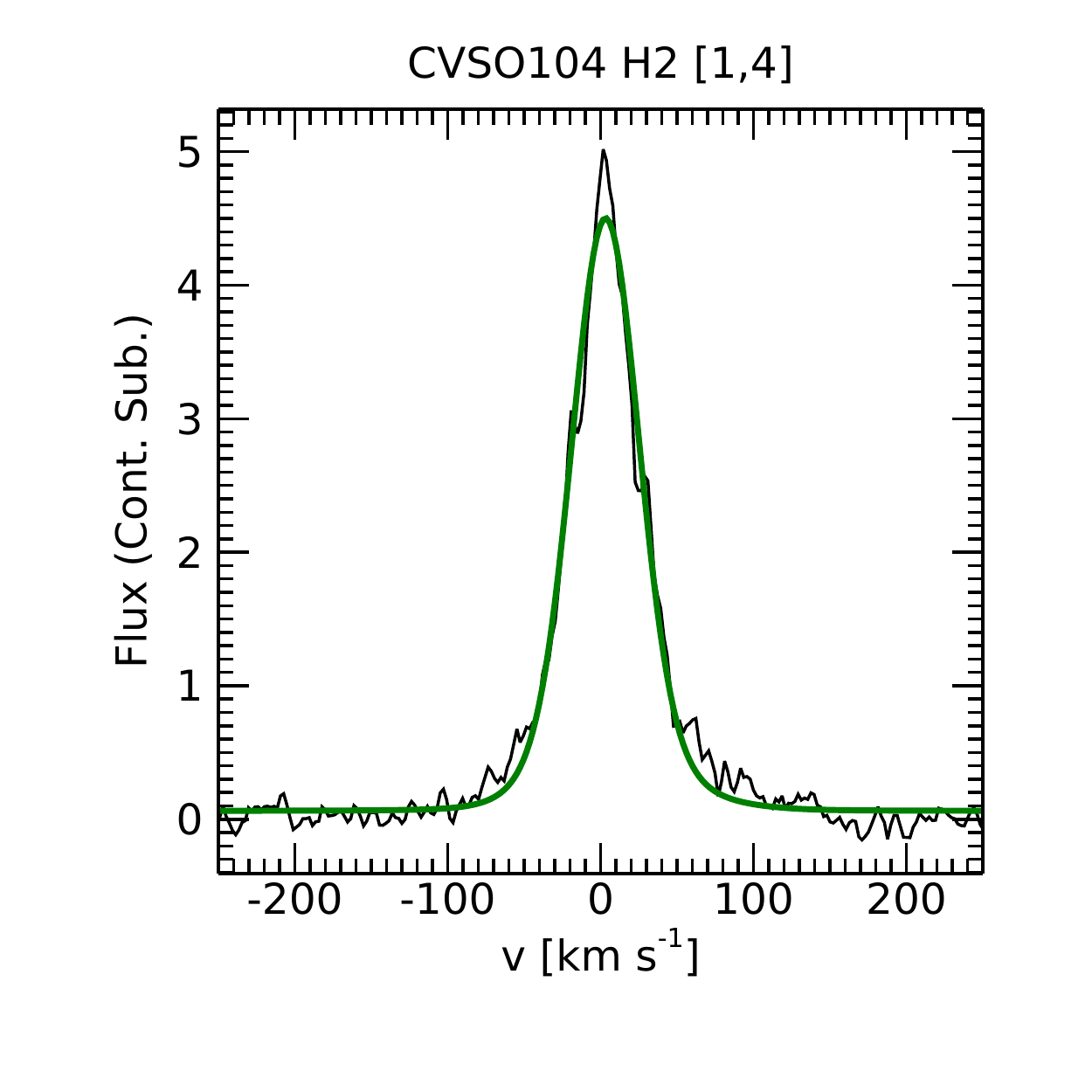}
\includegraphics[trim=20 40 0 0,width=.65\columnwidth, angle=0]{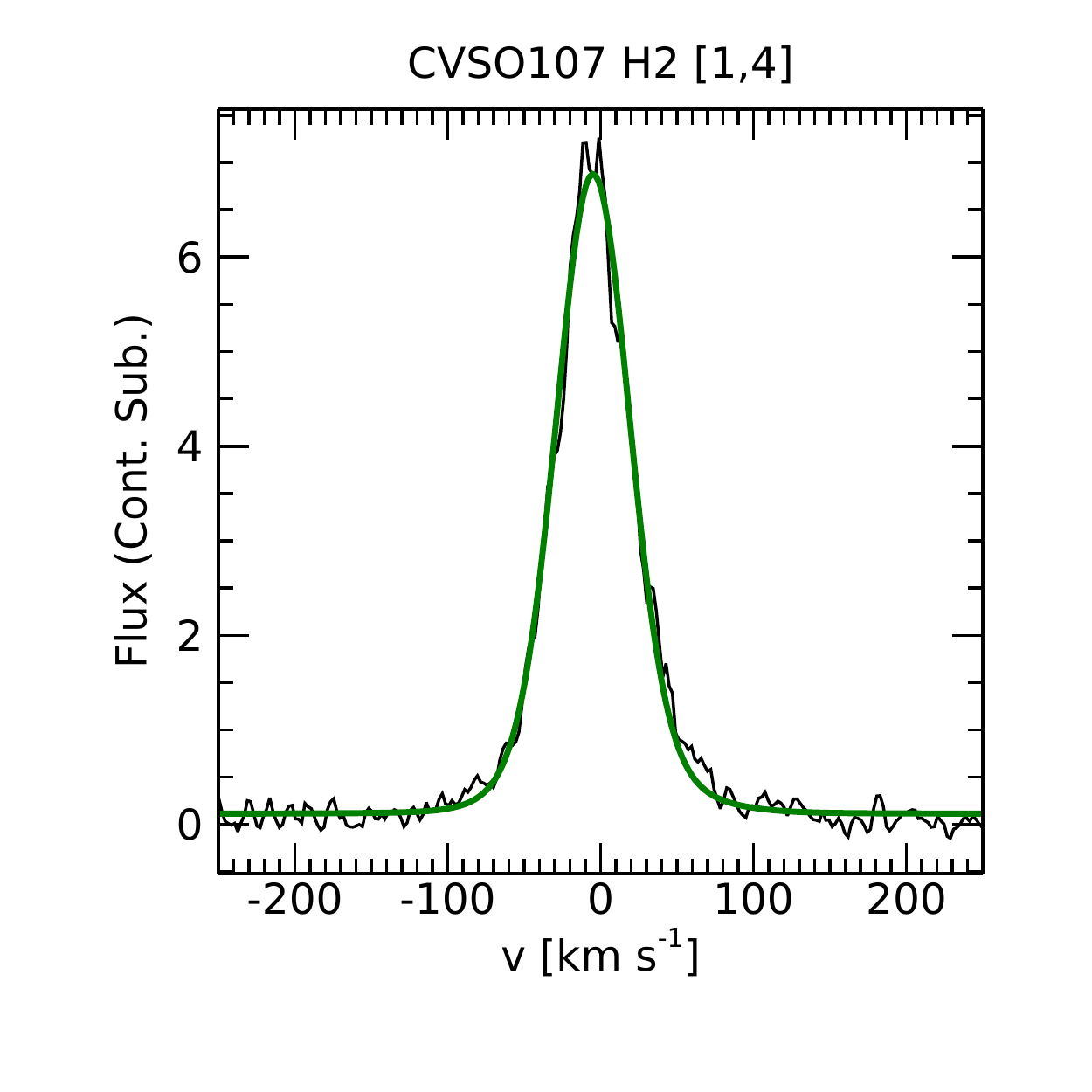}
\includegraphics[trim=20 40 0 0,width=.65\columnwidth, angle=0]{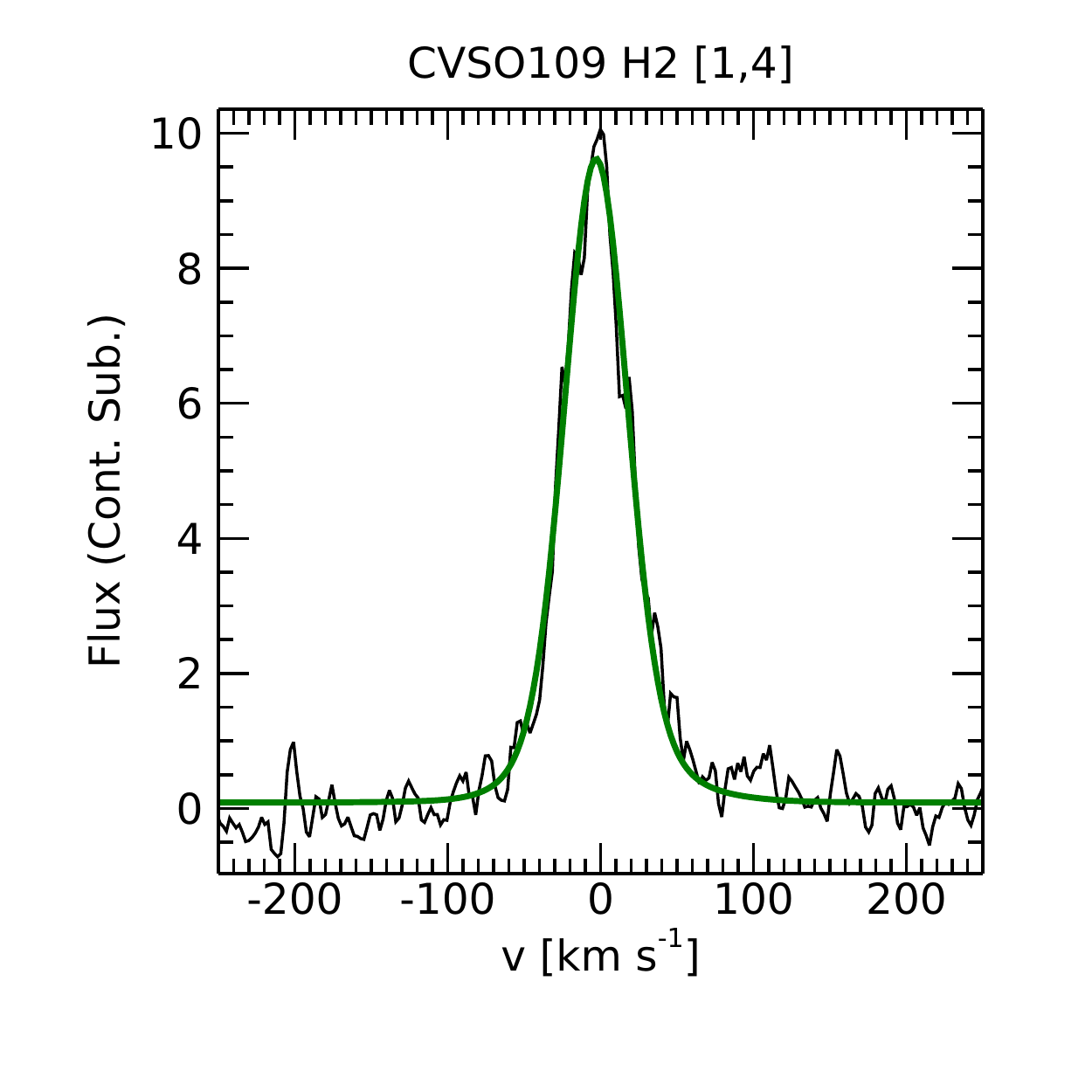}
\includegraphics[trim=20 40 0 0,width=.65\columnwidth, angle=0]{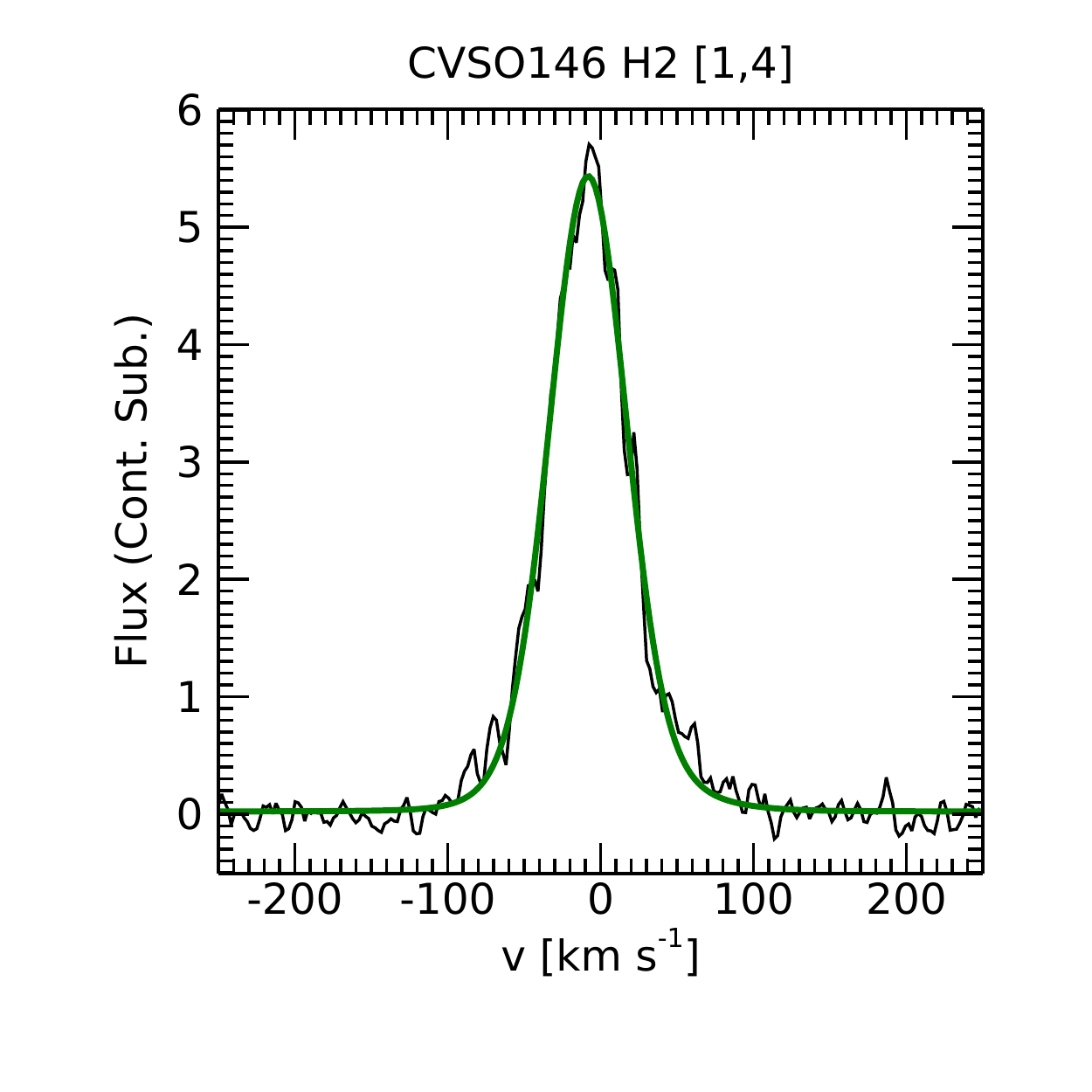}
\includegraphics[trim=20 40 0 0,width=.65\columnwidth, angle=0]{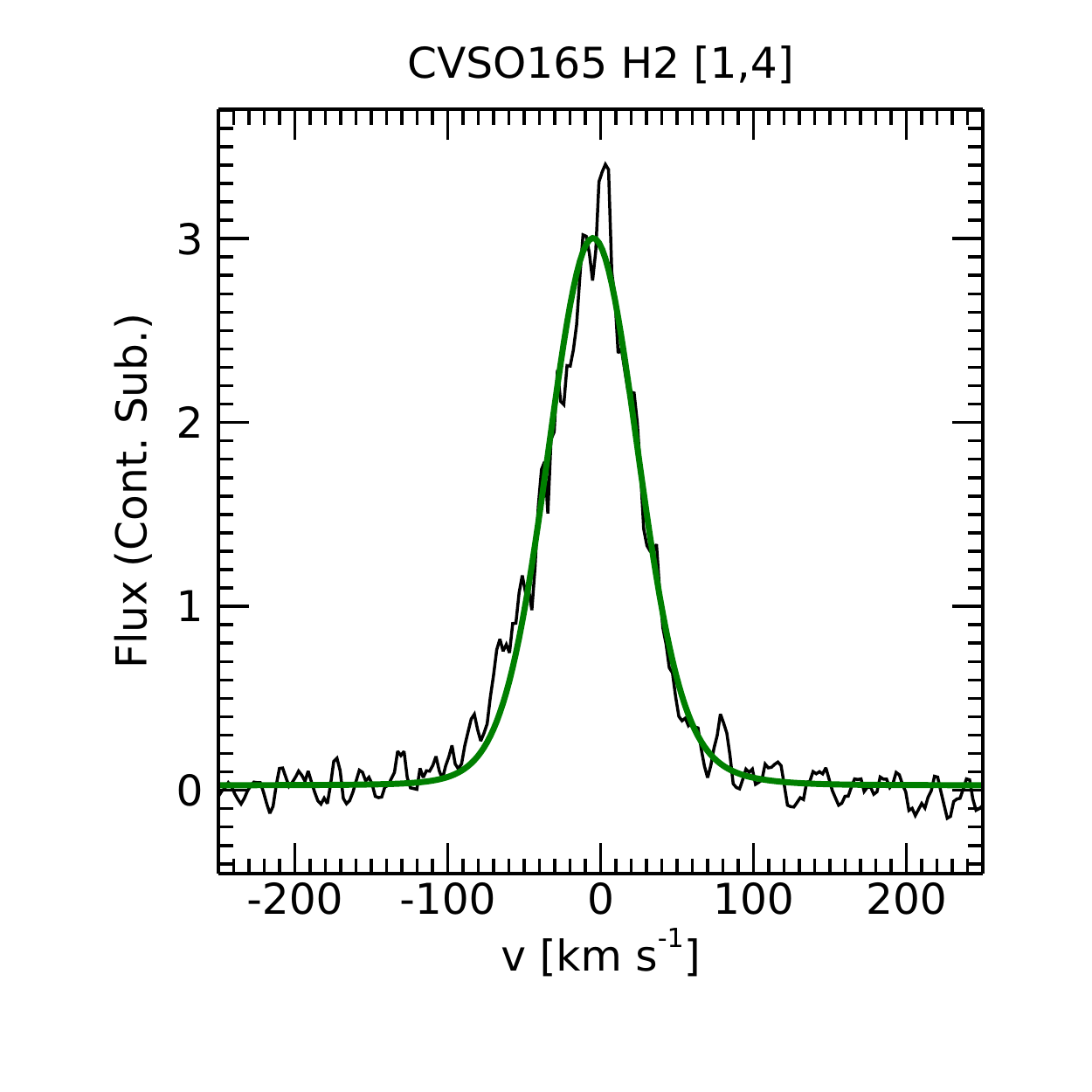}
\includegraphics[trim=20 40 0 0,width=.65\columnwidth, angle=0]{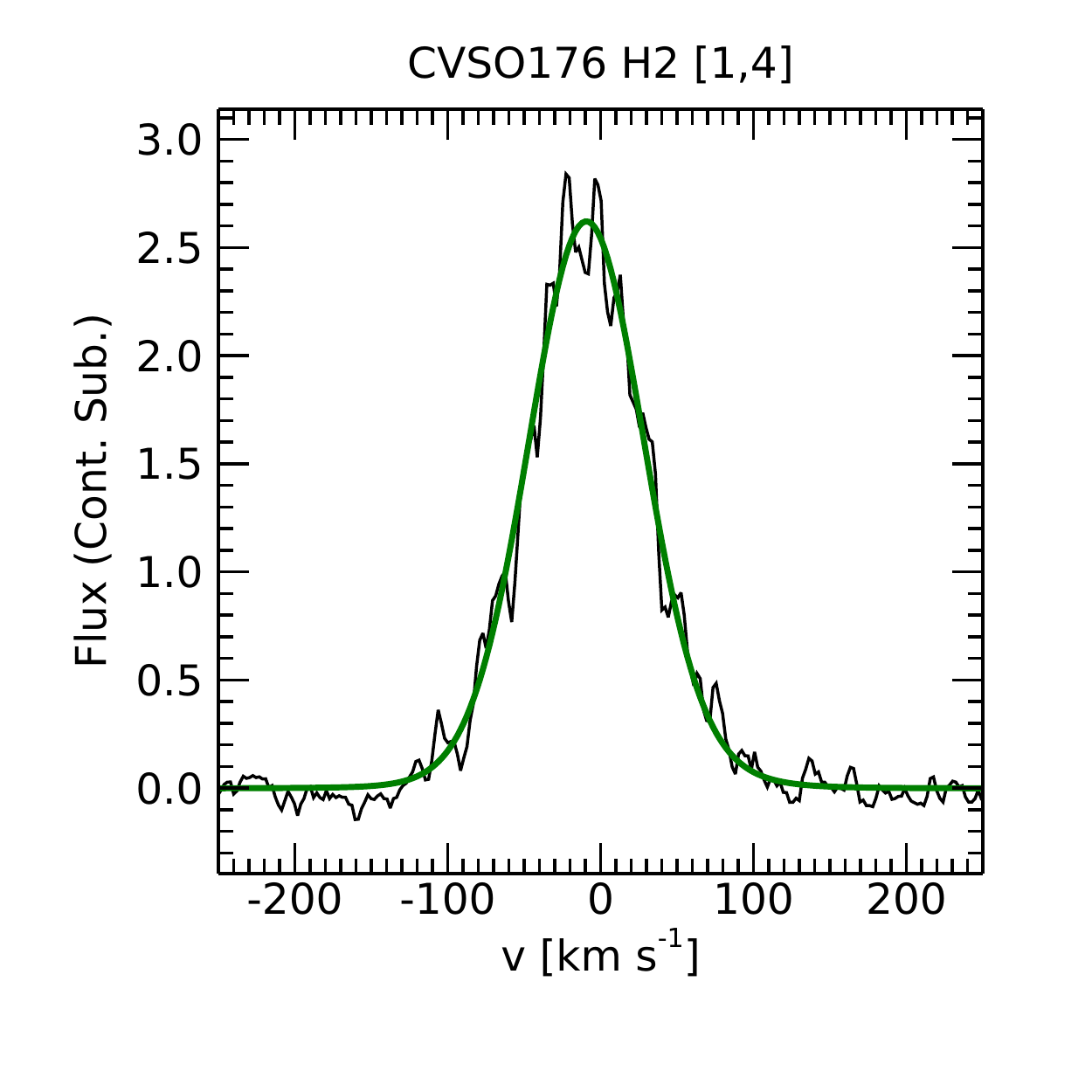}
\includegraphics[trim=20 40 0 0,width=.65\columnwidth, angle=0]{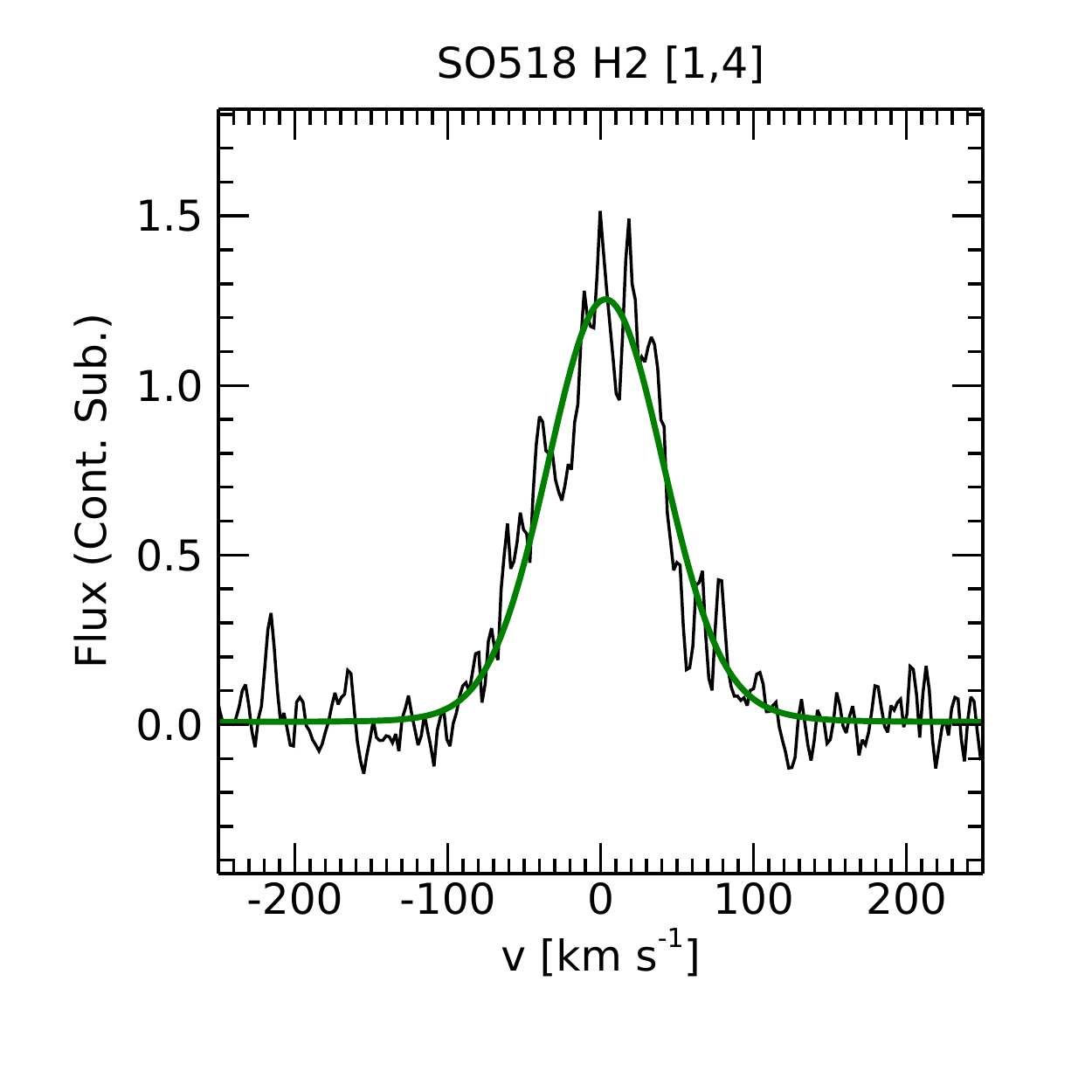}
\includegraphics[trim=20 40 0 0,width=.65\columnwidth, angle=0]{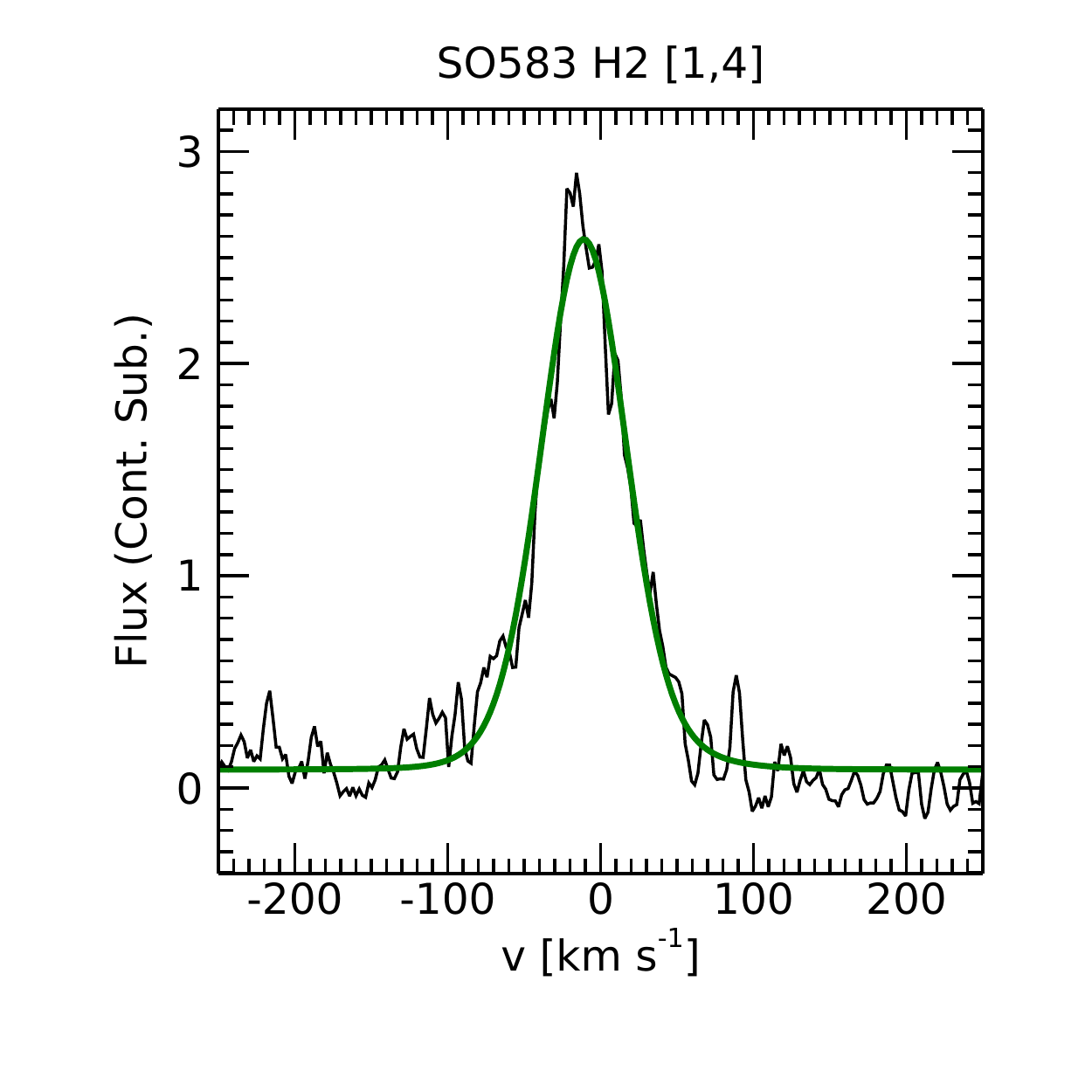}
\begin{center}\caption{\label{fig:complete_sample_H2_14}Continuum-subtracted UV H2 [1,4] averaged profiles. In green we plot the fit to the profile. Flux units are $\rm 10^{-15} ergs^{-1} cm^{-2} \AA^{-1}$. For each panel we indicate the target name and the line diagnostics.}\end{center}
\end{figure*}

\begin{figure*}[!h]
\includegraphics[trim=20 40 0 0,width=.65\columnwidth, angle=0]{Figures/Gauss_decomp_H2/CVSO58_H2_17_final_fit.pdf}
\includegraphics[trim=20 40 0 0,width=.65\columnwidth, angle=0]{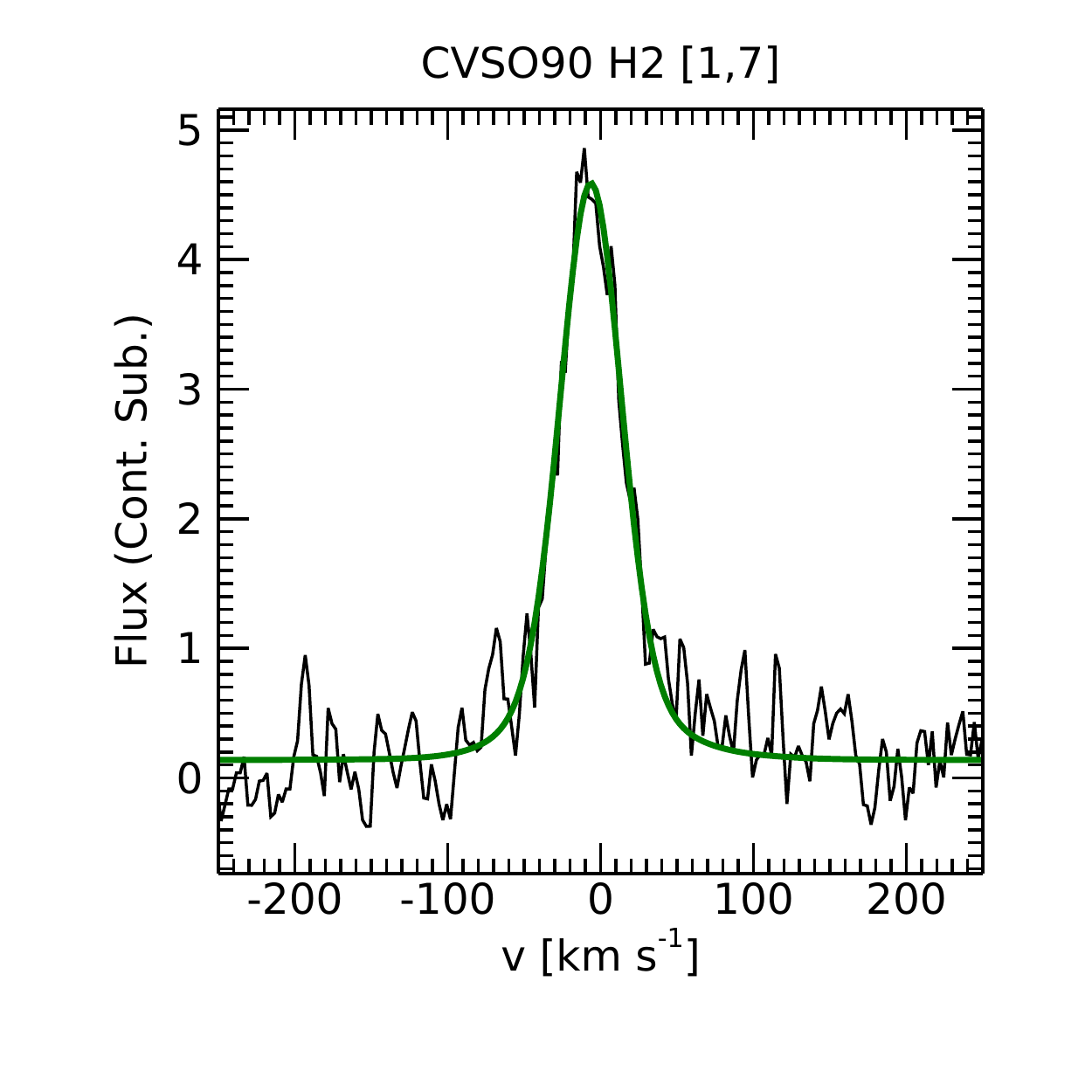}
\includegraphics[trim=20 40 0 0,width=.65\columnwidth, angle=0]{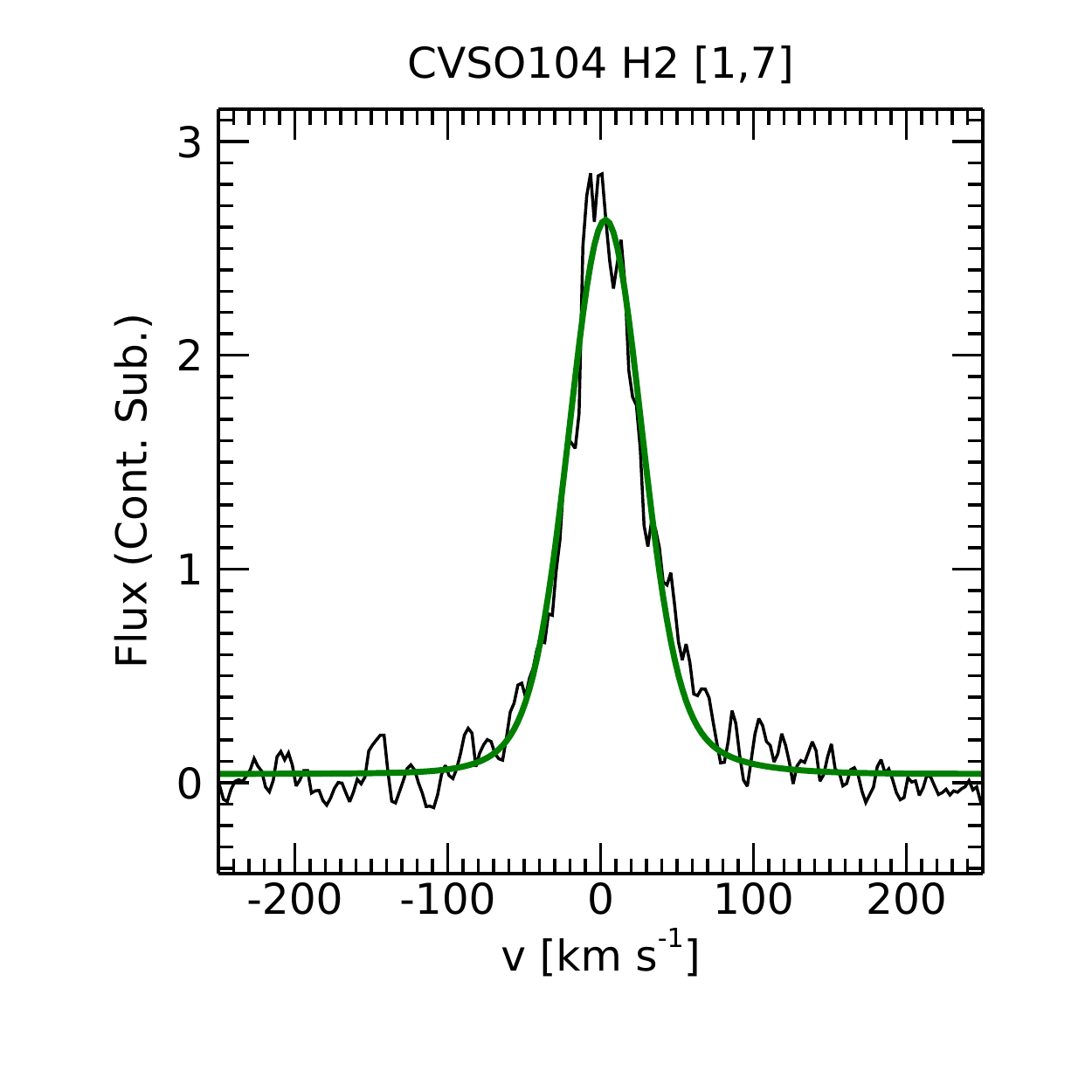}
\includegraphics[trim=20 40 0 0,width=.65\columnwidth, angle=0]{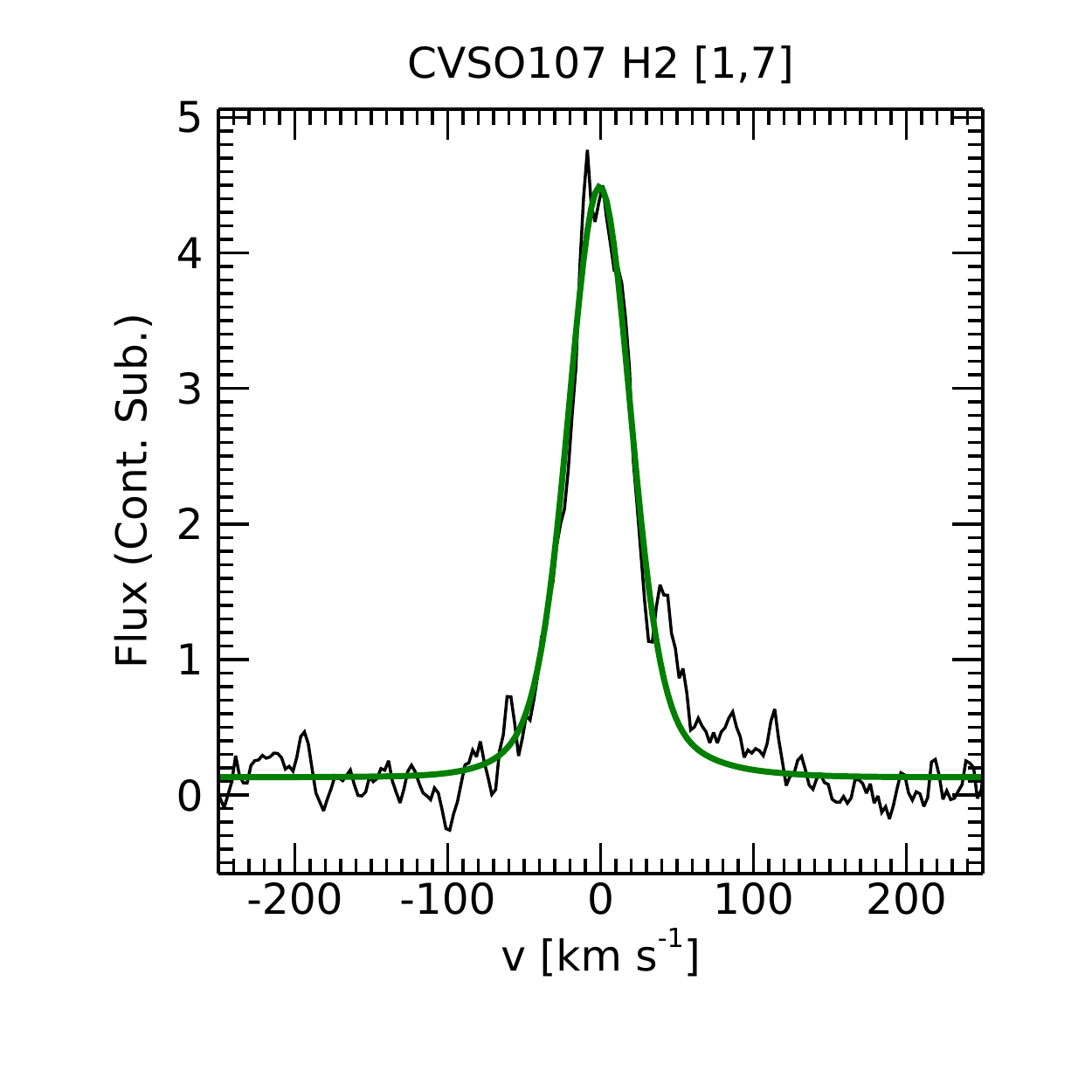}
\includegraphics[trim=20 40 0 0,width=.65\columnwidth, angle=0]{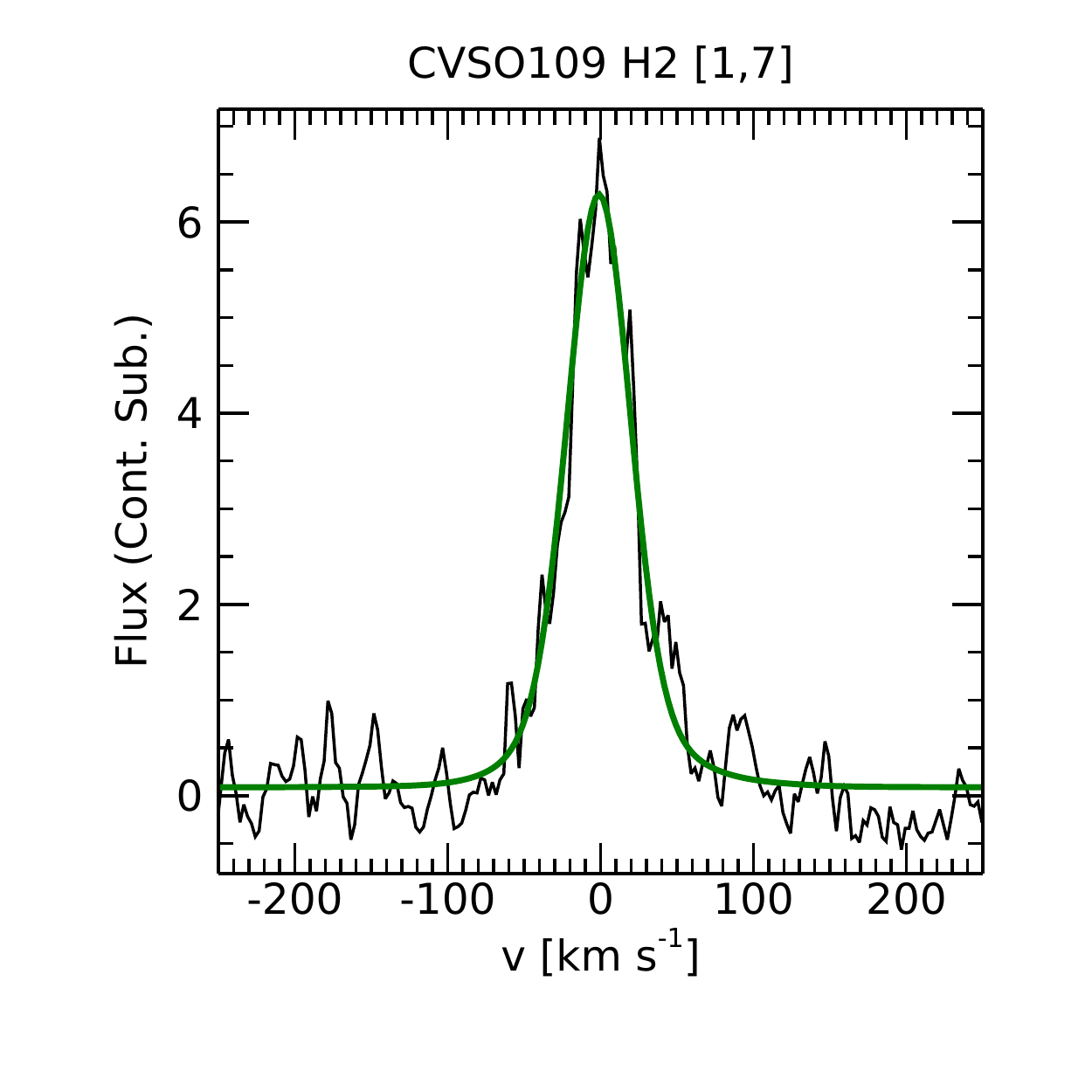}
\includegraphics[trim=20 40 0 0,width=.65\columnwidth, angle=0]{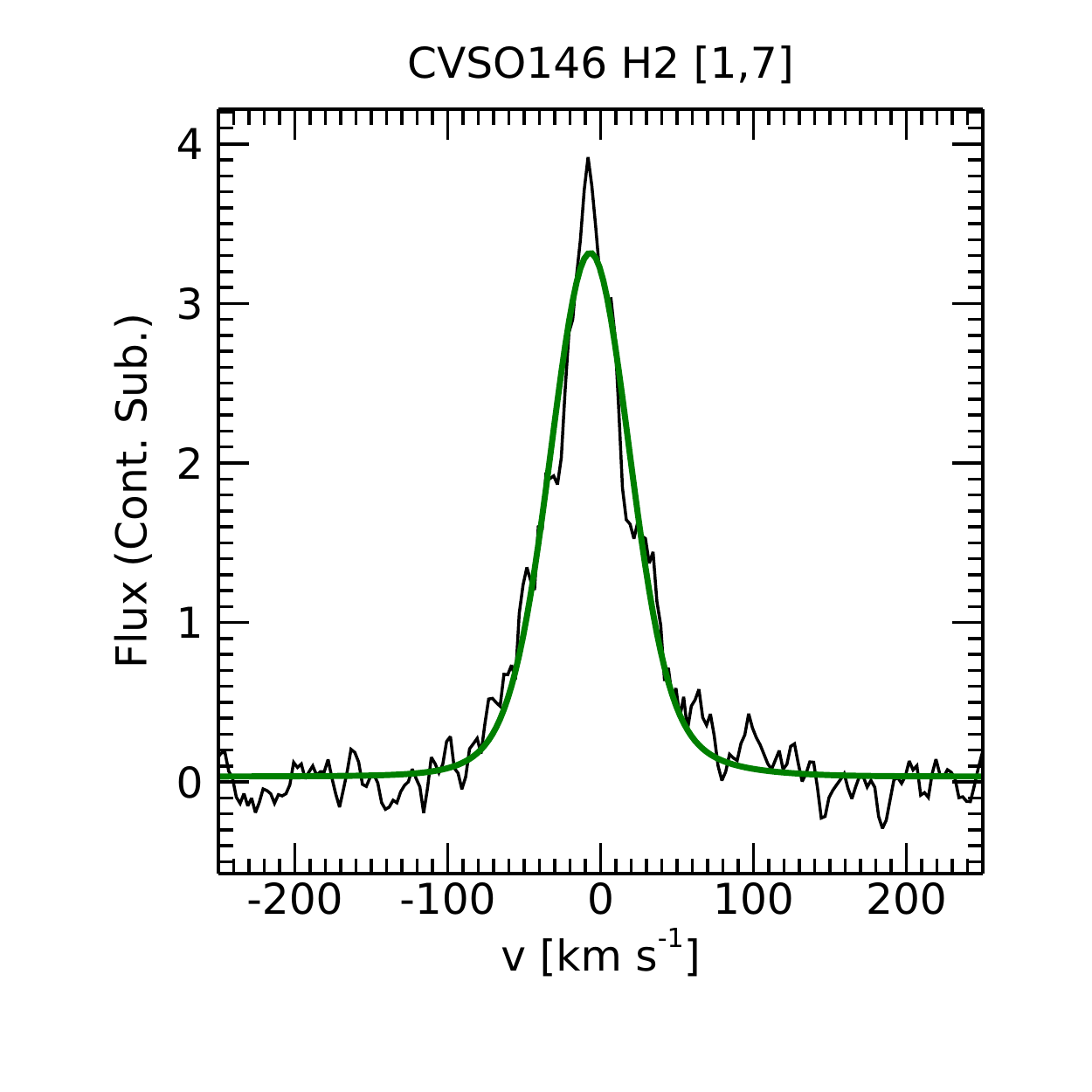}
\includegraphics[trim=20 40 0 0,width=.65\columnwidth, angle=0]{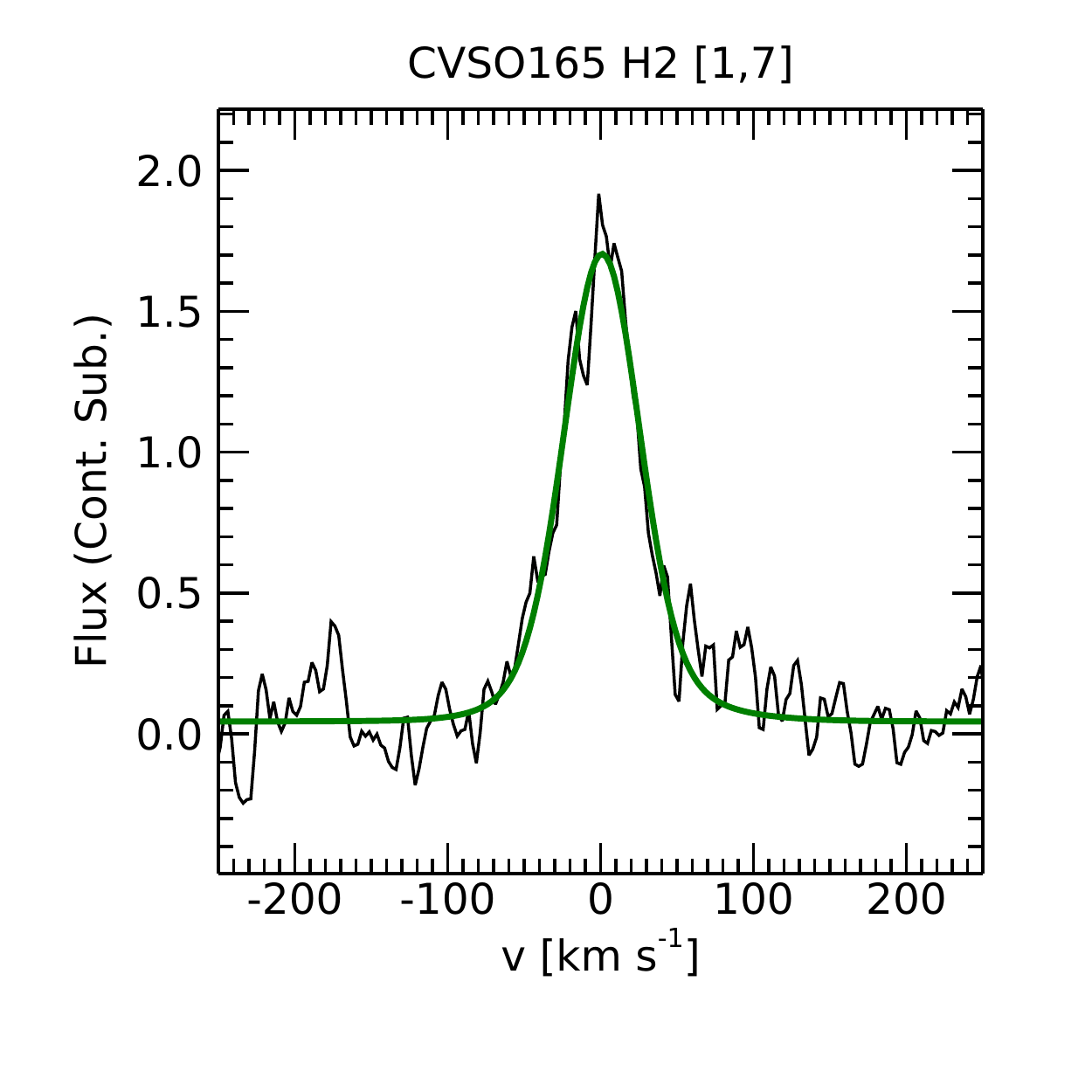}
\includegraphics[trim=20 40 0 0,width=.65\columnwidth, angle=0]{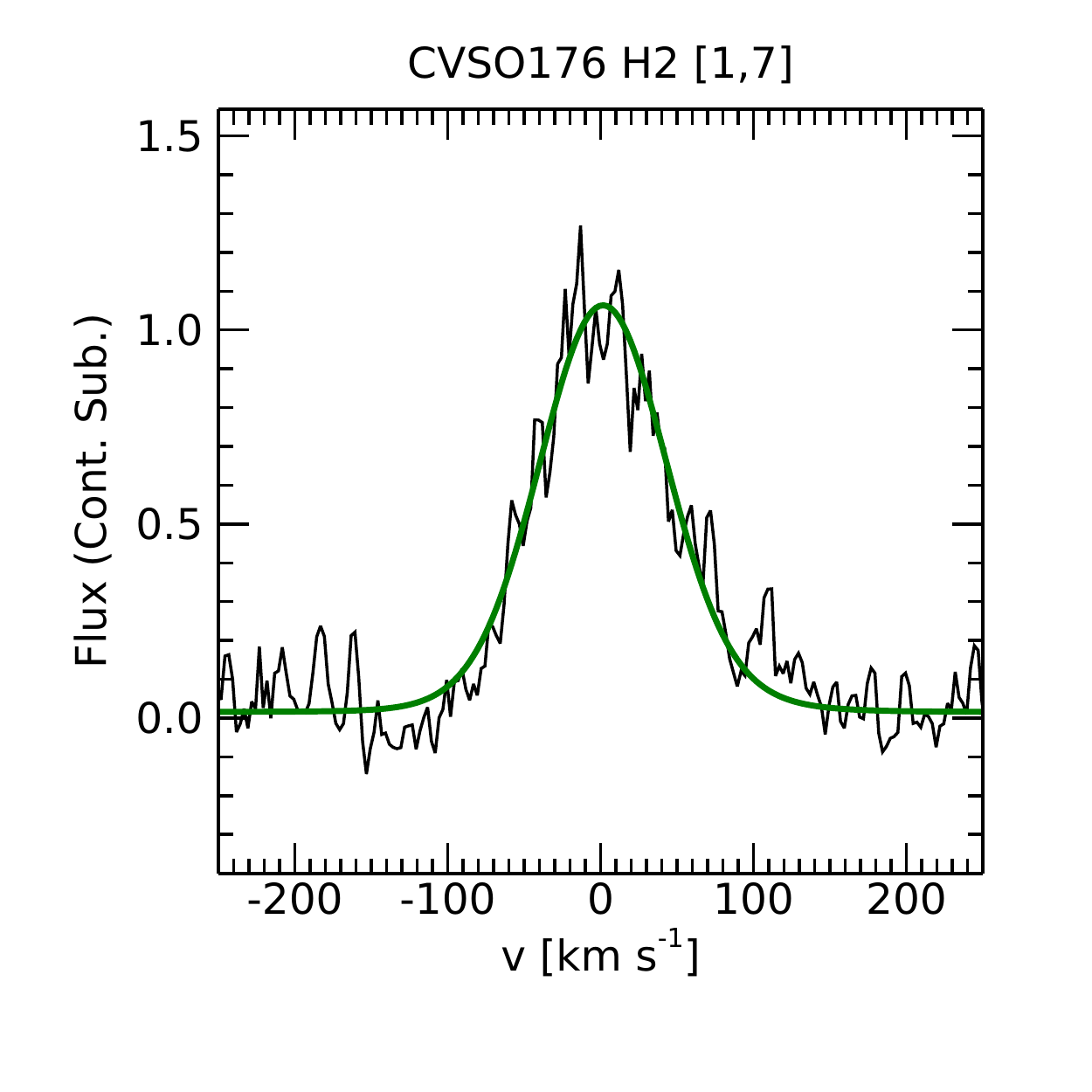}
\includegraphics[trim=20 40 0 0,width=.65\columnwidth, angle=0]{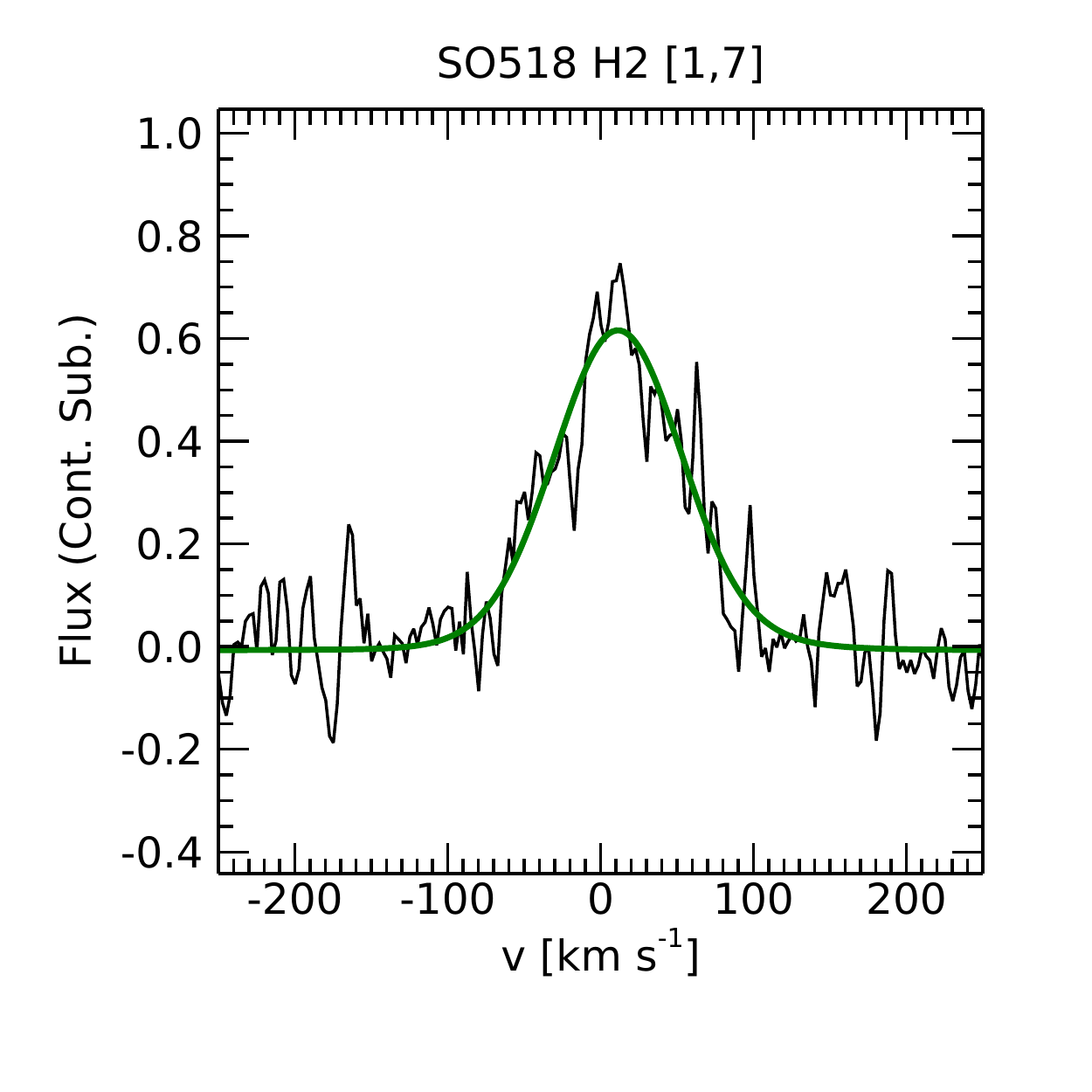}
\includegraphics[trim=20 40 0 0,width=.65\columnwidth, angle=0]{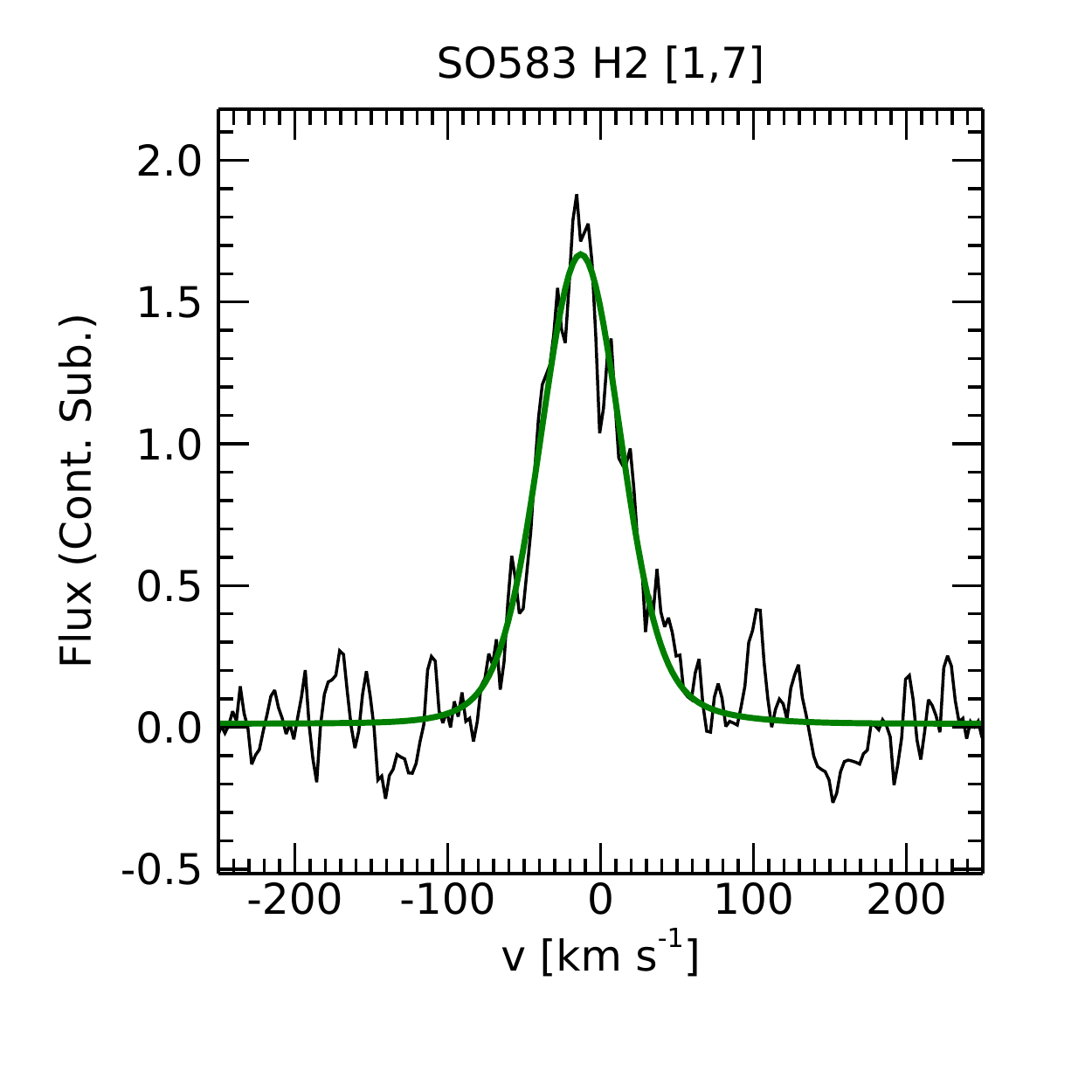}
\begin{center}\caption{\label{fig:complete_sample_H2_17}Continuum-subtracted UV H2 [1,7] averaged profiles. In green we plot the fit to the profile. Flux units are $\rm 10^{-15} ergs^{-1} cm^{-2} \AA^{-1}$. For each panel we indicate the target name and the line diagnostics.}\end{center}
\end{figure*}

\clearpage

\begin{table}[!ht]
\small
\center
\caption{\label{tab:GaussDecompRes_H2_14} Gaussian fit results for $\rm H_2$ [1,4] and $\rm H_2$ [1,7] lines.}
\begin{tabular}{lllllll}
\hline
SOURCE   & $\rm v_p^{(a)}\ (\pm err)$     & $\rm FWHM^{(b)}\ (\pm err)$    & $\rm Flux_0^{(c)}\ (\pm err)$     	& $\rm L_0^{(c)}\ (\pm err)$ & $\rm Flux^{(d)}\ (\pm err)$     			& $\rm L^{(d)}\ (\pm err)$	\\
         & [$\rm km$ $\rm s^{-1}$] & [$\rm km$ $\rm s^{-1}$] & [$\rm 10^{-14}$ $\rm erg$ $\rm s^{-1}$ $\rm cm^{-2}$]     & [$\rm 10^{-5}$ $\rm L_{\odot}$] & [$\rm 10^{-14}$ $\rm erg$ $\rm s^{-1}$ $\rm cm^{-2}$]     & [$\rm 10^{-5}$ $\rm L_{\odot}$] \\
\hline
\multicolumn{7}{c}{$\rm H_2$ [1,4]} \\
\hline
CVSO58	&   -2.42 (0.19) 	&	61.81 (0.57) & 18.74 (1.34) &  71.28 (5.11)   & 2.35 (0.17) &   8.94  (0.64) \\
CVSO90	&	-5.43 (0.14) 	& 	39.56 (0.44) & 7.84  (0.65) &  28.07 (2.32)   & 6.06 (0.50) &	21.71 (1.79) \\
CVSO104 &	3.97 (0.13)  	&   44.96 (0.40) & 4.10  (0.30) &  16.65 (1.20)   & 2.45 (0.18) &	9.95  (0.72) \\	
CVSO107 &	-4.24 (0.13) 	& 	49.45 (0.38) & 9.90  (0.52) &  33.72 (1.78)   & 4.57 (0.24) &	15.59 (0.82) \\
CVSO109 &	-2.23 (0.25) 	& 	41.21 (0.71) & 11.77 (0.86) &  58.79 (4.28)   & 9.10 (0.66) &	45.46 (3.31) \\		
CVSO146 &	-7.43 (0.13) 	& 	52.05 (0.39) & 17.48 (1.11) &  60.18 (3.82)   & 3.73 (0.23) &	12.85 (0.81) \\
CVSO165 &	-4.25 (0.22)	& 	61.81 (0.65) & 3.85  (0.29) &  19.22 (1.47)   & 2.30 (0.18) &	11.49 (0.88) \\
CVSO176 &   -8.67 (0.20) 	& 	82.42 (0.51) & 26.61 (2.06) &  75.99 (5.89)   & 1.99 (0.15) &	5.70  (0.44) \\
SO518   &    3.90 (0.66)	&   82.41 (1.57) & 21.86 (2.67) &  105.09 (12.86) & 1.63 (0.20) &	7.83  (0.95) \\	
SO583   &  -10.29 (0.35)	& 	58.18 (1.12) & 6.92  (0.66) &  32.01 (3.08)   & 2.47 (0.24) &	11.44 (1.10) \\
\hline
\multicolumn{7}{c}{$\rm H_2$ [1,7]} \\
\hline                        
CVSO58	&  -0.23 (0.38)	&  61.81 (0.99)  & 13.89 (1.16)	 &  52.83 (4.42)  & 1.87 (0.16) &	7.11  (0.60) \\
CVSO90	&  -5.49 (0.52)	&  38.03 (1.88)  & 6.16 (0.71)	 &	22.07 (2.55)  & 4.80 (0.56) &	17.22 (1.99) \\
CVSO104 &   4.15 (0.23)	&  47.10 (0.78)  & 2.95 (0.28)	 &	11.98 (1.15)  & 1.79 (0.17) &	7.28  (0.70) \\
CVSO107 &   0.27 (0.25)	&  38.04 (0.87)  & 7.11 (0.59)   &	24.25 (2.02)  & 3.37 (0.28) &	11.50 (0.96) \\
CVSO109 &  -0.16 (0.43)	&  39.56 (1.35)  & 9.54 (0.79)   &	47.70 (3.93)  & 7.44 (0.61) &	37.20 (3.07) \\
CVSO146 &  -5.92 (0.31)	&  52.05 (1.01)  & 13.39 (1.20)	 &  46.08 (4.15)  & 3.01 (0.27) &	10.37 (0.94) \\
CVSO165 &   1.84 (0.66)	&  49.45 (2.19)  & 2.82 (0.41)	 &	14.09 (2.07)  & 1.72 (0.25) &	8.57  (1.26) \\
CVSO176 &   2.44 (0.88)	&  89.90 (2.22)  & 15.45 (2.08)  &	44.13 (5.96)  & 1.27 (0.17) &	3.61  (0.50) \\
SO518   &  12.25 (1.65)	&  89.91 (3.85)  & 15.68 (2.39)	 &  75.38 (11.48) & 1.28 (0.20) &	6.16  (0.94) \\    
SO583   &  -12.06 (0.68) &  52.05 (1.86) & 5.52 (0.71)	 &  25.57 (3.30)  & 2.05 (0.26) &	9.47  (1.22) \\ 
 \hline
\hline
\end{tabular}
\begin{quotation}    
\textbf{Notes.} 

$\rm ^{(a)}$ The $\rm v_p$ errors reported in this Table are those resulting from the Gaussian decomposition procedure (Sect \ref{sec:data_analysis}). The estimated $\rm \sim 15$ $\rm km$ $\rm s^{-1}$ uncertainty due to the wavelength calibration of COS should be added in quadrature.

$\rm ^{(b)}$ Values corrected for the instrumental broadening with the appropriate COS LSF (see Sect. \ref{subsec:UV_spectra} for details)

$\rm ^{(c)}$ Values corrected for the extiction values Av computed in \citet{Manara2021} assuming the reddening law by \citet{Whittet2004} towards HD29647 and $R_v=3.1$.

$\rm ^{(d)}$ Values not corrected for the extiction.

\end{quotation}

\end{table}

\clearpage

\clearpage
\onecolumn
\begin{longtable}{lcccccc}
\caption{Gaussian decomposition of optical forbidden lines.}\label{tab:GaussDecompRes}\\ 
\hline
\small
Line    & $\rm v_p\ (\pm err)$     & $\rm FWHM\ (\pm err)$    & $\rm Flux_0\ (\pm err)$     & $\rm L_0\ (\pm err)$ & $\rm Flux^{(c)}\ (\pm err)$     			& $\rm L^{(c)}\ (\pm err)$	\\
        & [$\rm km$ $\rm s^{-1}$] & [$\rm km$ $\rm s^{-1}$] & [$\rm 10^{-15}$ $\rm erg$ $\rm s^{-1}$ $\rm cm^{-2}$]     & [$\rm 10^{-5}$ $\rm L_{\odot}$] & [$\rm 10^{-15}$ $\rm erg$ $\rm s^{-1}$ $\rm cm^{-2}$]     & [$\rm 10^{-5}$ $\rm L_{\odot}$] \\
\hline
\endfirsthead
\multicolumn{4}{c}%
{\tablename\ \thetable\ -- \textit{Continued from previous page}} \\
\hline
Line    & $\rm v_p\ (\pm err)$     & $\rm FWHM\ (\pm err)$    & $\rm Flux_0\ (\pm err)$     & $\rm L_0\ (\pm err)$ & $\rm Flux^{(c)}\ (\pm err)$     			& $\rm L^{(c)}\ (\pm err)$	\\
        & [$\rm km$ $\rm s^{-1}$] & [$\rm km$ $\rm s^{-1}$] & [$\rm 10^{-15}$ $\rm erg$ $\rm s^{-1}$ $\rm cm^{-2}$]     & [$\rm 10^{-5}$ $\rm L_{\odot}$]  & [$\rm 10^{-15}$ $\rm erg$ $\rm s^{-1}$ $\rm cm^{-2}$]     & [$\rm 10^{-5}$ $\rm L_{\odot}$] \\
\hline
\endhead
\hline \multicolumn{4}{r}{\textit{Continued on next page}} \\
\endfoot
\hline
\endlastfoot 
         \multicolumn{5}{c}{CVSO58}\\
\hline
$[$OI$]$630	    & -0.65 (0.13)     &  34.50 (0.54)    &   3.21 (0.02) &   1.22 (0.01) &   1.70 (0.02) &   0.65 (0.01) \\
                & -13.95 (1.62)    &  138.71 (5.14)   &   5.10 (0.08) &   1.94 (0.03) &   2.71 (0.08) &   1.03 (0.03) \\      
                & -118.10 (0.55)   &  53.75 (1.51)    &   1.76 (0.03) &   0.67 (0.01) &   0.94 (0.03) &   0.36 (0.01) \\
$[$OI$]$557	    &   3.08 (0.99)    &  66.14 (2.73)    &   1.23 (0.04) &   0.47 (0.02) &   0.59 (0.04) &   0.23 (0.02) \\
$[$NII$]$658	& -63.27 (4.78)    &  82.22 (6.29)    &   0.40 (0.03) &   0.15 (0.01) &   0.22 (0.03) &   0.08 (0.01) \\
                & -123.03 (0.69)   &  39.70 (1.69)    &   0.68 (0.01) &   0.26 (0.01) &   0.37 (0.01) &   0.14 (0.01) \\
$[$SII$]$673	& -27.35 (3.33)    &  117.34 (8.92)   &   0.67 (0.06) &   0.26 (0.02) &   0.38 (0.06) &   0.14 (0.02) \\
                & -120.34 (1.13)   &  31.32 (2.28)    &   0.37 (0.02) &   0.14 (0.01) &   0.21 (0.02) &   0.08 (0.01) \\
$[$SII$]$406	&  -3.43 (1.06)    &   48.18 (2.86)   &   1.82 (0.05) &   0.69 (0.02) &   0.63 (0.05) &   0.24 (0.02) \\
                & -113.093 (1.98)  &   39.43 (6.57)   &   0.79 (0.04) &   0.30 (0.01) &   0.27 (0.04) &   0.10 (0.01) \\      
\hline
         \multicolumn{5}{c}{CVSO90}\\
\hline
$[$OI$]$630	    &   141.88 (0.53)   &   44.18 (1.66)    &   0.67 (0.04) &   0.24 (0.01) &   0.62 (0.04) &   0.22 (0.01) \\
                &   -1.35  (0.07)   &   19.66 (0.21)    &   1.30 (0.02) &   0.47 (0.01) &   1.20 (0.02) &   0.43 (0.01) \\
                &   -32.97 (0.84)   &   180.13 (1.85)   &   5.12 (0.15) &   1.83 (0.05) &   4.73 (0.15) &   1.70 (0.05) \\     
                &   -134.80 (0.24)  &   44.36  (0.85)   &   1.40 (0.04) &   0.50 (0.01) &   1.30 (0.04) &   0.46 (0.01) \\
$[$OI$]$557	    &   -1.72 (0.31)    &   30.12 (1.04)    &   0.66 (0.03) &   0.24 (0.01) &   0.60 (0.03) &   0.22 (0.01) \\
                &   -10.99 (2.62)   & 118.05 (6.93)     &   0.71 (0.10) &   0.25 (0.04) &   0.65 (0.10) &   0.23 (0.04) \\
$[$NII$]$658	&   145.08 (0.58)   & 32.34 (4.37)      &   0.27 (0.02) &   0.10 (0.01) &   0.25 (0.02) &   0.09 (0.01) \\
                &   -44.08 (1.14)   & 32.92 (2.21)      &   0.13 (0.02) &   0.05 (0.01) &   0.12 (0.02) &   0.04 (0.01) \\
                &   -149.86 (0.51)  & 32.02 (0.91)      &   0.27 (0.02) &   0.09 (0.01) &   0.25 (0.02) &   0.09 (0.01) \\
$[$SII$]$673	&   142.30 (0.62)   &   25.42 (1.44)    &   0.20 (0.02) &   0.07 (0.01) &   0.18 (0.02) &   0.06 (0.01) \\
                &   -146.82 (1.17)  &   55.24 (2.37)    &   0.31 (0.04) &   0.11 (0.01) &   0.29 (0.04) &   0.10 (0.01) \\
$[$SII$]$406	&   234.49 (1.66)   &   48.93 (2.35)    &   0.60 (0.07) &   0.22 (0.03) &   0.53 (0.07) &   0.19 (0.03)	\\			
                &   -27.79 (3.57)   &   92.39 (8.32)    &   0.80 (0.14) &   0.29 (0.05) &   0.70 (0.14) &   0.25 (0.05) \\     
                &   -125.95 (1.57)  &   54.22 (2.64)    &   0.82 (0.08) &   0.29 (0.03) &   0.72 (0.08) &   0.26 (0.03) \\
\hline
         \multicolumn{5}{c}{CVSO104}\\
\hline
$[$OI$]$630	    &	-1.01 (0.37)    &   30.18 (2.38)    &   0.85 (0.03) &   0.34 (0.01) &   0.72 (0.03) &   0.29 (0.01) \\
                &   -3.02 (0.92)    &   93.49 (5.34)    &   2.13 (0.09) &   0.86 (0.04) &   1.82 (0.09) &   0.74 (0.04) \\
$[$OI$]$557	    &    -4.65 (1.25)   &   73.78 (3.41)    &   0.97 (0.08) &   0.39 (0.03) &   0.81 (0.08) &   0.33 (0.03) \\
$[$NII$]$658	&	ND				&	ND				&	< 0.65			&	< 0.26	&	            &		        \\ 	
$[$SII$]$673	&	ND				&	ND				&	< 0.62			&	< 0.25	&	            &		        \\ 
$[$SII$]$406	&    -4.21 (1.99)   &   38.36 (7.92)    &   0.26 (0.03) &   0.10 (0.01) &   0.20 (0.03) &   0.08 (0.01) \\
\hline                                                                                        	
         \multicolumn{5}{c}{CVSO107}\\                                                        	
\hline
$[$OI$]$630	    &   0.26 (0.10)    &   24.74 (0.48)    &   1.51 (0.01) &   0.51 (0.01) &   1.19 (0.01) &   0.41 (0.01) \\
                &   -7.93 (0.43)   &   102.03 (2.63)   &   3.43 (0.06) &   1.17 (0.02) &   2.70 (0.06) &   0.92 (0.02) \\   
                &   -105.89 (0.44) &   40.93 (0.91)    &   0.78 (0.02) &   0.26 (0.01) &   0.61 (0.02) &   0.21 (0.01) \\
$[$OI$]$557	    & -1.83 (1.07)     &   33.77 (3.03)    &   0.38 (0.02) &   0.13 (0.01) &   0.29 (0.02) &   0.10 (0.01) \\
                & 2.52 (4.64)      &   75.80 (12.54)   &   0.78 (0.04) &   0.27 (0.01) &   0.60 (0.04) &   0.20 (0.01) \\ 
$[$NII$]$658	&	ND			   &	ND			   &	< 1.11	   &	< 0.38	   &	           &		       \\
$[$SII$]$673	& -21.01 (2.84)    &   84.09 (8.22)    &   0.52 (0.05) &   0.18 (0.02) &   0.42 (0.05) &   0.14 (0.02) \\ 
                & -102.87 (1.35)   &   53.34 (2.79)    &   0.50 (0.03) &   0.17 (0.01) &   0.40 (0.03) &   0.14 (0.01) \\ 
$[$SII$]$406	&   4.91 (1.59)    &   34.86 (2.00)    &   0.97 (0.03) &   0.33 (0.01) &   0.65 (0.03) &   0.22 (0.01) \\
                &  -27.16 (4.17)   &   54.07 (8.58)    &   0.96 (0.05) &   0.33 (0.02) &   0.65 (0.05) &   0.22 (0.02) \\    
                &  -105.39 (0.97)  &   44.18 (2.43)    &   0.92 (0.04) &   0.31 (0.01) &   0.62 (0.04) &   0.21 (0.01) \\
\hline
         \multicolumn{5}{c}{CVSO109}\\
\hline
$[$OI$]$630	    &   0.48 (0.19)    &   24.89 (2.24)    &   2.15 (0.03) &   1.07 (0.02) &   1.99 (0.03) &   0.99 (0.02) \\
                &   1.57 (1.34)    &   72.14 (14.90)   &   2.36 (0.10) &   1.18 (0.05) &   2.18 (0.10) &   1.09 (0.05) \\
$[$OI$]$557	    &   5.00 (0.61)    &   40.26 (1.74)    &   1.36 (0.07) &   0.68 (0.04) &   1.25 (0.07) &   0.62 (0.04) \\
$[$NII$]$658	&	ND			   &	ND			   &	< 1.45	   &	< 0.72	   &	           &		       \\
$[$SII$]$673	&	ND			   &	ND			   &	< 1.06	   &	< 0.53	   &	           &		       \\
$[$SII$]$406	&  -1.26 (0.63)    &   37.46 (1.39)    &   1.58 (0.08) &   0.79 (0.04) &   1.39 (0.08) &   0.69 (0.04) \\
\hline
         \multicolumn{5}{c}{CVSO146}\\
\hline
$[$OI$]$630	    &  0.37 (0.20)   &   31.79 (0.39)    &   2.03 (0.04) &   0.70 (0.01) &   1.26 (0.04) &   0.43 (0.01) \\   
                & -0.01 (0.61)   &   119.69 (1.14)   &   5.18 (0.15) &   1.78 (0.05) &   3.22 (0.15) &   1.11 (0.05) \\				
$[$OI$]$557	    & 0.33 (0.86)    &   62.53 (1.91)    &   1.29 (0.06) &   0.44 (0.02) &   0.75 (0.06) &   0.26 (0.02) \\
$[$NII$]$658	&	ND			 &	ND				 &	< 0.85		 &	< 0.29		 &	             &		         \\ 	      
$[$SII$]$673	& -51.90 (1.81)  &   41.32 (6.26)    &   0.28 (0.04) &   0.10 (0.01) &   0.18 (0.04) &   0.06 (0.01) \\ 
$[$SII$]$406	&  1.94 (1.09)   &   19.97 (4.23)    &   0.61 (0.02) &   0.21 (0.01) &   0.27 (0.02) &   0.09 (0.01) \\	
                & -21.47 (8.76)  &   126.59 (19.19)  &   1.69 (0.16) &   0.58 (0.05) &   0.76 (0.16) &   0.26 (0.05) \\	
\hline
         \multicolumn{5}{c}{CVSO165}\\
\hline
$[$OI$]$630	    &  92.17 (1.16)   &   79.81 (2.37)   &   1.30 (0.08) &   0.65 (0.04) &   1.11 (0.08) &   0.55 (0.04) \\
                &  1.32 (0.40)    &   48.13 (1.56)   &   1.65 (0.05) &   0.83 (0.02) &   1.41 (0.05) &   0.70 (0.02) \\
                &  -80.29 (3.40)  &   112.87 (4.48)  &   1.28 (0.11) &   0.64 (0.05) &   1.09 (0.11) &   0.55 (0.05) \\
$[$OI$]$557	    &  3.30 (1.08)    &   66.19 (2.89)   &   0.60 (0.05) &   0.30 (0.02) &   0.50 (0.05) &   0.25 (0.02) \\
$[$NII$]$658	&	ND			  &	ND				 &	< 0.51		 &	< 0.25		 &	             & 	             \\ 	
$[$SII$]$673	&  120.30 (0.59)  &   17.58 (1.14 )  &   0.15 (0.01) &   0.08 (0.01) &               &               \\
                & -120.03 (1.21)  &   28.07 (4.75)   &   0.21 (0.02) &   0.10 (0.01) &               &               \\
$[$SII$]$406	&	ND			  &	ND				 &	< 1.85		 &	< 0.92		 &	             &	             \\ 			
\hline
         \multicolumn{5}{c}{CVSO176}\\
\hline
$[$OI$]$630	    &  1.22 (0.58)   &   68.70 (1.40)    &   1.63 (0.02) &   0.47 (0.01) &   0.74 (0.02) &   0.21 (0.01) \\
                & -97.58 (1.77)  &   32.51 (3.35)    &   0.16 (0.01) &   0.05 (0.01) &   0.07 (0.01) &   0.02 (0.01) \\
$[$OI$]$557	    &  5.66 (0.78)   &   70.40 (1.53)    &   0.75 (0.02) &   0.21 (0.01) &   0.30 (0.02) &   0.09 (0.01) \\
$[$NII$]$658	&	ND			 &	ND				 &	< 0.45		 &	< 0.13		 &	             &	             \\ 	
$[$SII$]$673	&	ND			 &	ND				 &	< 0.51		 &	< 0.15		 &	             &	             \\ 
$[$SII$]$406	& -9.87 (1.94)   &   52.15 (4.69)    &   0.97 (0.03) &   0.28 (0.01) &   0.26 (0.03) &   0.07 (0.01) \\
\hline
         \multicolumn{5}{c}{SO518}\\
\hline
$[$OI$]$630	    & 86.18 (4.12 )  &   93.17 (5.42)    &   0.73 (0.03) &   0.35 (0.01) &   0.33 (0.03) &   0.16 (0.01) \\
                & -17.35 (0.44)  &   84.68 (2.04)    &   4.95 (0.02) &   2.38 (0.01) &   2.24 (0.02) &   1.08 (0.01) \\
                & -79.58 (0.43)  &   44.47 (1.00)    &   1.38 (0.01) &   0.66 (0.01) &   0.62 (0.01) &   0.30 (0.01) \\
$[$OI$]$557	    & -11.86 (0.68)  &   100.22 (1.87)   &   1.11 (0.02) &   0.53 (0.01) &   0.45 (0.02) &   0.21 (0.01) \\                 
$[$NII$]$658	&  4.19 (0.37)   &   15.37 (0.83)    &   0.25 (0.01) &   0.12 (0.01) &   0.12 (0.01) &   0.06 (0.01) \\
                & -58.75 (4.57)  &   51.48 (11.98)   &   0.18 (0.02) &   0.09 (0.01) &   0.09 (0.02) &   0.04 (0.01) \\
$[$SII$]$673	& -51.95 (1.69)  &   67.47 (2.28)    &   0.64 (0.02) &   0.31 (0.01) &   0.31 (0.02) &   0.15 (0.01) \\				
                & -77.82 (0.36)  &   20.84 (1.38)    &   0.28 (0.01) &   0.13 (0.01) &   0.14 (0.01) &   0.06 (0.01) \\
$[$SII$]$406	&  28.74 (3.68)  &   153.78 (2.42)   &   1.20 (0.02) &   0.58 (0.01) &   0.32 (0.02) &   0.15 (0.01) \\ 	
                & -53.48 (1.10)  &   91.12 (1.89)    &   1.51 (0.01) &   0.72 (0.01) &   0.49 (0.01) &   0.19 (0.01) \\
\hline
         \multicolumn{5}{c}{SO583}\\
\hline
$[$OI$]$630   	&  2.17 (0.98)   &   16.23 (2.16)    &   0.84 (0.07) &   0.39 (0.03) &   0.61 (0.07) &   0.28 (0.03) \\
                & -7.76 (1.87)   &   106.50 (3.78)   &   6.23 (0.44) &   2.88 (0.21) &   4.54 (0.44) &   2.10 (0.21) \\
$[$OI$]$557	    & -37.50 (4.81)  &   90.12 (4.09)    &   1.52 (0.28) &   0.70 (0.13) &   1.06 (0.28) &   0.49 (0.13) \\
$[$NII$]$658	& -0.93 (0.57)   &   26.84 (1.59)    &   2.15 (0.11) &   1.00 (0.05) &   1.59 (0.11) &   0.74 (0.05) \\
$[$SII$]$673	& 31.60 (5.84)   &   148.39 (12.21)  &   3.51 (0.54) &   1.62 (0.25) &   2.62 (0.54) &   1.21 (0.25) \\
$[$SII$]$406	&	ND			 &	ND				 &	< 10.84		 &	< 5.02		 &	             &	             \\ 	
\hline
         \multicolumn{5}{c}{SO1153}\\
\hline
$[$OI$]$630	    &  -9.70 (0.05)   &   24.61 (0.14)    &   2.86 (0.01) &   1.36 (0.01) &   2.64 (0.01) &   1.26 (0.01)	\\     
                &  -41.94 (0.82)  &   148.23 (1.54)   &   4.62 (0.07) &   2.20 (0.03) &   4.27 (0.07) &   2.03 (0.03)	\\      
                &  -43.32 (0.15)  &   32.95 (0.31)    &   2.16 (0.01) &   1.03 (0.01) &   1.99 (0.01) &   0.95 (0.01)	\\
                &  -87.48 (0.08)  &   48.35 (0.28)    &   7.15 (0.02) &   3.40 (0.01) &   6.61 (0.02) &   3.14 (0.01)	\\
$[$OI$]$557	    & -9.24 (0.17)    &   32.42 (0.48)    &   0.46 (0.01) &   0.22 (0.01) &   0.42 (0.01) &   0.20 (0.01)	\\
                & -46.82 (1.82)   &   158.20 (4.19)   &   0.67 (0.05) &   0.32 (0.02) &   0.61 (0.05) &   0.29 (0.02)	\\
$[$NII$]$658	& -16.45 (0.18)   &   19.40 (0.36)    &   0.28 (0.01) &   0.13 (0.01) &   0.26 (0.01) &   0.12 (0.01)   \\
                & -89.34 (0.17)   &   37.31 (0.49)    &   0.63 (0.01) &   0.30 (0.01) &   0.58 (0.01) &   0.28 (0.01)   \\
$[$SII$]$673	& -62.10 (0.29)   &   71.59 (0.33)    &   2.57 (0.03) &   1.22 (0.01) &   2.39 (0.03) &   1.14 (0.01)	\\     
                & -88.80 (0.11)   &   25.75 (0.41)    &   0.69 (0.01) &   0.33 (0.01) &   0.64 (0.01) &   0.31 (0.01)	\\                
$[$SII$]$406	& -46.53 (1.97)   &   50.34 (2.52)    &   1.16 (0.03) &   0.55 (0.01) &   1.01 (0.03) &   0.48 (0.01)   \\
                & -90.98 (1.09)   &   49.93 (1.36)    &   1.99 (0.03) &   0.95 (0.01) &   1.74 (0.03) &   0.83 (0.01)   \\
\hline
\end{longtable} 
\begin{quotation}    
\textbf{Notes.} 

$\rm ^{\dagger}$ Nebular component.

$\rm ^{(a)}$ Values corrected for the instrumental broadening.

$\rm ^{(b)}$ Values corrected for the extiction. The Av values were derived in \citet{Manara2021} assuming the reddening law by \citet{Cardelli1989} and $R_v=3.1$.

$\rm ^{(c)}$ Values not corrected for the extiction.
\end{quotation}
\clearpage
\twocolumn

\end{appendix}

\end{document}